\documentclass[usenatbib]{aa}

\begin{document}

\title{The CARMENES search for exoplanets around M dwarfs}

\subtitle{Radial-velocity variations of active stars in visual-channel spectra}

    \author{L.\,Tal-Or\inst{1}
     \and M.\,Zechmeister\inst{1}
     \and A.\,Reiners\inst{1} 
     \and S.\,V.\,Jeffers\inst{1}
     \and P.\,Sch\"ofer\inst{1}
     \and A.\,Quirrenbach\inst{2}
     \and P.\,J.\,Amado\inst{3}      
     \and I.\,Ribas\inst{4}
     \and J.\,A.\,Caballero\inst{5}
     \and J.\,Aceituno\inst{3,6}
     \and F.\,F.\,Bauer\inst{1}
     \and V.\,J.\,S.\,B\'ejar\inst{7}
     \and S.\,Czesla\inst{8}
     \and S.\,Dreizler\inst{1}
     \and B.\,Fuhrmeister\inst{8}
     \and A.\,P.\,Hatzes\inst{9}
     \and E.\,N.\,Johnson\inst{1}
     \and M.\,K\"urster\inst{10}
     \and M.\,Lafarga\inst{4}
     \and D.\,Montes\inst{11}
     \and J.\,C.\,Morales\inst{4}
     \and S.\,Reffert\inst{2}      
     \and S.\,Sadegi\inst{2}
     \and W.\,Seifert\inst{2}
     \and D.\,Shulyak\inst{1}
}
   \institute{Institut f\"ur Astrophysik, Georg-August-Universit\"at, 
              Friedrich-Hund-Platz 1, 37077 G\"ottingen, Germany\\
              \email{levtalor@astro.physik.uni-goettingen.de}
         \and Landessternwarte, Zentrum f\"ur Astronomie der Universt\"at Heidelberg, 
              K\"onigstuhl 12, D-69117 Heidelberg, Germany
         \and Instituto de Astrof\'isica de Andaluc\'ia (IAA-CSIC), Glorieta de la Astronom\'ia s/n, 
              E-18008 Granada, Spain
         \and Institut de Ci\`encies de l’Espai (ICE, CSIC), Campus UAB, c/ de Can Magrans s/n, E-08193 Bellaterra, Barcelona, Spain 
              and Institut d’Estudis Espacials de Catalunya (IEEC), E-08034 Barcelona, Spain
         \and Centro de Astrobiolog\'ia (CSIC-INTA), ESAC campus, Camino Bajo del Castillo s/n, E-28692 Villanueva de la Ca\~nada, Madrid, Spain
         \and Centro Astron\'omico Hispano-Alem\'an (CSIC-MPG), 
              Observatorio Astron\'omico de Calar Alto, 
              Sierra de los Filabres-04550 G\'ergal, Almer\'ia, Spain
         \and Instituto de Astrof\'sica de Canarias, V\'ia L\'actea s/n, 38205 La Laguna, 
              Tenerife, Spain, and Departamento de Astrof\'isica, Universidad de La Laguna, 
              38206 La Laguna, Tenerife, Spain
         \and Hamburger Sternwarte, Gojenbergsweg 112, D-21029 Hamburg, Germany
         \and Th\"uringer Landessternwarte Tautenburg, Sternwarte 5, D-07778 Tautenburg, Germany
         \and Max-Planck-Institut f\"ur Astronomie,
              K\"onigstuhl 17, D-69117 Heidelberg, Germany
         \and Departamento de F\'isica de la Tierra y Astrof\'isica \& UPARCOS-UCM (Unidad de F\'isica de Part\'iculas y del Cosmos de la UCM), Facultad de Ciencias F\'isicas, Universidad Complutense de Madrid, E-28040 Madrid, Spain
              }

\date{Received 27 Nov 2017; accepted 28 Feb 2018}
\abstract
{Previous simulations predicted the activity-induced radial-velocity (RV) variations of M dwarfs to range from $\sim1$\,cm\,s$^{-1}$ to $\sim1$\,km\,s$^{-1}$, depending on various stellar and activity parameters.
}
{We investigate the observed relations between RVs, stellar activity, and stellar parameters of M dwarfs by analyzing CARMENES high-resolution visual-channel spectra ($0.5$--$1$\,$\mu$m), which were taken within the CARMENES RV planet survey during its first $20$ months of operation.}
{During this time, $287$ of the CARMENES-sample stars were observed at least five times. From each spectrum we derived a relative RV and a measure of chromospheric H$\alpha$ emission. In addition, we estimated the chromatic index (CRX) of each spectrum, which is a measure of the RV wavelength dependence.
}
{Despite having a median number of only $11$ measurements per star, we show that the RV variations of the stars with RV scatter of $>10$\,m\,s$^{-1}$ and a projected rotation velocity $v \sin{i}>2$\,km\,s$^{-1}$ are caused mainly by activity. We name these stars `active RV-loud stars' and find their occurrence to increase with spectral type: from\,$\sim3\%$ for early-type M dwarfs (M$0.0$--$2.5$\,V) through\,$\sim30\%$ for mid-type M dwarfs (M$3.0$--$5.5$\,V) to\,$>50\%$ for late-type M dwarfs (M$6.0$--$9.0$\,V). Their RV-scatter amplitude is found to be correlated mainly with $v \sin{i}$.
For about half of the stars, we also find a linear RV--CRX anticorrelation, which indicates that their activity-induced RV scatter is lower at longer wavelengths. For most of them we can exclude a linear correlation between RV and H$\alpha$ emission.
}
{Our results are in agreement with simulated activity-induced RV variations in M dwarfs. The RV variations of most active RV-loud M dwarfs are likely to be caused by dark spots on their surfaces, which move in and out of view as the stars rotate.}
\keywords{stars: late-type -- stars: activity -- stars: rotation -- techniques: radial velocities}
\authorrunning{Tal-Or et al. (2018)}
\titlerunning{Radial-velocity variations of active stars in visual-channel spectra}
\maketitle

\section{Introduction}
\label{sec1}

Precise radial-velocity (PRV) surveys of well-defined stellar samples constitute an important pathway to characterize their inner planetary systems \citep[e.g.,][]{WinnFabrycky2015}. The first PRV surveys focused mainly on solar-type stars \citep*[e.g.,][]{Campbell1988}, and largely overlooked M dwarfs, the most abundant type of stars in our Galaxy. There were two main reasons for doing so. First, M dwarfs are intrinsically fainter than solar-type stars, particularly in visible light, where most PRV instruments were operating. Second, a large fraction of M dwarfs are fast rotating and magnetically active, two phenomena that are linked to each other \citep[e.g.,][]{Reiners2012,Newton2017,Jeffers2018arXiv}. Fast rotation and activity complicate the detection of planetary-induced RV signals \citep[e.g.,][]{SaarDonahue1997,Hatzes2002,Reiners2010a,Barnes2011}.

Recently the focus of PRV surveys has shifted slightly towards M dwarfs. The shift is driven by the desire to complement our picture of planetary occurrence rates around different type stars \citep[e.g.,][]{Howard2012,DressingCharbonneau2013}, and by the ability of detecting low-mass habitable-zone planets around nearby stars with RV instruments of $\sim1$\,m\,s$^{-1}$ accuracy \citep[e.g.,][]{Anglada-Escude2016,Ribas2016}. In the previous decade \citet*{Zechmeister2009b} monitored the RVs of $40$ M dwarfs with UVES at the VLT \citep{Dekker2000}. The largest PRV survey of M dwarfs to date was carried out by \citet{Bonfils2013}, who monitored $102$ southern hemisphere stars with HARPS \citep{mayor03}. A few additional surveys are either being planned or executed \citep{Fischer2016}. However, to use M-dwarf PRV surveys to their full extent, there is a need for efficient tools to distinguish between orbital- and activity-induced RV signals. Understanding the ways in which rotation and activity alter the measured RVs of M dwarfs is essential in developing such tools.

Stellar RVs are measured mainly from photospheric absorption lines. There are several activity-related phenomena that can cause variations in M-dwarf spectra. Inhomogeneities of the stellar photosphere, such as spots and faculae that corotate with the star, alter the shapes of absorption lines by changing the brightness and spectral appearance of parts of the stellar disk \citep[e.g.][]{SaarDonahue1997}. Since these features are often accompanied by strong magnetic fields, they also cause Zeeman splitting of spectral lines \citep[e.g.,][]{Semel1989,Donati2006Sci,Reiners2007}. Moreover, magnetically active regions suppress the convective blueshift of cool stars, leading to a net redshift of their spectra \citep[e.g.,][]{Gray2009}. In addition, photospheric magnetic activity can induce chromospheric features such as filaments and plages, the latter being associated with line emission. The extreme cases of magnetically induced chromospheric activity are strong flares that could alter the entire spectral appearance of a star \citep[e.g.,][]{Fuhrmeister2008}.

To understand the impact of these phenomena on measured M-dwarf RVs, two complementary efforts are being made. The first is simulating the effects via forward modeling the spectra and RVs of active rotating stars. The second is measuring the RVs of active stars and trying to correlate the observed variations with various activity indicators, which could be derived either from the spectra themselves or from auxiliary simultaneous measurements.

Regarding forward modeling, \citet{Reiners2010a} and \citet{Barnes2011} simulated the impact of active regions, such as cool spots, on RV measurements of M dwarfs over the wavelength range of $0.5$--$2.2$\,$\mu$m for a wide variety of possible combinations of star and spot parameters. They concluded that the flux contrast between the active regions and the unperturbed photosphere will induce RV-variation amplitudes that can range from $\sim1$\,cm\,s$^{-1}$ to $\sim1$\,km\,s$^{-1}$. These amplitudes depend on the specific parameters of the star and its spots, such as the spot-coverage fraction, the projected rotation velocity ($v \sin{i}$) of the star, its spectral type (SpT), and the spot-to-photosphere temperature contrast. For instance, they predicted higher RV amplitudes for higher spot-to-photosphere temperature contrasts ($\sim$\,$1000$\,K), while for low contrasts ($\sim$\,$200$\,K), they predicted a stronger decrease of RV amplitudes when going from observations at $\sim0.5$\,$\mu$m to observations at $\sim1.0$\,$\mu$m.

\citet{Reiners2013} estimated the additional impact that the Zeeman effect would have on active-star RVs. They concluded that in the wavelength range of $0.5$--$2.3$\,$\mu$m it could be as large as the flux-contrast effect, depending on the strength of the magnetic fields that accompany the active regions. In
contrast to RV variations caused by the flux-contrast effect of low temperature-contrast spots, the amplitude of RV variations caused by the Zeeman effect grows with wavelength. However, the actual impact of the effect on M-dwarf spectra is highly uncertain, mainly due to our limited knowledge of Land{\'e} $g$ factors of molecular levels.

Convective-blueshift suppression by active regions would also impact RVs \citep[e.g.,][]{Meunier2010}. The impact of such convection suppression on solar-type RVs was studied in detail \citep[e.g.,][]{Jeffers2014,Dumusque2014SOAP,Reiners2016,Meunier2017}, and it was shown that it could reach a few times $10$\,m\,s$^{-1}$. Assuming simple scaling relations, \citet{Kurster2003} estimated the impact of the suppression on M-dwarf RVs to be smaller by
an order of magnitude.

Contrary to photospheric surface inhomogeneities, the direct effect of chromospheric emission variations on RVs has not yet
been numerically addressed. This is most likely due to the relative difficulty of simulating chromospheric processes \citep{Reiners2009c}.

With regard to observations, several investigations have been published on activity-induced RV variations of M dwarfs, and on using different indicators to associate the variations with activity. The most widely used activity indicators in these works were variations in the average absorption-line shape, measured via the different moments of the cross-correlation function \citep[CCF, e.g.,][]{Queloz2001}, and variations of chromospheric emission lines, such as Ca\,{\sc ii}\,H\&K and H$\alpha$. In addition, simultaneous (or contemporaneous) photometric time series were used to determine rotational periods ($P_{\rm rot}$), and to trace activity-related phenomena \citep[e.g.,][]{Bonfils2007,Anglada-Escude2016}.

\citet{Kurster2003} reported a correlation between Barnard's
star RVs and variations in the H$\alpha$ line. \citet{Bonfils2007} used photometric time series, as well as the chromospheric Ca\,{\sc ii}\,H\&K and H$\alpha$ emission lines, to attribute the $35$-day RV signal of GJ\,$674$ to stellar activity. \citet{Bonfils2013} also reported detecting a few periodic RV signals that correlated with at least one of their activity indicators: Ca\,{\sc ii}\,H\&K, H$\alpha$, and the different moments of the CCF. Unfortunately, \citet{Bonfils2013} specifically excluded from their sample rapid rotators ($v \sin{i}>6.5$\,km\,s$^{-1}$), which would naturally be the most active stars in their volume-limited sample. \citet{Barnes2014} reported RV measurements of 15 M$5$--$9$\,V stars, with many of them presenting activity-induced signals of up to $\sim100$\,m\,s$^{-1}$. Despite their low number and short time-span of observations ($3$--$5$ RVs per star, gathered in one week), they found significant correlations between RV-scatter amplitude, $v \sin{i}$, and chromospheric H$\alpha$ activity. Finally, \citet{Suarez2017} reported detecting periodic activity-induced RV signals with amplitudes of $1$--$10$\,m\,s$^{-1}$ in 18 M$0$--$5$\,V stars. Their results relied on several years
of monitoring and some tens to hundreds RVs per star. They found a significant correlation of RV scatter with the chromospheric activity indicators Ca\,{\sc ii}\,H\&K and H$\alpha$. However, only slowly rotating stars, with $20$\,d$<P_{\rm rot}<100$\,d, were observed for this work.

The direct effect of strong chromospheric emission on RVs was reported only for {a few} M dwarfs. \citet{Reiners2009c} described a strong flare event in the M$6$\,V star CN Leo, which was detected via a $0.9$\,dex enhancement of H$\alpha$ emission. The flare was accompanied by an RV deviation of a few hundred m\,s$^{-1}$. However, flares with $<0.4$\,dex changes in H$\alpha$ emission did not result in RV deviations at the $10$\,m\,s$^{-1}$ level. \citet{Barnes2014} reported tentative evidence for $\sim20$\,m\,s$^{-1}$ RV excursions that might have been caused by flares on the M$5.5$\,V star Proxima Centauri, but with a weak correlation with changes in H$\alpha$ emission.

Out of the few dedicated PRV instruments that specifically aim at surveying nearby M dwarfs to detect their planets, CARMENES was the first to be on sky \citep[][and references therein]{Quirrenbach2016SPIE,Fischer2016}. Its sample is composed of $>300$ nearby northern hemisphere M-dwarfs, and covers the full range of M-dwarf SpT (M$0$--$9$\,V). Fast-rotating and active stars were not explicitly excluded from the survey \citep{Reiners2017a}. Taking advantage of this large and highly diverse sample, we use the CARMENES visual-channel spectra to empirically investigate the relation between M-dwarf RVs, their rotation, and their activity. In particular, we focus on stars with $v \sin{i}>2$\,km\,s$^{-1}$ and activity-induced RV variations of $>10$\,m\,s$^{-1}$, which we call `active RV-loud stars'.

The CARMENES survey is briefly presented in Sect. \ref{sec2}. Section \ref{sec3} explains our data reduction, and RV and activity measurements. In Sect. \ref{sec4} we present our sample of active RV-loud stars. Using this sample, we investigate the observed connection between RVs, stellar rotation, and stellar activity in Sect. \ref{sec5}, and we summarize our findings in Sect. \ref{sec6}.

\section{CARMENES survey}
\label{sec2}

The contents of the CARMENES-survey input catalog \citep[Carmencita,][]{Caballero2016}, are described by \citet{Alonso-Floriano2015}, \citet{Cortes-Contreras2017}, and \citet{Jeffers2018arXiv}. Carmencita contains detailed information on about $2200$ M dwarfs brighter than $J=11.5$\,mag that can be observed from Calar Alto (i.e., $\delta \gtrsim -23$\,deg). Using the compiled information, the stars were prioritized, and $324$ stars were selected for the CARMENES-survey sample. 

The selected stars for the CARMENES sample were presented by \citet{Reiners2017a}, who listed several stellar parameters for each survey star, including information on their $P_{\rm rot}$, $v \sin{i}$, and H$\alpha$ emission. Stars with known stellar companions at less than $5$\,arcsec separation were excluded from the survey \citep{Cortes-Contreras2017}, but stars with known planetary systems were not \citep[e.g.,][]{Trifonov2018}. Although rapid rotation and strong magnetic activity degrade RV precision, stars with high $v \sin{i}$ values or strong H$\alpha$ emission were not excluded from the survey.

Following a successful commissioning of the CARMENES instrument in October--December 2015, the CARMENES survey started in January 2016 \citep{Quirrenbach2016SPIE}. More than $7000$ M-dwarf spectra were recorded during the first 20 months of operation. The median number of measurements per star was $11$, $166$ stars were observed more than $10$ times, and $60$ stars were observed more than $30$ times.

Despite providing high-quality spectra \citep[e.g.,][]{Reiners2017a}, during its first months of operations, the near-infrared channel (NIR) suffered from several systematic issues that prevented us from reaching an RV accuracy comparable to the visual channel (VIS). Some of these systematic issues are still under investigation. For this reason, only VIS spectra are used for this paper.

\section{RV and activity measurements}
\label{sec3}

All raw frames collected for the CARMENES survey are extracted and wavelength-calibrated with CARACAL, the CARMENES data-reduction pipeline \citep[e.g.,][]{Caballero2016SPIE}. The CARACAL extraction procedure is based mainly on the REDUCE package \citep{Piskunov2002} and on the flat-relative optimal-extraction algorithm \citep{Zechmeister2014}. Nightly wavelength calibrations are based on the spectra of three different hollow-cathode lamps: Th-Ne, U-Ne, and U-Ar \citep[e.g.,][]{LovisPepe2007,Redman2011,Sarmiento2014,AmmlerGuenther2017}, and on a Fabry-P{\'e}rot etalon spectrum \citep{Schaefer2012}. Calibration spectra are taken at the beginning and end of each night. The wavemap for each night, that is, the wavelength value assigned to each pixel of the extracted frames of that night, is produced using the method described by \citet*{Bauer2015}. The sub-pixel continuous drifts of both instrument channels are monitored by taking Fabry-P{\'e}rot etalon spectra simultaneously with stellar spectra through the additional fibers available in both channels \citep{Quirrenbach2016SPIE}.

\subsection{RV measurements}

Radial velocities are measured from the CARMENES spectra with SERVAL \citep{Zechmeister2018}, which is an iterative least-squares fitting algorithm similar to TERRA \citep{AngladaEscudeButler2012}. For each star SERVAL creates a template spectrum to fit its individual observations by coadding all CARMENES spectra of that star. Then SERVAL measures individual RVs for each echelle order, and the final RV for each observation is taken as the weighted average of the individual-order RVs.

We currently require a minimum of five spectra to produce a template spectrum for a given star and to calculate its RVs. By September 2017, all $324$ CARMENES-sample stars were observed at least once \citep{Reiners2017a}, but only $287$ of them fulfilled the $n_{\rm RV}\geq5$ criterion.

\begin{figure}
\resizebox{\hsize}{!}
{\includegraphics{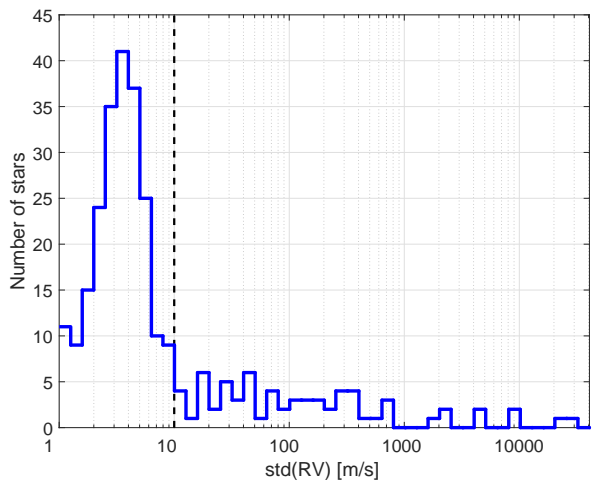}}
\caption{Histogram of std(RV) per star of the $287$ stars that were observed at least five times before September 2017. The RVs used for this plot were derived from the CARMENES VIS spectra with SERVAL \citep{Zechmeister2018}. The dashed line is the adopted boundary between RV-quiet and RV-loud stars at std(RV) $=10$\,m\,s$^{-1}$.}
\label{fig1}
\end{figure}

Beyond true RV excursions caused by stellar orbital motions, there are several additional sources of RV scatter. First, RV-measurement precision is limited by photon noise and by the amount of RV information in the observed stellar spectrum \citep[e.g.,][]{Bouchy2001,Reiners2010a}. Second, instrumental systematic effects might cause sub-pixel shifts of the spectra that look similar to Doppler shifts. Not all systematic effects can be calibrated out \citep[e.g.,][]{Halverson2016}. Third, stellar activity can cause apparent RV signals, as discussed above.

\citet{Reiners2017a} showed that an RV precision of $\lesssim1$\,m\,s$^{-1}$ can be achieved in the VIS for CARMENES-sample stars with $v \sin{i}\lesssim2$\,km\,s$^{-1}$, except for a few faint and very late-type M dwarfs (M$8.0$--$9.5$\,V). The median internal RV uncertainty that we measured from the VIS spectra taken before September 2017 was $\sim1.3$\,m\,s$^{-1}$. The deviation from $\sim1.0$\,m\,s$^{-1}$ can be explained by the fact that the survey includes $75$ stars with $v \sin{i}>2$\,km\,s$^{-1}$ and some faint M$8.0$--$9.5$\,V stars.

\citet{Trifonov2018} investigated the VIS RV accuracy of seven CARMENES-sample stars with known planetary systems, and the possible existence of systematic effects. They showed that despite our careful wavelength-calibration and drift-correction processes, which were shown to be precise at $<1$\,m\,s$^{-1}$ level \citep*{Bauer2015}, the measured RVs still contain nightly zero-point offsets (NZPs) at $\sim3$\,m\,s$^{-1}$ level. The origin of these NZPs is still being investigated. Nevertheless, when the RVs are corrected for the NZPs, the instrument performance is comparable to other PRV instruments, such as HARPS \citep{mayor03} and HIRES \citep{Vogt1994HIRES}. Since the NZP uncertainties, which are added in quadrature to the internal RV uncertainties, are typically $\sim1.0$\,m\,s$^{-1}$, the median NZP-corrected RV uncertainty that we measured was $\sim1.7$\,m\,s$^{-1}$.

Figure \ref{fig1} shows a histogram of RV standard-deviation values, std(RV), of the stars that were observed at least five times before September 2017. To avoid possible biases from outliers and small-number statistics, we estimated std(RV) as $1.48$ times the median absolute deviation around the median \citep[e.g.,][]{Rousseeuw1993}. The distribution peaks at $3$--$4$\,m\,s$^{-1}$, with tails extending out to $\sim1$\,m\,s$^{-1}$ and $\sim10$\,m\,s$^{-1}$. Hence, the RV scatter is larger than the typical uncertainties for most of the survey stars. This excess RV scatter can be explained by the presence of activity-induced RV variations and by reflex orbital motions caused by companion objects.

We here focus on the stars with std(RV) greater than $10$\,m\,s$^{-1}$. We name them `RV-loud stars', and by September 2017, we had identified $67$ such stars in the CARMENES sample. We expect the scatter of the RV-loud stars to be dominated by stellar activity and by possible orbital motions, but not by systematic measurement errors.

\subsection{Rotation and activity measurements}

Stellar rotation of M dwarfs, the amount and nature of active regions on their photospheres, and the level of chromospheric emission are three quantities that are closely related to each other \citep[e.g.,][]{Reiners2007}. In order to estimate their impact on strong RV variability, we need to measure each of them independently.

Using CARMENES spectra, \citet{Reiners2017a} derived $v \sin{i}$ values for $75$ out of the $324$ CARMENES-sample stars. Given the instrument resolution ($94,600$ in the VIS), reliable values could only be derived for stars with $v \sin{i}\geq2$\,km\,s$^{-1}$. For the rest of the stars the value $v \sin{i}=2$\,km\,s$^{-1}$ can be used as an upper limit. Another independent measure of rotation is $P_{\rm rot}$, which is usually determined from photometric time series. Unfortunately, we do not have $P_{\rm rot}$ values for many of the CARMENES-sample stars, including some of the fast rotators \citep{Caballero2016,Diez-Alonso2017,Reiners2017a,Jeffers2018arXiv}.

Two indicators that may be related to photospheric activity levels were defined by \citet{Zechmeister2018} and are measured by SERVAL from each spectrum: the chromatic index (CRX) and the differential line width (dLW). The CRX is a measure of the RV--$\log{\lambda}$ correlation. For each observation, SERVAL fits a straight line to the scatter plot of its individual-order RVs vs $\log{\lambda}$. The slope of this line is defined as the CRX. The dLW indicates differential changes in the line widths of the observation compared to the template. Unfortunately, these two indicators are not direct measures of the spot-coverage fraction or the average spot-to-photosphere temperature contrast. However, for the wavelength range covered by CARMENES VIS, both \citet{Reiners2010a} and \citet{Barnes2011} predicted a strong RV chromaticity to be created by cool spots on the surface of rotating M dwarfs, particularly for low spot-to-photosphere temperature contrast (e.g., $\sim200$\,K). In addition, \citet{Zechmeister2018} showed that dLW variations are correlated with the rotational phase of the active star YZ CMi.

Additional indicators of photospheric activity that we measure from CARMENES spectra are variations in the different CCF moments: the full width at half-maximum (FWHM), bisector inverse slope (BIS), and CCF contrast \citep[e.g.,][]{Reiners2018}. Such variations can be related to changes in the distribution of inhomogeneities on the stellar surface \citep[e.g.,][]{Herrero2016}. Since M-dwarf visible-light spectra contain a large number of blended molecular lines \citep[e.g.,][]{Reiners2017a}, their CCF analysis is not as straightforward as it is for solar-type stars \citep[e.g.,][]{Astudillo-Defru2015,SuarezMascareno2017}. The situation is even more complicated for fast-rotating M dwarfs. We are currently extending our CCF analysis of M-dwarf spectra to also include fast rotators.

Several indicators of chromospheric activity are routinely measured from each CARMENES spectrum \citep[e.g.,][]{Jeffers2018arXiv}. They include measurements of excess emission in H$\alpha$, Na D$_1$\,and\,D$_2$, and the Ca\,{\sc ii} infrared triplet, among other lines. The measurements are carried out by estimating the pseudo-equivalent width (pEW) of each line after subtracting from the observed star spectrum a template spectrum of an inactive star of a similar SpT. For instance, the pEW(H$\alpha$) values are measured in the wavelength region [$6559.6,6569.6$\,$\AA$], after normalizing it in the regions [$6550.0,6555.0$\,$\AA$] and [$6576.0,6581.0$\,$\AA$]. This is similar to the methods described by \citet[][]{Young1989} and \citet[][]{Montes1995}. We also translate pEW(H$\alpha$) measurements into H$\alpha$ luminosity relative to the bolometric one ($\log{L_{{\rm H}\alpha}/L_{\rm bol}}$) using the relation
\begin{equation}
\frac{L_{{\rm H}\alpha}}{L_{\rm bol}}=-{\rm pEW(H\alpha)}\cdot\chi(T_{\rm eff}),
\label{eq1}
\end{equation}
where $T_{\rm eff}$ is the effective temperature of the star, which is estimated by translating its SpT to $T_{\rm eff}$ with the relations of \citet{Pecaut2013}. The $\chi$ values are taken from \citet{ReinersBasri2008}. The SpT and an average $\log{L_{{\rm H}\alpha}/L_{\rm bol}}$ value for each CARMENES-sample star can be found in \citet{Reiners2017a}.

For the purpose of investigating RV--rotation and RV--activity correlations in this paper, we focus on the following four quantities: $v \sin{i}$, CRX, dLW, and $\log{L_{{\rm H}\alpha}/L_{\rm bol}}$. We do not use $P_{\rm rot}$ as a measure of rotation, not only because we do not have it for all of our stars, but also because we wish to limit ourselves to quantities that can be determined directly from the spectra. We use only $\log{L_{{\rm H}\alpha}/L_{\rm bol}}$ as a chromospheric-activity indicator, both because H$\alpha$ is the line that is most widely used for that purpose \citep[e.g.][]{Jeffers2018arXiv} and because our preliminary results show that the variations in chromospheric emission can be measured more accurately in H$\alpha$ than in the other lines.

\section{CARMENES sample of active RV-loud stars}
\label{sec4}

The goal of this paper is to study the main common features of activity-induced RV variations on a sample of stars, rather than making an in-depth analysis of individual active stars. To do so, we selected a subsample of stars for which we could attribute the RV scatter to rotation and/or activity with high confidence. Only stars that fulfilled the following criteria were selected for the study:
\begin{itemize}
 \item no orbital RV variations of $>10$\,m\,s$^{-1}$ due to companions, either known or suspected;
 \item std(RV) $>10$\,m\,s$^{-1}$;
 \item number of RVs per star $n_{\rm RV}>10$;
 \item $v \sin{i}>2$\,km\,s$^{-1}$.
\end{itemize} 
The reasoning behind these criteria is detailed below.

First, we removed from the sample $17$ RV-loud stars whose RV scatter is dominated by orbital motion and not by stellar activity: $4$ known planetary systems \citep{Trifonov2018}, and another $13$ spectroscopic binaries that passed our initial survey filters \citep[e.g.,][]{Cortes-Contreras2017,Jeffers2018arXiv}, but were discovered at an early stage of the CARMENES survey. Most of the new binaries have std(RV)\,$>1$\,km\,s$^{-1}$, and they will be addressed in a forthcoming paper.

Next, we excluded the $220$ RV-quiet stars with std(RV)$<10$\,m\,s$^{-1}$ because separating orbital RV signals from activity-induced ones close to the measurement accuracy limit with a small number of RVs is a difficult task. With m\,s$^{-1}$ instruments, the separation of signals smaller than $10$\,m\,s$^{-1}$ typically relies on periodogram analysis, comparing significant RV frequencies with significant frequencies of the different activity indicators \citep[e.g.,][]{Bonfils2007,Suarez2017}. However, for a small number of data points and a low RV-scatter to RV-uncertainty ratio, there is a low probability of finding any significant periodogram peaks \citep[e.g.,][]{Zechmeister2009a}. With a median number of $11$ RVs per star and a median RV uncertainty of $\sim1.7$\,m\,s$^{-1}$, we decided to exclude from our analysis here all RV-quiet stars. We postpone the investigation of their activity-induced RV signals to after we have acquired enough RVs per star to compute and analyze meaningful periodograms.

\begin{figure}
\resizebox{\hsize}{!}
{\includegraphics{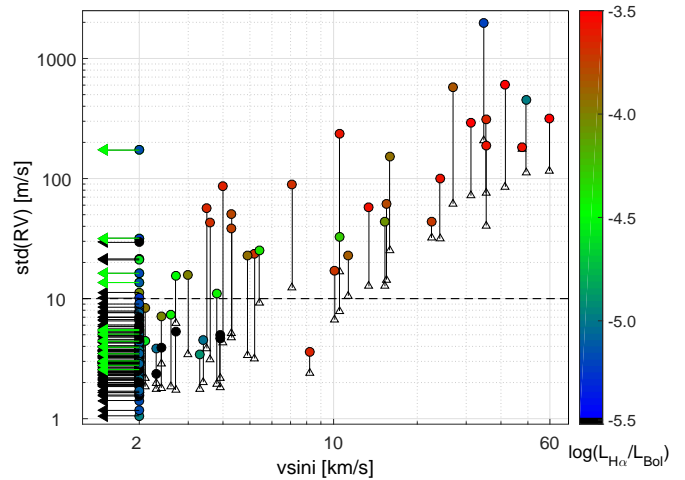}}
\caption{CARMENES VIS std(RV)--$v \sin{i}$ scatter plot for the $166$ stars with $n_{\rm RV}>10$. The points are color-coded according to their H$\alpha$ luminosity. Stars with only an upper limit of $v \sin{i}\lesssim2$\,km\,s$^{-1}$ are marked with arrows: green arrows for host stars of known planetary systems, and black arrows for stars with no detected planets. The open triangles are the median RV uncertainties ($\overline{\delta{\rm RV}}$) of the stars with $v \sin{i}>2$\,km\,s$^{-1}$. 
For clarity, each triangle is connected by a solid line to its point. The dashed line is the adopted boundary between RV-quiet and RV-loud stars at std(RV) $=10$\,m\,s$^{-1}$.
}
\label{fig2}
\end{figure}

The std(RV) of the remaining $50$ RV-loud stars ranges from $0.01$ to $2.0$ km\,s$^{-1}$. Thus, these stars represent the high end of the expected activity-induced RV scatter in M dwarfs \citep[e.g.,][]{Barnes2011}. However, even with a high ratio of the RV-scatter to RV-uncertainty, our experience shows that it is hard to distinguish between orbital and activity-induced RV signals for $n_{\rm RV}\leq10$. With such a low $n_{\rm RV}$, it is possible to confirm the compatibility of the RV variations with a known period \citep[e.g.,][]{TalOr2015}, but it is impossible to reliably determine the parameters of a signal of an unknown period \citep[e.g.,][]{Halbwachs2014sf2a}. For that reason, we excluded another $14$ RV-loud stars with $n_{\rm RV}\leq10$, and left them for future analysis.

Applying the criteria above in a different order, we selected the remaining $36$ RV-loud stars from the $166$ stars that were observed at least $11$ times by September 2017. Using the published occurrence rates of planets around M dwarfs \citep{Fressin2013,DressingCharbonneau2013}, \citet{Garcia-Piquer2017} estimated the exoplanet yield of the CARMENES survey. They concluded that the largest RV semi-amplitudes that will be discovered by CARMENES should be $\sim30$\,m\,s$^{-1}$, and that the CARMENES sample should include $\sim10$ planets that exert RV semi-amplitudes of $10$--$30$\,m\,s$^{-1}$ on their hosts. Hence, a handful of such systems could be contained in this subsample of $166$ stars with $n_{\rm RV}>10$.

\begin{table*}
\caption{Main properties of the active RV-loud sample. Values are taken from Carmencita and \citet{Reiners2017a}.}
\begin{centering}
\begin{tabular}{llccccccc}
\hline
\hline
 Karmn & Name   & $\alpha$ & $\delta$ & $J$ & SpT & $P_{\rm rot}$ & $v \sin{i}$ & log \\
            &      &  (J2000)  & (J2000) & (mag) &      & (d) & (km\,s$^{-1}$) & $L_{{\rm H}\alpha}/L_{\rm bol}$ \\
\hline
J01033+623 & V388 Cas & 01:03:21.49 & +62:21:57.2 & $8.61$ & M5.0\,V & $ 1.06$ & $10.5$ & $-3.57$ \\ 
J01352-072 & Barta 161 12 & 01:35:14.05 & -07:12:52.8 & $8.96$ & M4.0\,V &  ...   & $59.8$ & $-3.50$ \\ 
J02088+494 & G 173-039 & 02:08:54.00 & +49:26:52.0 & $8.42$ & M3.5\,V &   ...  & $24.1$ & $-3.56$ \\ 
J03473-019 & G 080-021 & 03:47:23.54 & -01:58:24.1 & $7.80$ & M3.0\,V & $ 3.88$ & $ 5.2$ & $-3.69$ \\ 
J04472+206 & RX J0447.2+2038 & 04:47:12.35 & +20:38:09.3 & $9.38$ & M5.0\,V &  ...  & $47.6$ & $-3.55$ \\ 
J05365+113 & V2689 Ori & 05:36:30.98 & +11:19:39.2 & $6.13$ & M0.0\,V & $12.04$ & $ 3.8$ & $-4.45$ \\ 
J06000+027 & G 099-049 & 06:00:03.86 & +02:42:22.9 & $6.91$ & M4.0\,V & $ 1.81$ & $ 4.9$ & $-3.93$ \\ 
J07446+035 & YZ CMi & 07:44:39.79 & +03:33:01.4 & $6.58$ & M4.5\,V & $ 2.78$ & $ 4.0$ & $-3.62$ \\ 
J08298+267 & DX Cnc & 08:29:48.11 & +26:46:24.5 & $8.23$ & M6.5\,V & $ 0.46$ & $10.5$ & $-4.31$ \\ 
J09449-123 & G 161-071 & 09:44:53.84 & -12:20:53.8 & $8.50$ & M5.0\,V &  ...  & $31.2$ & $-3.37$ \\ 
J12156+526 & StKM 2-809 & 12:15:39.56 & +52:39:08.9 & $8.59$ & M4.0\,V &   ...  & $35.3$ & $-3.64$ \\ 
J12189+111 & GL Vir & 12:18:57.94 & +11:07:37.3 & $8.53$ & M5.0\,V & $ 0.49$ & $15.5$ & $-3.79$ \\ 
J14173+454 & RX J1417.3+4525 & 14:17:22.17 & +45:25:45.7 & $9.47$ & M5.0\,V &  ...  & $15.9$ & $-3.94$ \\ 
J15218+209 & OT Ser & 15:21:53.01 & +20:58:41.7 & $6.61$ & M1.5\,V & $ 3.38$ & $ 4.3$ & $-3.75$ \\ 
J15499+796 & LP 022-420 & 15:49:53.86 & +79:39:53.9 & $9.72$ & M5.0\,V &  ...  & $26.9$ & $-3.83$ \\ 
J16313+408 & G 180-060 & 16:31:18.57 & +40:51:56.8 & $9.46$ & M5.0\,V & $ 0.51$ & $ 7.1$ & $-3.68$ \\ 
J16555-083 & vB 8 & 16:55:34.36 & -08:23:54.8 & $9.78$ & M7.0\,V &  ...  & $ 5.4$ & $-4.40$ \\ 
J16570-043 & LP 686-027 & 16:57:06.24 & -04:21:02.3 & $7.97$ & M3.5\,V & $ 1.21$ & $10.1$ & $-3.71$ \\ 
J17338+169 & 1RXS J173353.5+165515 & 17:33:53.02 & +16:55:10.7 & $8.89$ & M5.5\,V & $ 0.27$ & $41.5$ & $-3.51$ \\ 
J18022+642 & LP 071-082 & 18:02:17.11 & +64:15:38.1 & $8.54$ & M5.0\,V & $ 0.28$ & $11.3$ & $-3.86$ \\ 
J18189+661 & LP 071-165 & 18:18:58.45 & +66:11:25.6 & $8.74$ & M4.5\,V &  ...  & $15.3$ & $-4.07$ \\ 
J18356+329 & LSR J1835+3259 & 18:35:37.80 & +32:59:41.6 & $10.27$ & M8.5\,V &  ...  & $49.2$ & $-5.01$ \\ 
J18498-238 & V1216 Sgr & 18:49:50.08 & -23:50:13.3 & $6.22$ & M3.5\,V & $ 2.87$ & $ 3.0$ & $-4.01$ \\ 
J19169+051S & V1298 Aql (vB 10) & 19:16:56.97 & +05:08:39.8 & $9.91$ & M8.0\,V &  ...  & $ 2.7$ & $-4.51$ \\ 
J19255+096 & LSPM J1925+0938 & 19:25:30.99 & +09:38:19.3 & $11.21$ & M8.0\,V &  ...  & $34.7$ & $-5.25$ \\ 
J19511+464 & G 208-042 & 19:51:09.61 & +46:29:04.5 & $8.59$ & M4.0\,V & $ 0.59$ & $22.5$ & $-3.72$ \\ 
J20093-012 & 2M J20091824-0113377 & 20:09:18.18 & -01:13:44.1 & $9.40$ & M5.0\,V &  ...  & $ 4.3$ & $-3.77$ \\ 
J22012+283 & V374 Peg & 22:01:13.59 & +28:18:25.5 & $7.63$ & M4.0\,V & $ 0.45$ & $35.4$ & $-3.57$ \\ 
J22468+443 & EV Lac & 22:46:48.71 & +44:19:55.4 & $6.11$ & M3.5\,V & $ 4.38$ & $ 3.5$ & $-3.65$ \\ 
J22518+317 & GT Peg & 22:51:54.18 & +31:45:14.4 & $7.70$ & M3.0\,V & $ 1.64$ & $13.4$ & $-3.55$ \\ 
J23548+385 & RX J2354.8+3831 & 23:54:51.27 & +38:31:34.8 & $8.94$ & M4.0\,V & $ 4.76$ & $ 3.6$ & $-3.68$ \\ 
\hline
\end{tabular}
\label{tab1}
\end{centering}
\end{table*}

Figure \ref{fig2} presents the std(RV)--$v \sin{i}$ scatter plot of these $166$ stars. The points are color-coded according to their H$\alpha$ luminosity. In addition, hosts of known planetary systems are marked with green arrows. 
Above the $10$\,m\,s$^{-1}$ boundary, which we adopted to separate RV-quiet from RV-loud stars, stars with $v \sin{i}$ values above and below $2$\,km\,s$^{-1}$ seem to form two distinct groups. While the stars with $v \sin{i}>2$\,km\,s$^{-1}$ typically have strong H$\alpha$ emission and no known planets, the group of stars with $v \sin{i}\lesssim2$\,km\,s$^{-1}$ have weaker H$\alpha$ emission and a few known planets. If planetary occurrence rates are not sensitive to $v \sin{i}$ values, we expect most of the stars with orbital RV semi-amplitudes of $10$--$30$\,m\,s$^{-1}$ to have $v \sin{i}\lesssim2$\,km\,s$^{-1}$, and that only one or two of these stars will have $v \sin{i}>2$\,km\,s$^{-1}$. Therefore, to minimize the chances of including in the selected sample stars whose RV scatter is dominated by planets as yet
undiscovered, we excluded the five stars with std(RV) of $10$--$30$\,m\,s$^{-1}$ but $v \sin{i}\lesssim2$\,km\,s$^{-1}$. The RV scatter of some of these stars could still be dominated by activity, but to be extra careful, we leave this question for future analysis. 

We are thus left with $31$ stars, which we name the `CARMENES sample of active RV-loud stars'. We do not claim that these stars do not have planetary systems, but that most likely the RV scatter of these stars is dominated by activity rather than by orbital motions. We find this sufficient to use this sample for investigating the main features of activity-induced RVs.

Table \ref{tab1} lists the main properties of our active RV-loud stars. Their SpT distribution shows that most of the stars in the sample ($24/31$) are mid-type M dwarfs (M$3.0$--$5.5$\,V). There are only two early-type M dwarfs (M$0.0$--$2.5$\,V), and only five late-type M dwarfs (M$6.0$--$9.0$\,V) in the sample. However, the fraction of active RV-loud stars in the whole CARMENES sample increases monotonically with SpT. In the subsample of $166$ stars with $n_{\rm RV}>10$, the fractions of early-, mid-, and late-type active RV-loud stars are $2/70$ ($3$\,\%), $24/88$ ($27$\,\%), and $5/8$ ($63$\,\%), respectively.

\section{Observed connection between RV scatter, stellar rotation, and stellar activity}
\label{sec5}

\begin{table*}
\caption{CARMENES VIS results for the active RV-loud sample.
}
\begin{centering}
\begin{tabular}{llcccccc}
\hline
\hline
\noalign{\smallskip}
Karmn & Name   & $n_{\rm RV}$ & $\overline{\delta{\rm RV}}$ & std(RV) & RV--CRX & RV--dLW & RV--$\log{L_{{\rm H}\alpha}/L_{\rm bol}}$\\
    &  & & (m\,s$^{-1}$) & (m\,s$^{-1}$) & log(p-value) & log(p-value) & log(p-value) \\
\hline
\noalign{\smallskip}
J01033+623 & V388 Cas & 12 & $16.8$ & $ 237\pm 48$ & $-5.6$ & $-0.4$ & $-0.4$ \\ 
J01352-072 & Barta 161 12 & 11 & $115.6$ & $ 315\pm 67$ & $-2.8$ & $-1.1$ & $-2.2$ \\ 
J02088+494 & G 173-039 & 17 & $31.5$ & $ 100\pm 17$ & $-0.3$ & $-0.5$ & $-0.1$ \\ 
J03473-019 & G 080-021 & 11 & $3.2$ & $  24\pm  5$ & $-0.1$ & $-1.0$ & $-0.4$ \\ 
J04472+206 & RX J0447.2+2038 & 11 & $177.1$ & $ 182\pm 39$ & $-0.1$ & $-0.1$ & $-0.3$ \\ 
J05365+113 & V2689 Ori & 31 & $1.9$ & $  11\pm  1$ & $-0.3$ & $-1.2$ & $-1.9$ \\ 
J06000+027 & G 099-049 & 14 & $3.4$ & $  23\pm  4$ & $-2.7$ & $-0.2$ & $-0.4$ \\ 
J07446+035 & YZ CMi & 44 & $4.3$ & $  87\pm  9$ & $-13.8$ & $-0.2$ & $-0.3$ \\ 
J08298+267 & DX Cnc & 13 & $7.9$ & $  33\pm  6$ & $-0.6$ & $-0.3$ & $-0.1$ \\ 
J09449-123 & G 161-071 & 11 & $72.0$ & $ 290\pm 62$ & $-0.7$ & $-0.7$ & $-0.6$ \\ 
J12156+526 & StKM 2-809 & 12 & $75.6$ & $ 314\pm 64$ & $-1.8$ & $-0.9$ & $-0.2$ \\ 
J12189+111 & GL Vir & 12 & $14.4$ & $  61\pm 13$ & $-1.8$ & $-0.1$ & $-0.1$ \\ 
J14173+454 & RX J1417.3+4525 & 12 & $25.3$ & $ 152\pm 31$ & $-0.4$ & $-0.4$ & $-1.4$ \\ 
J15218+209 & OT Ser & 36 & $5.2$ & $  39\pm  5$ & $-6.0$ & $-0.2$ & $-0.3$ \\ 
J15499+796 & LP 022-420 & 14 & $61.6$ & $ 577\pm109$ & $-0.3$ & $-0.3$ & $-1.4$ \\ 
J16313+408 & G 180-060 & 12 & $12.3$ & $  89\pm 18$ & $-1.7$ & $-0.3$ & $-0.3$ \\ 
J16555-083 & vB 8 & 104 & $9.2$ & $  25\pm  2$ & $-0.8$ & $-0.7$ & $-0.8$ \\ 
J16570-043 & LP 686-027 & 13 & $6.7$ & $  17\pm  3$ & $-0.1$ & $-0.6$ & $-0.5$ \\ 
J17338+169 & 1RXS J173353.5+165515 & 12 & $85.3$ & $ 604\pm123$ & $-1.5$ & $-0.6$ & $-0.3$ \\ 
J18022+642 & LP 071-082 & 14 & $10.5$ & $  23\pm  4$ & $-0.8$ & $-0.3$ & $-0.3$ \\ 
J18189+661 & LP 071-165 & 12 & $12.7$ & $  44\pm  9$ & $-1.7$ & $-0.2$ & $-0.2$ \\ 
J18356+329 & LSR J1835+3259 & 14 & $112.9$ & $ 449\pm 85$ & $-2.7$ & $-2.8$ & $-0.3$ \\ 
J18498-238 & V1216 Sgr & 25 & $3.4$ & $  16\pm  2$ & $-0.6$ & $-0.3$ & $-0.0$ \\ 
J19169+051S & V1298 Aql (vB 10) & 31 & $6.3$ & $  16\pm  2$ & $-1.3$ & $-0.2$ & $-0.4$ \\ 
J19255+096 & LSPM J1925+0938 & 15 & $208.1$ & $1986\pm363$ & $-6.9$ & $-0.5$ & $-0.7$ \\ 
J19511+464 & G 208-042 & 13 & $32.0$ & $  44\pm  9$ & $-1.1$ & $-0.1$ & $-0.1$ \\ 
J20093-012 & 2M J20091824-0113377 & 12 & $4.7$ & $  51\pm 10$ & $-5.0$ & $-0.5$ & $-0.7$ \\ 
J22012+283 & V374 Peg & 12 & $40.6$ & $ 190\pm 39$ & $-0.6$ & $-0.0$ & $-0.0$ \\ 
J22468+443 & EV Lac & 74 & $3.8$ & $  57\pm  5$ & $-16.0$ & $-0.0$ & $-0.5$ \\ 
J22518+317 & GT Peg & 12 & $12.7$ & $  58\pm 12$ & $-1.5$ & $-0.0$ & $-0.0$ \\ 
J23548+385 & RX J2354.8+3831 & 13 & $3.1$ & $  43\pm  8$ & $-5.1$ & $-0.6$ & $-0.1$ \\ 
\hline
\end{tabular}
\label{tab2}
\end{centering}
\end{table*}

\subsection{Correlations of std(RV) with global spectral parameters}

\begin{figure*}
\minipage{0.33\textwidth}
{\includegraphics[width=\linewidth]{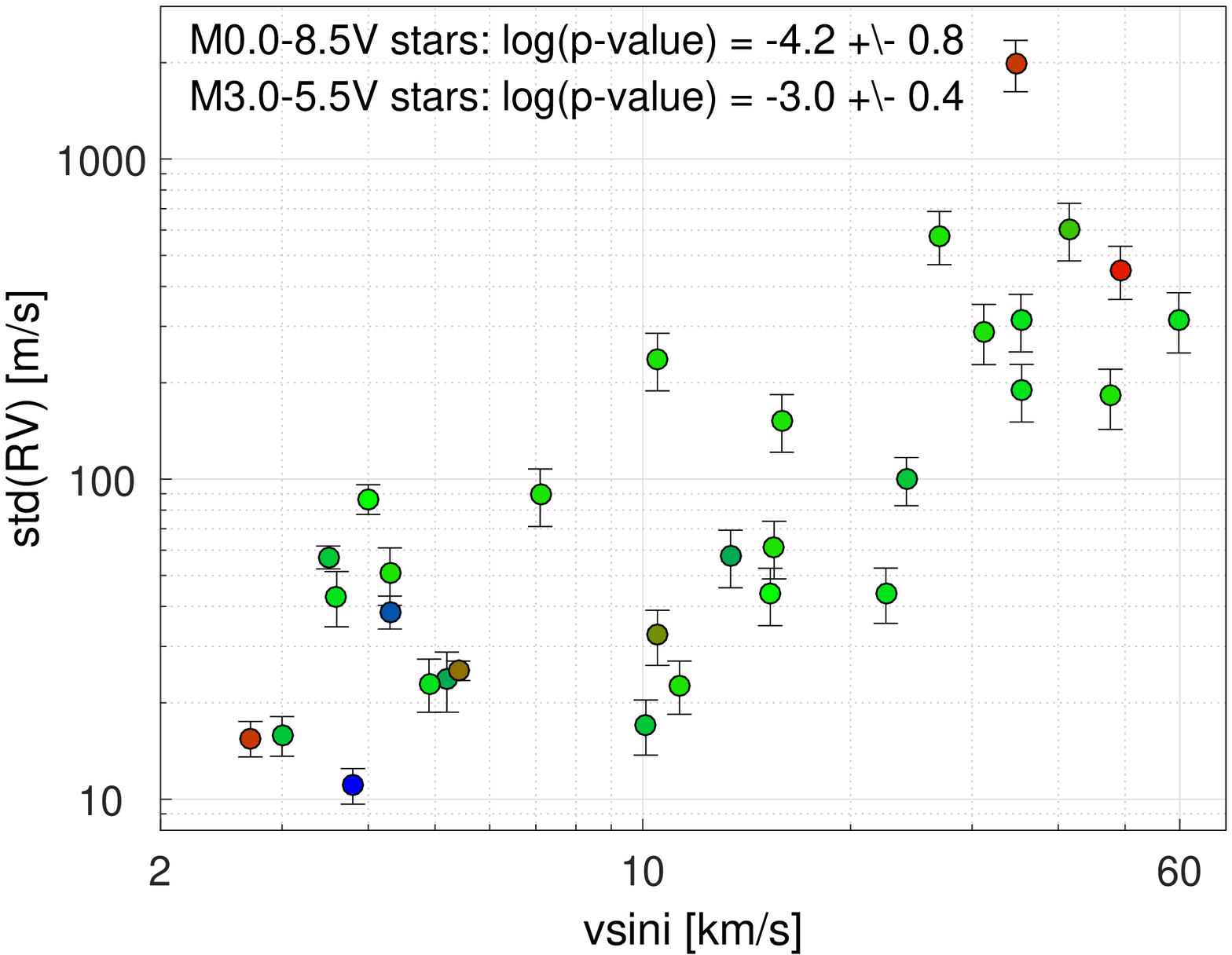}}
\endminipage\hfill
\minipage{0.33\textwidth}
{\includegraphics[width=\linewidth]{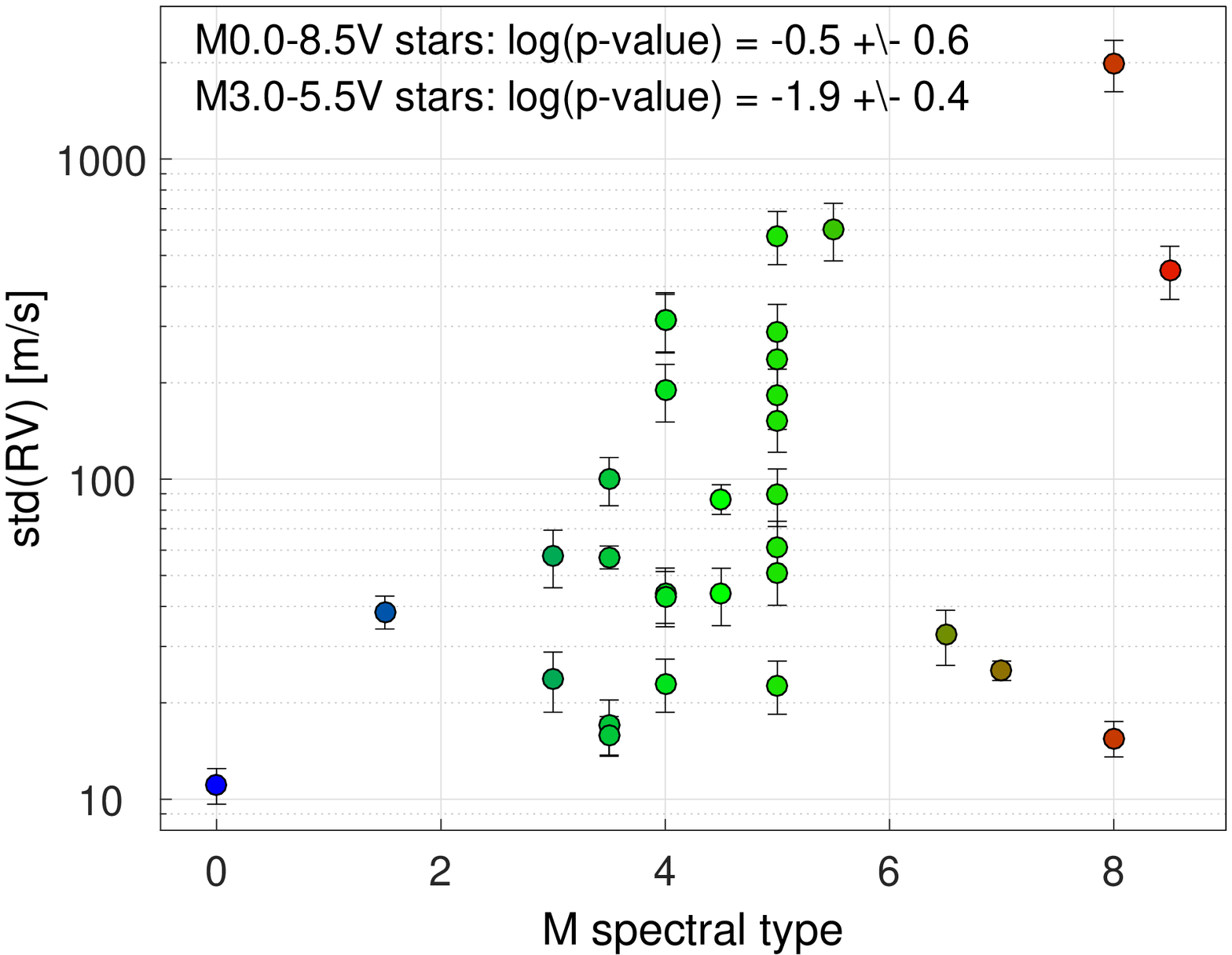}}
\endminipage\hfill
\minipage{0.33\textwidth}
{\includegraphics[width=\linewidth]{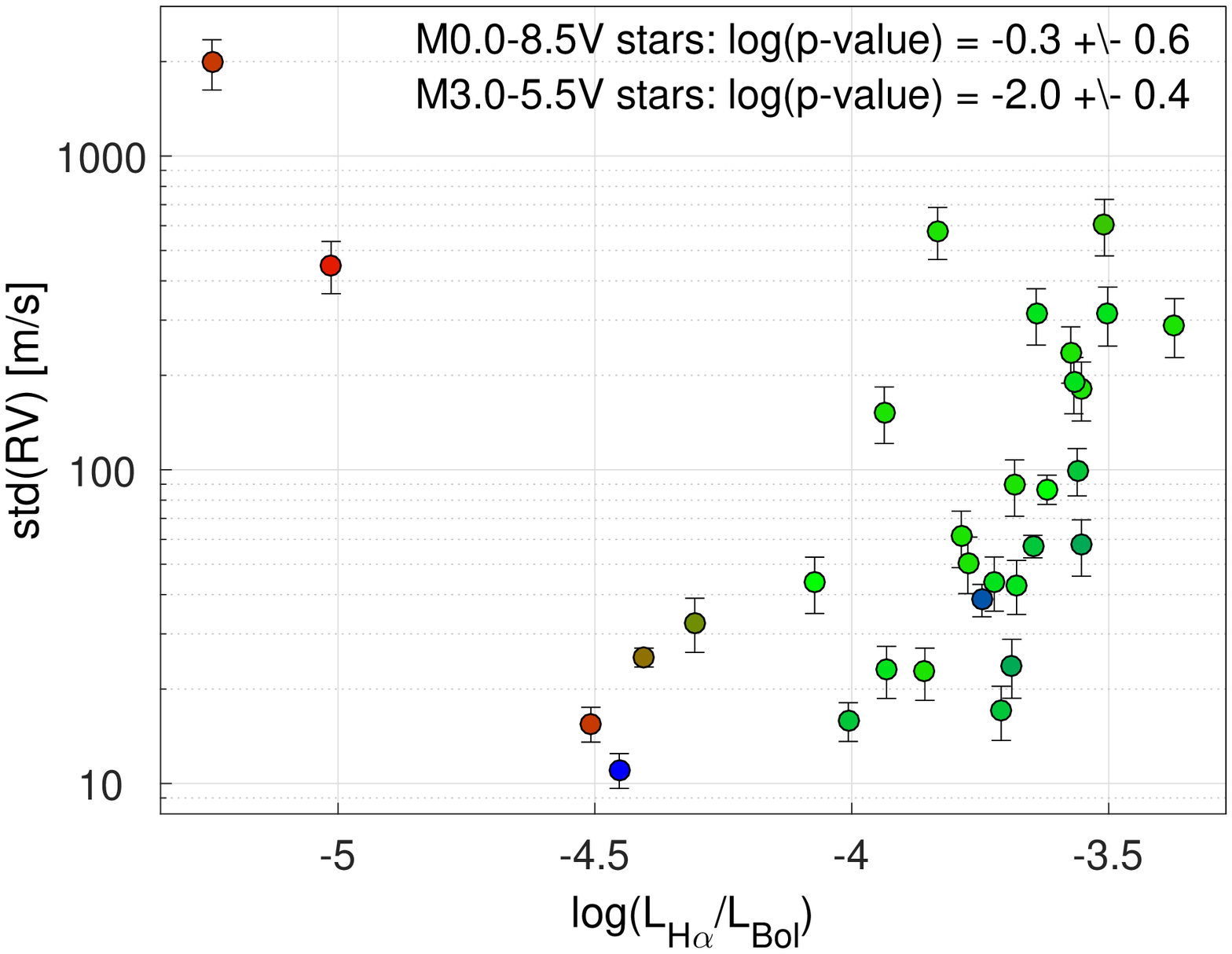}}
\endminipage
\caption{Correlating std(RV) of the active RV-loud stars with three of their spectral parameters -- {\it left}: $v \sin{i}$; {\it middle}: SpT; and {\it right}: $\log{L_{{\rm H}\alpha}/L_{\rm bol}}$. The points are color-coded according to SpT in all three plots, from M$0$ in blue to M$9$ in red. Correlation significances, as measured by the p(F$_{\rm test}$) values, are shown in the insets. A log(p-value)\,$<-2.3$ means a significant correlation, while a log(p-value) between $-1.3$ and $-2.3$ means a marginal correlation.
}
\label{fig3}
\end{figure*}

Table \ref{tab2} shows the $n_{\rm RV}$ and std(RV) values, as well as the median RV uncertainties ($\overline{\delta{\rm RV}}$), of the active RV-loud stars, which were measured from their CARMENES VIS spectra. As a measure of std(RV) uncertainty we took std(RV)($2$\,$n_{\rm RV})^{-0.5}$ \citep{Press2002}.

To understand the mechanisms driving the observed large RV scatter, we first searched for correlations between std(RV) and three of the parameters shown in Table \ref{tab1}: SpT, $v \sin{i}$, and $\log{L_{{\rm H}\alpha}/L_{\rm bol}}$. These three parameters were chosen because they were previously suggested to influence the RV-scatter amplitude \citep[e.g.,][]{Barnes2011}. Unfortunately, we do not have indicators of the photospheric level of activity, from which we could estimate the spot-coverage fraction, or the average spot-to-photosphere temperature contrast. The chromospheric H$\alpha$ emission can be used as a proxy to the photospheric activity level only if the two are correlated -- a non-trivial assumption for M dwarfs \citep[e.g.,][]{Reiners2007,Reiners2010b}.

To estimate linear correlation significance, we used the p(F$_{\rm test}$) value \citep{Fisher1925} of fitting a straight line to the data. The default p-value threshold frequently used for claiming a statistically significant correlation is $0.05$ (or log(p-value)\,$=-1.3$). Recently, \citet{Benjamin2018} proposed that it should be redefined to $0.005$ (log(p-value)\,$=-2.3$), and that findings with $0.005<$\,p-value\,$<0.05$ should be labeled as having ``suggestive evidence''. We follow these criteria.

Figure \ref{fig3} shows the std(RV)--$v \sin{i}$, std(RV)--SpT, and std(RV)--$\log{L_{{\rm H}\alpha}/L_{\rm bol}}$ correlation plots. The points are color-coded according to SpT. In each plot we estimated two linear correlation significances: one for the whole active RV-loud sample ($31$ stars), and one for the subsample of mid-type M dwarfs ($24$ stars). To prevent the difference in sample size from affecting the p-values, we bootstrapped the sample with $1000$ iterations, selecting each time a random subsample of $21$ stars for the p-value estimation. The final p-values were taken as the medians of these $1000$ iterations, and the p-value uncertainties were tuned to cover the central $68.2\%$ of the distribution. The resulting p-values are shown in the plot insets.

There is a significant std(RV)--$v \sin{i}$ correlation in our active RV-loud sample. There are no std(RV)--$\log{L_{{\rm H}\alpha}/L_{\rm bol}}$ or std(RV)--SpT correlations for the whole sample, but there is suggestive evidence for such correlations when we restrict our sample to only the mid-type M dwarfs.

The std(RV)--$v \sin{i}$ correlation suggests that the spectral line broadening that accompanies stellar rotation plays an important role in driving strong RV scatter in active M dwarfs. In order to verify that the loss of RV information content due to the spectral line broadening is not, by itself, sufficient to explain the observed RV scatter and that it is the combination of activity and rotation that drives it, we performed the following test: we repeated the correlation analysis shown in Fig. \ref{fig3} after replacing the std(RV) values with the excess RV scatter of each star, which we defined as the quadratic difference $({\rm std(RV)}^{2} - \overline{\delta{\rm RV}}^{2})^{0.5}$. 
We found this substitution to slightly weaken the correlation significances presented in Fig. \ref{fig3}, but not to change the presented qualitative results. For instance, the log(p-value) of the linear correlation between excess RV scatter and $v \sin{i}$ for the full sample of active RV-loud stars was found to be $-3.3$. This indicates a weaker correlation than the std(RV)--$v \sin{i}$ one, for which we found log(p-value) $=-4.2$, but it is still significant.

The marginal std(RV)--$\log{L_{{\rm H}\alpha}/L_{\rm bol}}$ correlation for the mid-type M dwarfs could be explained in three ways: (1) a higher chromospheric emission directly drives a higher RV scatter; (2) the level of chromospheric emission is correlated with the level of photospheric activity, but the spots are the ones that drive RV scatter; and (3) this correlation is an indirect outcome of the std(RV)--$v \sin{i}$ correlation, and the rotation-activity correlation for M dwarfs \citep[e.g.,][]{Reiners2007,Newton2017,Jeffers2018arXiv}. The last explanation could also hold true for the weak std(RV)--SpT correlation seen for the mid-type M dwarfs in Fig. \ref{fig3}, which could be an indirect outcome of the rotation-SpT correlation in this range of spectral types \citep[e.g.,][]{Reiners2012}.

The above findings are in agreement with the predictions of \citet{Reiners2010a} and \citet{Barnes2011}. Their simulations suggest that the large scatter around the std(RV)--$v \sin{i}$ correlation seen in the left panel of Fig. \ref{fig3} could be caused by the spot characteristics of the active stars, such as their spot coverage fraction, and their spot-to-photosphere temperature contrast. 
We also recall that an additional source of scatter could be true orbital motions of these stars due to their as yet undiscovered planetary systems.

\subsection{Correlations of RVs with spectral activity indicators}

\begin{figure}
\resizebox{\hsize}{!}
{\includegraphics{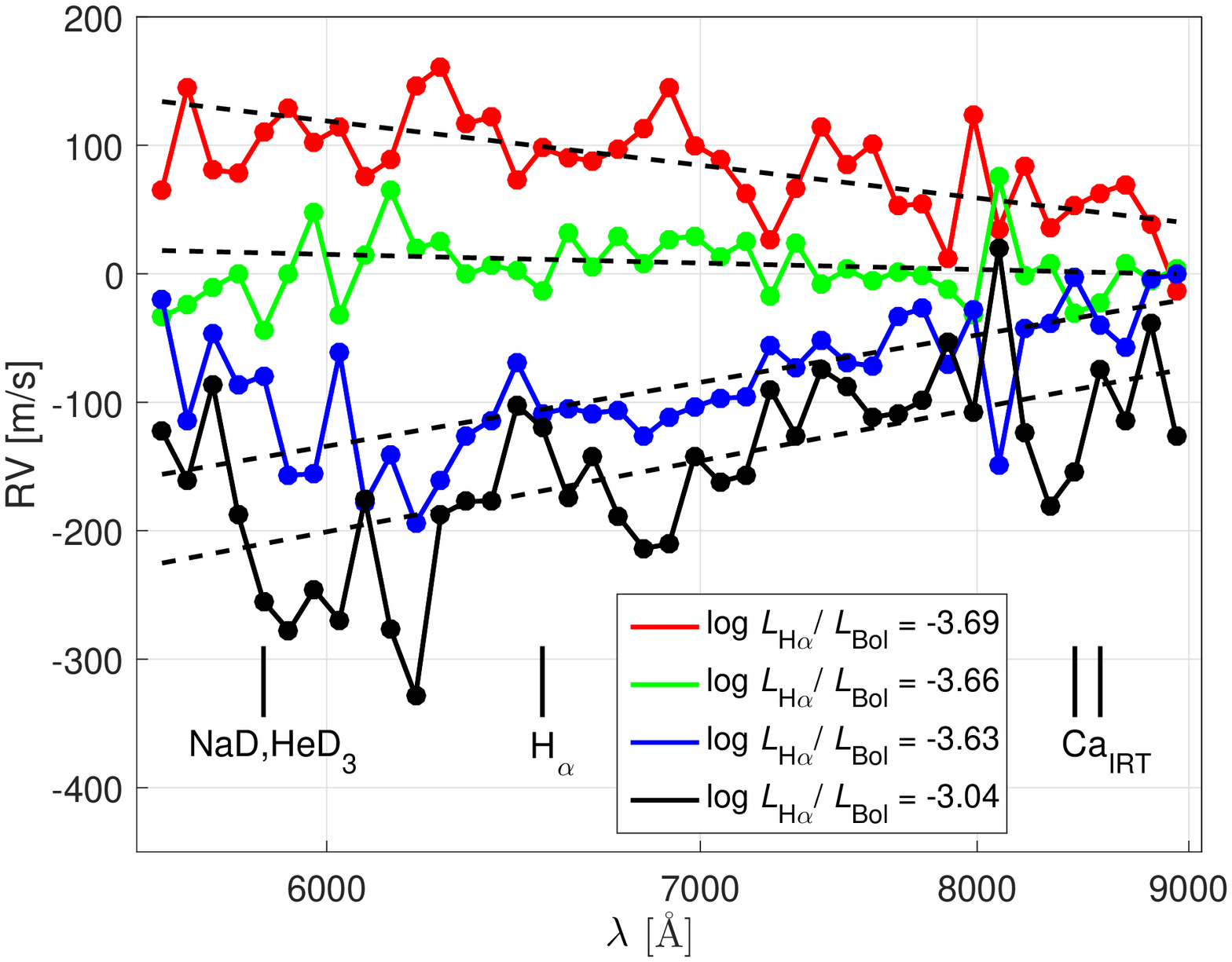}}
{\includegraphics[width=84mm]{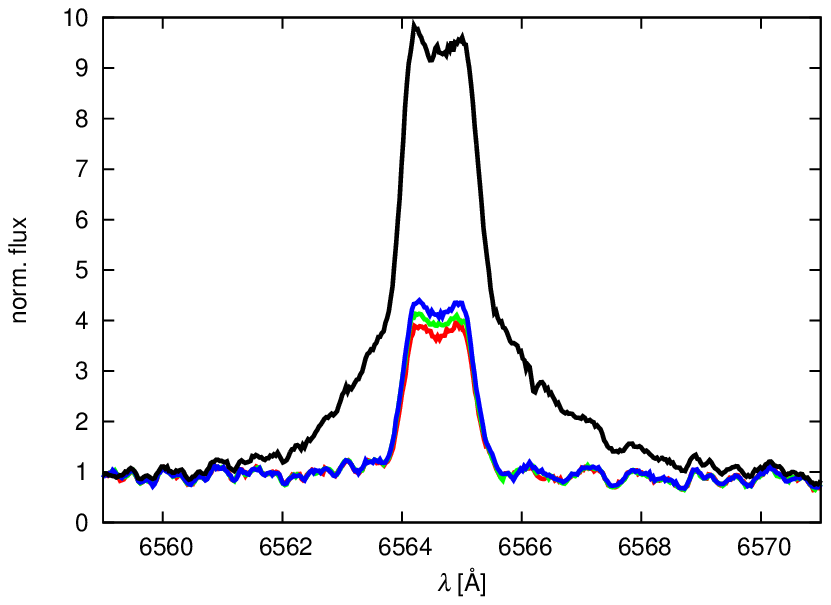}}
\caption{CARMENES VIS measurements of J22468+443 (EV Lac). {\it Top}: Order-by-order RVs from four representative observations. Red, green, and blue show three observations with close-to-median H$\alpha$ emission. Black plots an observation with extremely high H$\alpha$ emission, taken during a strong flare event \citep{Fuhrmeister2018arXiv}. The corresponding $\log{L_{{\rm H}\alpha}/L_{\rm bol}}$ values are given in the inset. Dashed lines show the best-fit straight lines to the RV--order scatter plots of each observation. Individual-order RV uncertainties are not shown in the plot for clarity, but they are on the order of $10$--$30$\,m\,s$^{-1}$, depending on the RV-information content and the S/N in each order. The markers at the bottom specify the locations of the indicated chromospheric emission lines. {\it Bottom}: Parts of the VIS spectra from these four observations, centered on the H$\alpha$ line. The colors are the same as in the top panel.
}
\label{fig4}
\end{figure}

\begin{figure*}
\minipage{0.33\textwidth}
{\includegraphics[width=\linewidth]{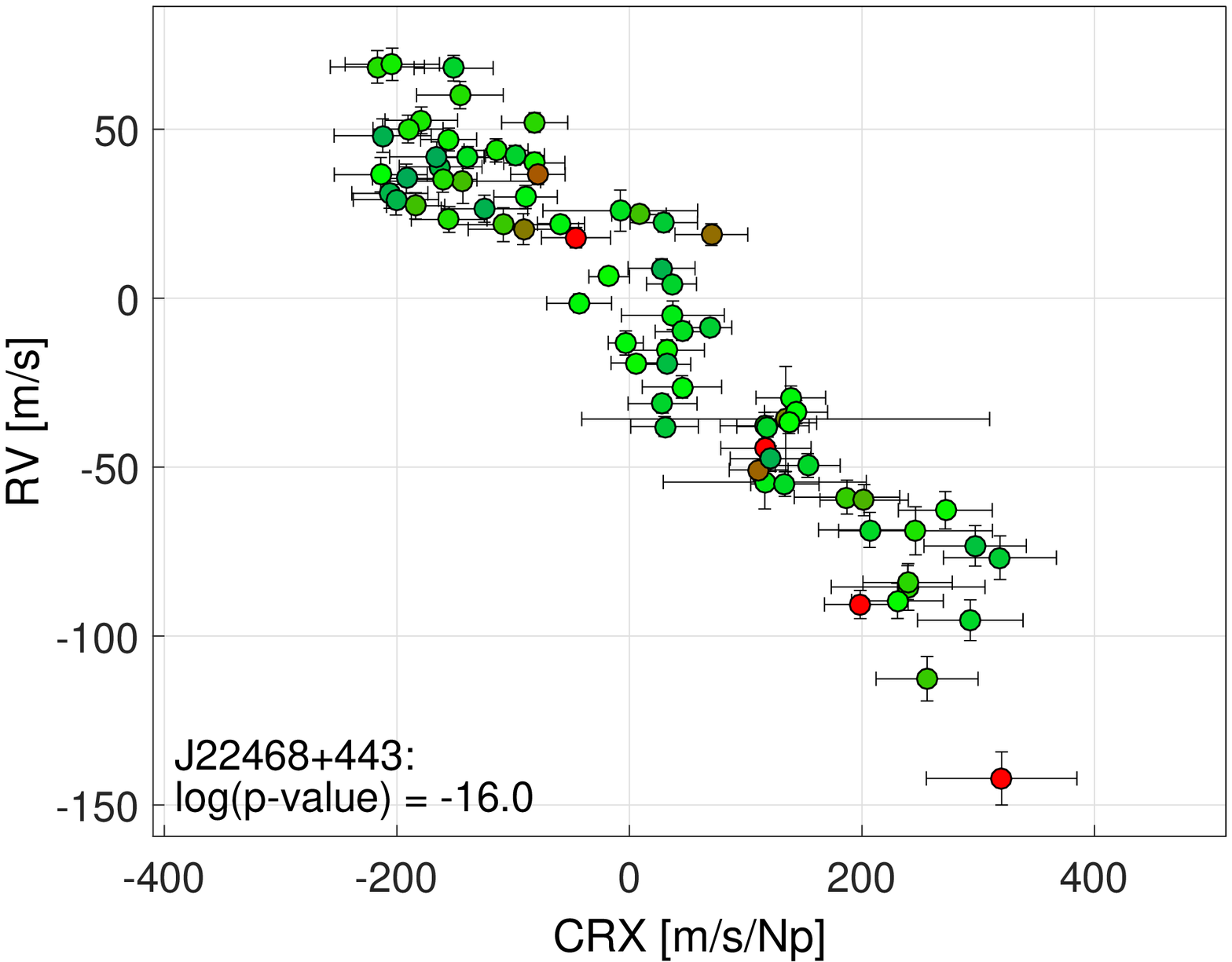}}
\endminipage\hfill
\minipage{0.33\textwidth}
{\includegraphics[width=\linewidth]{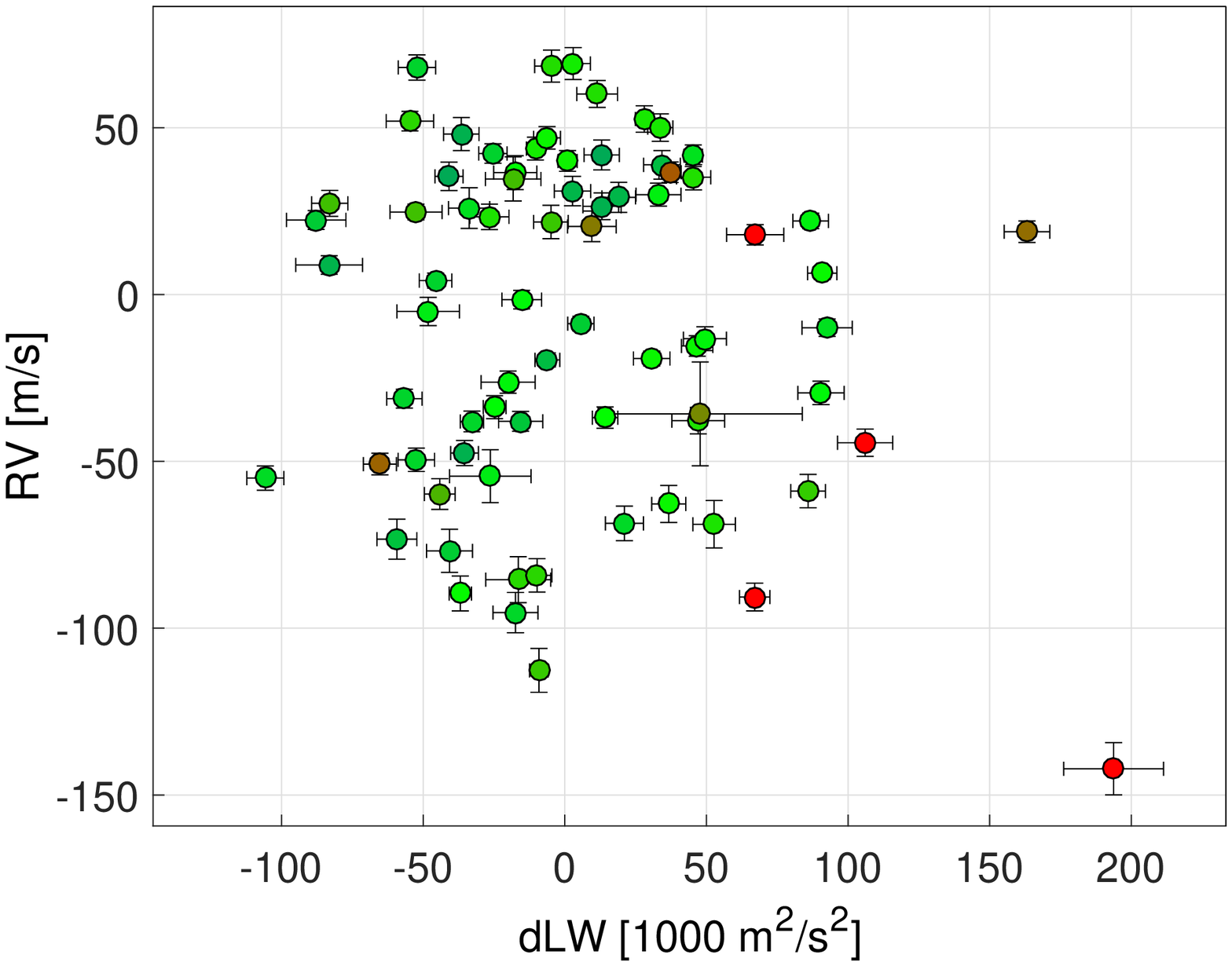}}
\endminipage\hfill
\minipage{0.33\textwidth}
{\includegraphics[width=\linewidth]{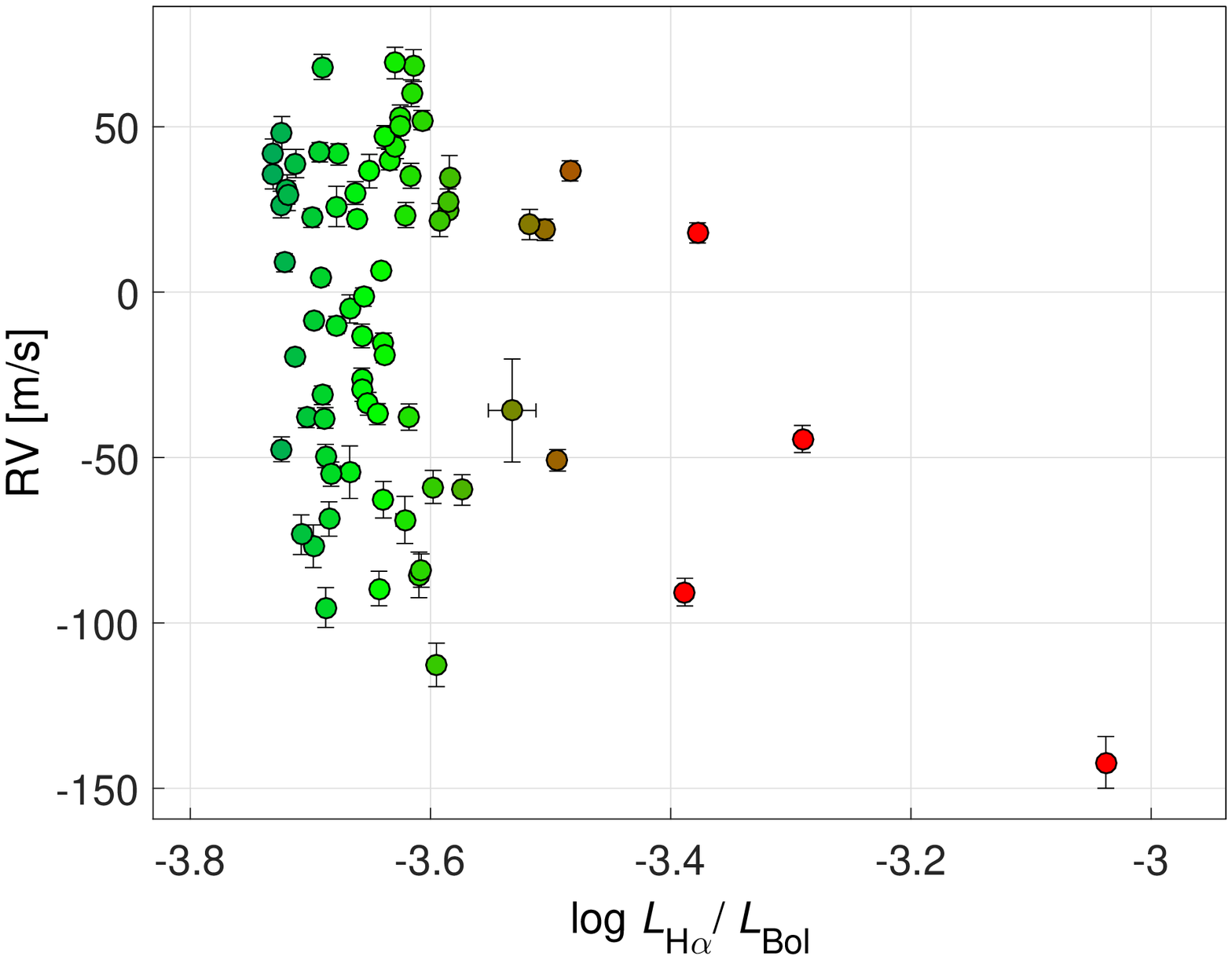}}
\endminipage
\caption{CARMENES VIS measurements of EV Lac. RV--CRX ({\it left}), RV--dLW ({\it middle}), and RV--$\log{L_{{\rm H}\alpha}/L_{\rm bol}}$ ({\it right}) scatter plots. Points are color-coded according to their $\log{L_{{\rm H}\alpha}/L_{\rm bol}}$ value, with H$\alpha$ enhancement of $>0.25$\,dex colored red. The CARMENES star id (Karmn) and the log(p-value) of the linear RV--CRX correlation are shown in the left panel. At least eight epochs of enhanced chromospheric activity can be seen via their higher H$\alpha$ emission, with the highest one belonging to a strong flare event. The outliers at the bottom of the left and middle panels belong to this event (see Fig. \ref{fig4} for its H$\alpha$ profile).
}
\label{fig5}
\end{figure*}

After searching for correlations between std(RV) and activity-related stellar parameters of the active RV-loud sample as a whole, we moved on to investigate individual stars of the sample for correlations between their stellar activity indicators and their measured RVs. Finding a parametric function that could translate measurable spectral activity indicators into RV deviations is a great challenge, but at the same time, it holds the promise to enable the modeling of RV time series with a combined model of planetary- and activity-induced signals. Moreover, such a function would obviate simultaneous auxiliary measurements to quantify activity-induced RVs, such as multi-band photometric observations. Unfortunately, no such a function has been proposed so far for M dwarfs. Lacking a physical model to guide our spectral data exploration, we searched our active RV-loud stars for linear correlations between their varying RVs and three of the spectral activity indicators that we have chosen to investigate in this paper: CRX, dLW, and $\log{L_{{\rm H}\alpha}/L_{\rm bol}}$.

To best illustrate the process, we show here a few representative plots for one of the most frequently observed stars in our sample: J22468+443 (EV Lac). In order to demonstrate the CRX definition, we show in the top panel of Fig. \ref{fig4} the order-by-order RVs from four representative observations of EV Lac: three observations with close-to-median H$\alpha$ emission but different CRX values, and one observation with extremely high H$\alpha$ emission. The latter observation was taken during a strong flare event \citep{Fuhrmeister2018arXiv}. The slopes of the best-fit lines to these RV--order scatter plots are related to the CRX values of these four observations \citep{Zechmeister2018}. The locations of activity-induced emission lines are marked at the top panel of Fig. \ref{fig4}. The bottom panel of Fig. \ref{fig4} shows the corresponding CARMENES spectra from these four observations, centered on the H$\alpha$ line.

The upper panel of Fig. \ref{fig4} shows that the RVs of EV Lac are strongly wavelength dependent. This dependence could be fit by a linear RV--$\log{\lambda}$ relation in virtually all observations, which means significant CRX values. The observation made during the strong flare event (black lines in Fig. \ref{fig4}) also showed a linear RV--$\log{\lambda}$ dependence, but its median RV was lower by $\sim50$\,m\,s$^{-1}$ from an observation with a similar CRX value, but a normal H$\alpha$ emission (blue lines in Fig. \ref{fig4}). Hence, the flare affects all order RVs in a similar fashion, introducing a similar deviation in all orders.

Figure \ref{fig5} shows the RV--CRX, RV--dLW, and RV--$\log{L_{{\rm H}\alpha}/L_{\rm bol}}$ scatter plots for EV Lac. Similar scatter plots for the other $30$ active RV-loud stars are presented in Fig. \ref{figA1}. As with the correlations between std(RV) and global stellar parameters, we again used the p(F$_{\rm test}$)
values of the best-fit lines to these scatter plots to estimate the RV--CRX, RV--dLW, and RV--$\log{L_{{\rm H}\alpha}/L_{\rm bol}}$ linear-correlation significances for all $31$ active RV-loud stars. The calculated p-values are listed in Table \ref{tab2}.

For EV Lac, there is a strong linear RV--CRX anticorrelation, while there are no RV--dLW or RV--$\log{L_{{\rm H}\alpha}/L_{\rm bol}}$ correlations. The RV--CRX anticorrelation, also referred to as a negative chromaticity \citep{Zechmeister2018}, indicates that the RV scatter decreases toward longer wavelength. This can also be seen in Fig. \ref{fig4}. This behavior is in agreement with the predictions of \citet{Reiners2010a} and \citet{Barnes2011} for a relatively low spot-to-photosphere temperature contrast.

\begin{figure}
\resizebox{\hsize}{!}
{\includegraphics{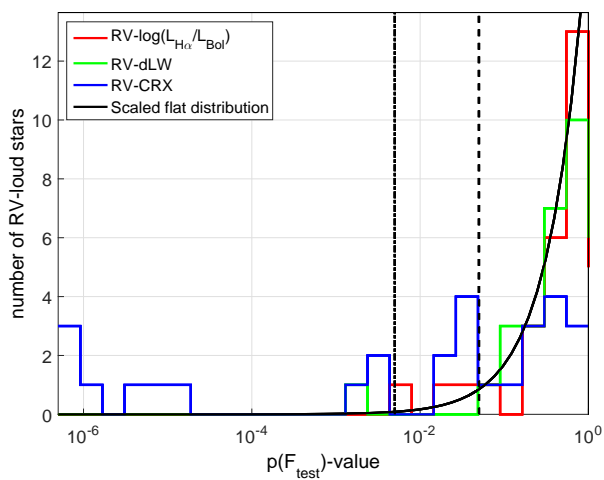}}
\caption{Histograms of p(F$_{\rm test}$) values of fitting a straight line to the RV--CRX, RV--dLW, and RV--$\log{L_{{\rm H}\alpha}/L_{\rm bol}}$ scatter plots for the stars in Table \ref{tab1}. The solid black curve represents a uniform distribution between $0$ and $1$ scaled so that its integral is equal the number of RV-loud stars. The dashed vertical line is the critical p-value for suggestive evidence ($0.05$), and the dash-dotted vertical line is the critical p-value for significant evidence ($0.005$). The histogram is truncated at $5\cdot10^{-7}$. Lower p-values are contained in the left-most column.
}
\label{fig6}
\end{figure}

Fig. \ref{fig6} shows the RV--CRX, RV--dLW, and RV--$\log{L_{{\rm H}\alpha}/L_{\rm bol}}$ linear-correlation p-value distributions on a log scale. The three distributions are compared to a uniform distribution between 0 and 1, which is the expected p-value distribution for the case of no correlation between these pairs of parameters \citep[e.g.,][]{Hung1997}. The deviation of the RV--CRX p-value distribution from a uniform one indicates that these two quantities are correlated for a large fraction of our active RV-loud stars. Using the criteria of \citet{Benjamin2018}, the RV--CRX correlation is significant for $10/31$ stars, and marginal for another $7/31$ stars. For comparison, the RV--dLW correlation is significant for only one star (namely LSR J1835+3259), and there is a marginal RV--$\log{L_{{\rm H}\alpha}/L_{\rm bol}}$ correlation for four stars only. Moreover, all $10$ stars with a significant RV--CRX correlation have a negative chromaticity. These findings are in agreement with the predictions of \citet{Reiners2010a} and \citet{Barnes2011}. The lack of RV chromaticity in about half of our active RV-loud stars can be explained by one (or more) of the following:
\begin{itemize}
\item complex spot patterns, with multiple spots whose chromatic contribution nearly cancels out;
\item higher spot-to-photosphere temperature contrasts;
\item strong magnetic fields associated with their spots, which counteract the flux-contrast RV chromaticity via the Zeeman splitting effect \citep{Reiners2013}.
\end{itemize}

The lack of linear RV--$\log{L_{{\rm H}\alpha}/L_{\rm bol}}$ or RV--dLW correlations does not exclude the possibility of more complicated relations. For instance, \citet{Bonfils2007} showed a `closed-loop' relation between RVs and H$\alpha$-index measurements of GJ\,$674$. In addition, \citet{Zechmeister2018} showed for YZ CMi that the position of each data point in the RV--dLW plane could depend on the rotational phase. Such interesting relations should probably be tested with a larger sample of active stars with known $P_{\rm rot}$. In addition, we should try reproducing such behavior with numeric simulations.

\section{Summary and future work}
\label{sec6}

During the first 20 months of the CARMENES survey, we observed all $324$ CARMENES sample stars \citep{Reiners2017a}. From their VIS spectra, we derived precise RVs for the $287$ stars that fulfilled the $n_{\rm RV}\geq5$ criterion \citep{Zechmeister2018}, with a median $\overline{\delta{\rm RV}}$ of $\sim1.7$\,m\,s$^{-1}$. Similarly to previous PRV surveys of M dwarfs \citep[e.g.,][]{Bonfils2013}, we found the std(RV) distribution of the observed stars to peak at $3$--$4$\,m\,s$^{-1}$. Currently, one of the main challenges of the CARMENES consortium is to explain the excess RV scatter for each star with a well-established model comprised of orbital motion and stellar activity.

Among the $287$ stars with $n_{\rm RV}\geq5$, we detected $67$ stars with std(RV) of $>$\,$10$\,m\,s$^{-1}$. Of these, we selected for the active RV-loud sample $31$ stars that have $n_{\rm RV}>10$, $v \sin{i}>2$\,km\,s$^{-1}$ and no known or suspected low-mass companions in orbit around them. This sample downsizing was done to ensure that the RV scatter of the stars in the sample is dominated by activity and not by orbital motion. Nevertheless, it is the largest and most comprehensive sample of active RV-loud M dwarfs investigated to date with a PRV instrument. We found the fraction of active RV-loud stars to increase with spectral type: from $\sim3\%$ for early-type M dwarfs through $\sim30\%$ for mid-type M dwarfs to $>50\%$ for late-type M dwarfs. This finding is consistent with the increase in the fraction of fast-rotating and active M dwarfs with spectral type \citep[e.g.,][]{Reiners2012,Jeffers2018arXiv}.

While investigating the relations between std(RV) and other stellar parameters of the active RV-loud stars, we found significant evidence for a positive std(RV)--$v \sin{i}$ correlation in all M-dwarf spectral types. Suggestive evidence for positive std(RV)--$\log{L_{{\rm H}\alpha}/L_{\rm bol}}$ and std(RV)--SpT correlations was found only for the mid-type M dwarfs. In addition, we have shown that the spectral line broadening caused by the rapid rotation of these stars is not sufficient to explain the observed RV scatter, and that the excess RV scatter is most likely caused by the additional photospheric activity that accompanies the rapid rotation. Our findings are in agreement with predictions from numeric simulations \citep[e.g.,][]{Reiners2010a,Barnes2011}.

Furthermore, we searched the active RV-loud sample for correlations between each star's RVs and three of its activity indicators: CRX, dLW, and $\log{L_{{\rm H}\alpha}/L_{\rm bol}}$. We found a significant RV--CRX correlation for $\sim30\%$ of the stars, and a marginal correlation for another $\sim20\%$ of them. For most of these stars we found a negative chromaticity, which means that the RV scatter decreases with wavelength. Comparing our findings with the predictions made by \citet{Reiners2010a}, \citet{Barnes2011}, and \citet{Reiners2013}, we see that such behavior can most likely be created by active regions on the surface of the rotating stars. The lack of RV chromaticity for about half of our active RV-loud sample can be explained by complex spot patterns, higher spot-to-photosphere temperature contrasts, or strong magnetic fields associated with their spots.

We found no significant linear RV--dLW or RV--$\log{L_{{\rm H}\alpha}/L_{\rm bol}}$ correlations for the vast majority of the active RV-loud sample. However, we did not exclude the possibility of more complicated relations between these two pairs of observables, such as the `closed-loop' relations that were found by \citet{Bonfils2007} and \citet{Zechmeister2018}, for example.

Periodogram analyses could reveal common periods for the different activity indicators and RVs. For instance, \citet{Suarez2017} used such period matching to attribute to activity $1$--$10$\,m\,s$^{-1}$ signals of $18$ M dwarfs. Moreover, matching the periods with photometrically measured rotation periods (or their harmonics) can help distinguishing between planetary- and activity-induced RV signals. However, for an accurate derivation of orbital parameters from RV time series of active stars, period discrimination is not enough. For this, one would have to model the activity-induced signals together with the orbital solution. Revealing the exact relations between spectral activity indicators and RVs would greatly help in such modeling. Deepening the understanding of such relations is on the agenda of the CARMENES consortium.

To this end, we do not have a direct measure of the spot-coverage fraction, or the average spot-to-photosphere temperature contrast -- two important parameters for understanding activity-induced RVs of M dwarfs \citep[e.g.,][]{Barnes2011}. We are working on identifying spectral indicators that will enable estimating these two quantities \citep[e.g.,][]{Herrero2016}. For example, finding a correlation between the different CCF moments and RV deviations of active stars would be an important step toward modeling activity-induced RVs.

In the coming few years, several novel instruments promise to push down RV-measurement precision towards $0.1$--$0.3$\,m\,s$^{-1}$ \citep[e.g.,][]{ESPRESSO,Halverson2016}. Such precision would mean that activity-induced RV variations are expected to be larger than the RV uncertainties for most of the M dwarfs, most of the time. Under these circumstances, extending our picture of planetary occurrence rates to lower-mass planets and lower-mass stars will require a paradigm shift, from treating activity-induced RV variations as an additional noise source \citep[e.g.,][]{Wright2005} to modeling RV time series with both orbital- and activity-induced signals \citep[e.g.,][]{Hebrard2016,Barnes2017a}. Farther in the future, one could even consider forward modeling the spectra themselves with the same ingredients.

\begin{acknowledgements}
We thank the anonymous referee for their helpful comments and suggestions. We also thank the dedicated CAHA observers and technical staff for ensuring that the CARMENES measurements over the last two years were of high quality. CARMENES is an instrument for the Centro Astron\'omico Hispano-Alem\'an de Calar Alto (CAHA, Almer\'{\i}a, Spain).   CARMENES is funded by the German Max-Planck-Gesellschaft (MPG), the Spanish Consejo Superior de Investigaciones Cient\'{\i}ficas (CSIC), the European Union through FEDER/ERF FICTS-2011-02 funds, and the members of the CARMENES Consortium (Max-Planck-Institut f\"ur Astronomie, Instituto de Astrof\'{\i}sica de Andaluc\'{\i}a, Landessternwarte K\"onigstuhl, Institut de Ci\`encies de l'Espai, Insitut f\"ur Astrophysik G\"ottingen, Universidad Complutense de Madrid, Th\"uringer Landessternwarte Tautenburg, Instituto de Astrof\'{\i}sica de Canarias, Hamburger Sternwarte, Centro de Astrobiolog\'{\i}a and Centro Astron\'omico Hispano-Alem\'an), with additional contributions by the Spanish Ministry of Economy, the German Science Foundation through the Major Research Instrumentation Programme and DFG Research Unit FOR2544 ``Blue Planets around Red Stars'', the Klaus Tschira Stiftung, the states of Baden-W\"urttemberg and Niedersachsen, and by the Junta de Andaluc\'{\i}a. L.T. also thanks Tsevi Mazeh and Shay Zucker from Tel-Aviv University for their useful comments on this work.
\end{acknowledgements}

\bibliographystyle{aa}
\bibliography{LevTalOr.bib}

\begin{appendix}
\section{Figures}
\label{app1}

\begin{figure*}
\minipage{0.33\textwidth}
{\includegraphics[width=\linewidth]{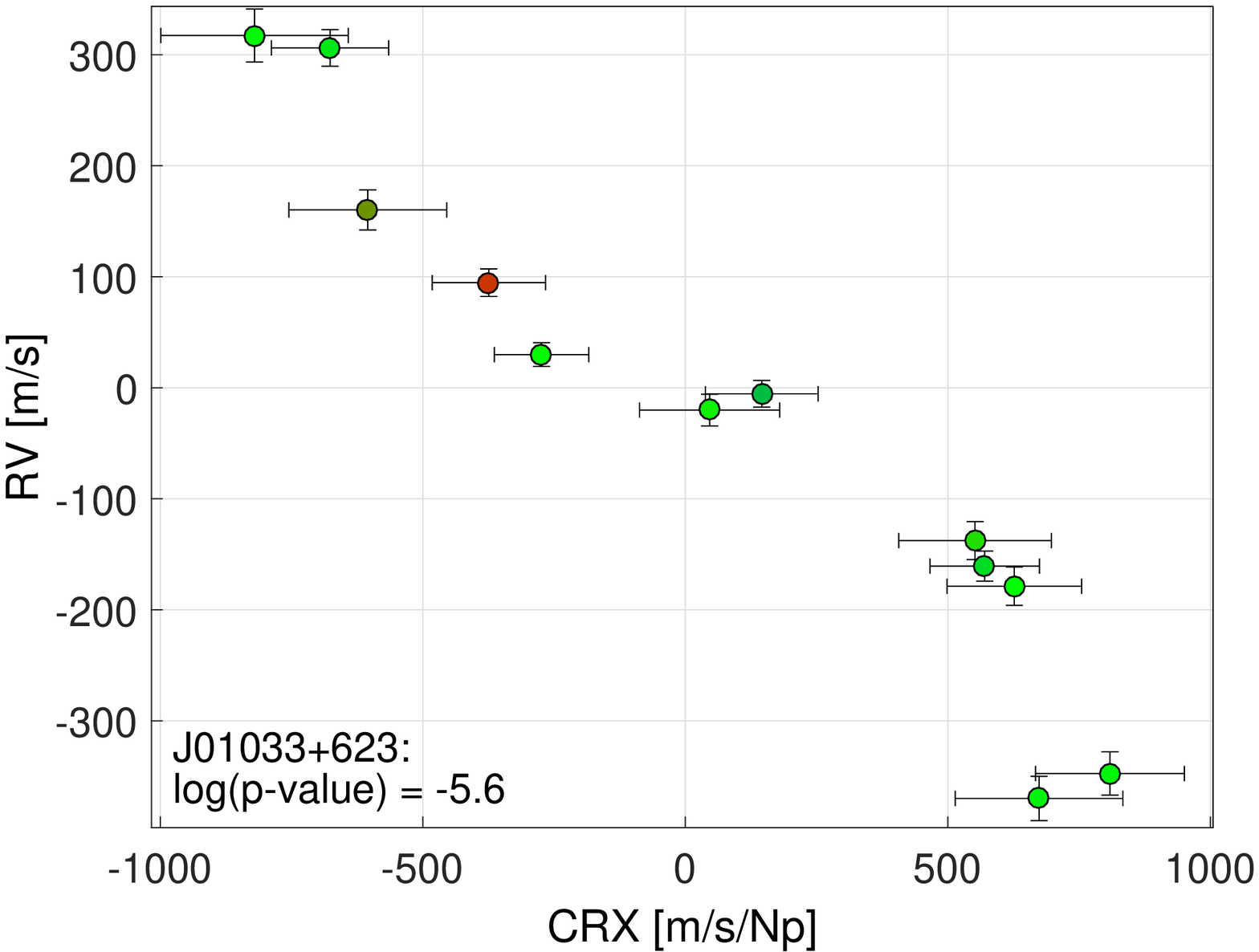}}
\endminipage\hfill
\minipage{0.33\textwidth}
{\includegraphics[width=\linewidth]{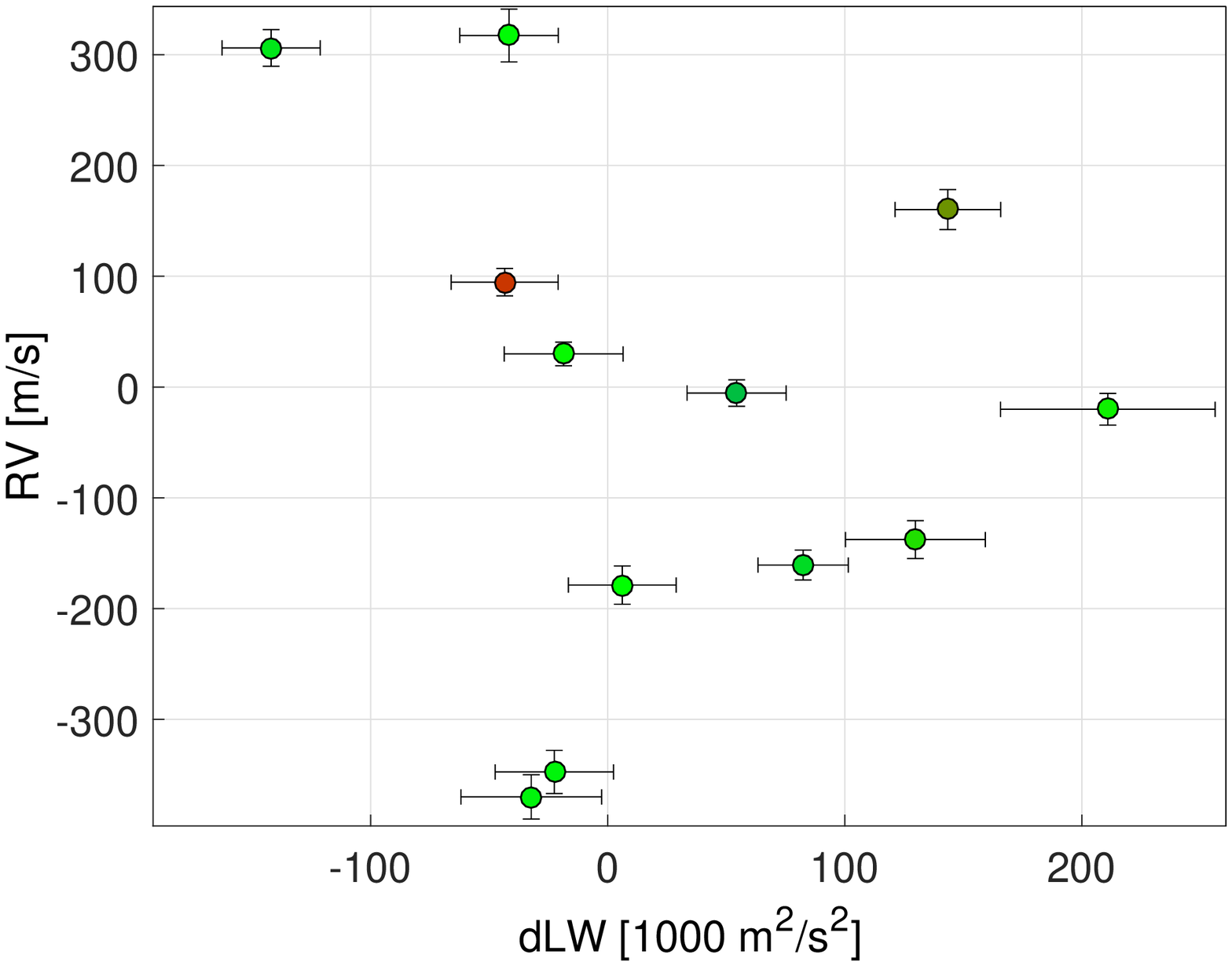}}
\endminipage\hfill
\minipage{0.33\textwidth}
{\includegraphics[width=\linewidth]{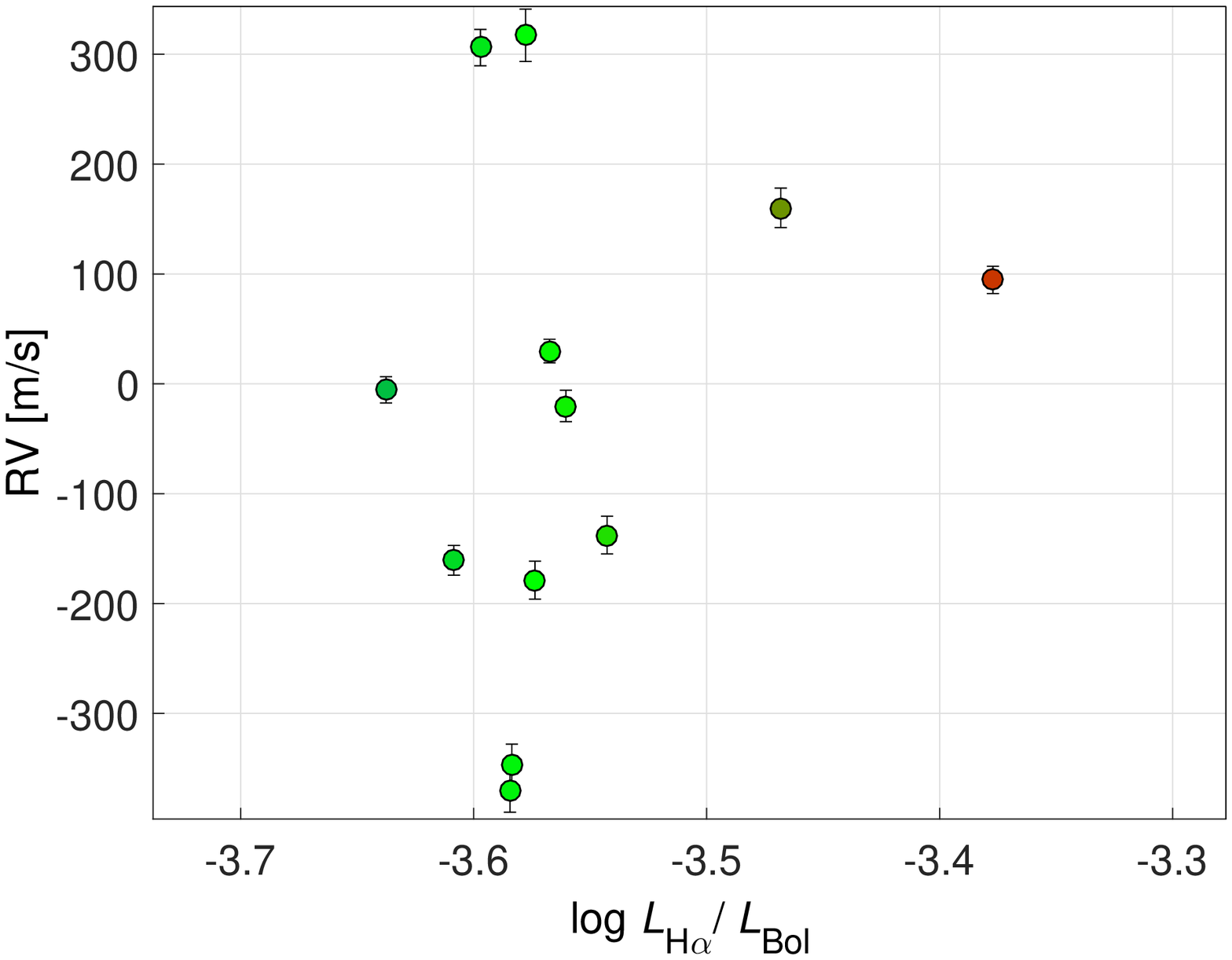}}
\endminipage


\minipage{0.33\textwidth}
{\includegraphics[width=\linewidth]{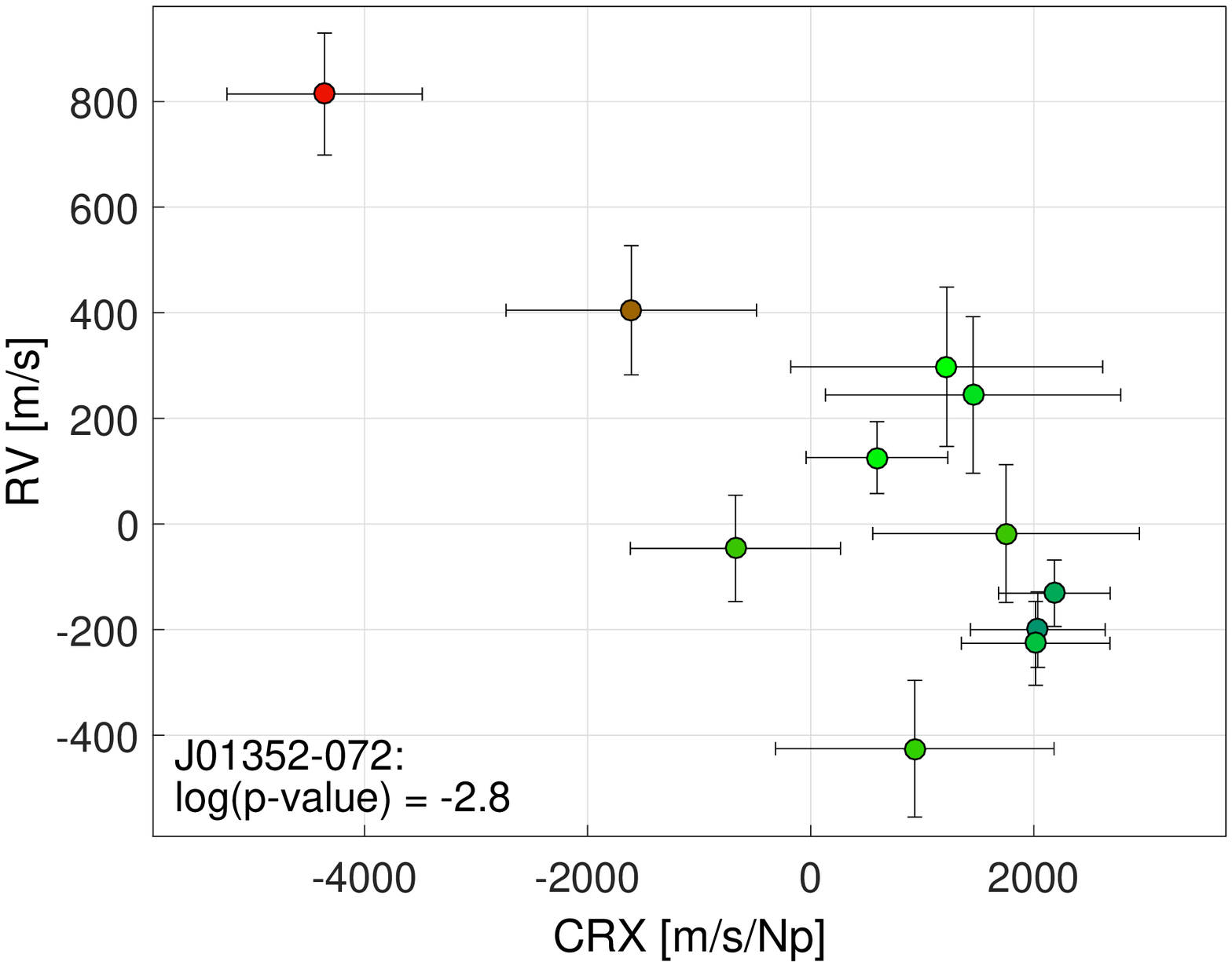}}
\endminipage\hfill
\minipage{0.33\textwidth}
{\includegraphics[width=\linewidth]{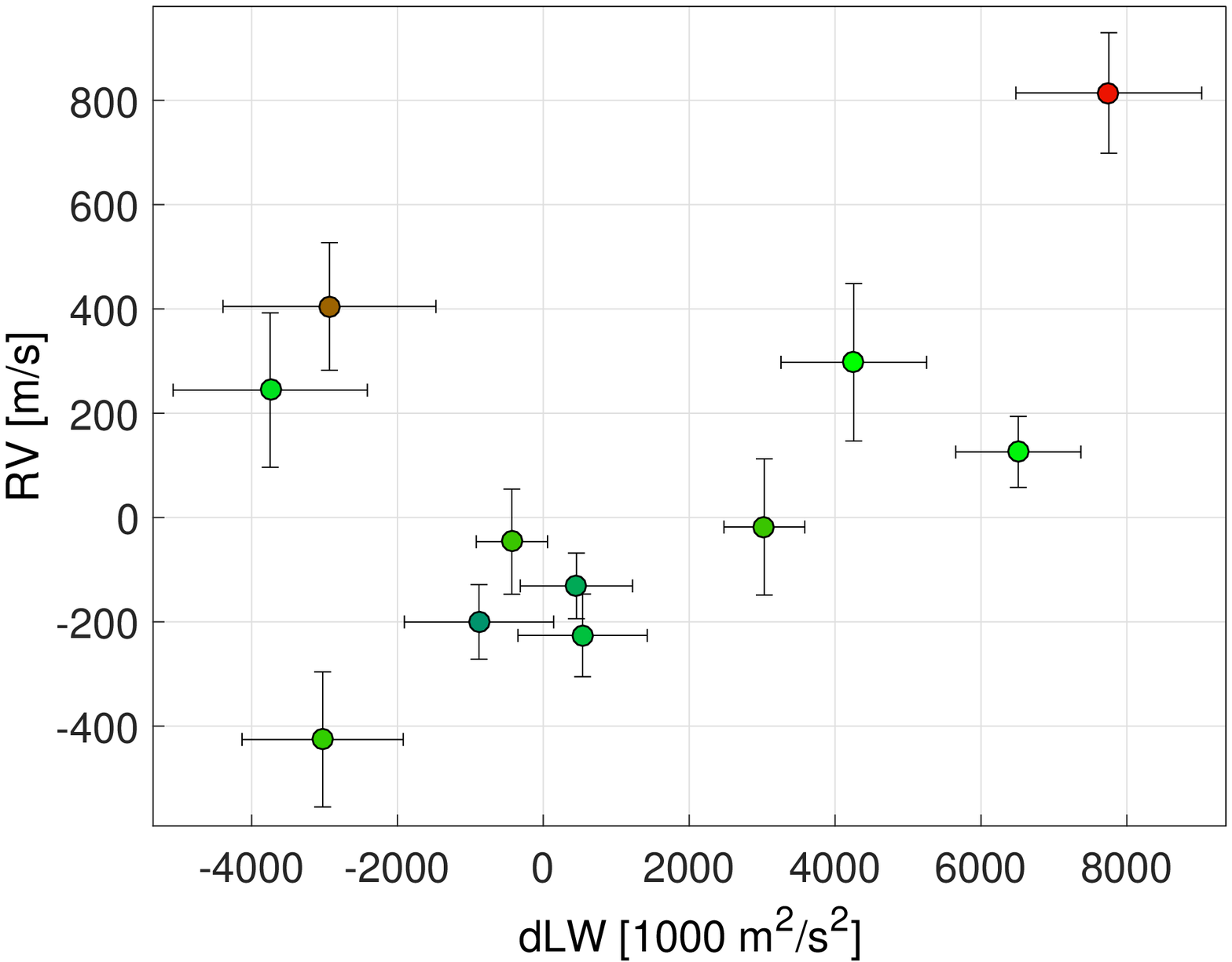}}
\endminipage\hfill
\minipage{0.33\textwidth}
{\includegraphics[width=\linewidth]{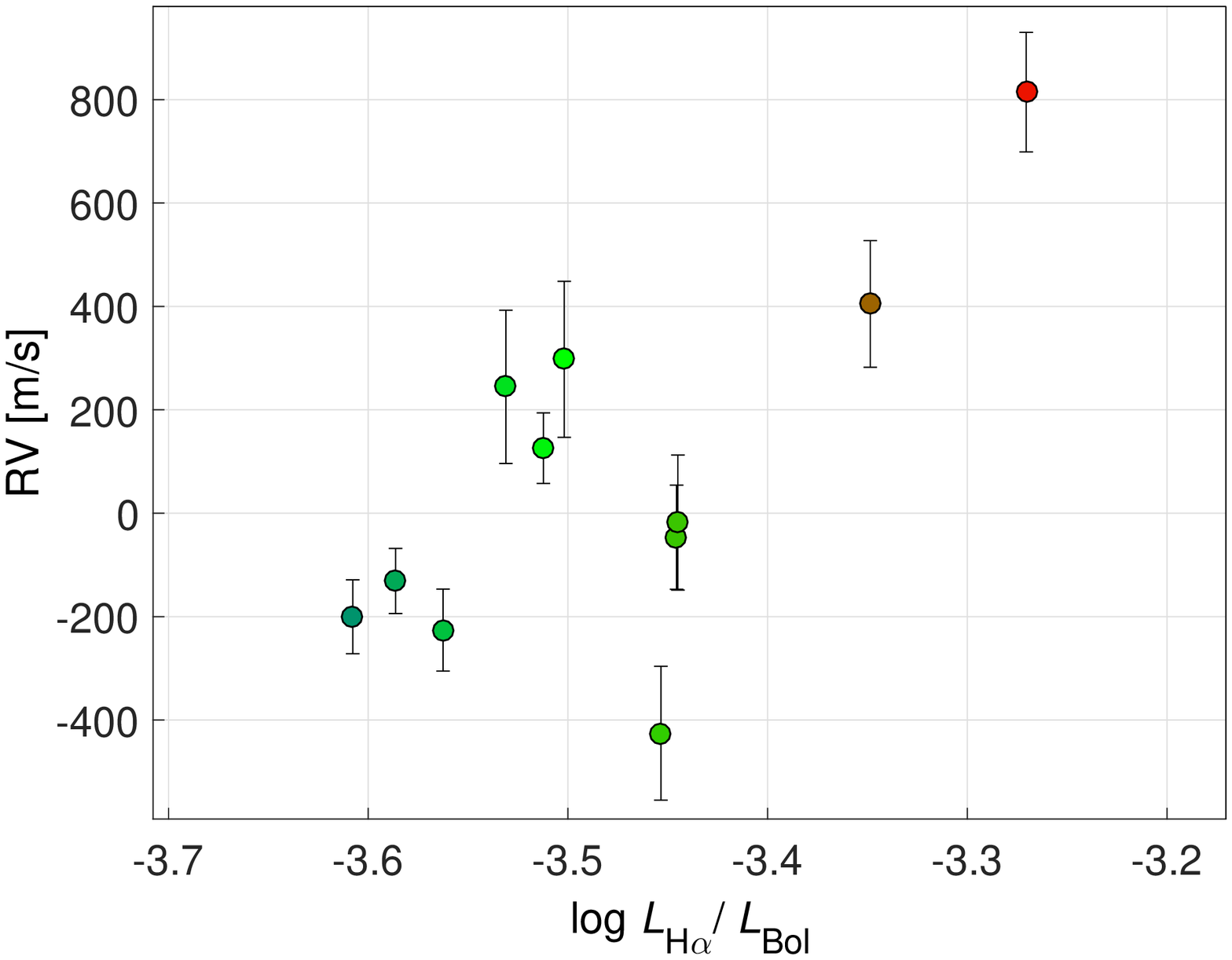}}
\endminipage


\minipage{0.33\textwidth}
{\includegraphics[width=\linewidth]{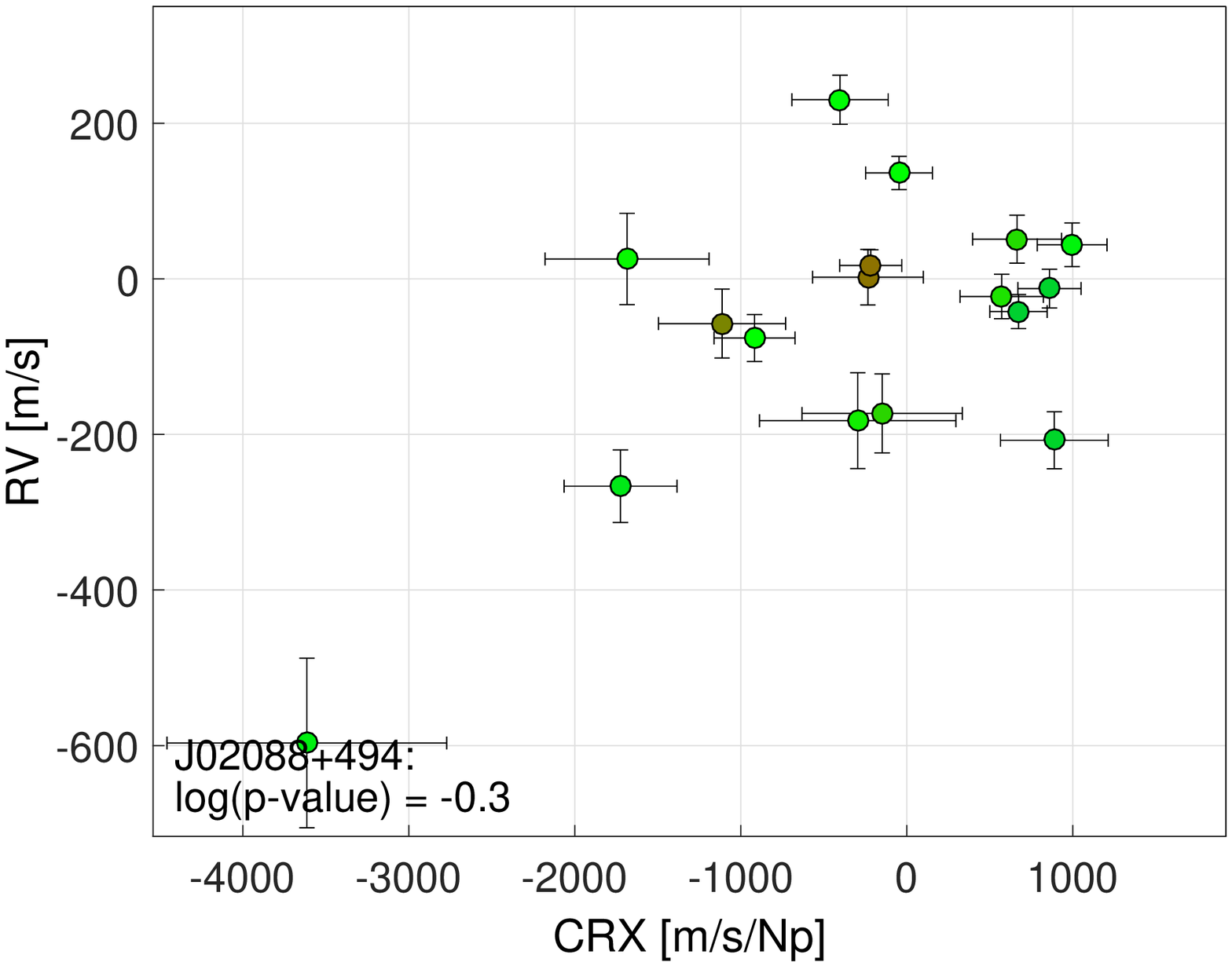}}
\endminipage\hfill
\minipage{0.33\textwidth}
{\includegraphics[width=\linewidth]{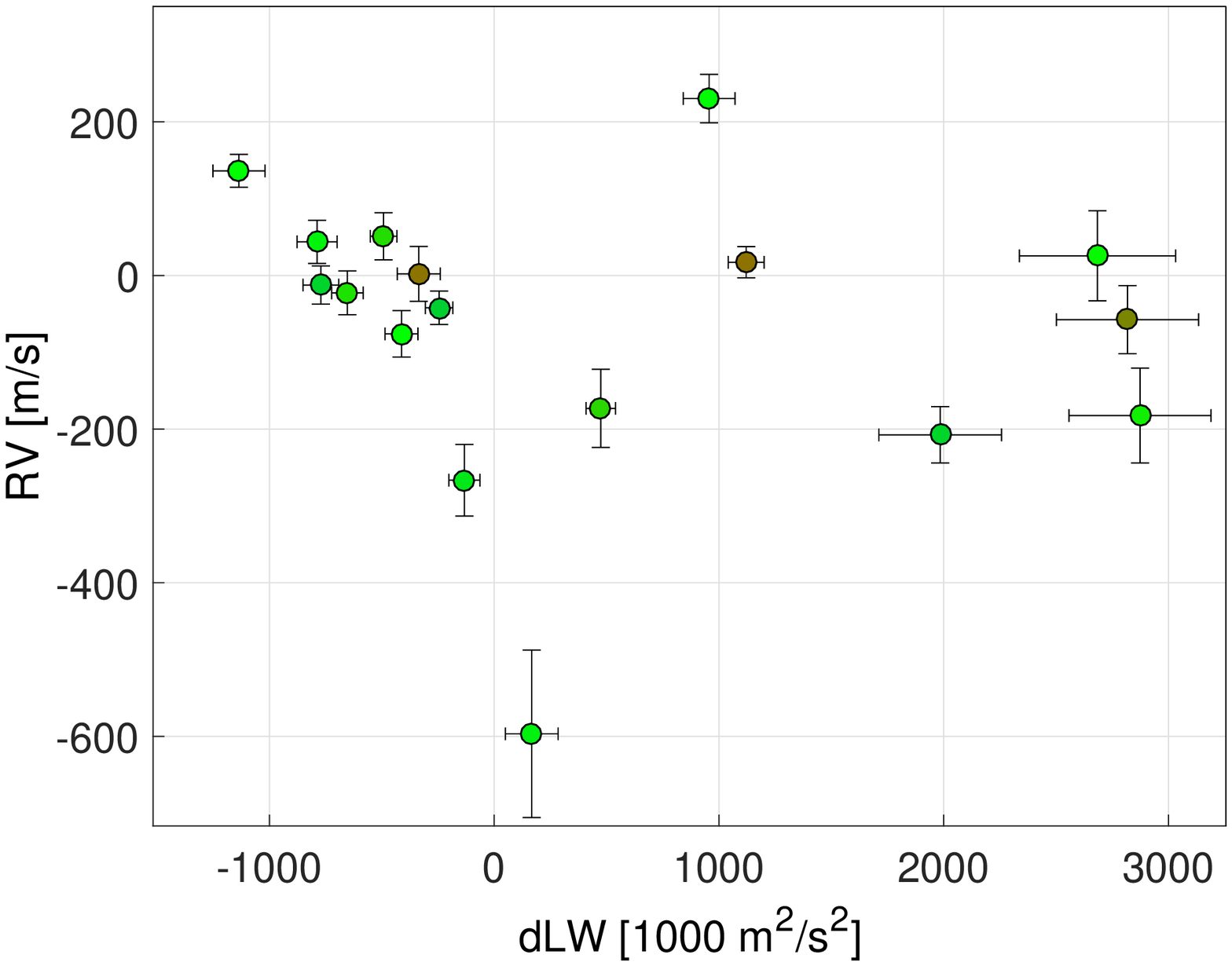}}
\endminipage\hfill
\minipage{0.33\textwidth}
{\includegraphics[width=\linewidth]{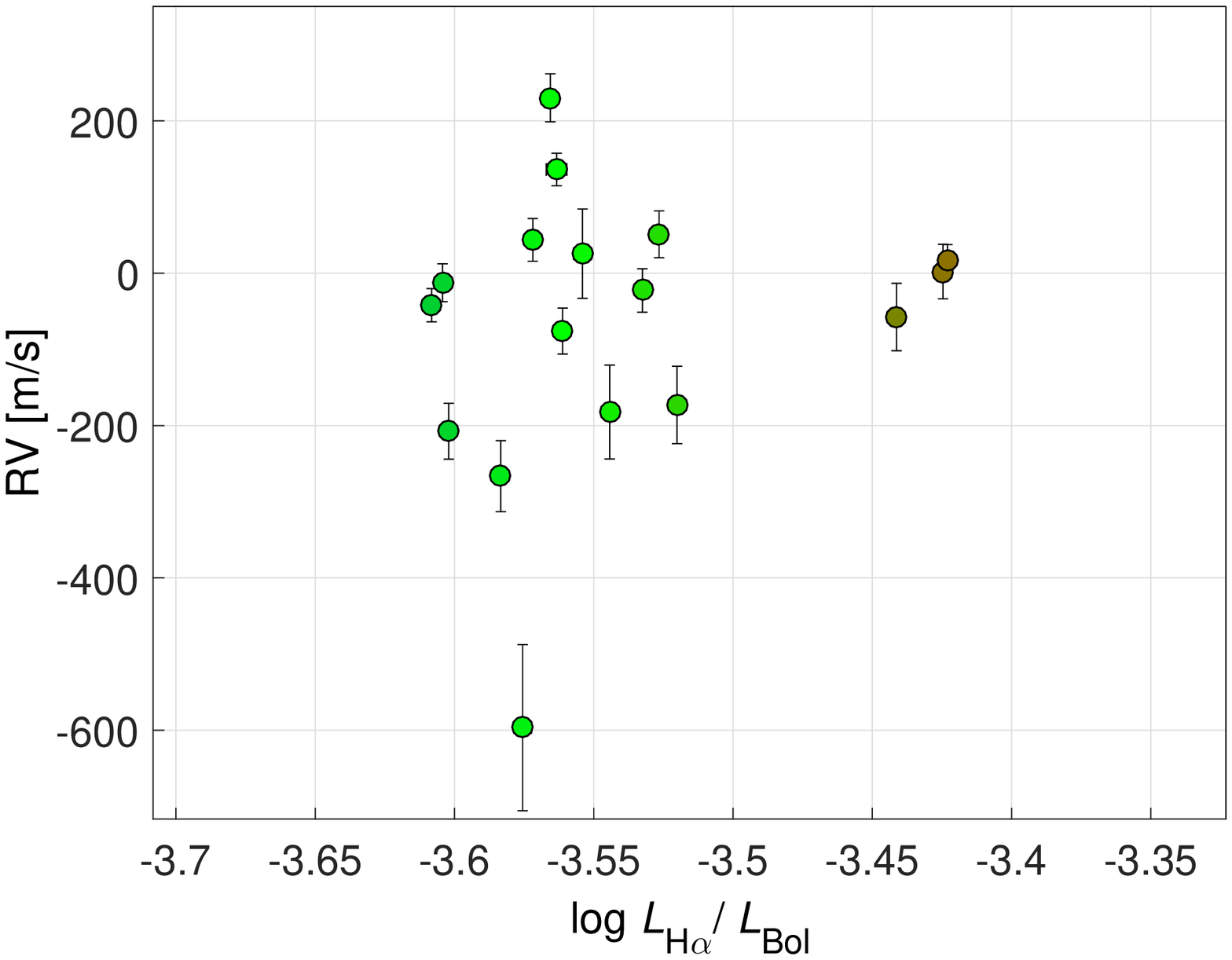}}
\endminipage


\minipage{0.33\textwidth}
{\includegraphics[width=\linewidth]{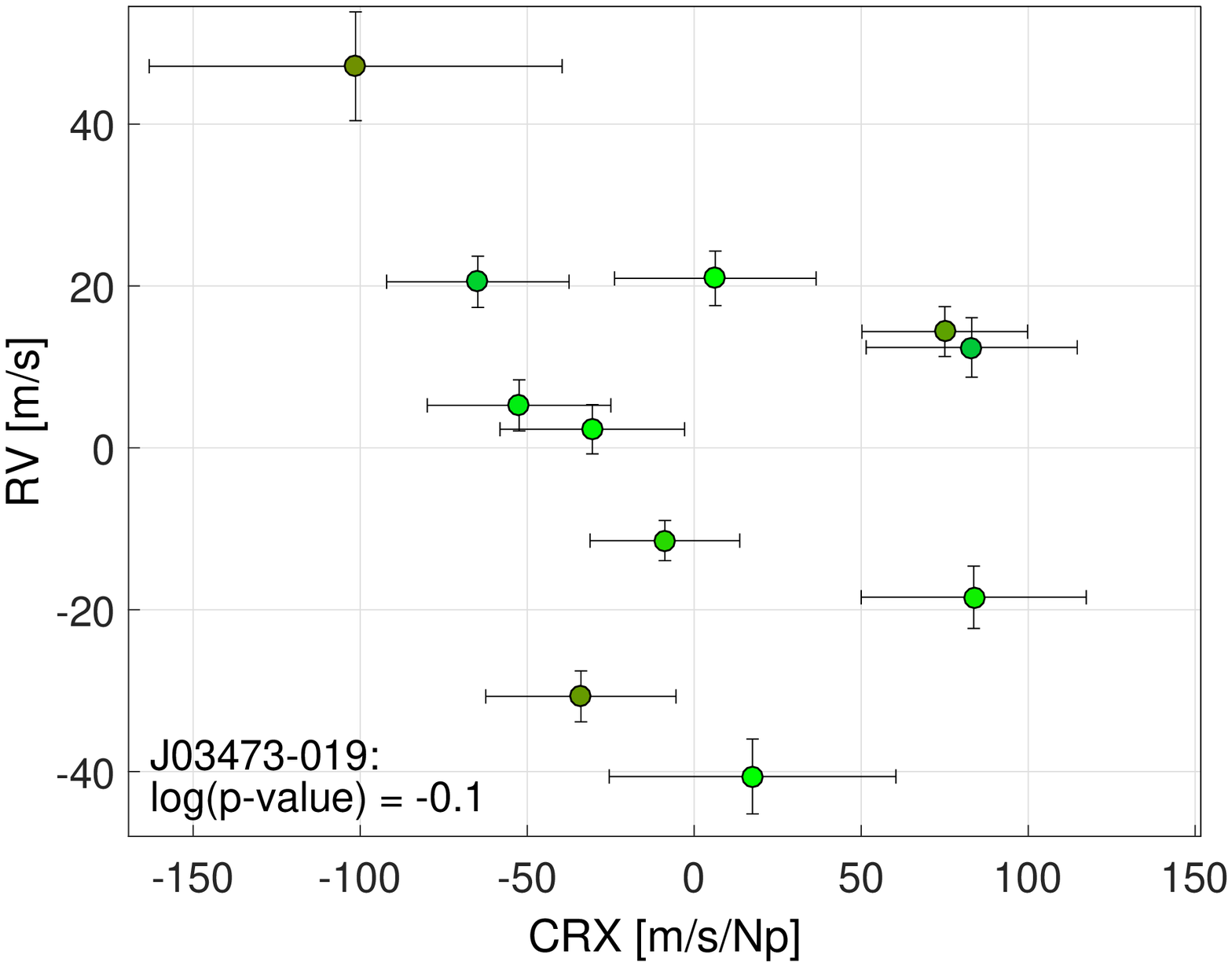}}
\endminipage\hfill
\minipage{0.33\textwidth}
{\includegraphics[width=\linewidth]{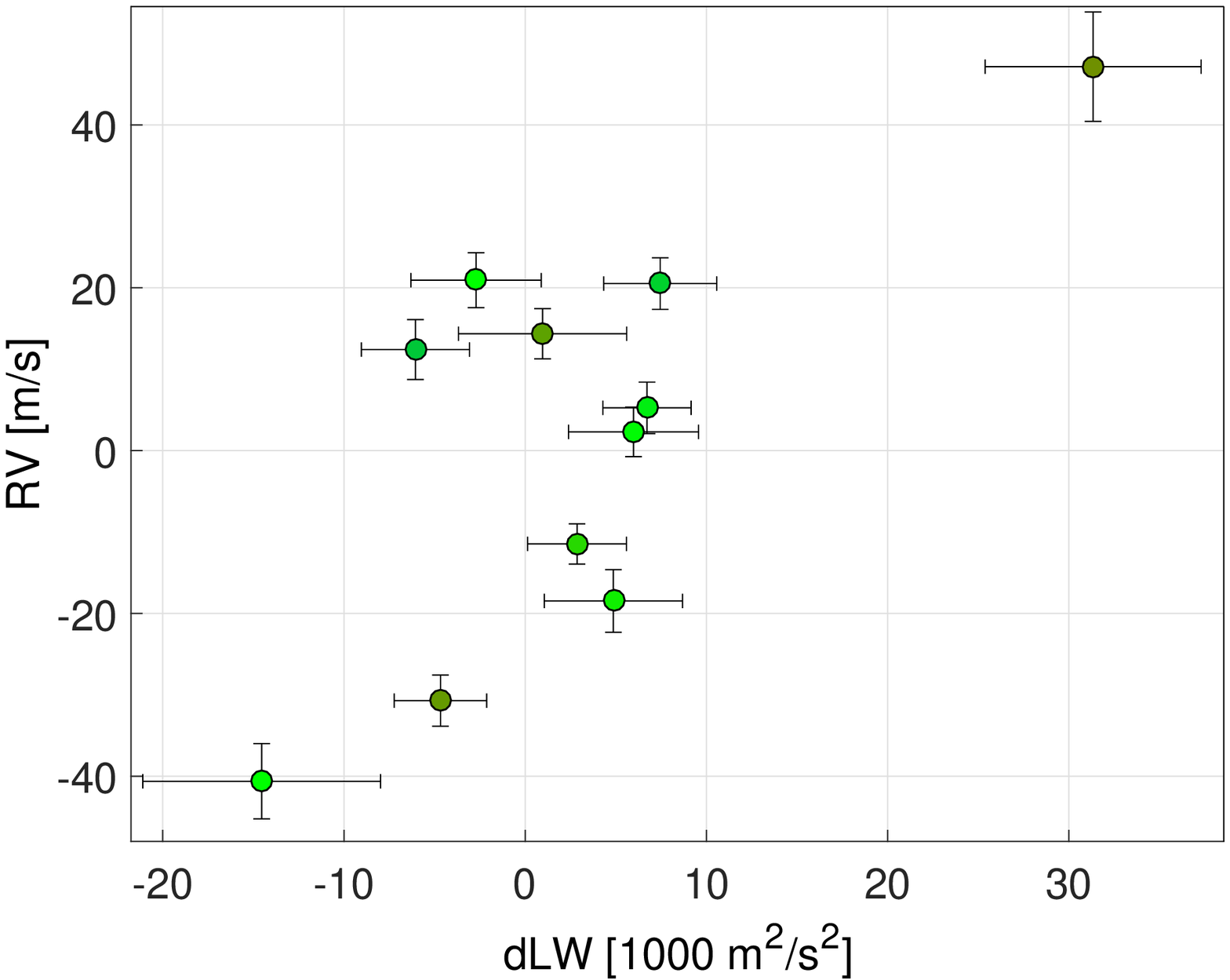}}
\endminipage\hfill
\minipage{0.33\textwidth}
{\includegraphics[width=\linewidth]{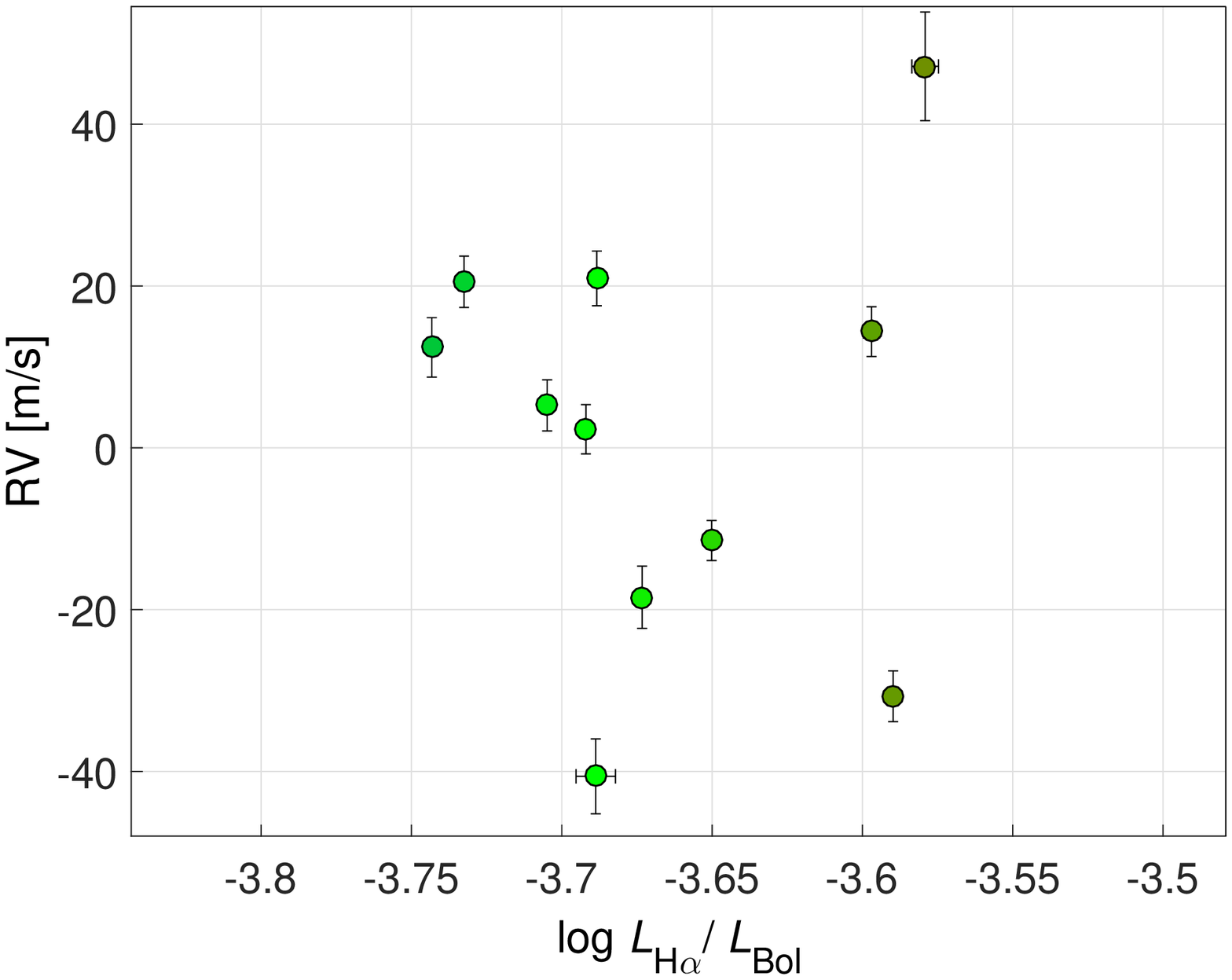}}
\endminipage


\minipage{0.33\textwidth}
{\includegraphics[width=\linewidth]{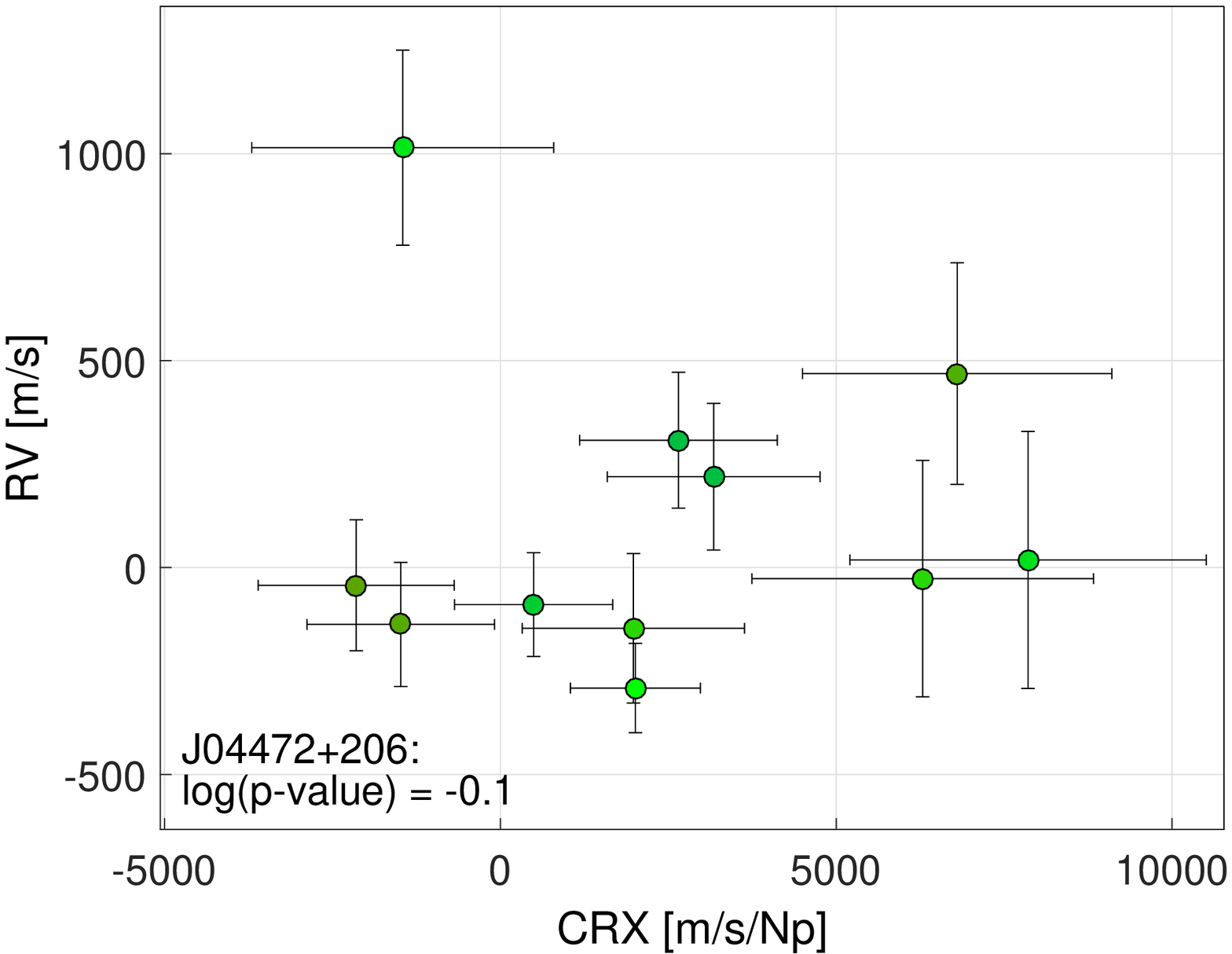}}
\endminipage\hfill
\minipage{0.33\textwidth}
{\includegraphics[width=\linewidth]{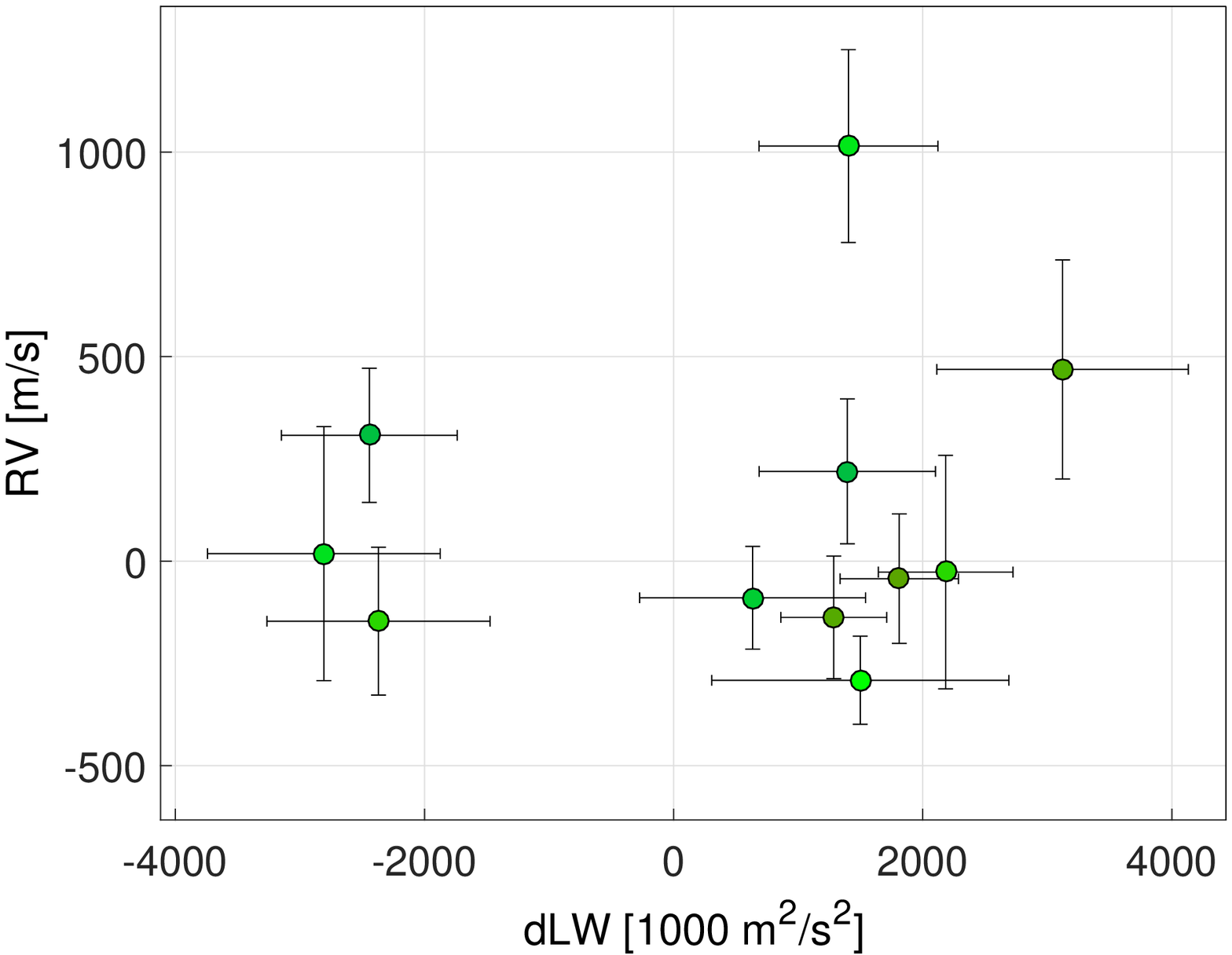}}
\endminipage\hfill
\minipage{0.33\textwidth}
{\includegraphics[width=\linewidth]{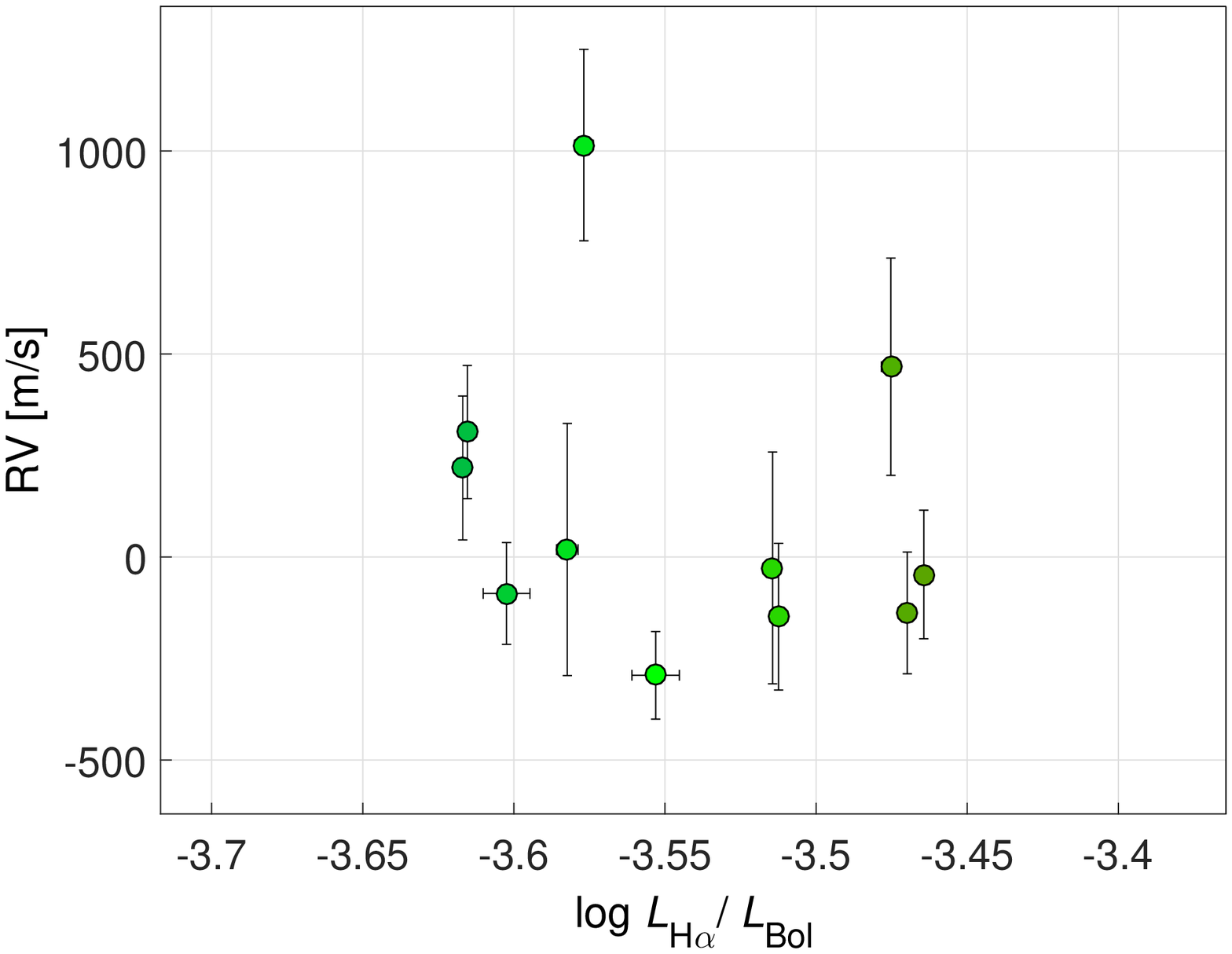}}
\endminipage
\caption{Similar plots to Fig. \ref{fig5} -- one plot per RV-loud star.}
\label{figA1}
\end{figure*}

\addtocounter{figure}{-1}

\begin{figure*}[!htp]
\minipage{0.33\textwidth}
{\includegraphics[width=\linewidth]{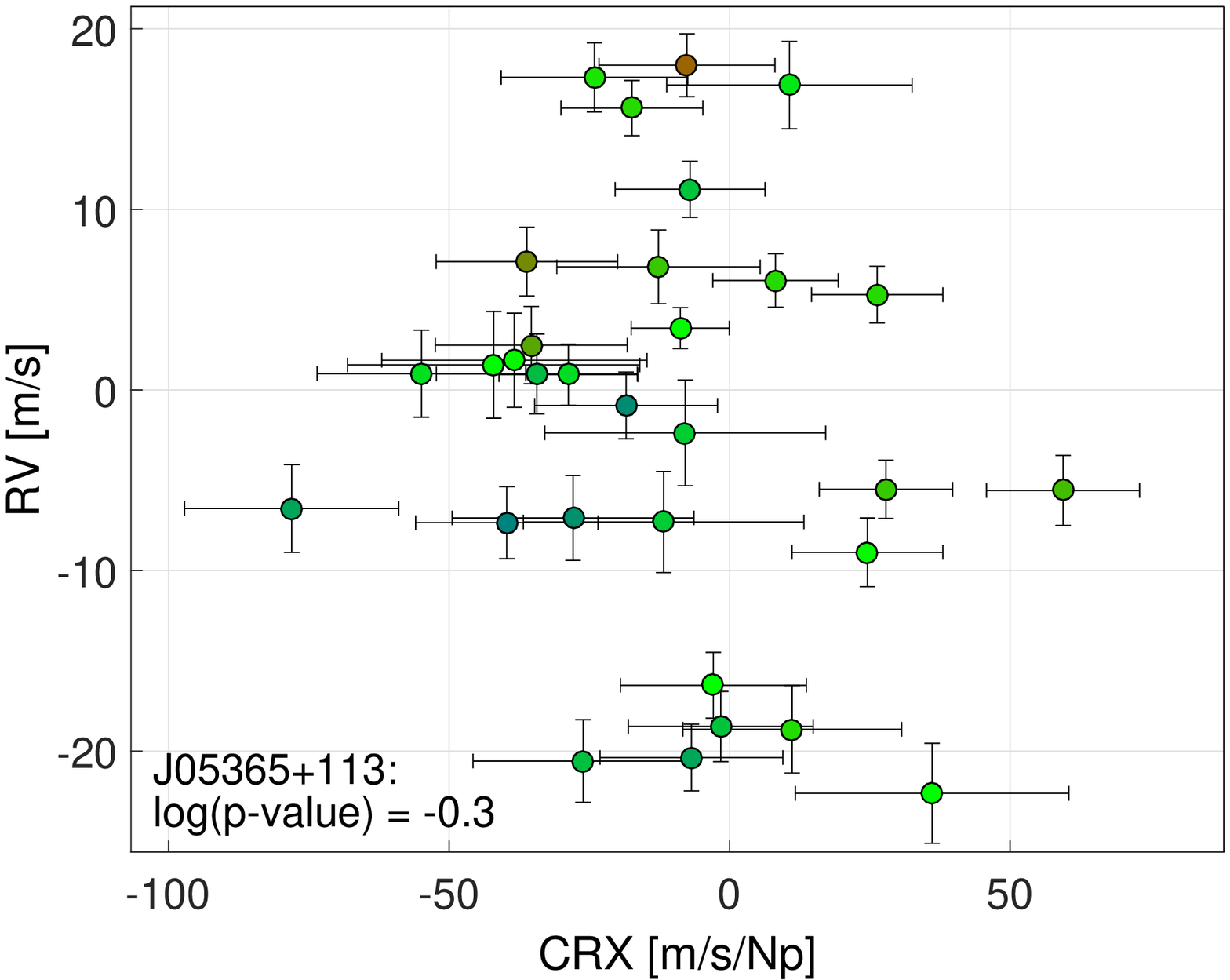}}
\endminipage\hfill
\minipage{0.33\textwidth}
{\includegraphics[width=\linewidth]{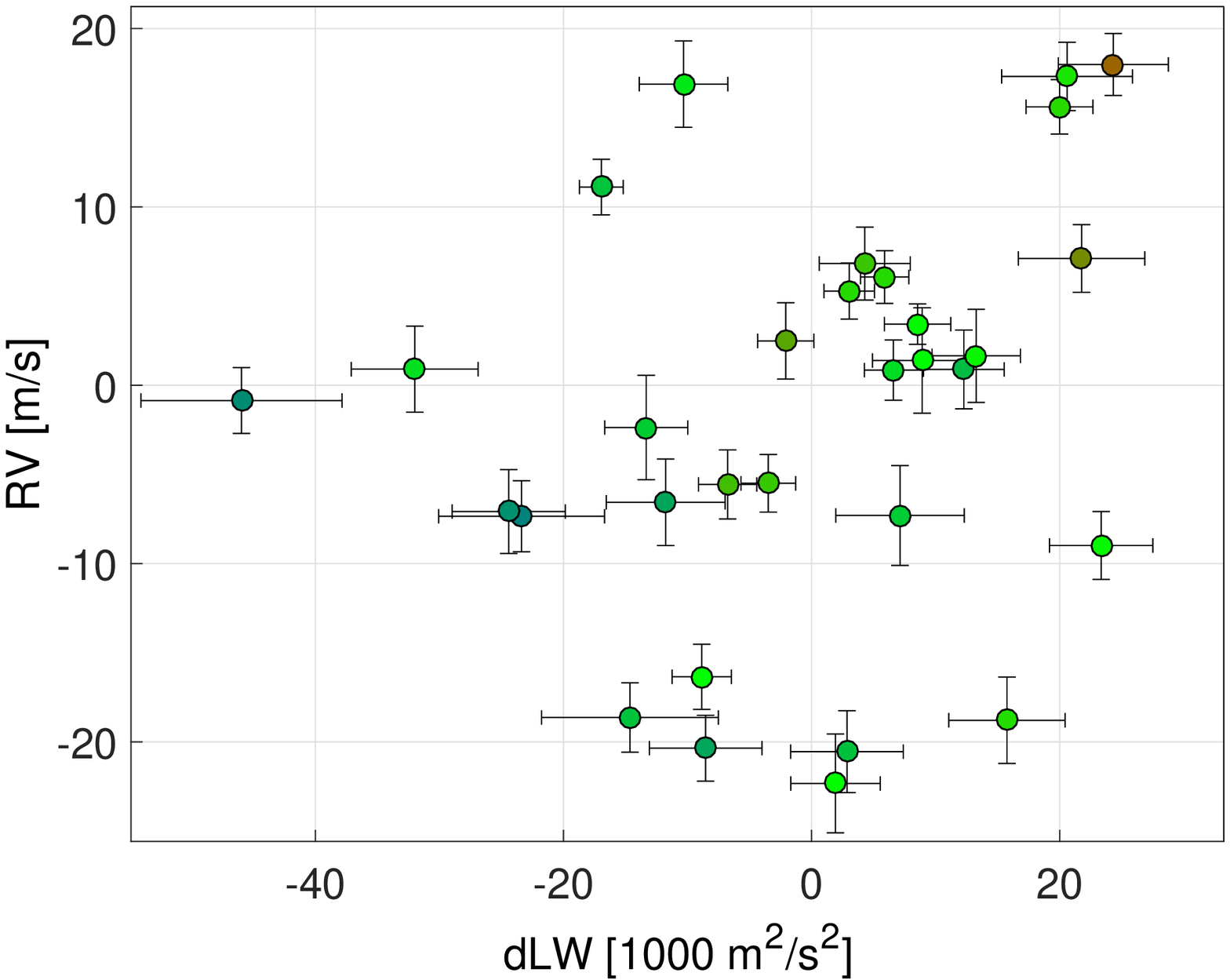}}
\endminipage\hfill
\minipage{0.33\textwidth}
{\includegraphics[width=\linewidth]{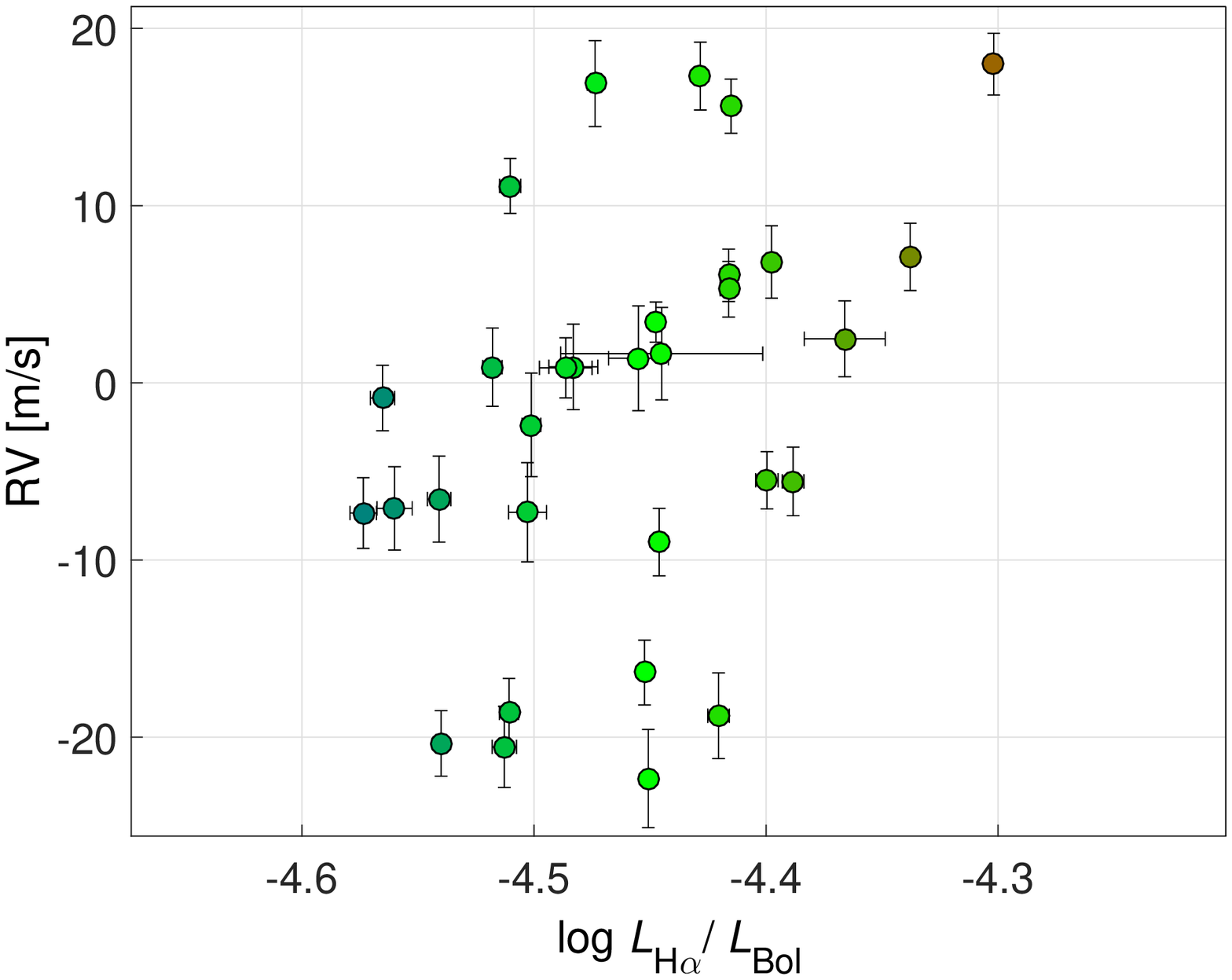}}
\endminipage


\minipage{0.33\textwidth}
{\includegraphics[width=\linewidth]{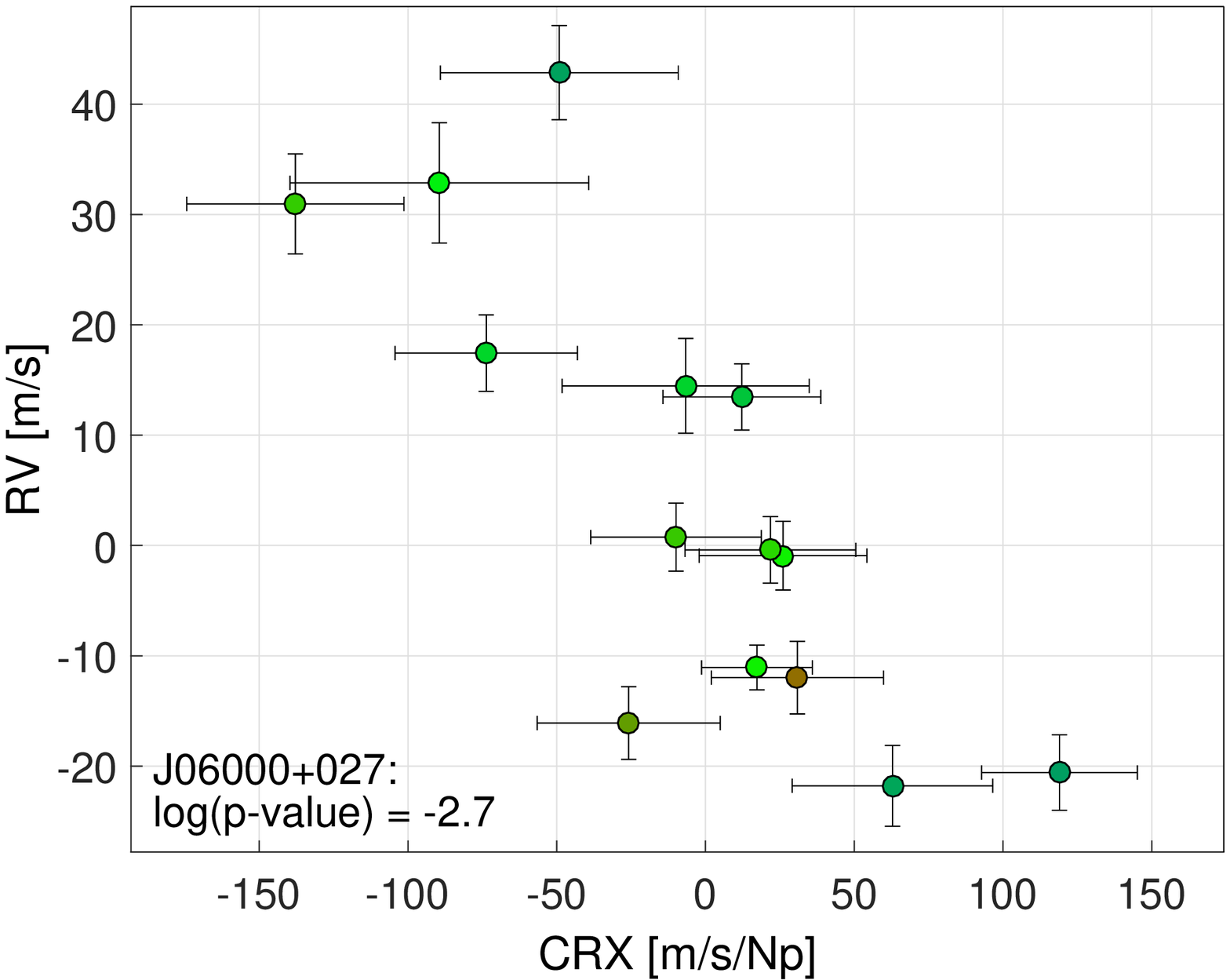}}
\endminipage\hfill
\minipage{0.33\textwidth}
{\includegraphics[width=\linewidth]{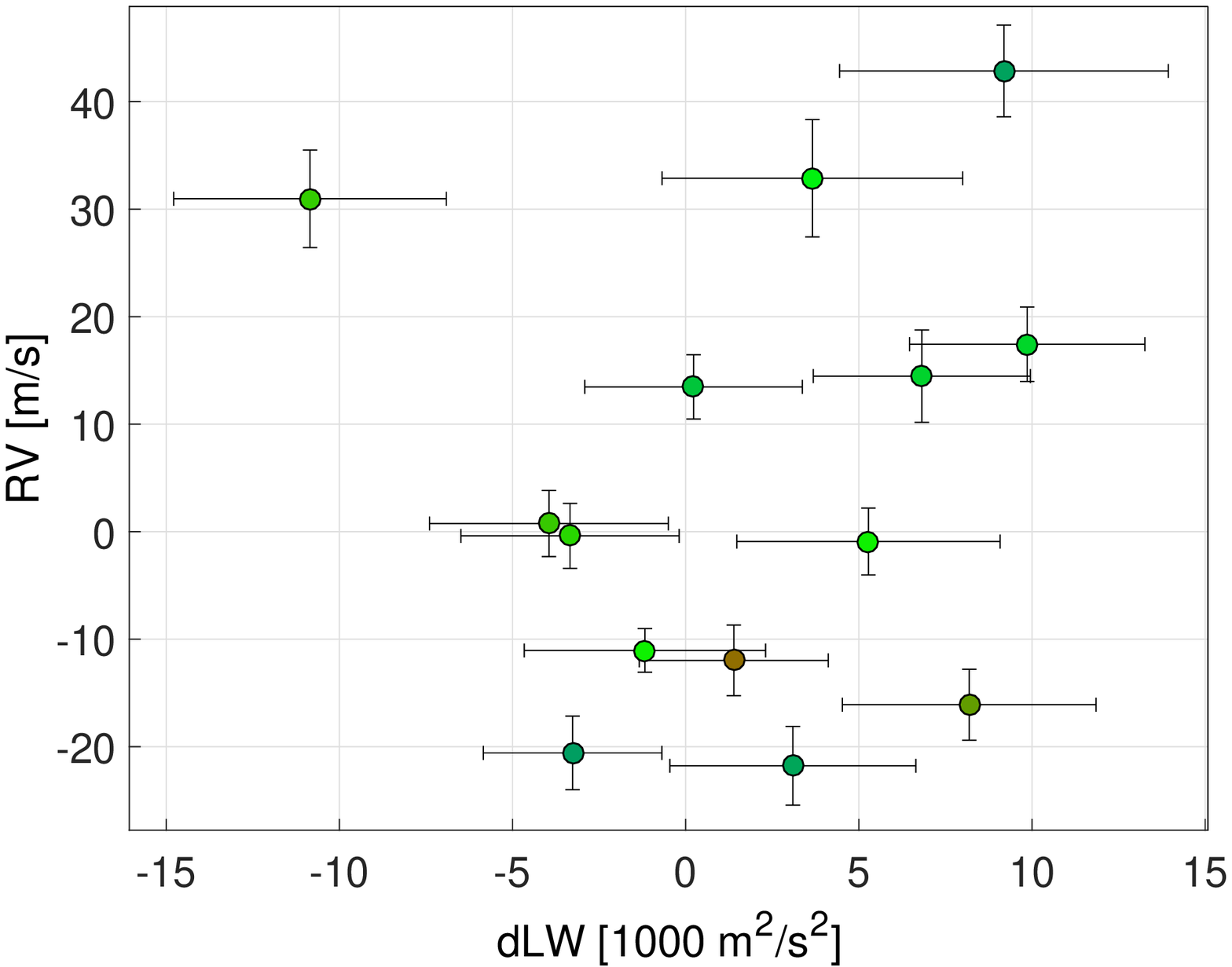}}
\endminipage\hfill
\minipage{0.33\textwidth}
{\includegraphics[width=\linewidth]{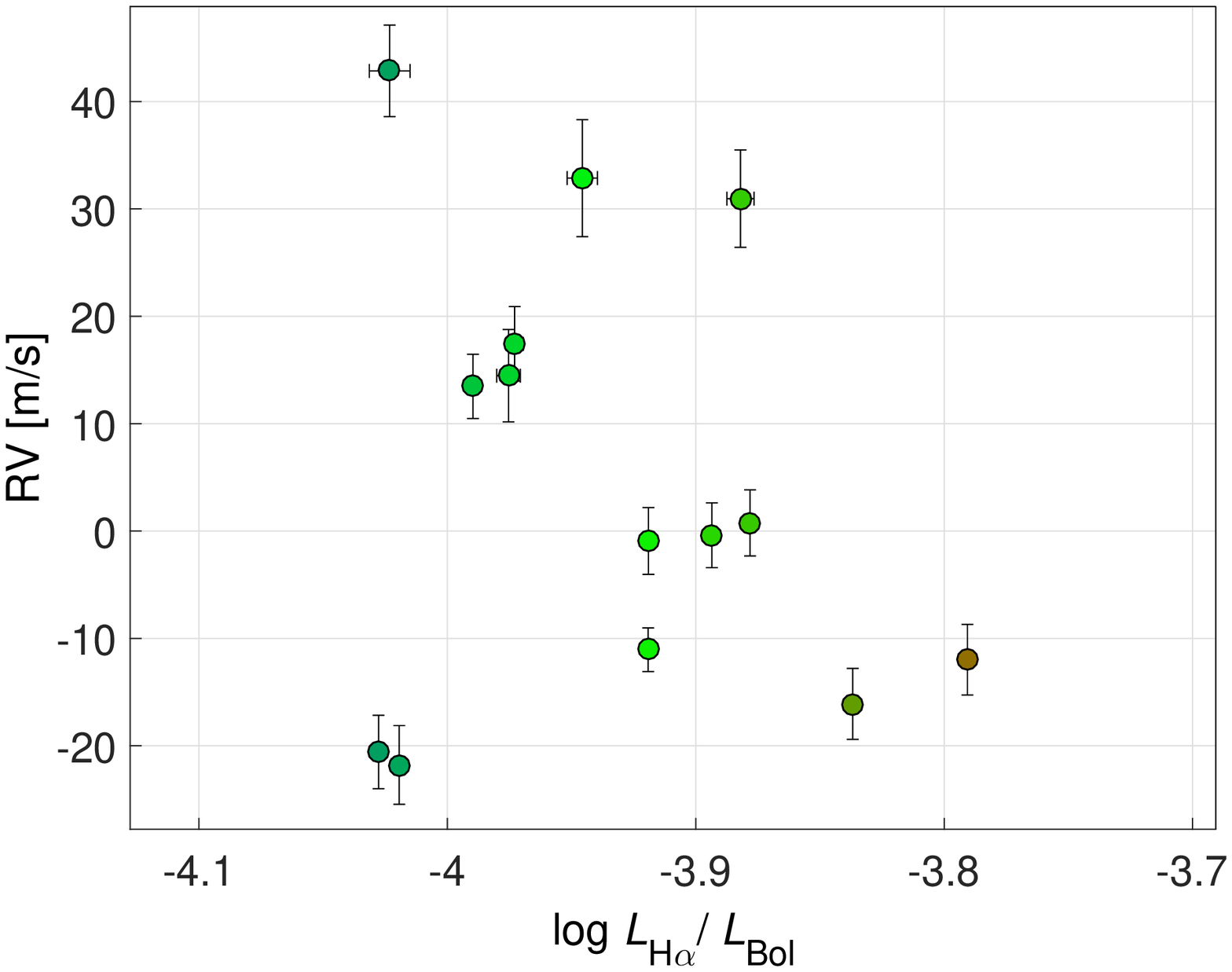}}
\endminipage


\minipage{0.33\textwidth}
{\includegraphics[width=\linewidth]{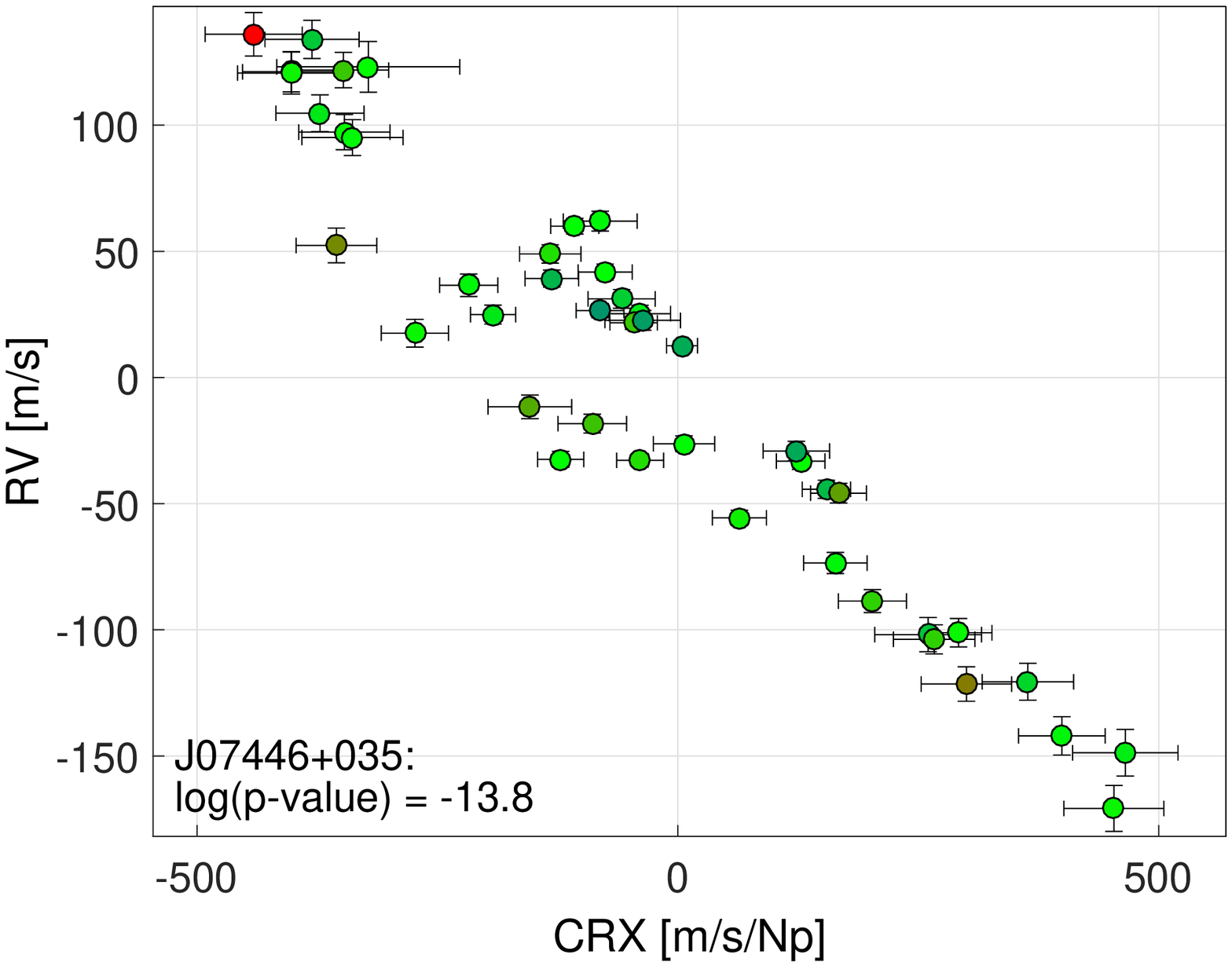}}
\endminipage\hfill
\minipage{0.33\textwidth}
{\includegraphics[width=\linewidth]{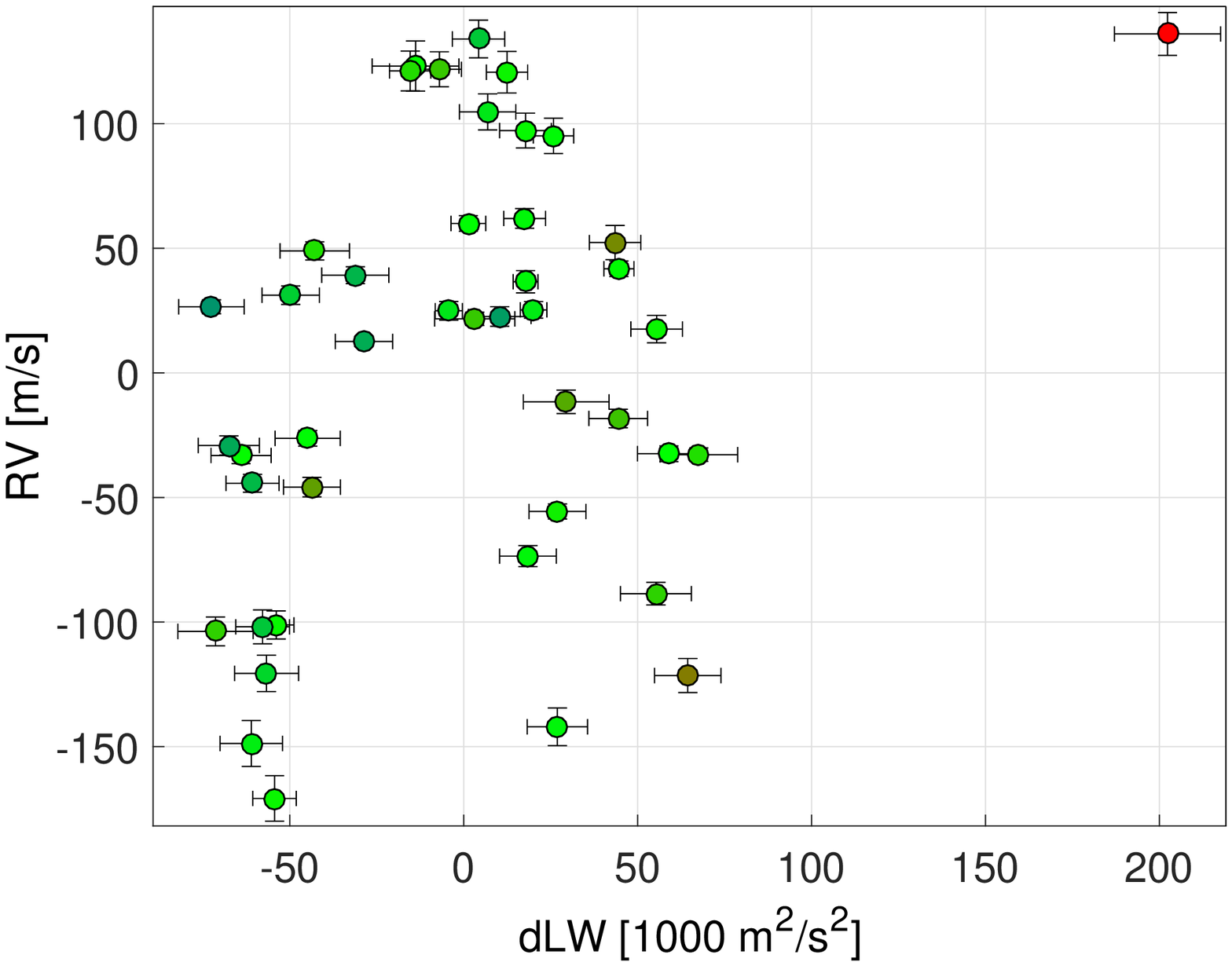}}
\endminipage\hfill
\minipage{0.33\textwidth}
{\includegraphics[width=\linewidth]{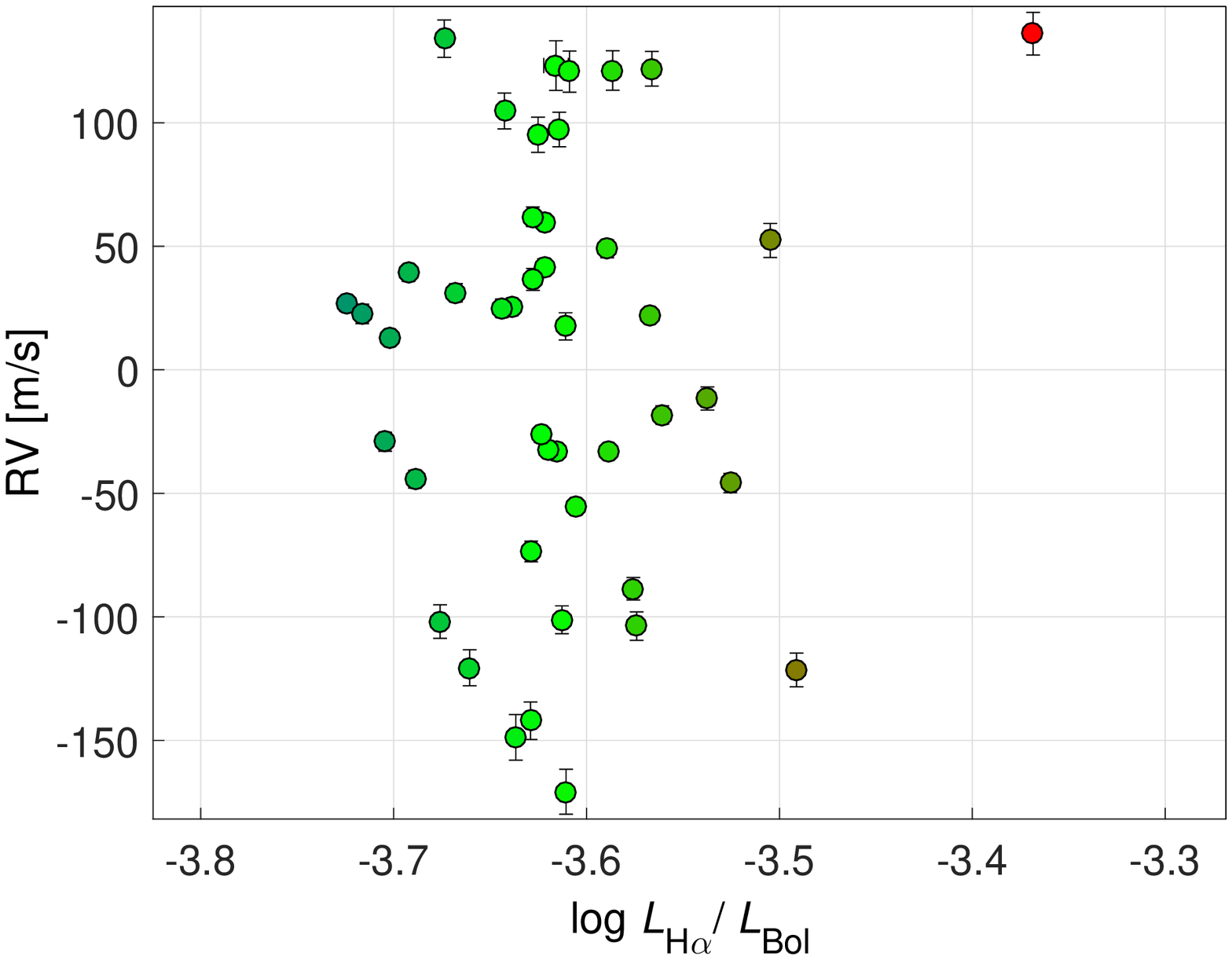}}
\endminipage


\minipage{0.33\textwidth}
{\includegraphics[width=\linewidth]{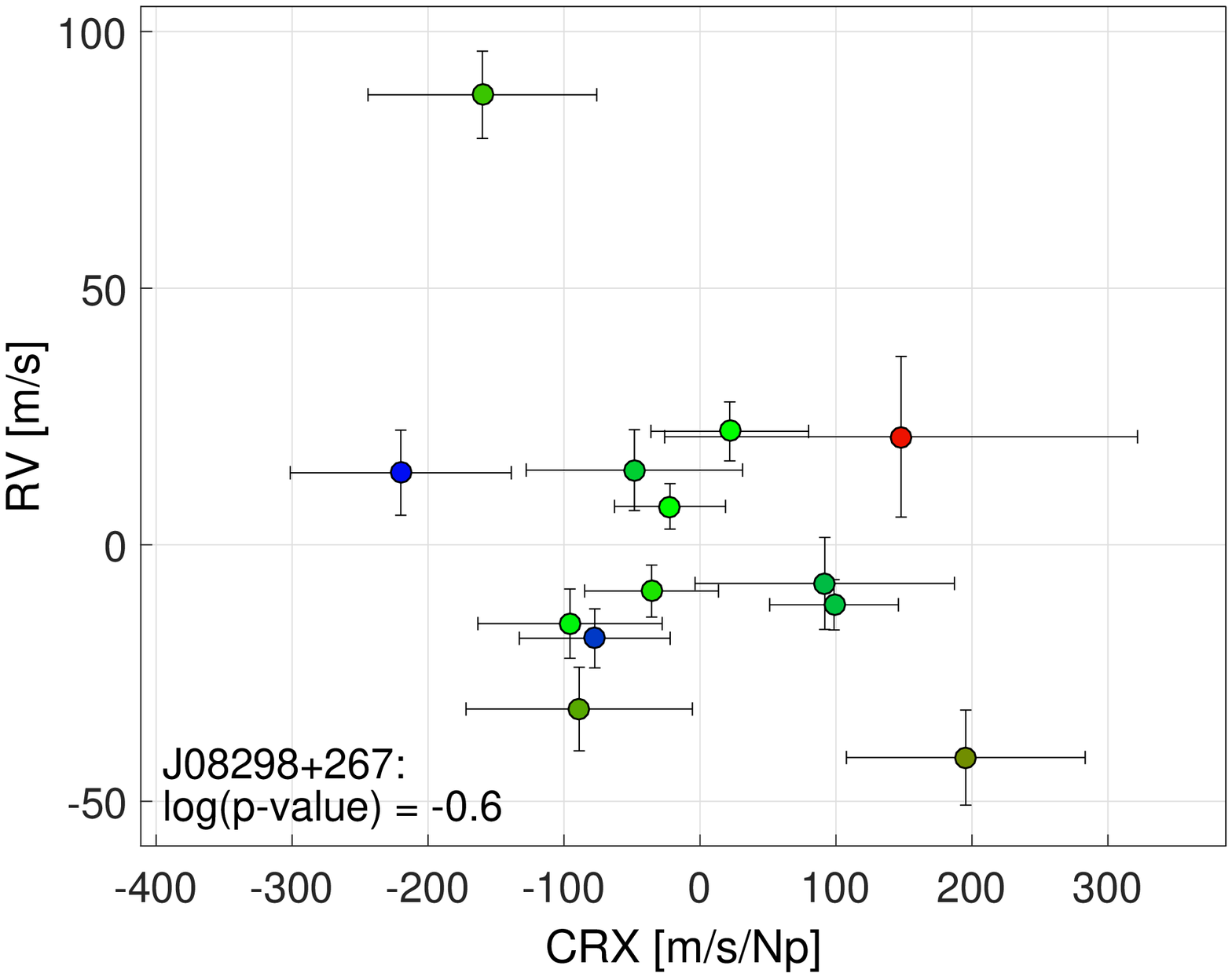}}
\endminipage\hfill
\minipage{0.33\textwidth}
{\includegraphics[width=\linewidth]{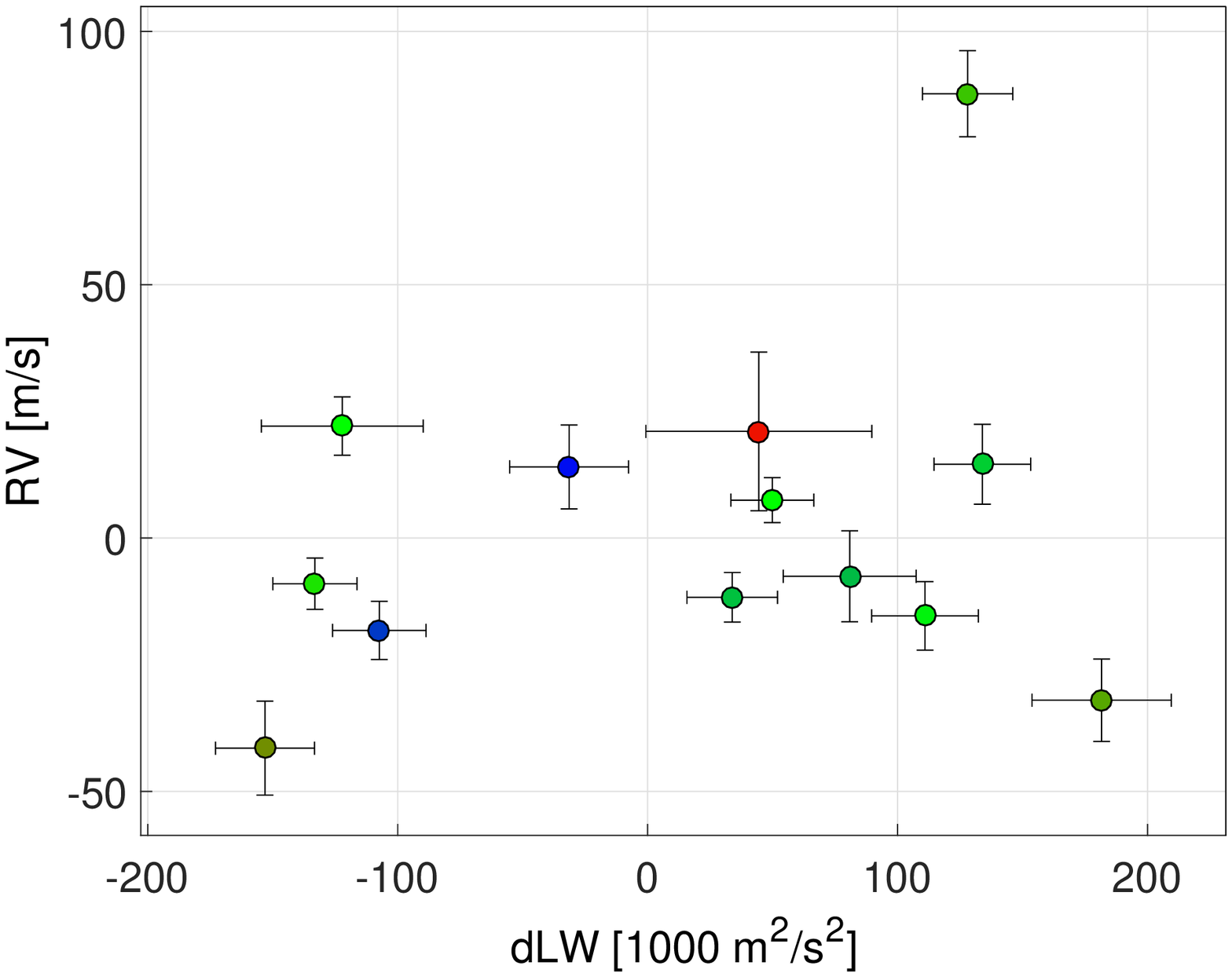}}
\endminipage\hfill
\minipage{0.33\textwidth}
{\includegraphics[width=\linewidth]{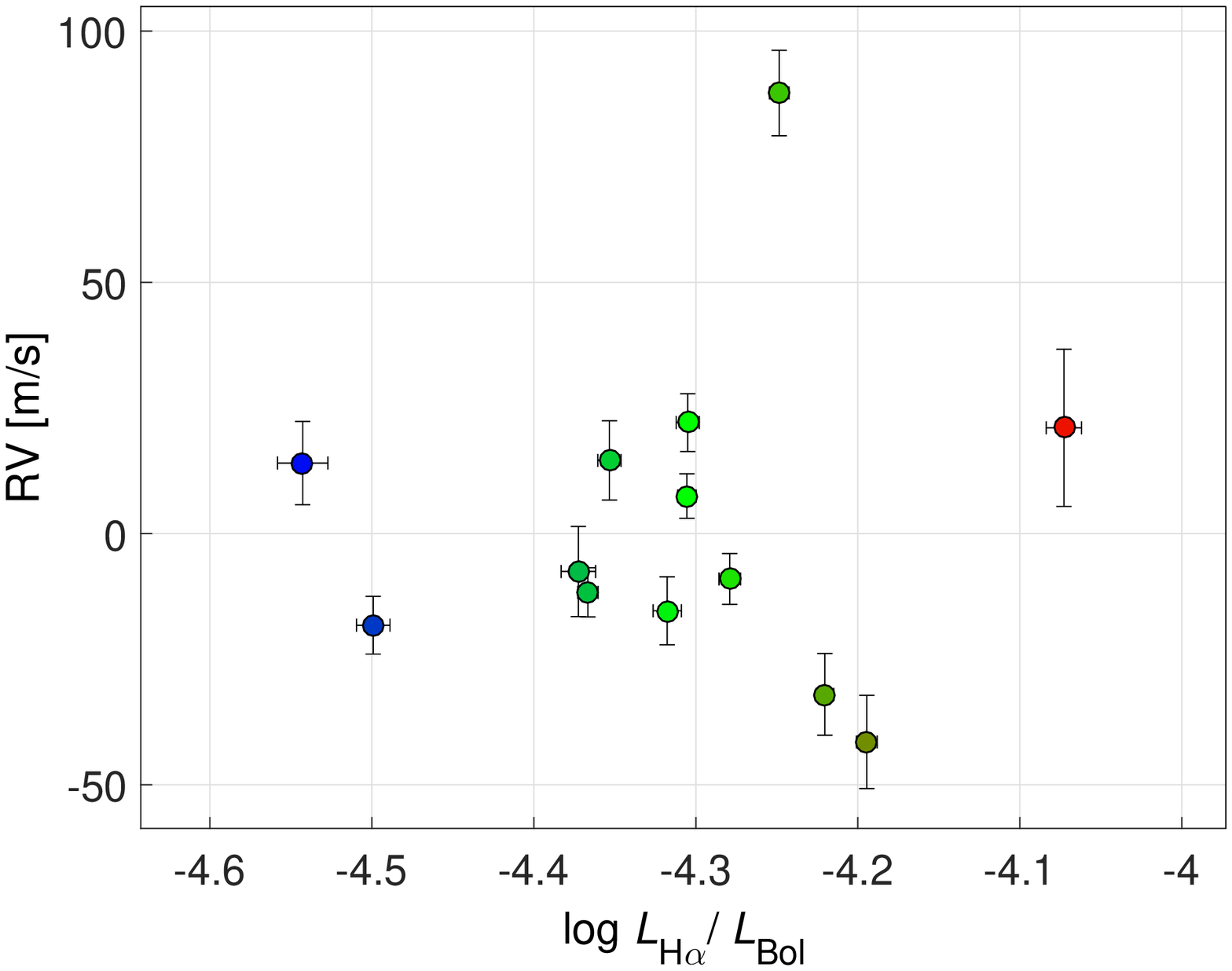}}
\endminipage


\minipage{0.33\textwidth}
{\includegraphics[width=\linewidth]{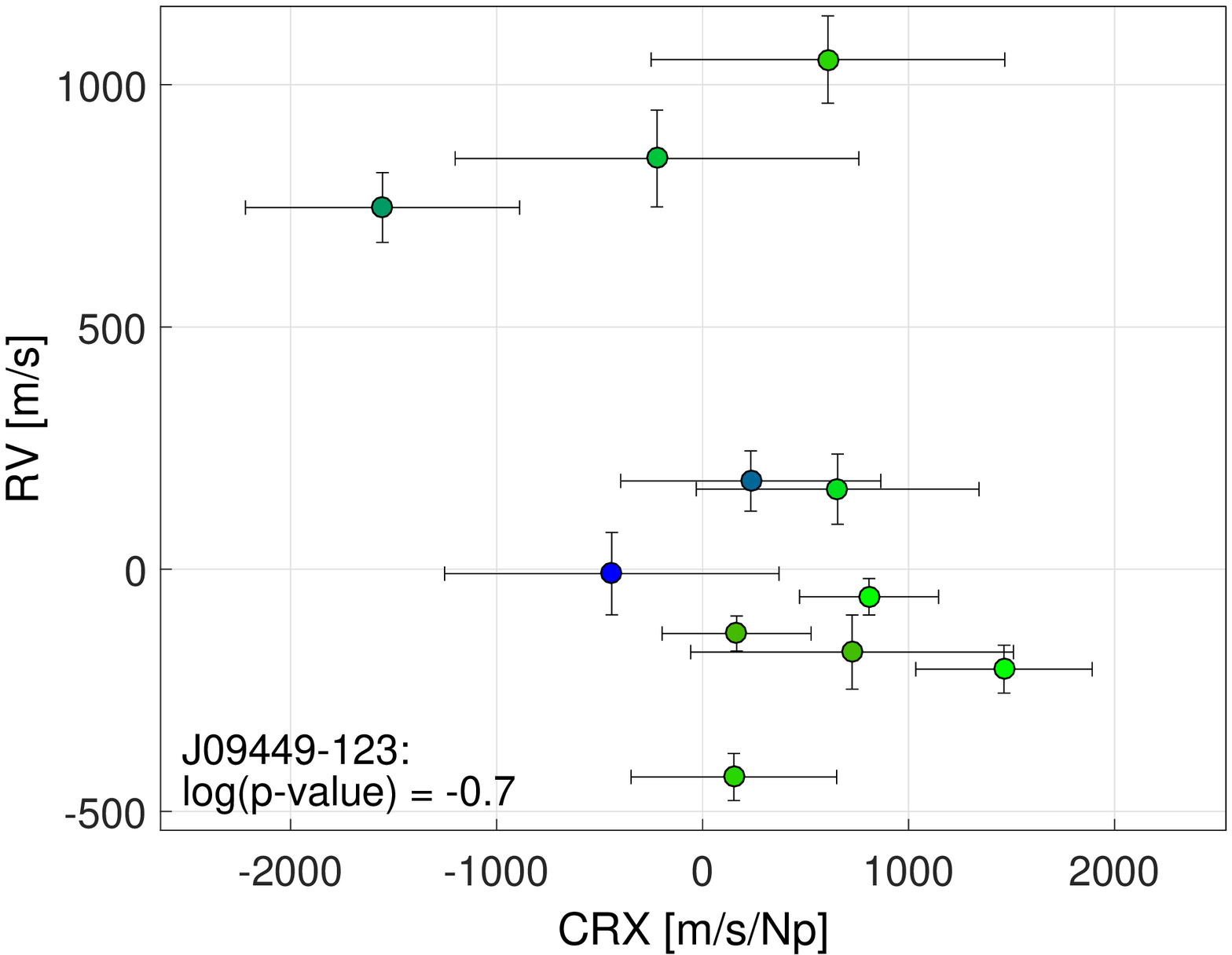}}
\endminipage\hfill
\minipage{0.33\textwidth}
{\includegraphics[width=\linewidth]{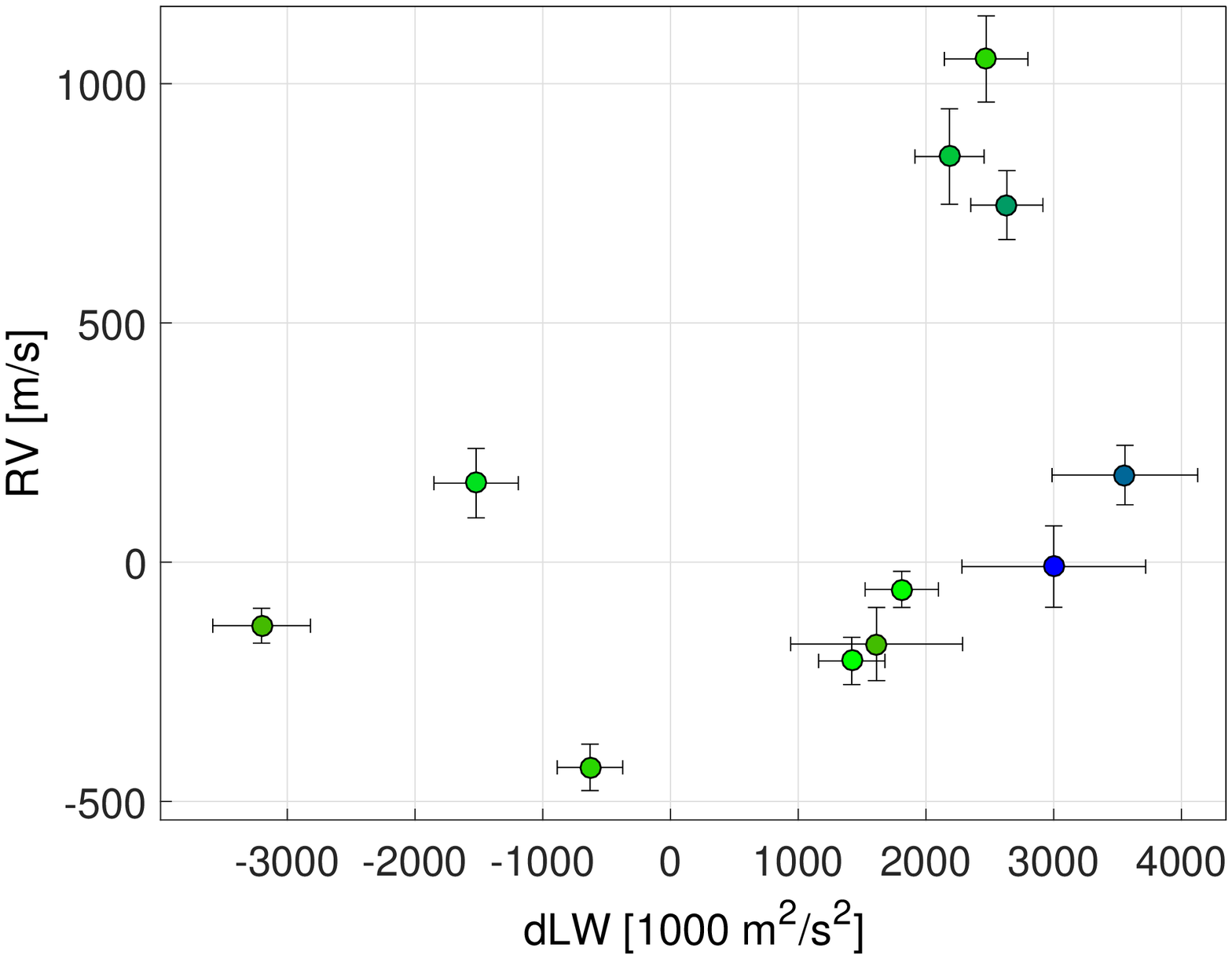}}
\endminipage\hfill
\minipage{0.33\textwidth}
{\includegraphics[width=\linewidth]{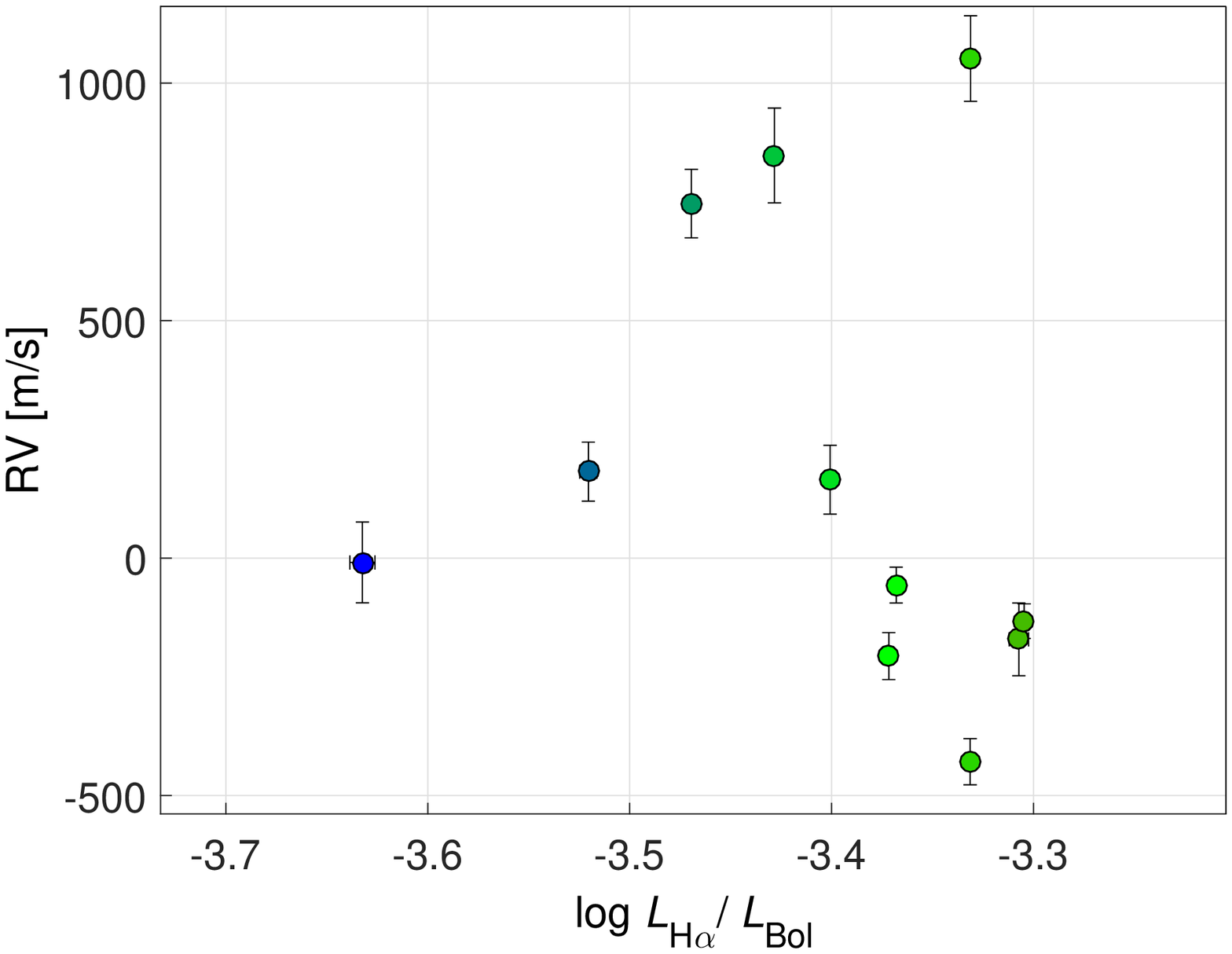}}
\endminipage
\caption{Continued.}
\label{figA1}
\end{figure*}

\addtocounter{figure}{-1}

\begin{figure*}[!htp]
\minipage{0.33\textwidth}
{\includegraphics[width=\linewidth]{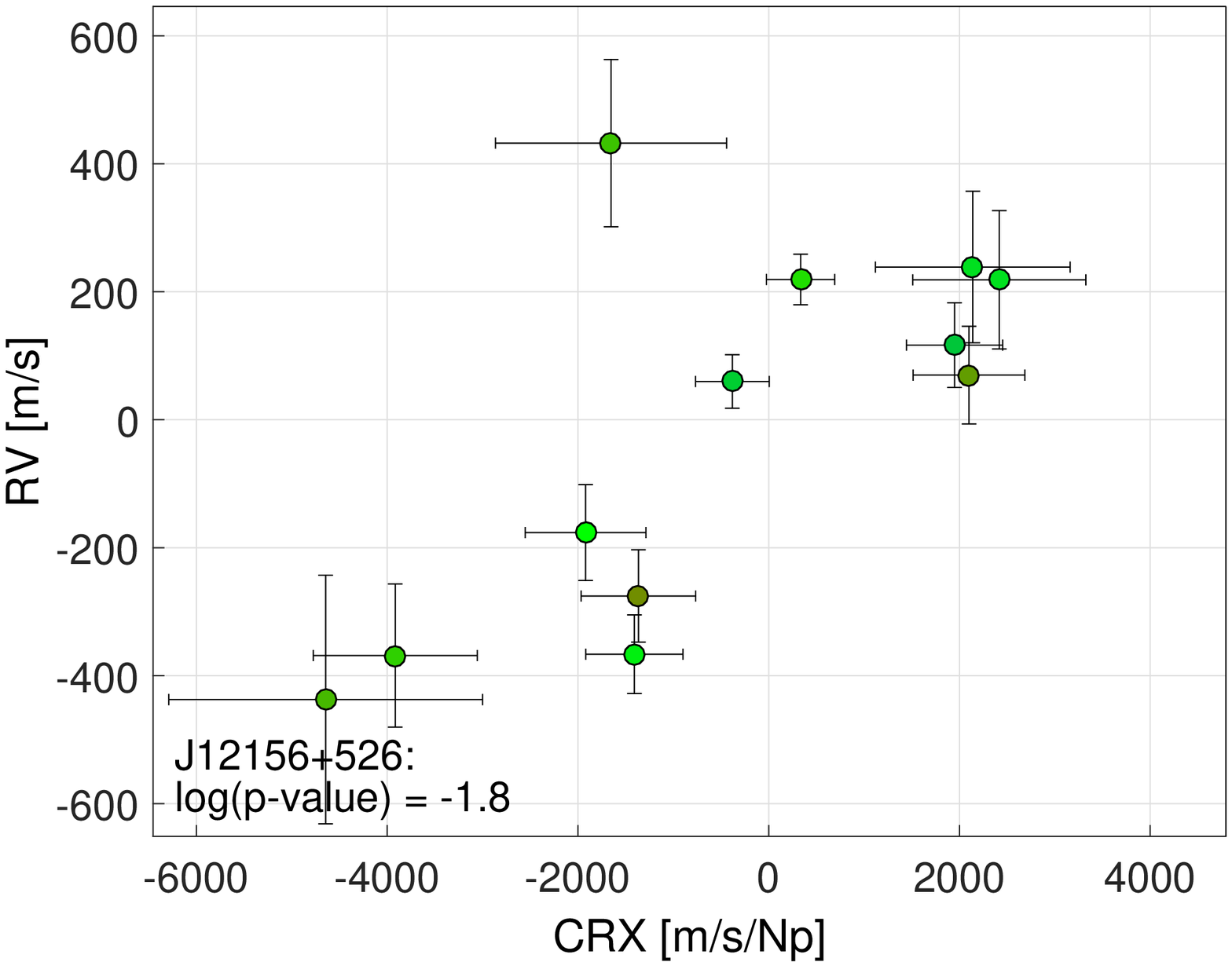}}
\endminipage\hfill
\minipage{0.33\textwidth}
{\includegraphics[width=\linewidth]{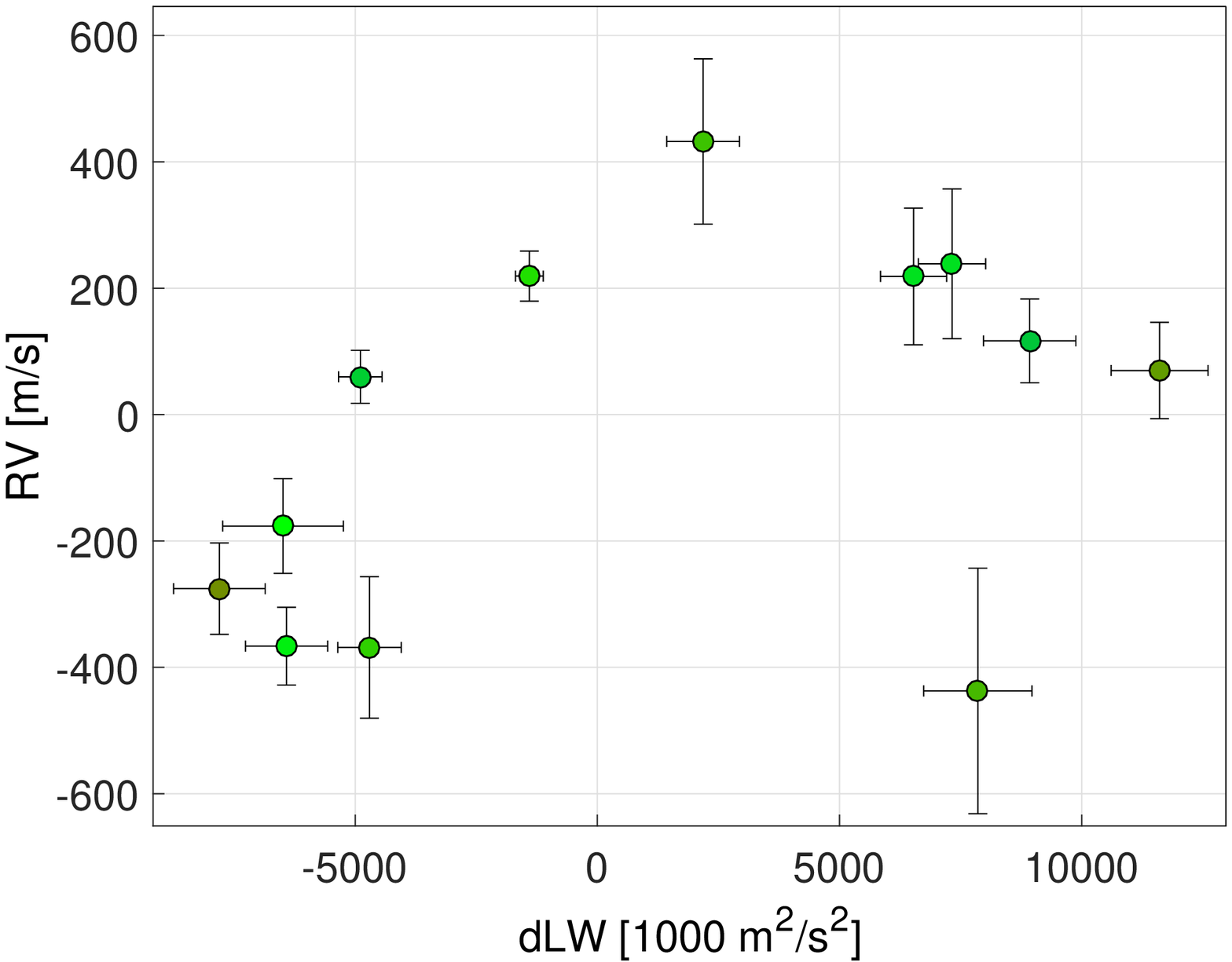}}
\endminipage\hfill
\minipage{0.33\textwidth}
{\includegraphics[width=\linewidth]{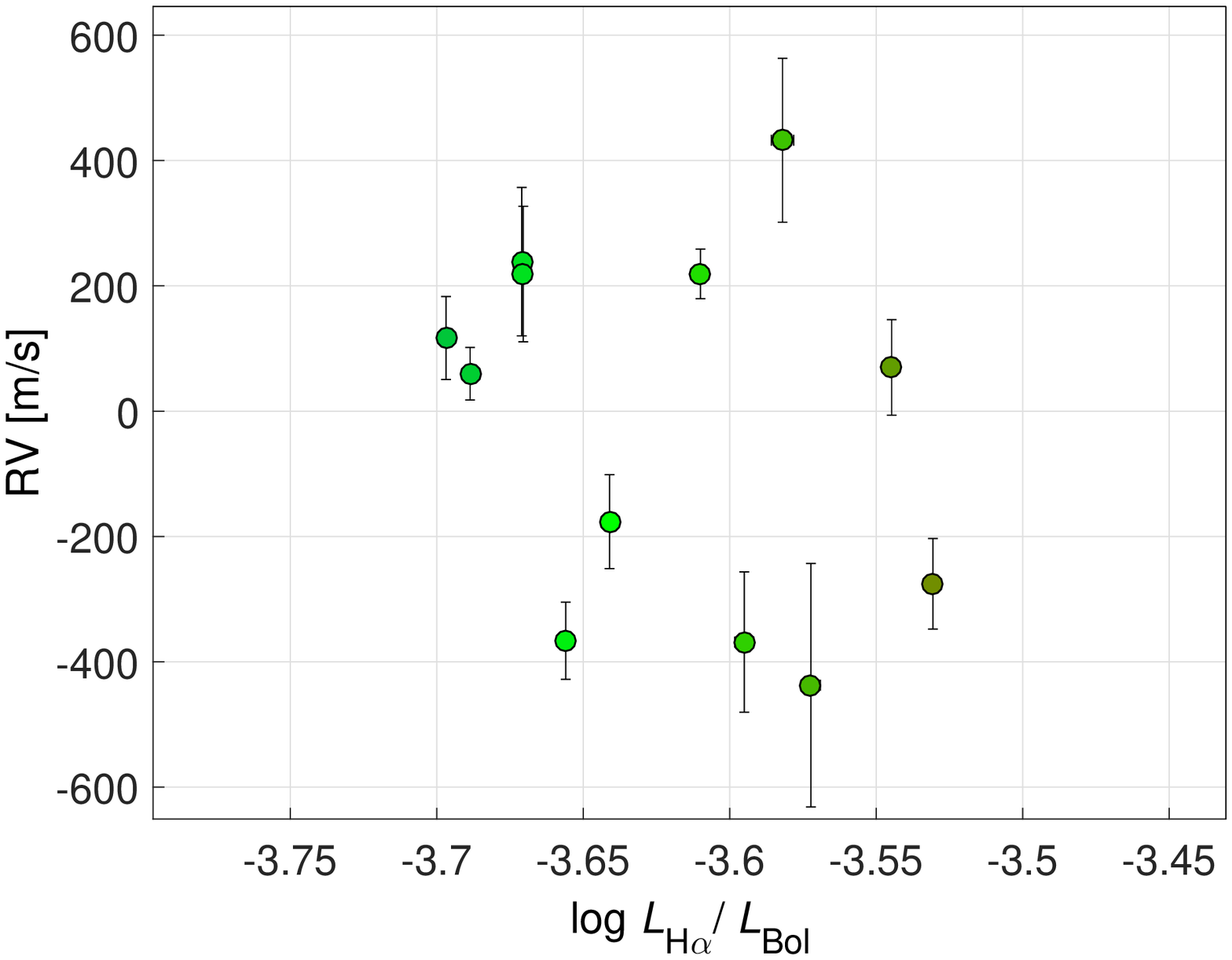}}
\endminipage


\minipage{0.33\textwidth}
{\includegraphics[width=\linewidth]{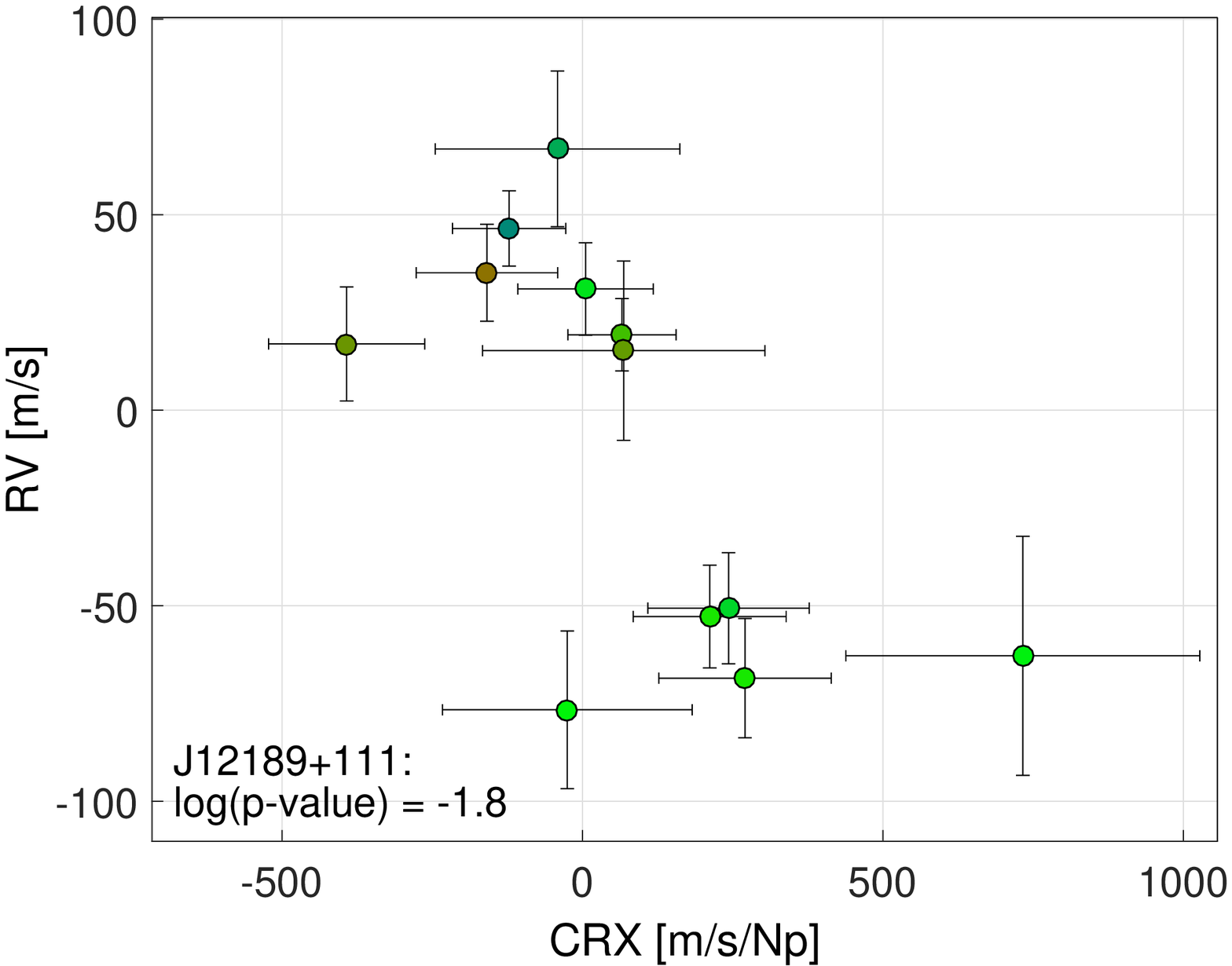}}
\endminipage\hfill
\minipage{0.33\textwidth}
{\includegraphics[width=\linewidth]{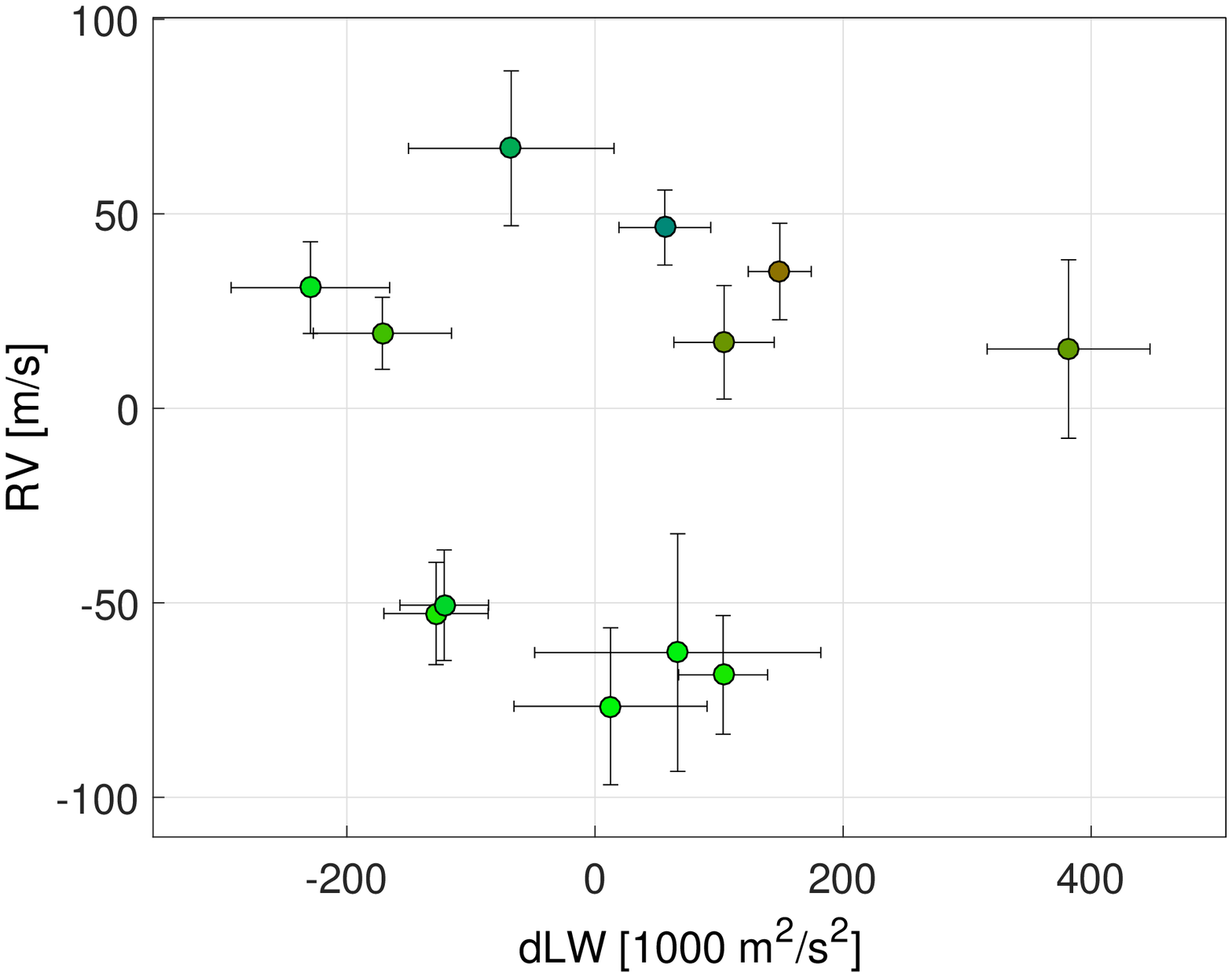}}
\endminipage\hfill
\minipage{0.33\textwidth}
{\includegraphics[width=\linewidth]{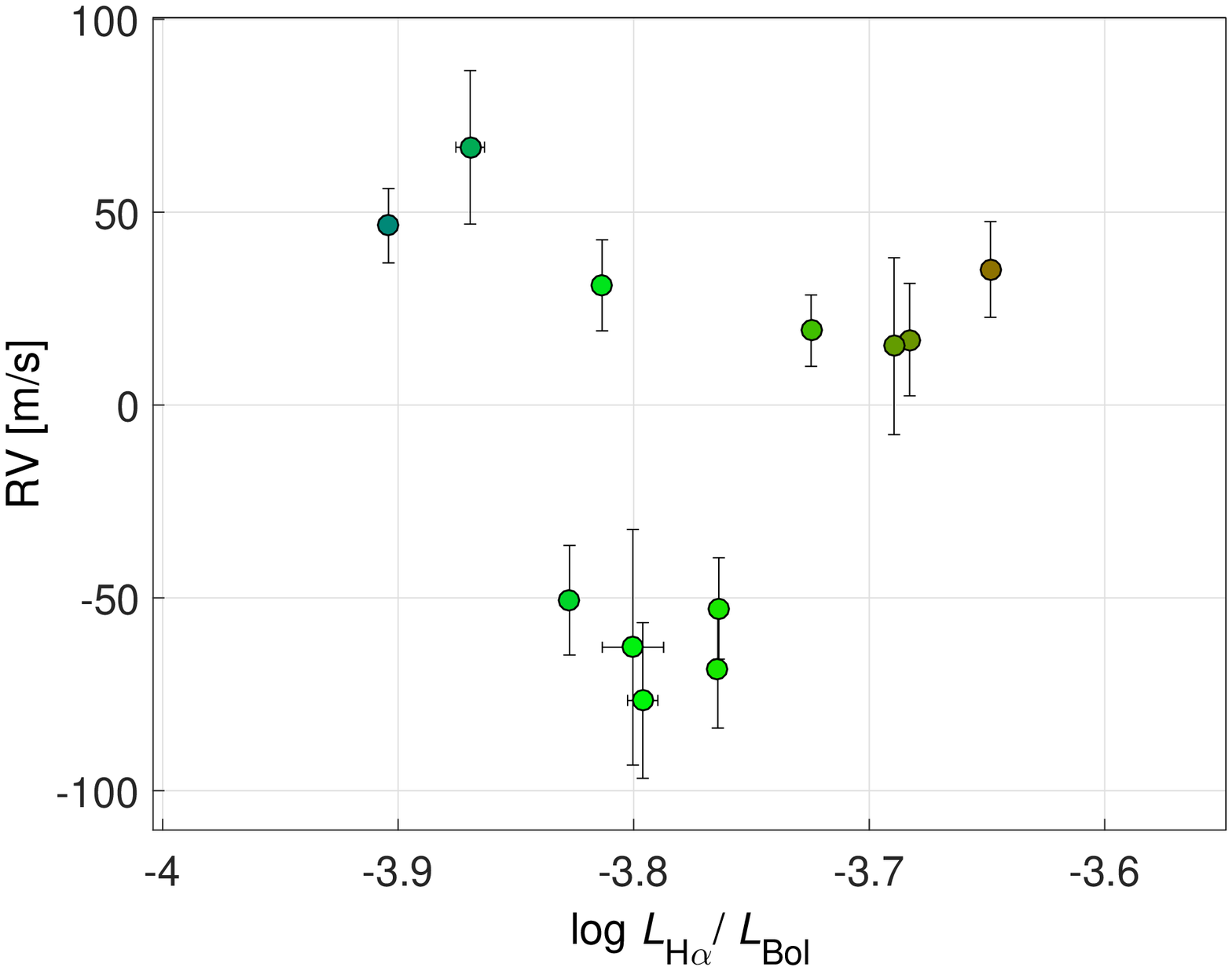}}
\endminipage


\minipage{0.33\textwidth}
{\includegraphics[width=\linewidth]{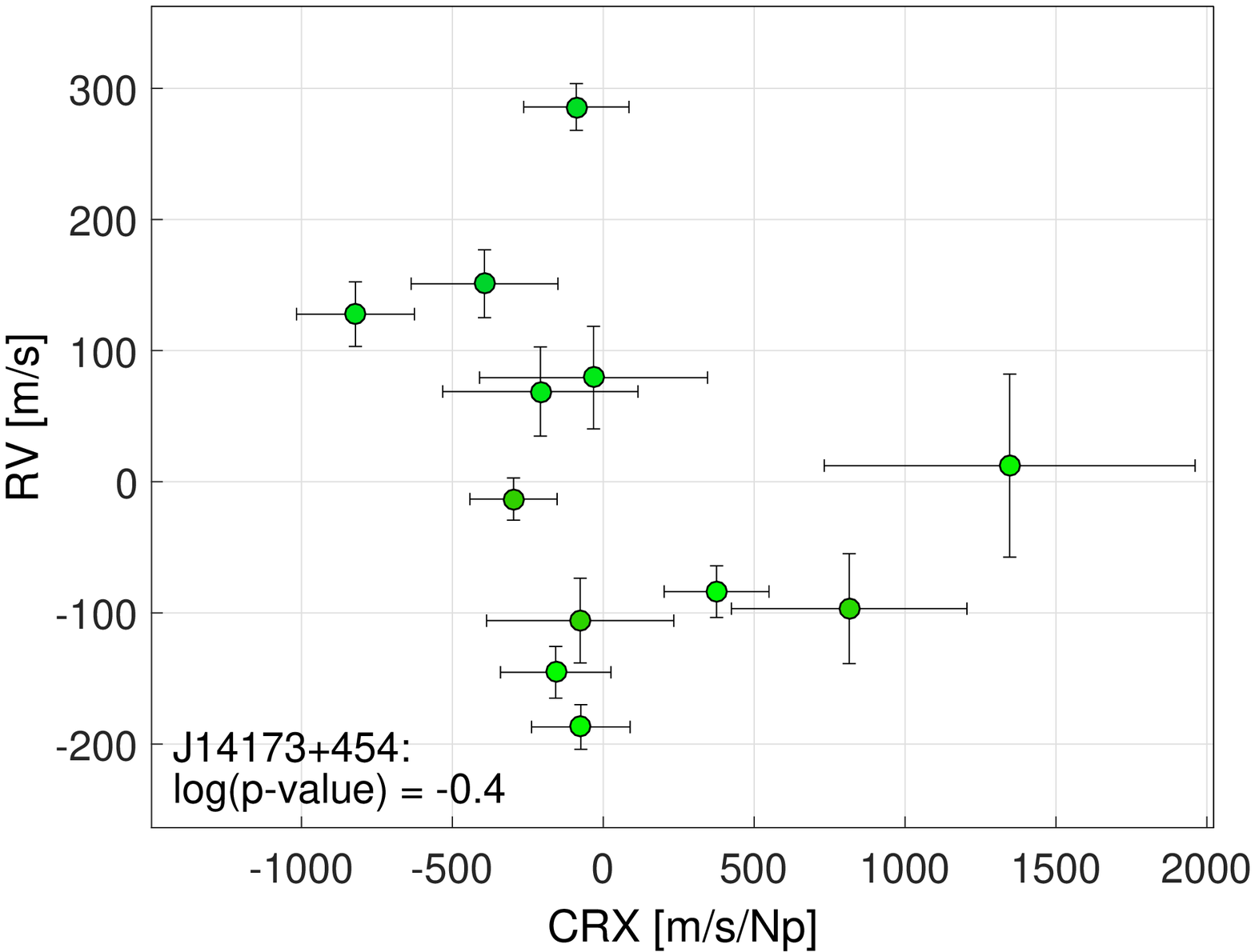}}
\endminipage\hfill
\minipage{0.33\textwidth}
{\includegraphics[width=\linewidth]{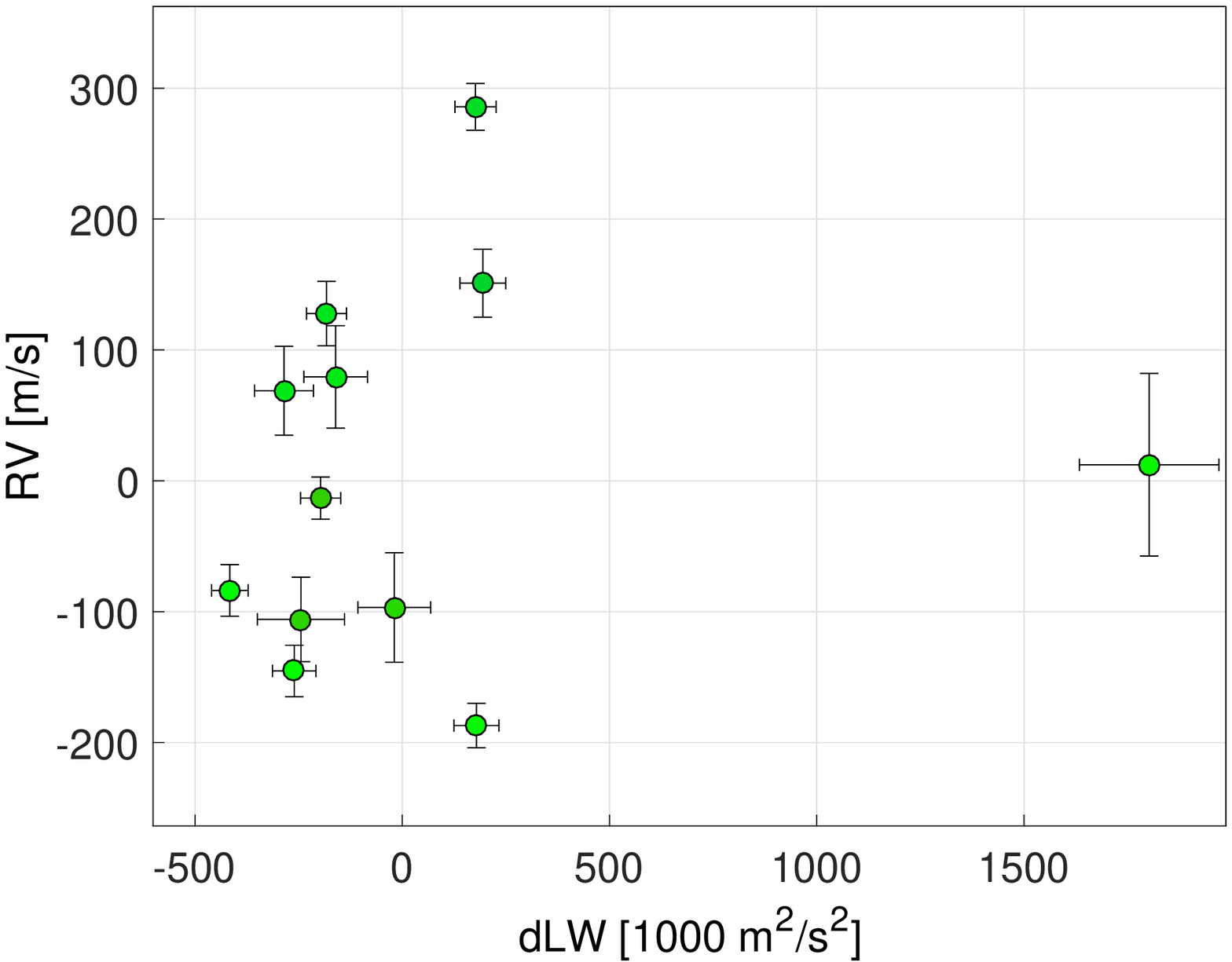}}
\endminipage\hfill
\minipage{0.33\textwidth}
{\includegraphics[width=\linewidth]{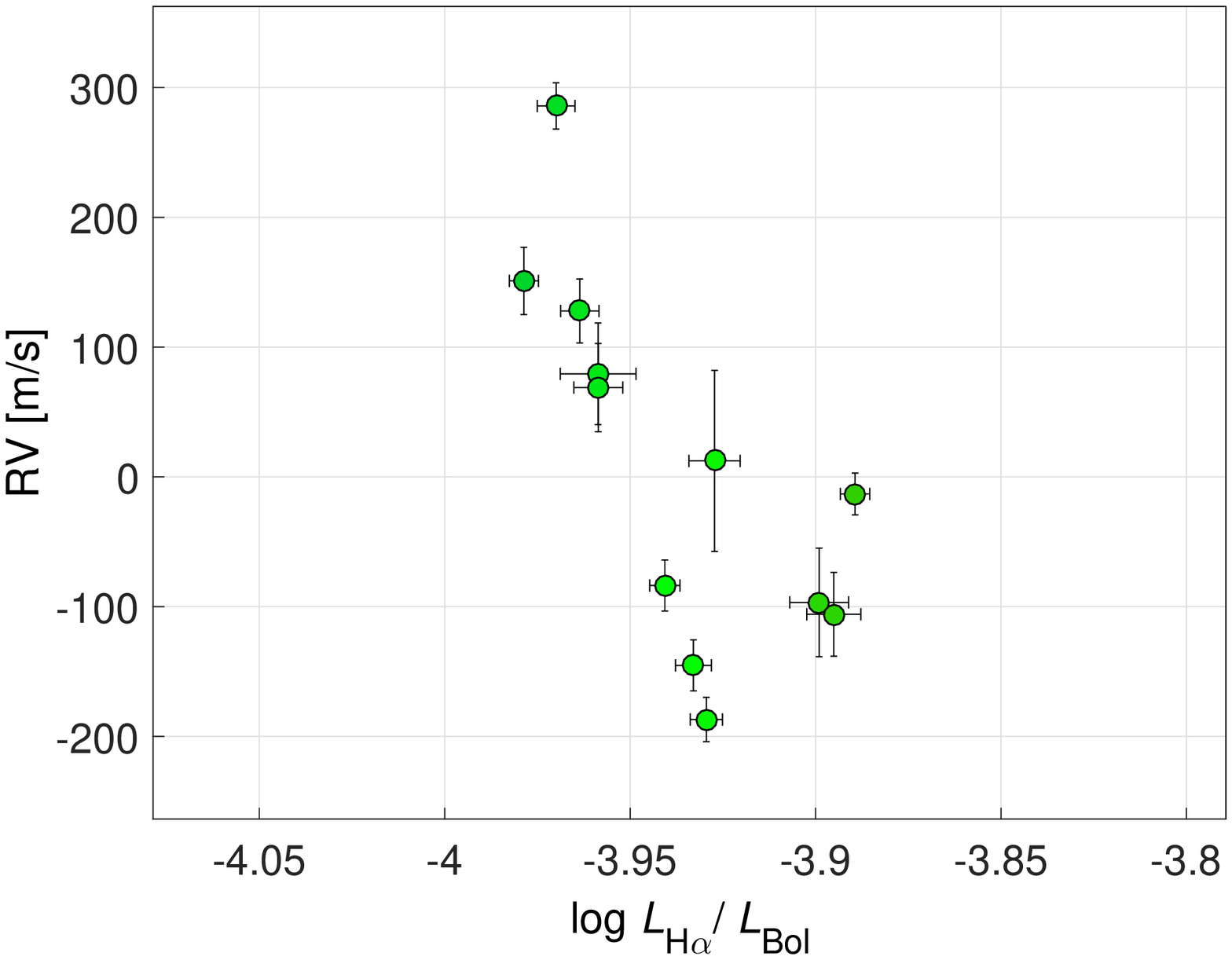}}
\endminipage


\minipage{0.33\textwidth}
{\includegraphics[width=\linewidth]{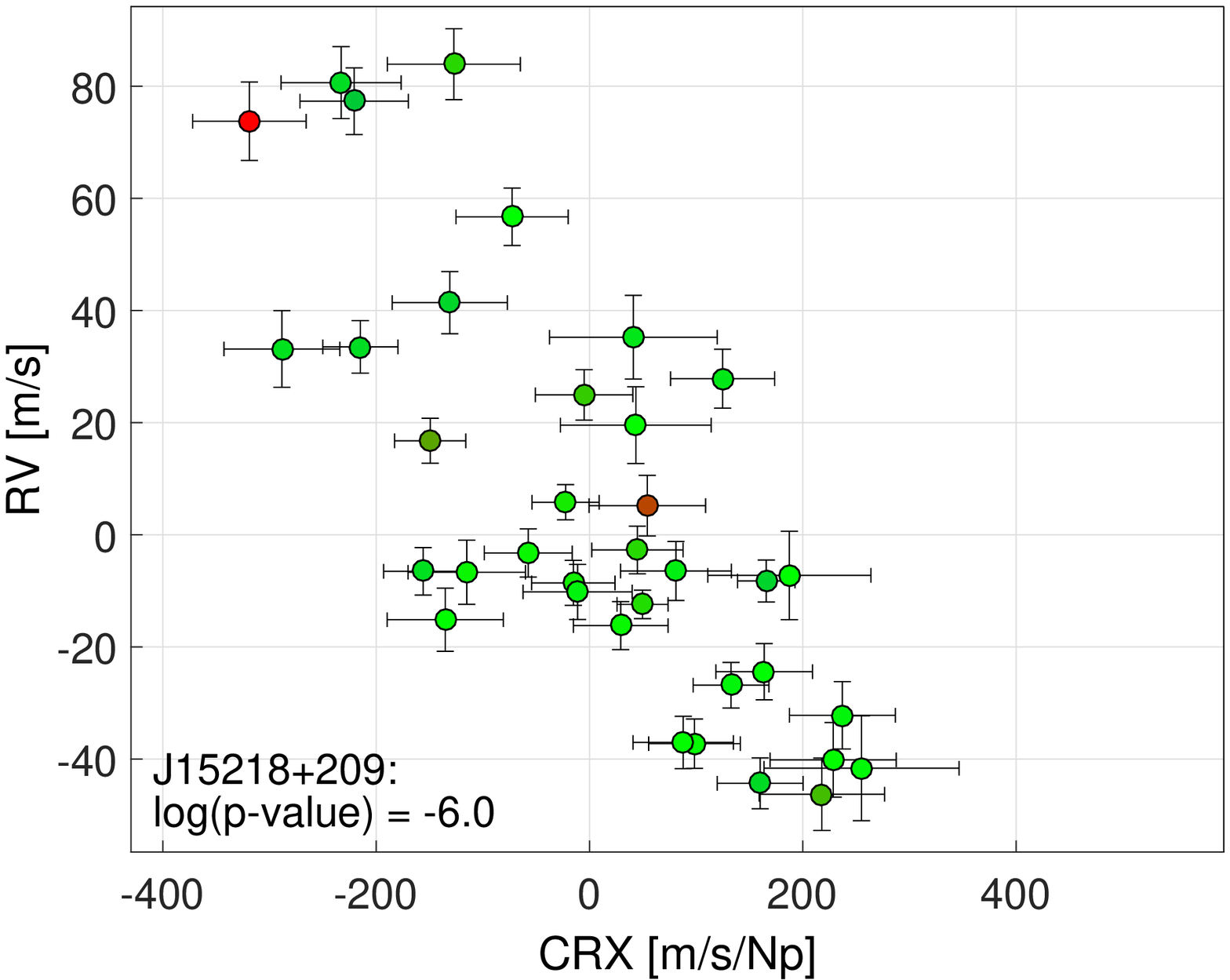}}
\endminipage\hfill
\minipage{0.33\textwidth}
{\includegraphics[width=\linewidth]{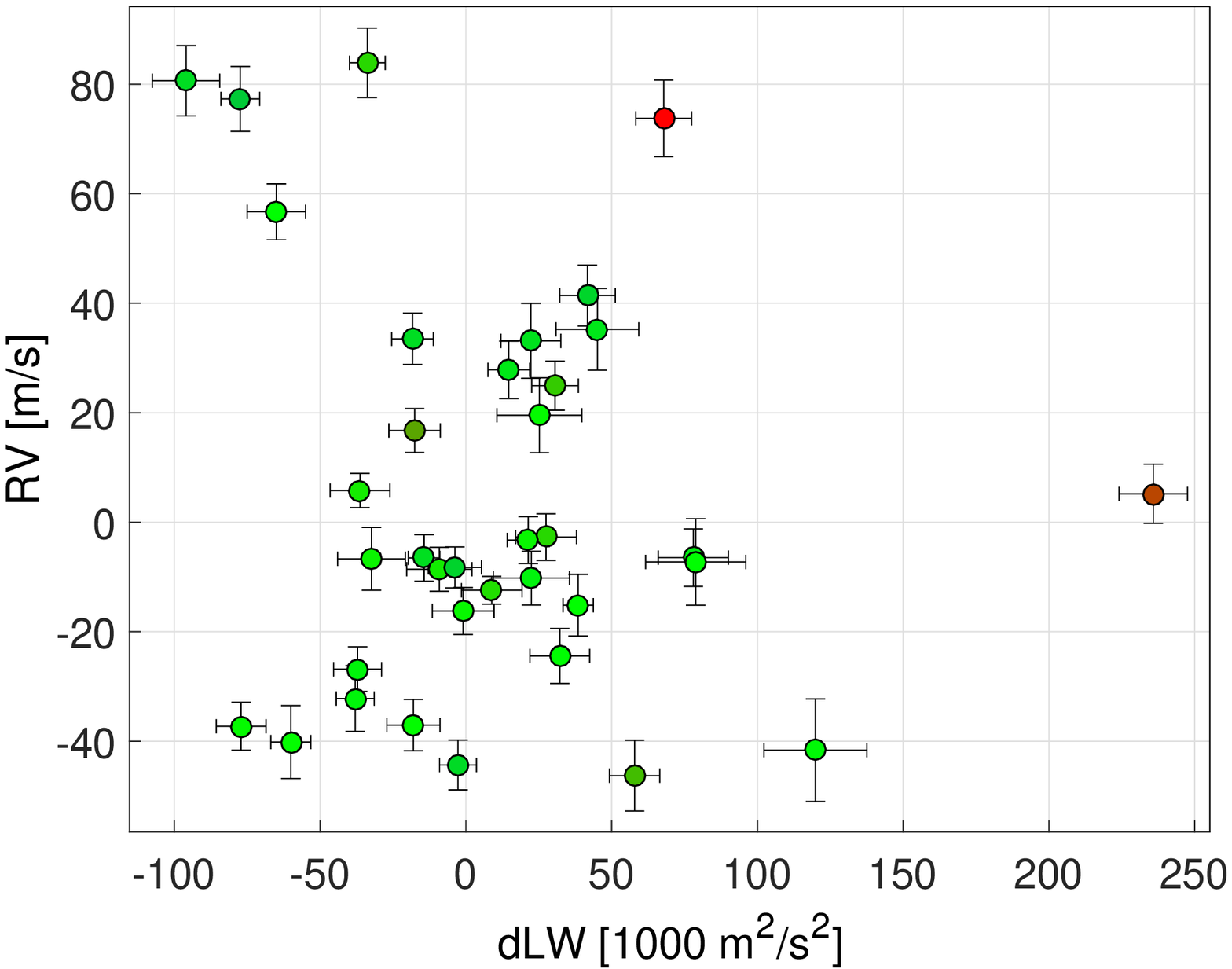}}
\endminipage\hfill
\minipage{0.33\textwidth}
{\includegraphics[width=\linewidth]{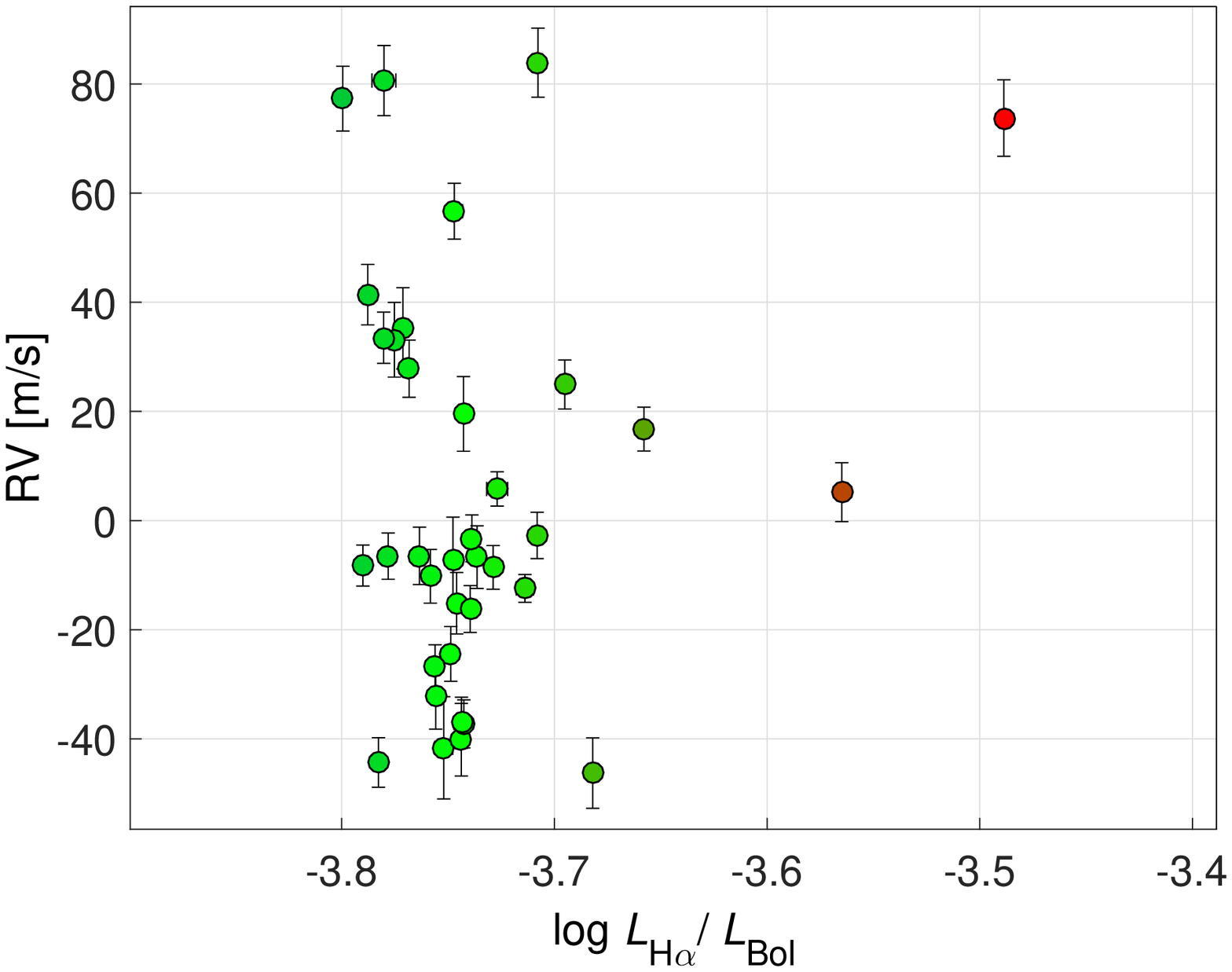}}
\endminipage


\minipage{0.33\textwidth}
{\includegraphics[width=\linewidth]{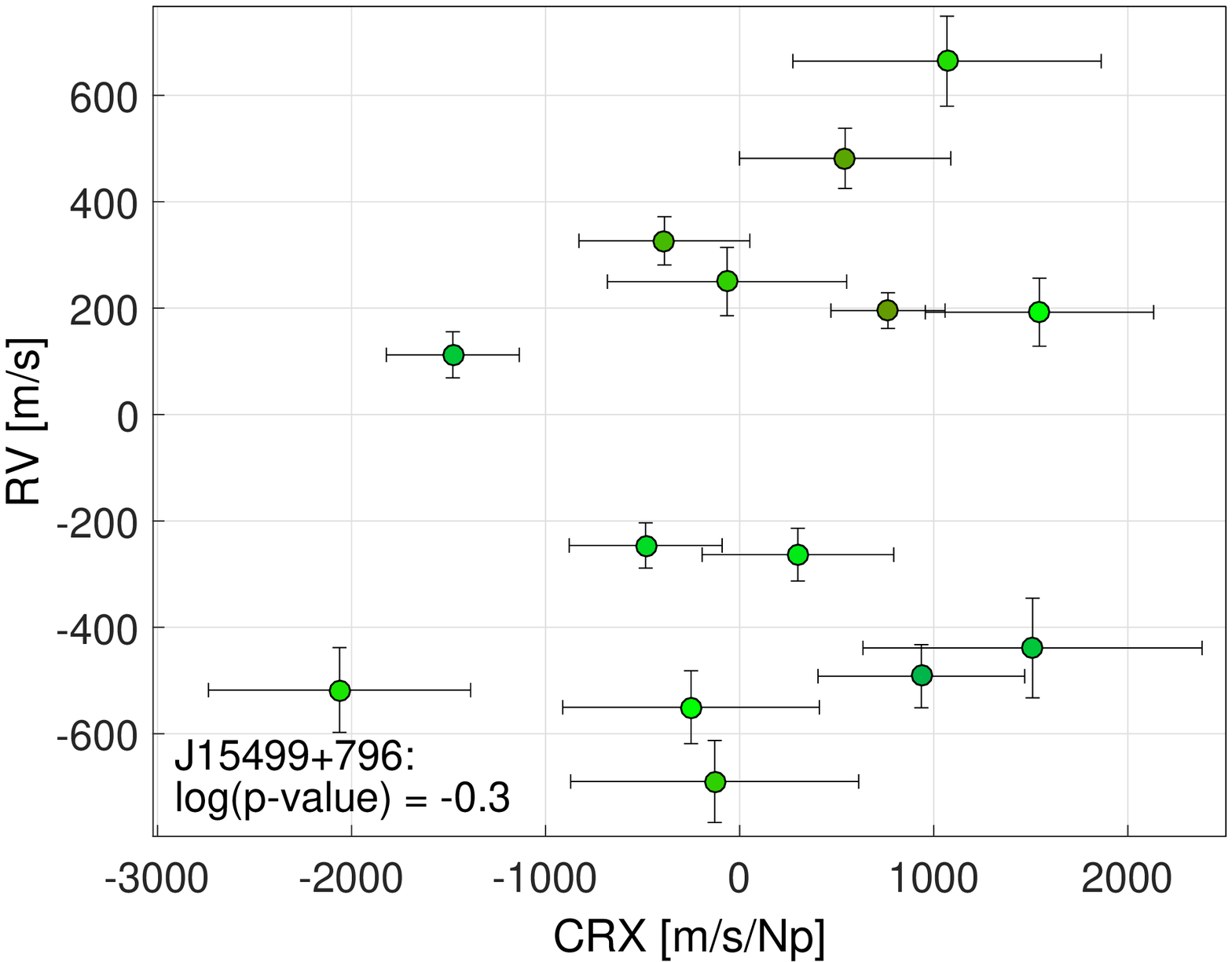}}
\endminipage\hfill
\minipage{0.33\textwidth}
{\includegraphics[width=\linewidth]{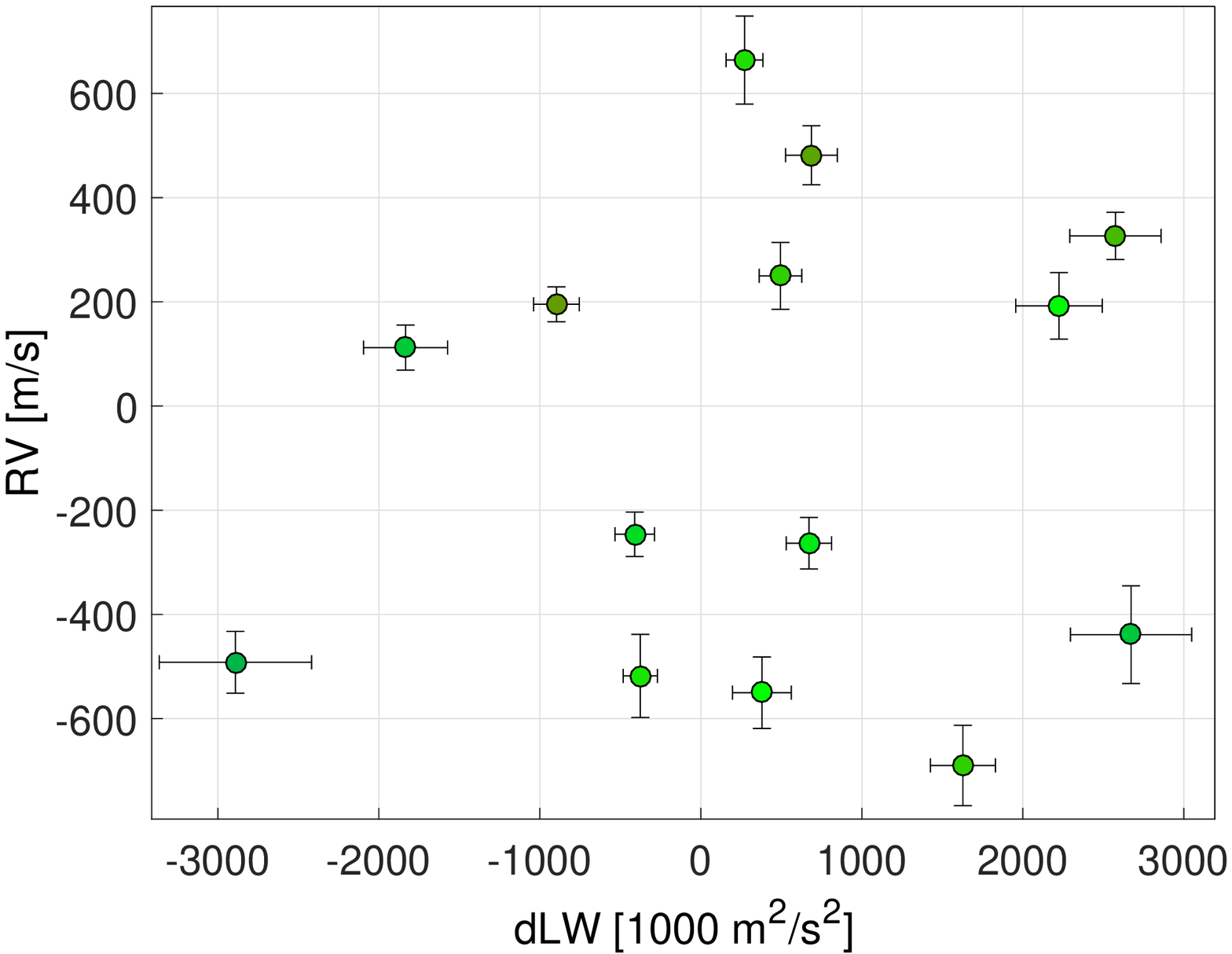}}
\endminipage\hfill
\minipage{0.33\textwidth}
{\includegraphics[width=\linewidth]{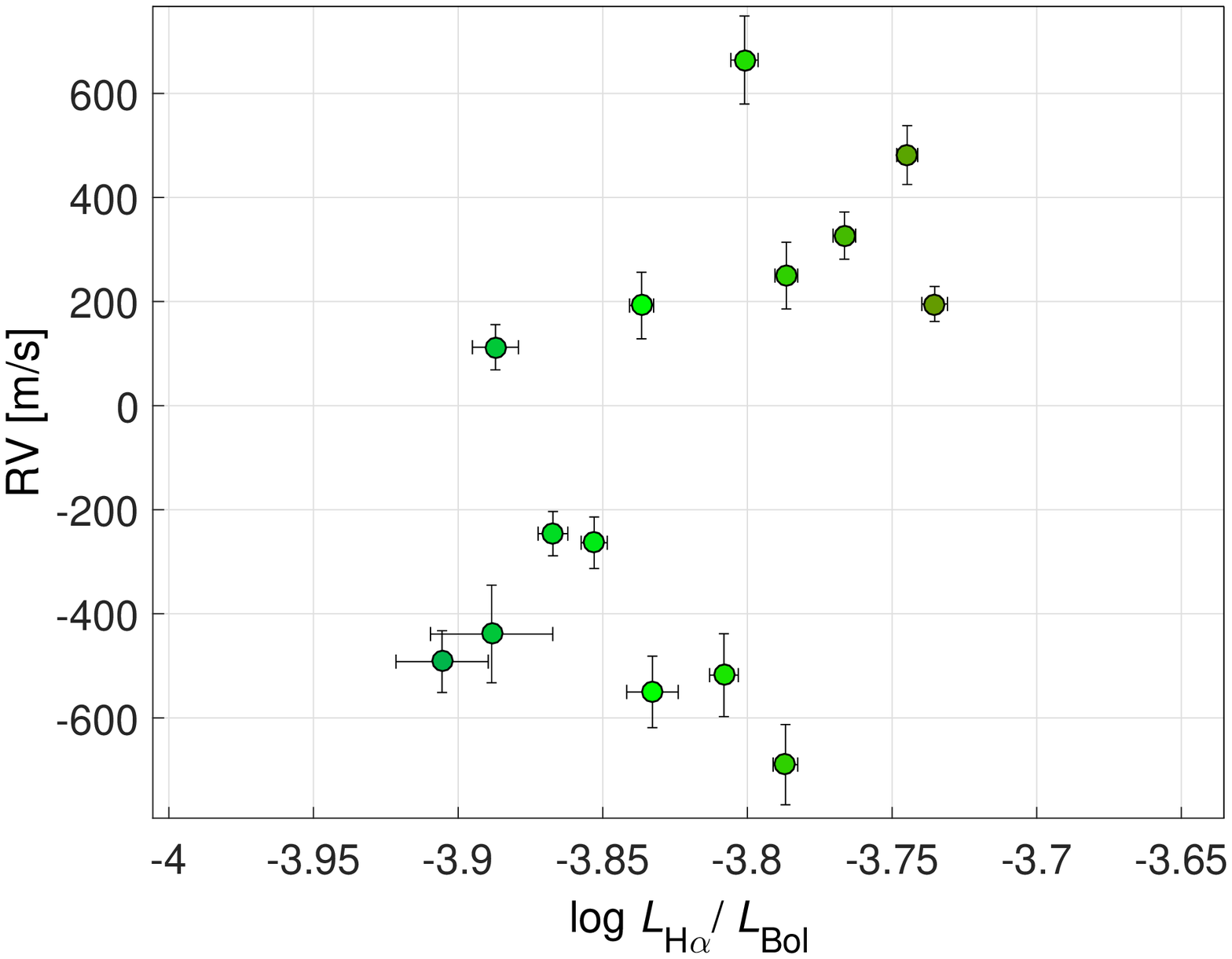}}
\endminipage
\caption{Continued.}
\label{figA1}
\end{figure*}

\addtocounter{figure}{-1}

\begin{figure*}[!htp]
\minipage{0.33\textwidth}
{\includegraphics[width=\linewidth]{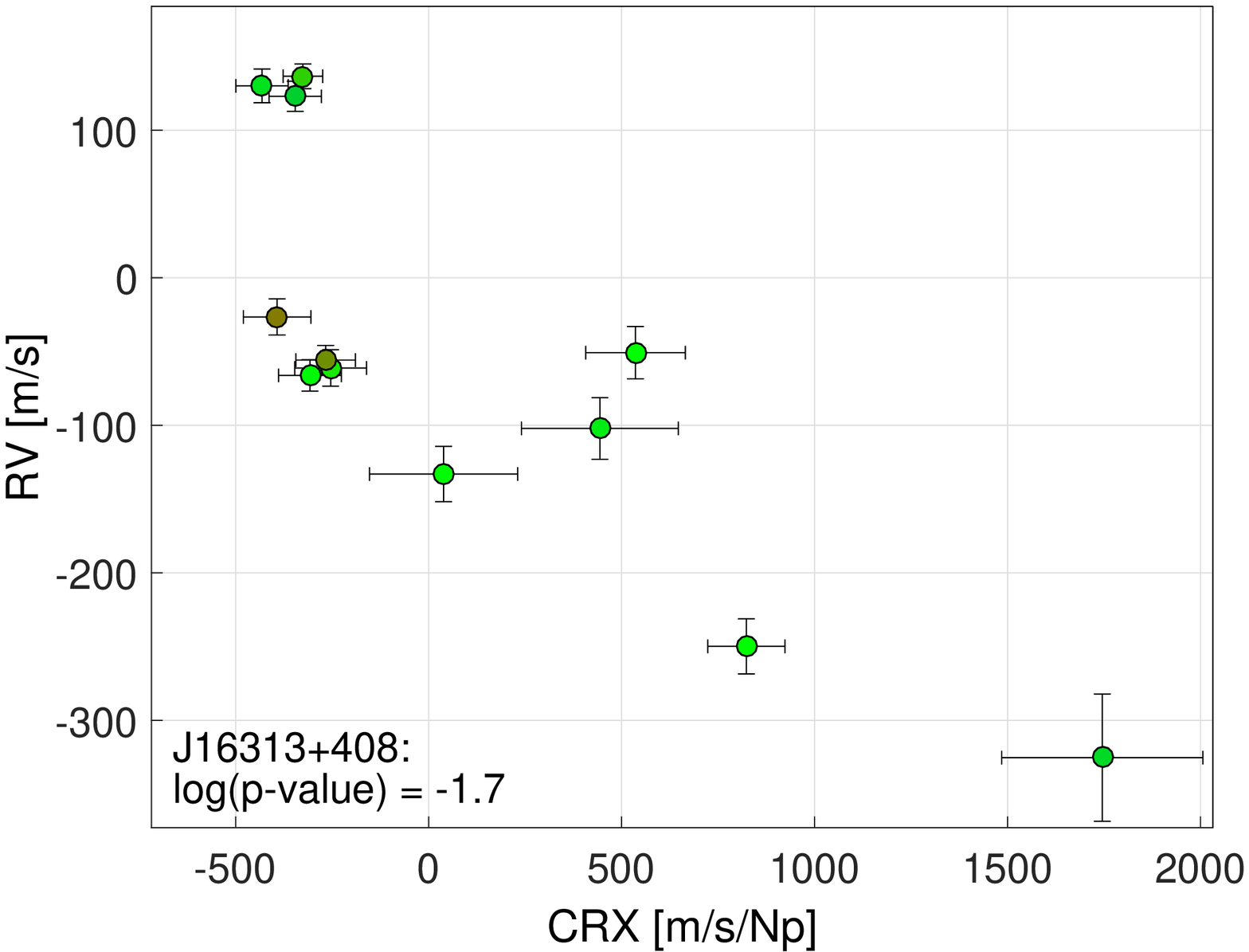}}
\endminipage\hfill
\minipage{0.33\textwidth}
{\includegraphics[width=\linewidth]{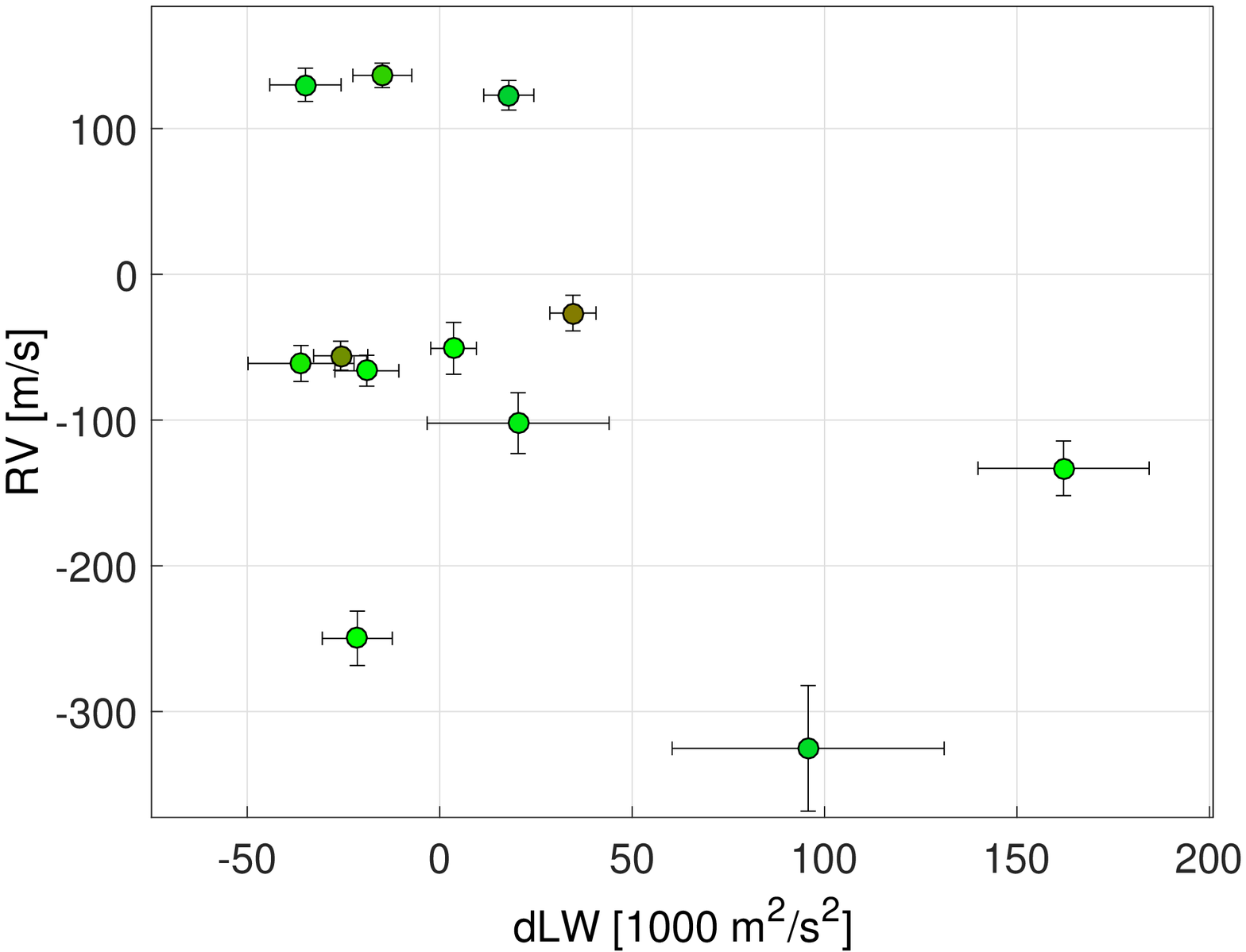}}
\endminipage\hfill
\minipage{0.33\textwidth}
{\includegraphics[width=\linewidth]{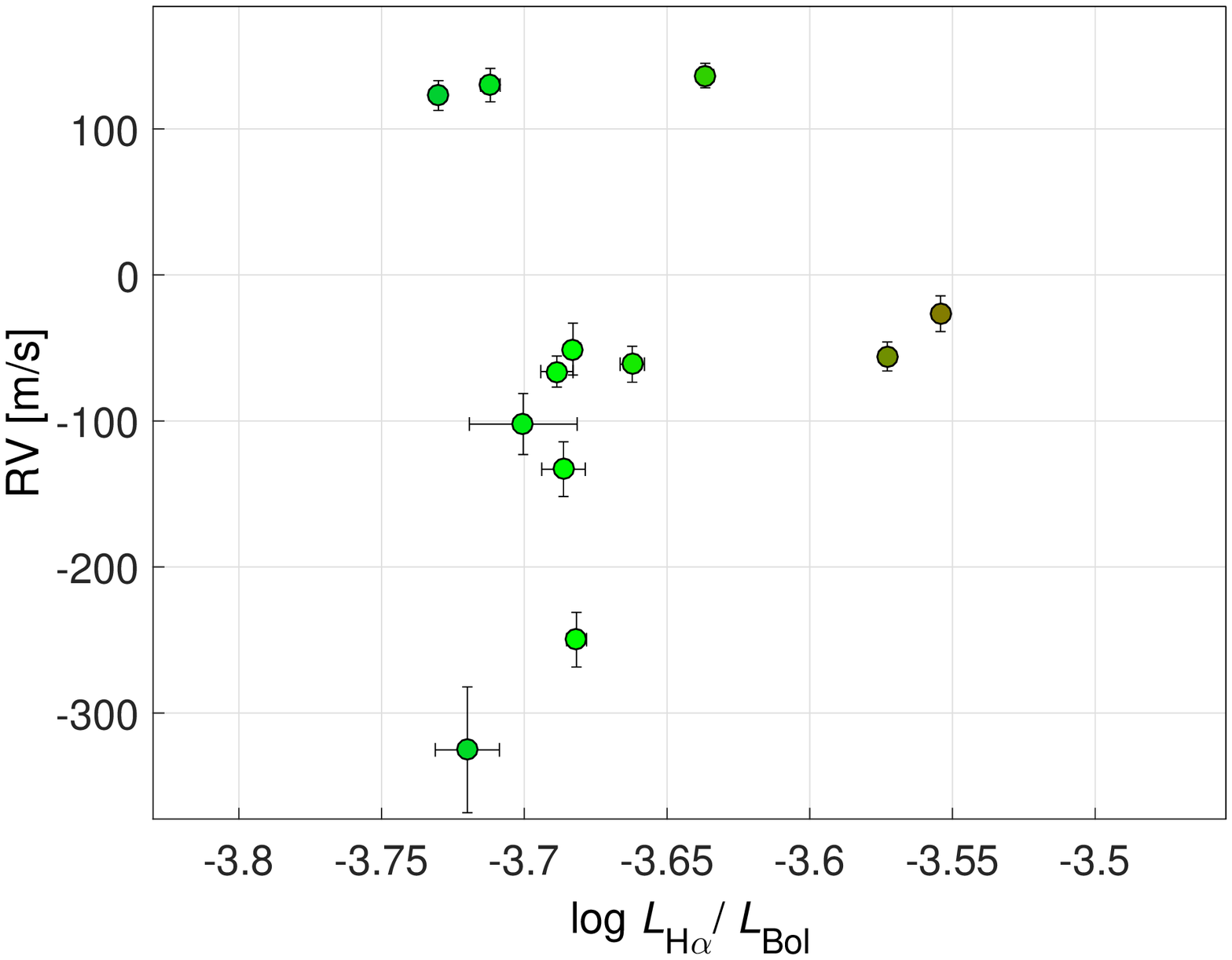}}
\endminipage


\minipage{0.33\textwidth}
{\includegraphics[width=\linewidth]{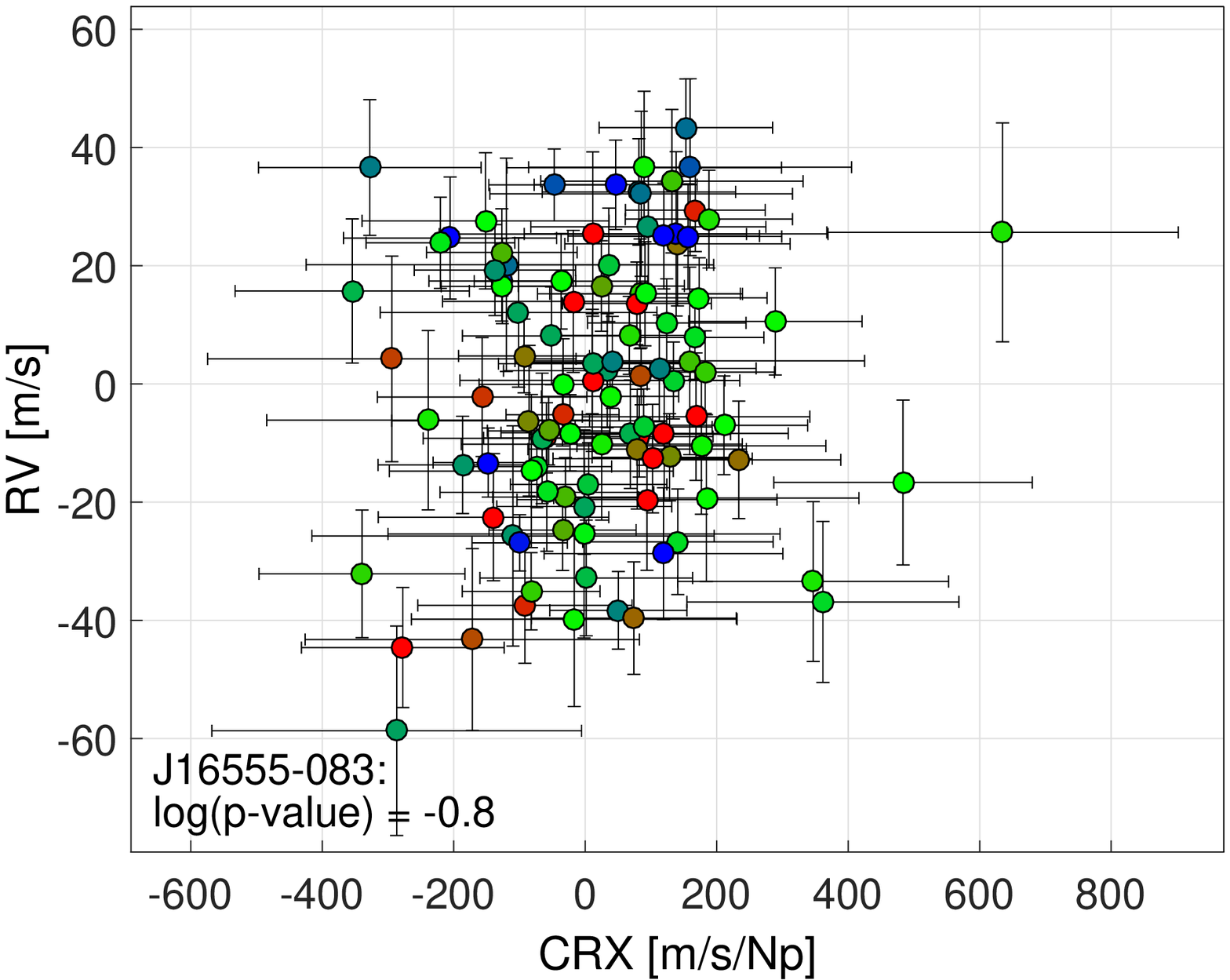}}
\endminipage\hfill
\minipage{0.33\textwidth}
{\includegraphics[width=\linewidth]{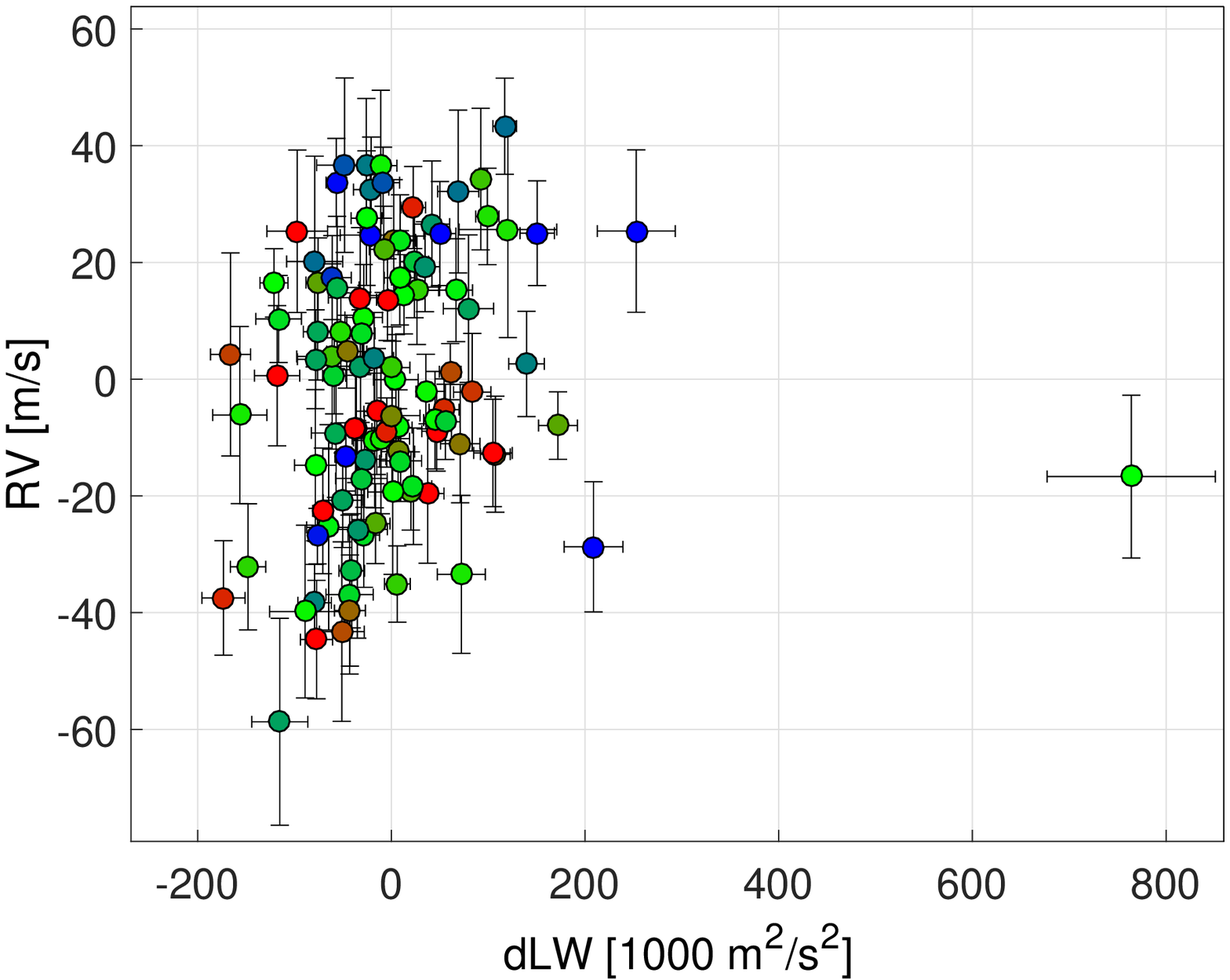}}
\endminipage\hfill
\minipage{0.33\textwidth}
{\includegraphics[width=\linewidth]{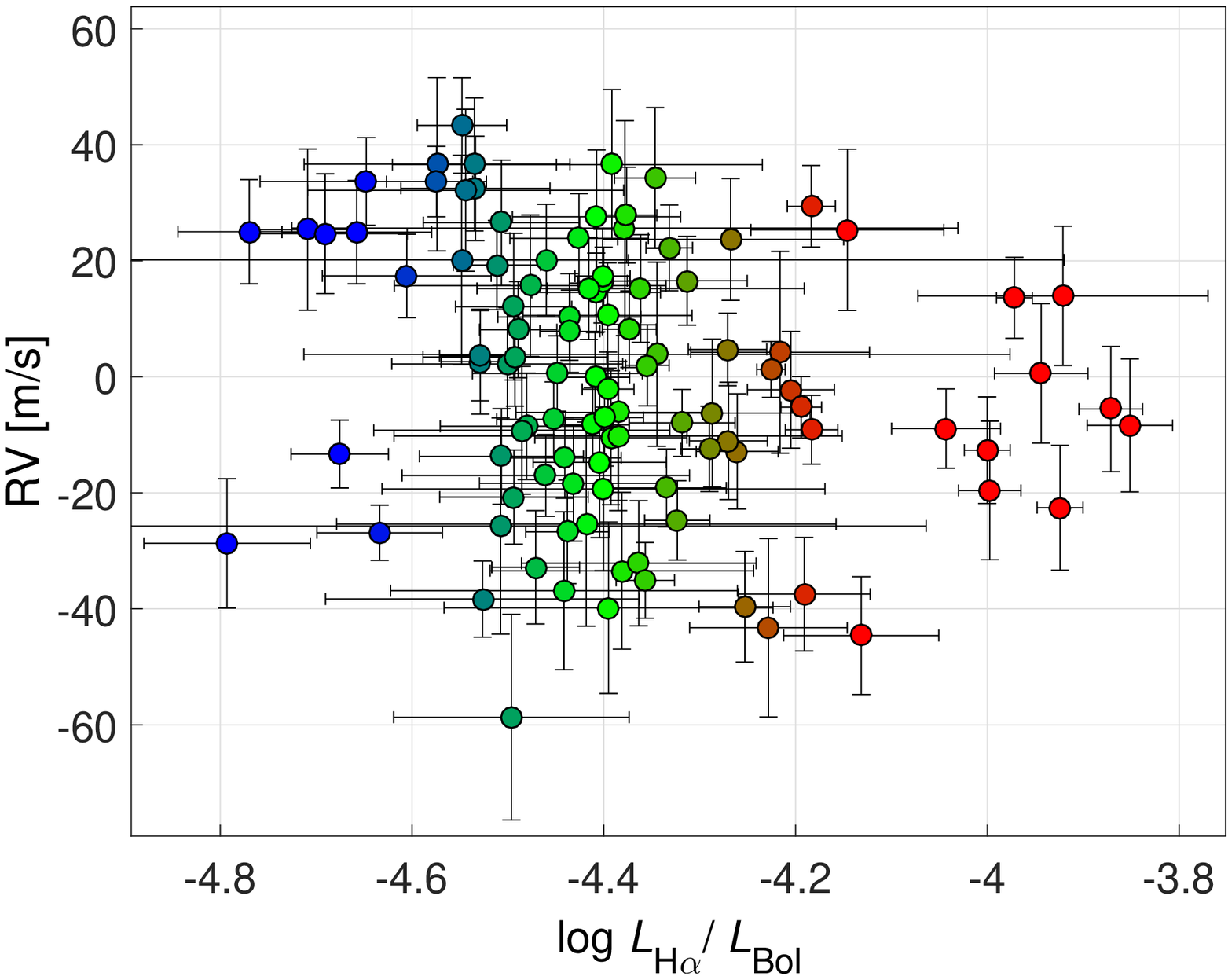}}
\endminipage


\minipage{0.33\textwidth}
{\includegraphics[width=\linewidth]{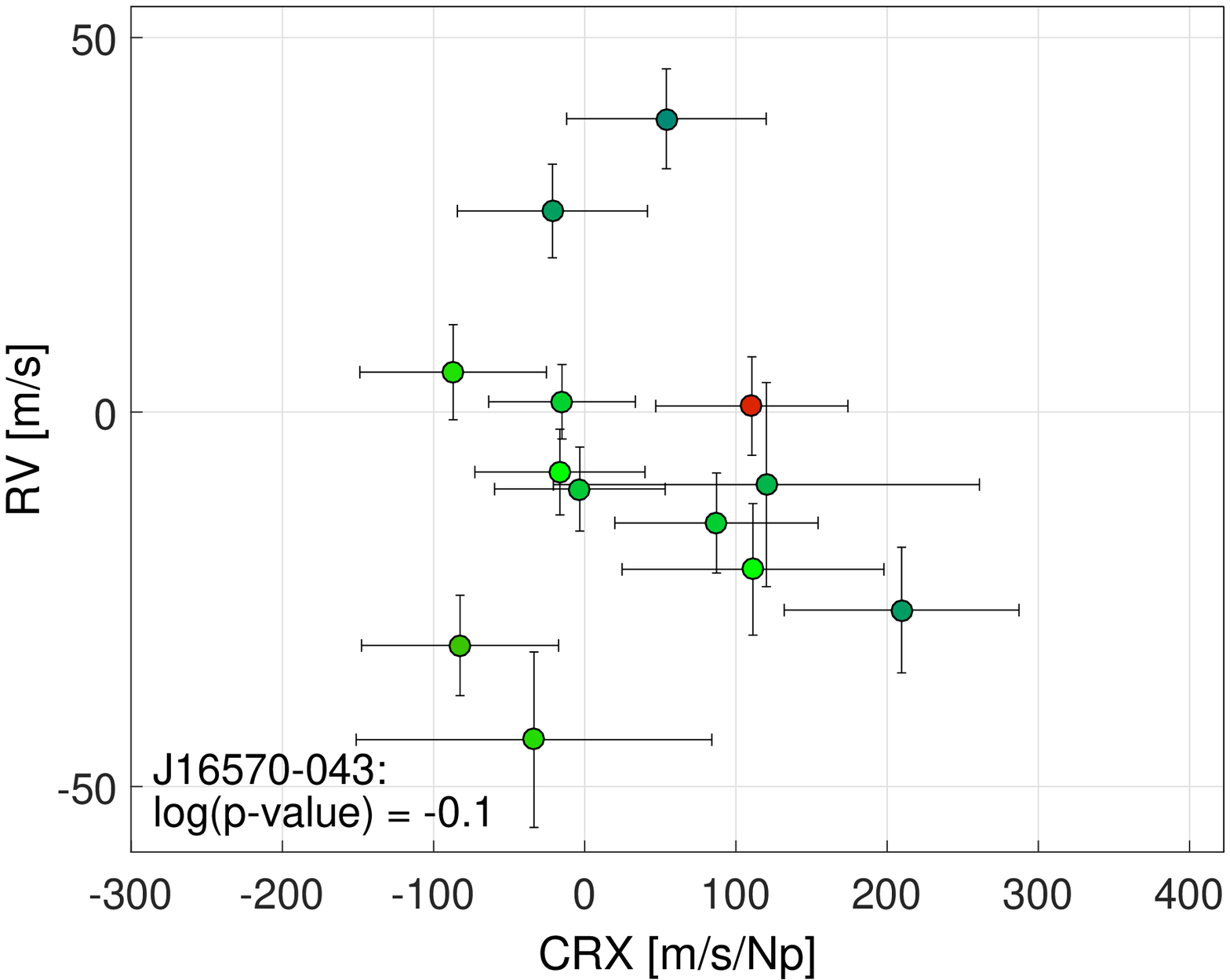}}
\endminipage\hfill
\minipage{0.33\textwidth}
{\includegraphics[width=\linewidth]{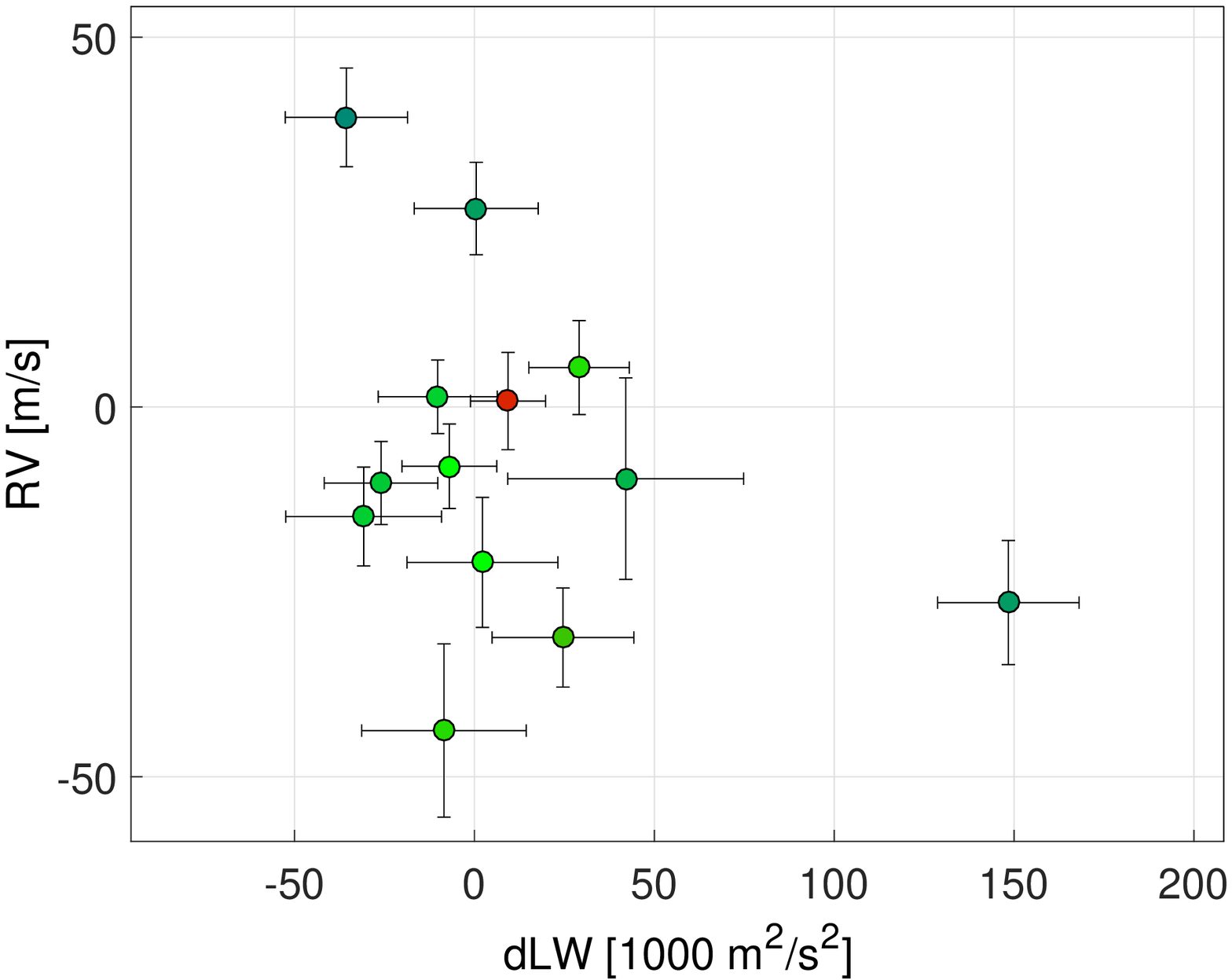}}
\endminipage\hfill
\minipage{0.33\textwidth}
{\includegraphics[width=\linewidth]{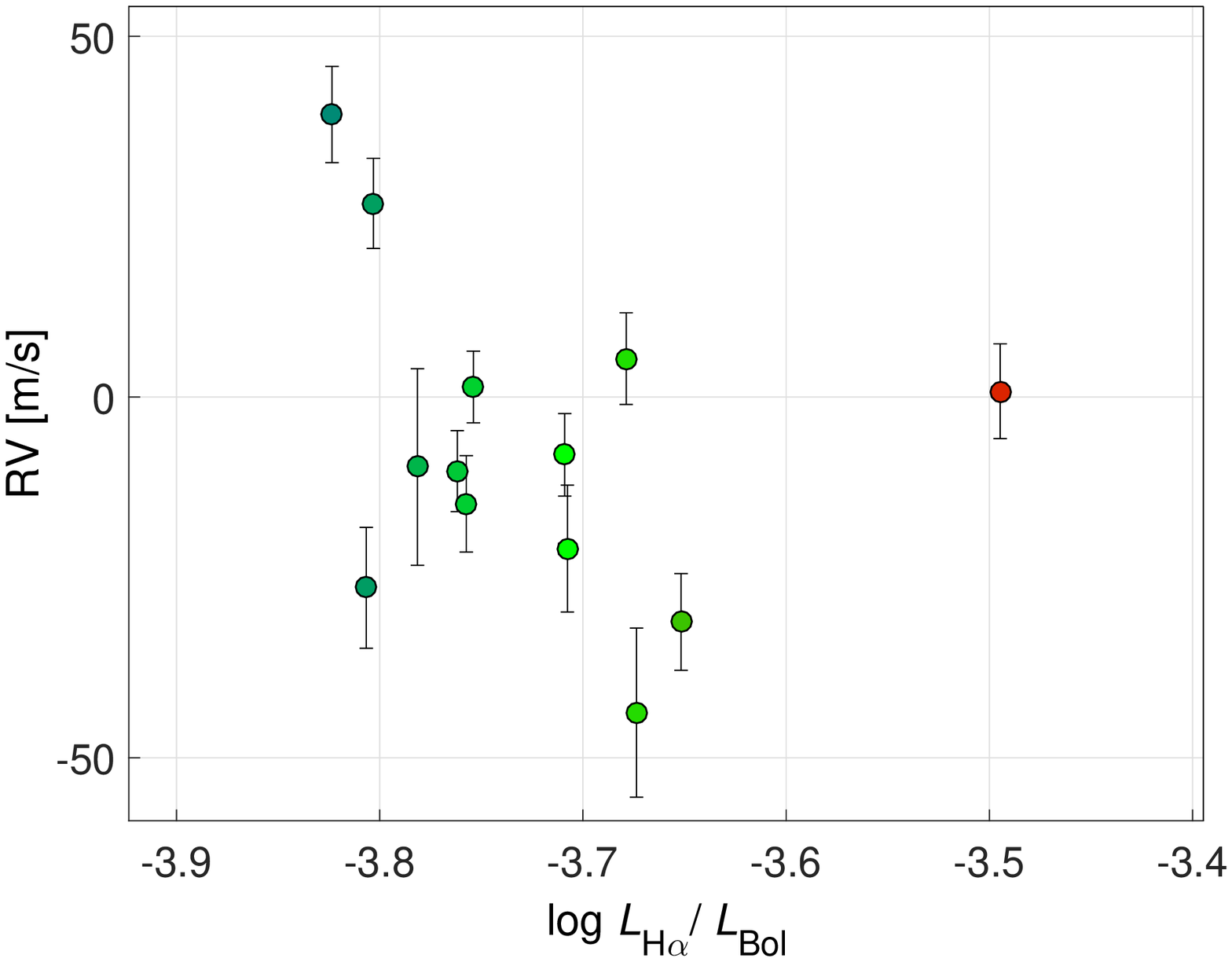}}
\endminipage


\minipage{0.33\textwidth}
{\includegraphics[width=\linewidth]{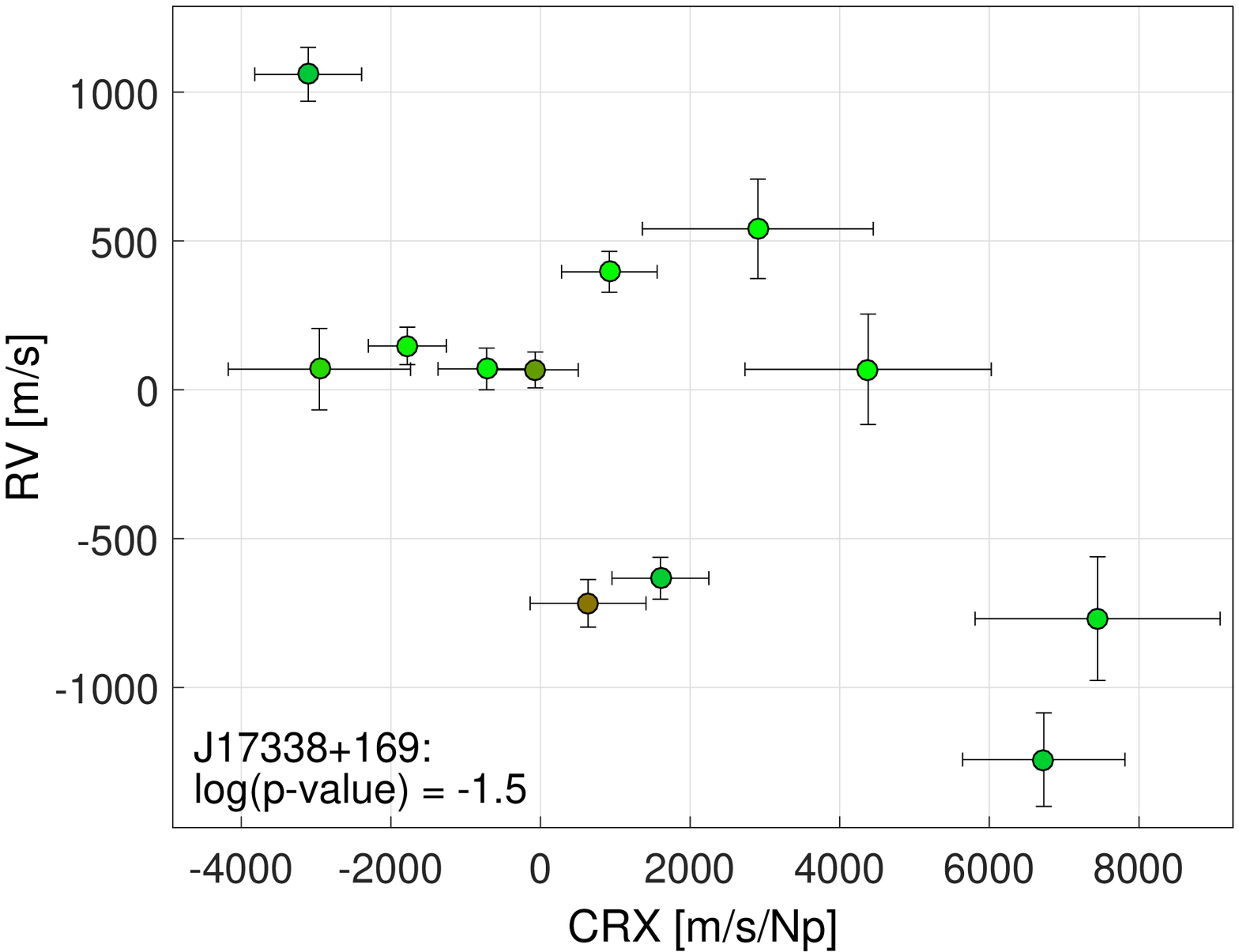}}
\endminipage\hfill
\minipage{0.33\textwidth}
{\includegraphics[width=\linewidth]{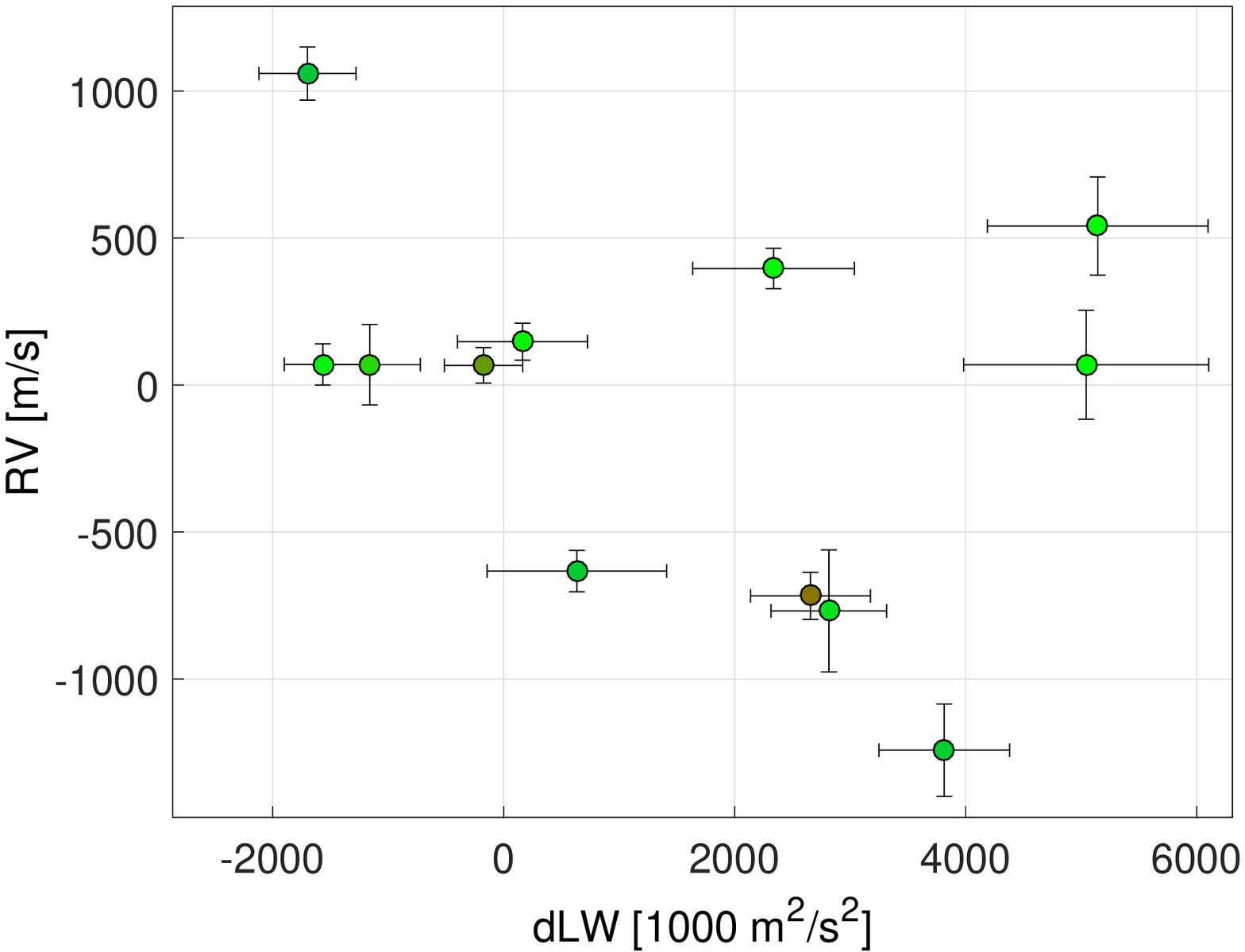}}
\endminipage\hfill
\minipage{0.33\textwidth}
{\includegraphics[width=\linewidth]{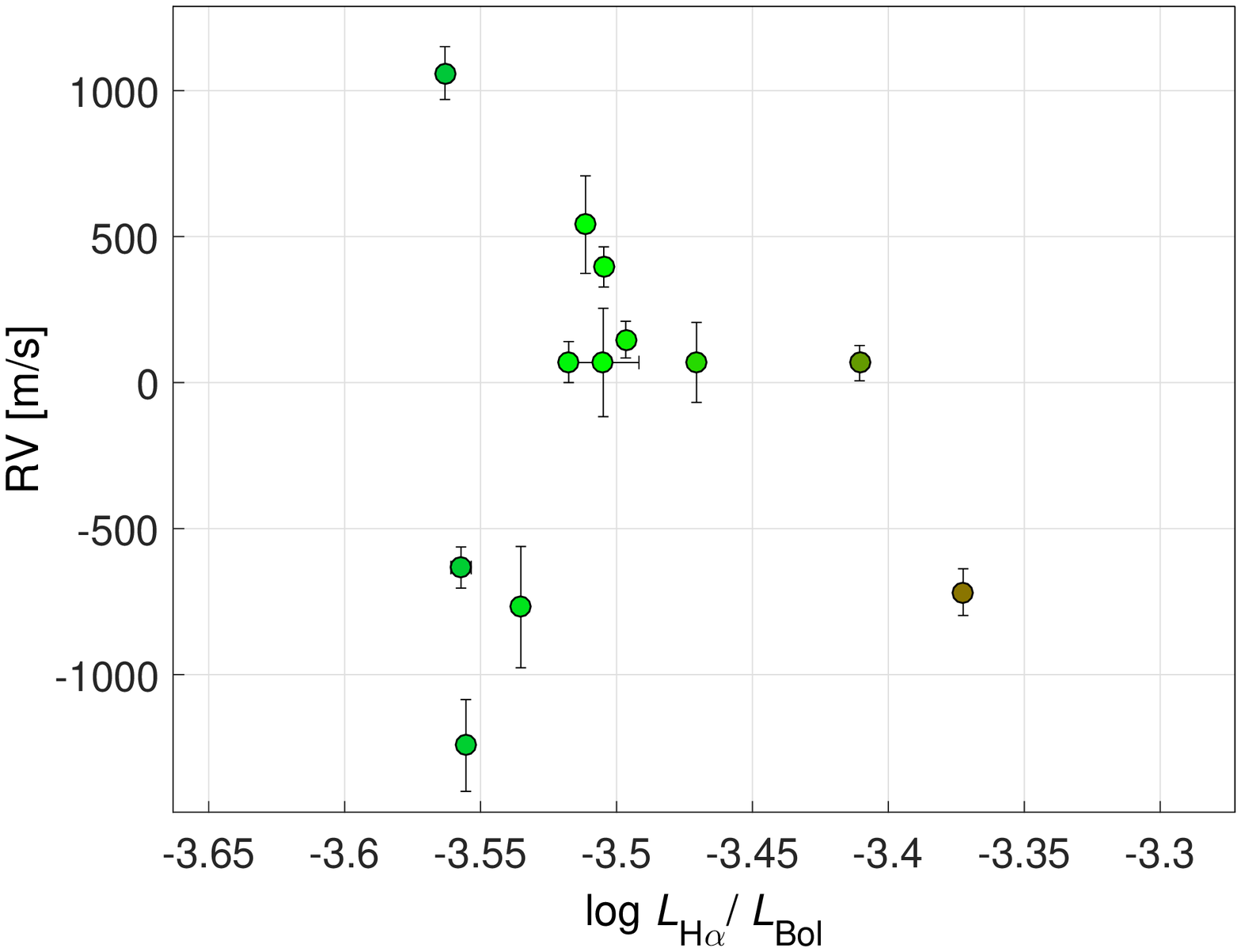}}
\endminipage


\minipage{0.33\textwidth}
{\includegraphics[width=\linewidth]{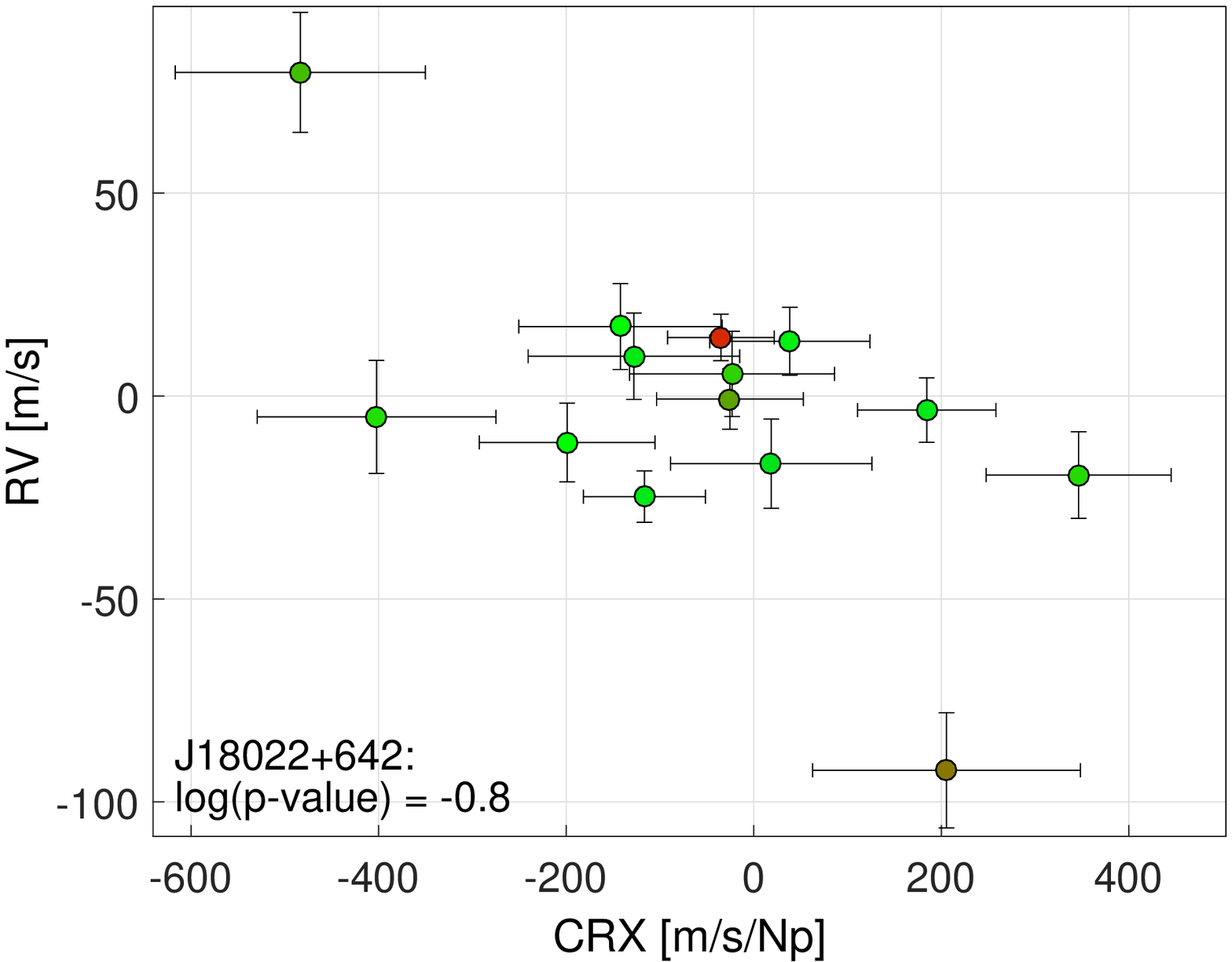}}
\endminipage\hfill
\minipage{0.33\textwidth}
{\includegraphics[width=\linewidth]{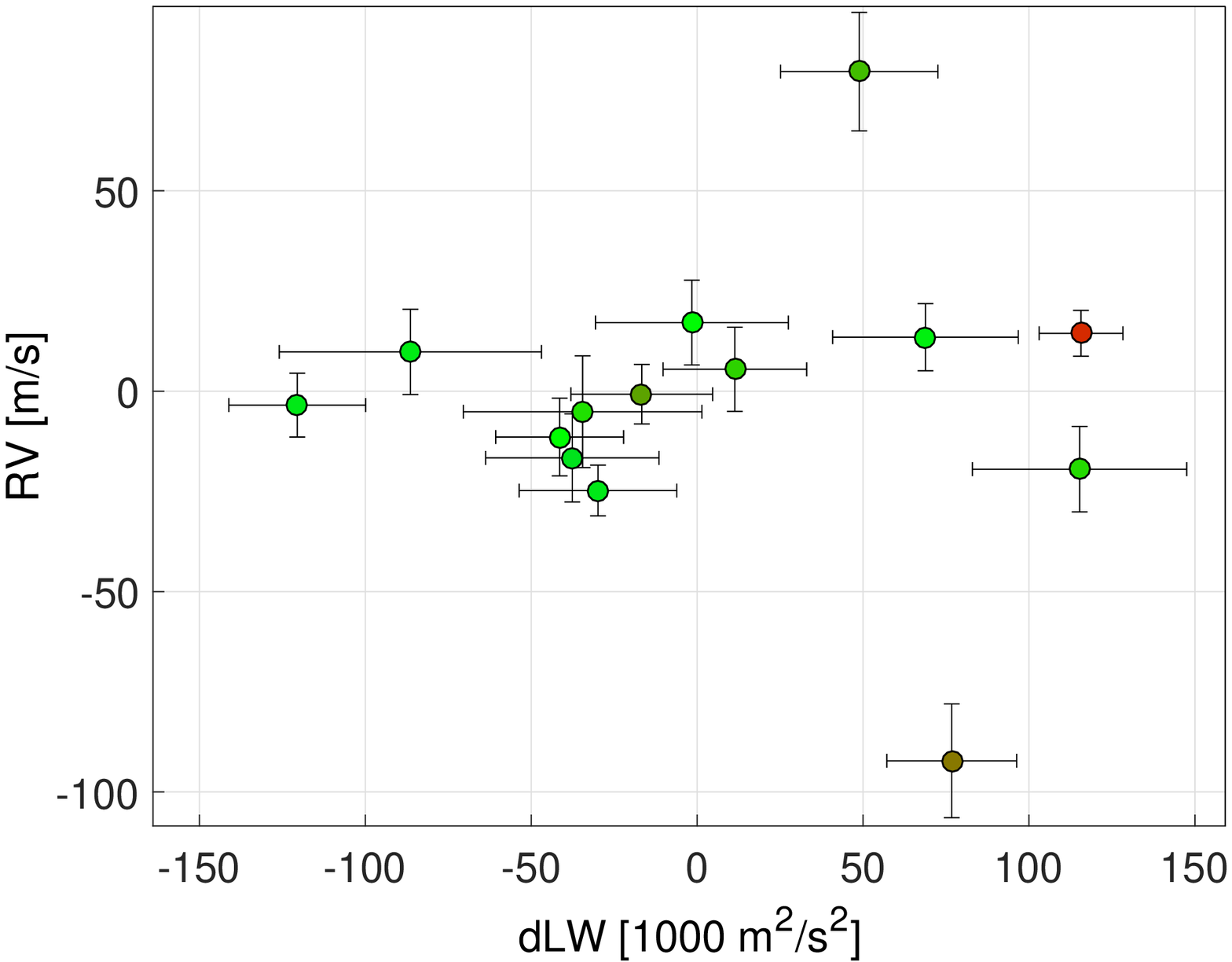}}
\endminipage\hfill
\minipage{0.33\textwidth}
{\includegraphics[width=\linewidth]{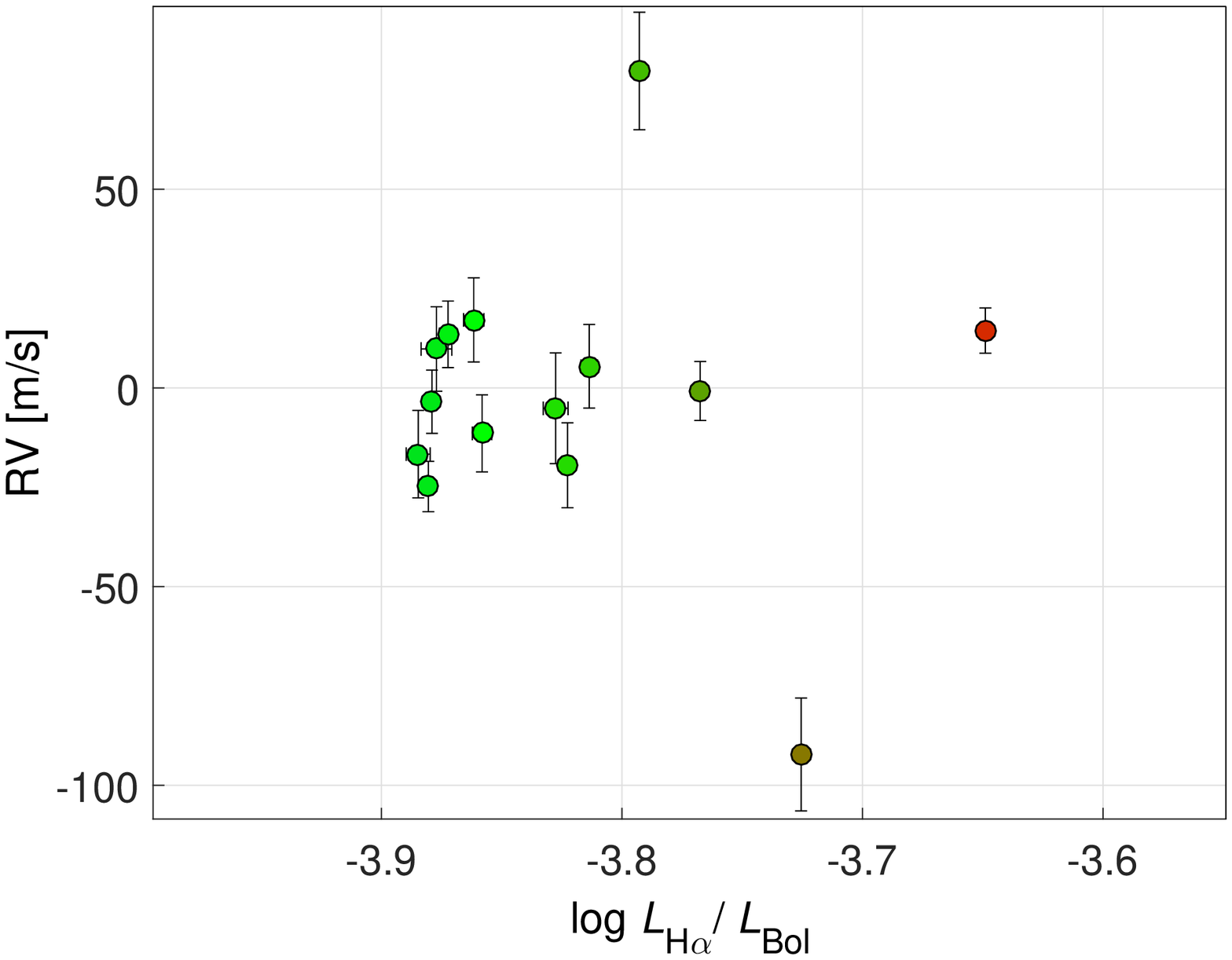}}
\endminipage
\caption{Continued.}
\label{figA1}
\end{figure*}

\addtocounter{figure}{-1}

\begin{figure*}[!htp]
\minipage{0.33\textwidth}
{\includegraphics[width=\linewidth]{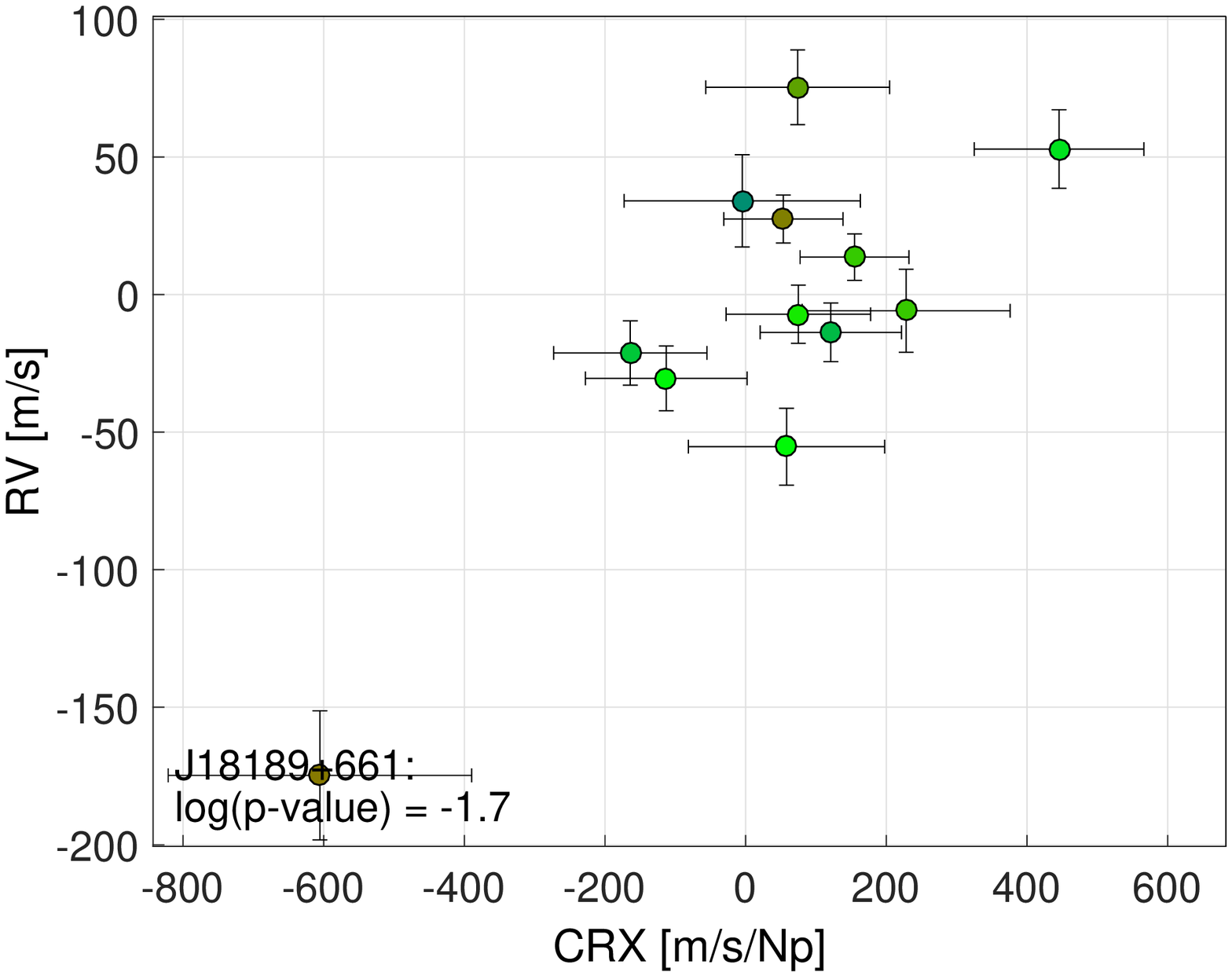}}
\endminipage\hfill
\minipage{0.33\textwidth}
{\includegraphics[width=\linewidth]{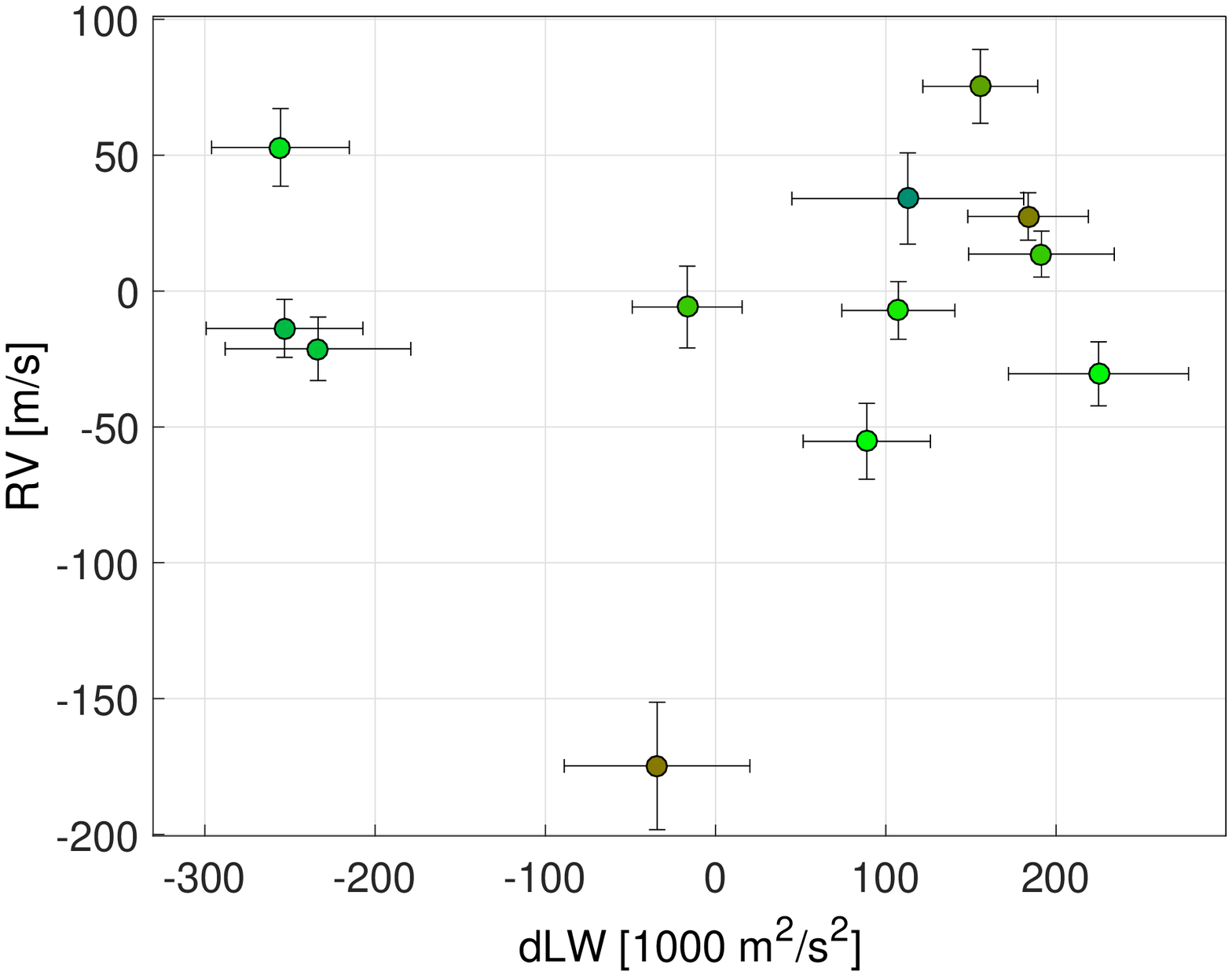}}
\endminipage\hfill
\minipage{0.33\textwidth}
{\includegraphics[width=\linewidth]{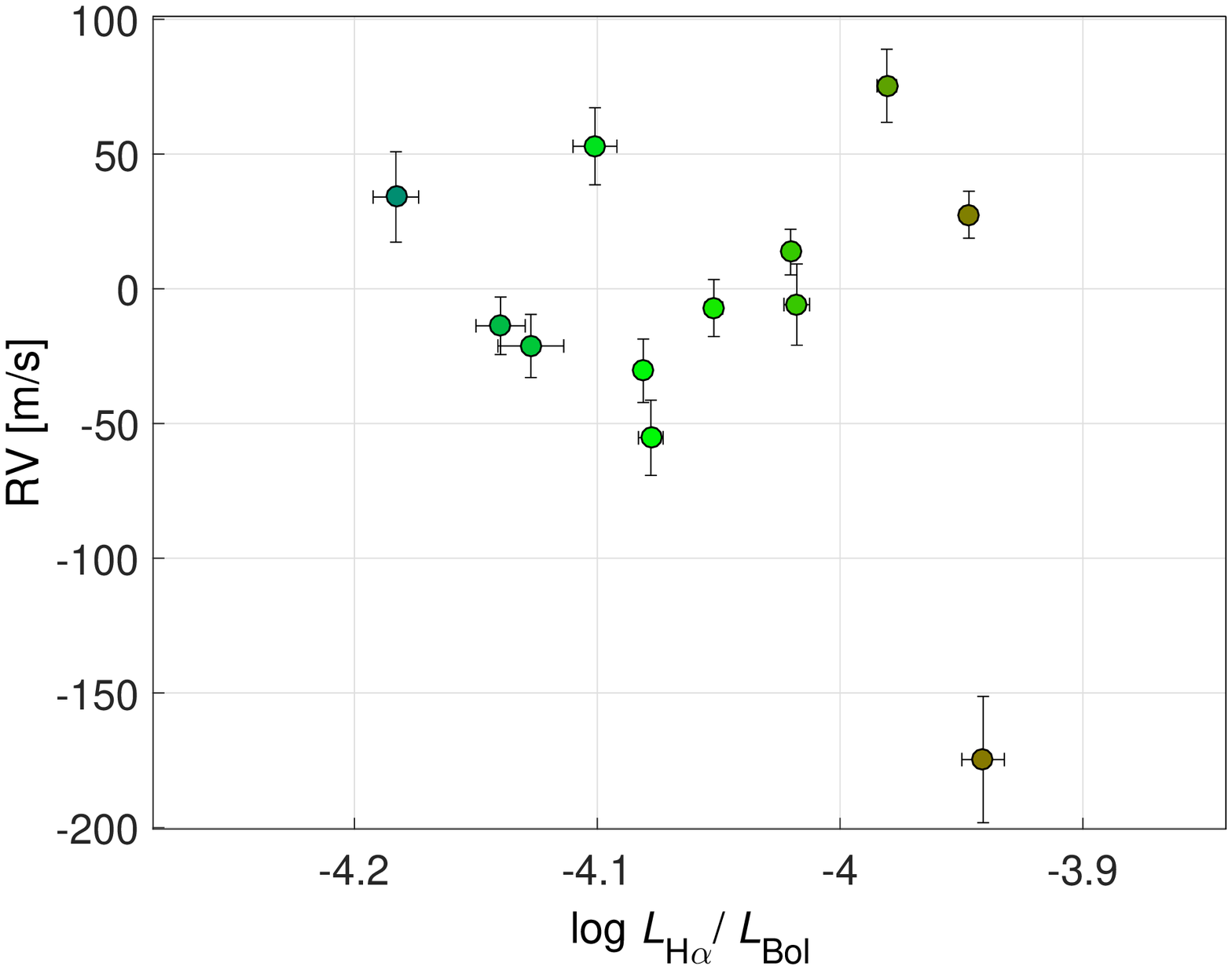}}
\endminipage


\minipage{0.33\textwidth}
{\includegraphics[width=\linewidth]{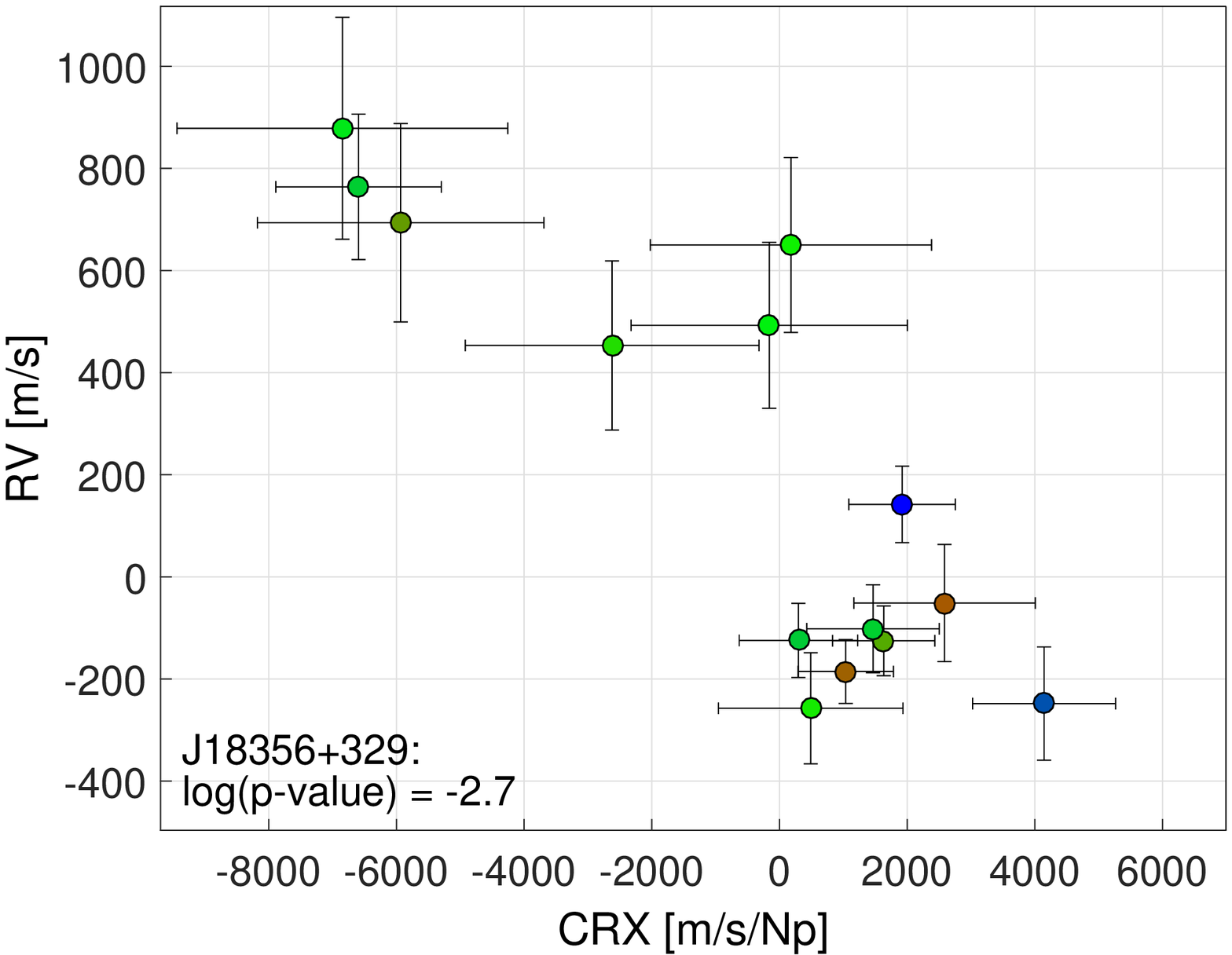}}
\endminipage\hfill
\minipage{0.33\textwidth}
{\includegraphics[width=\linewidth]{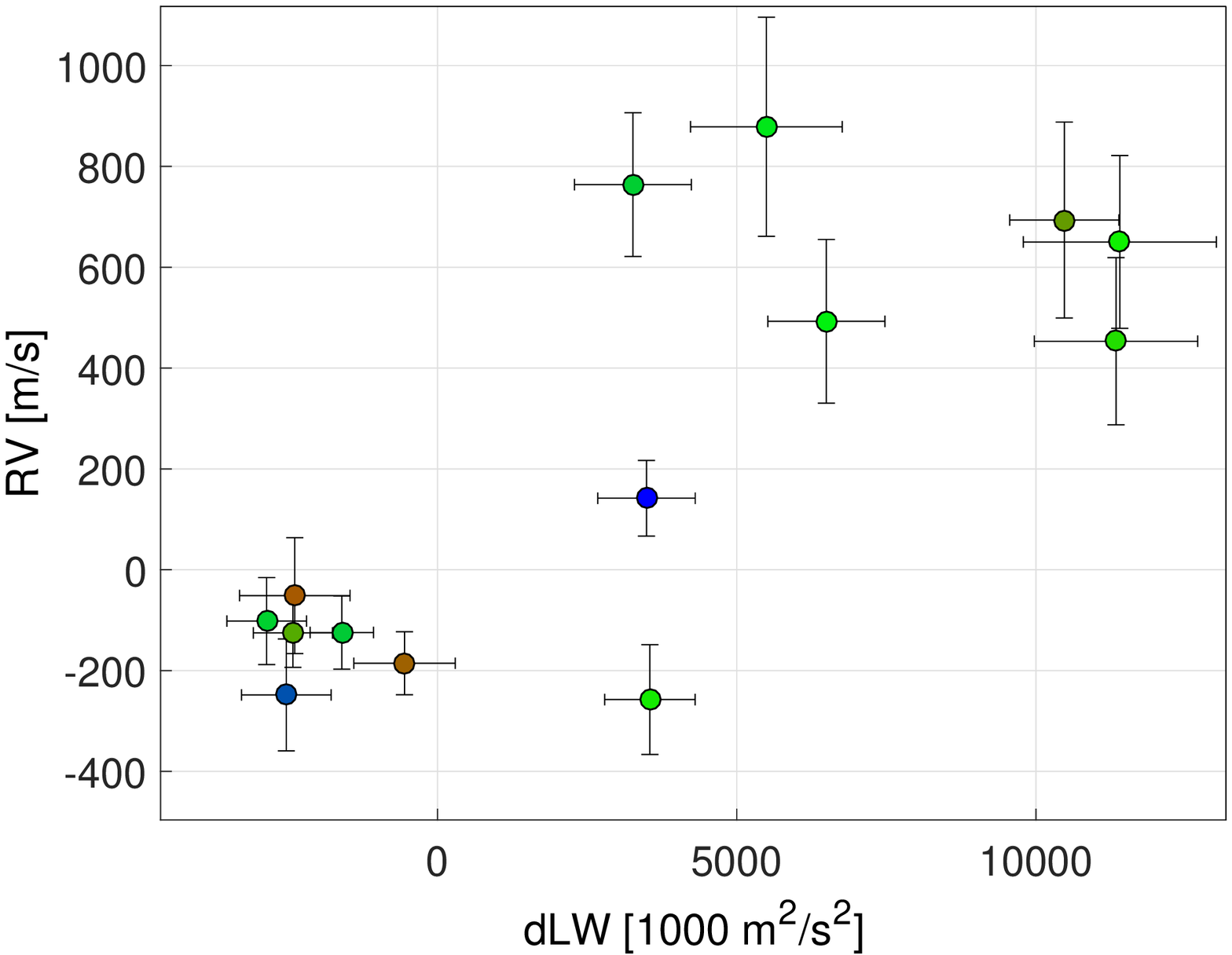}}
\endminipage\hfill
\minipage{0.33\textwidth}
{\includegraphics[width=\linewidth]{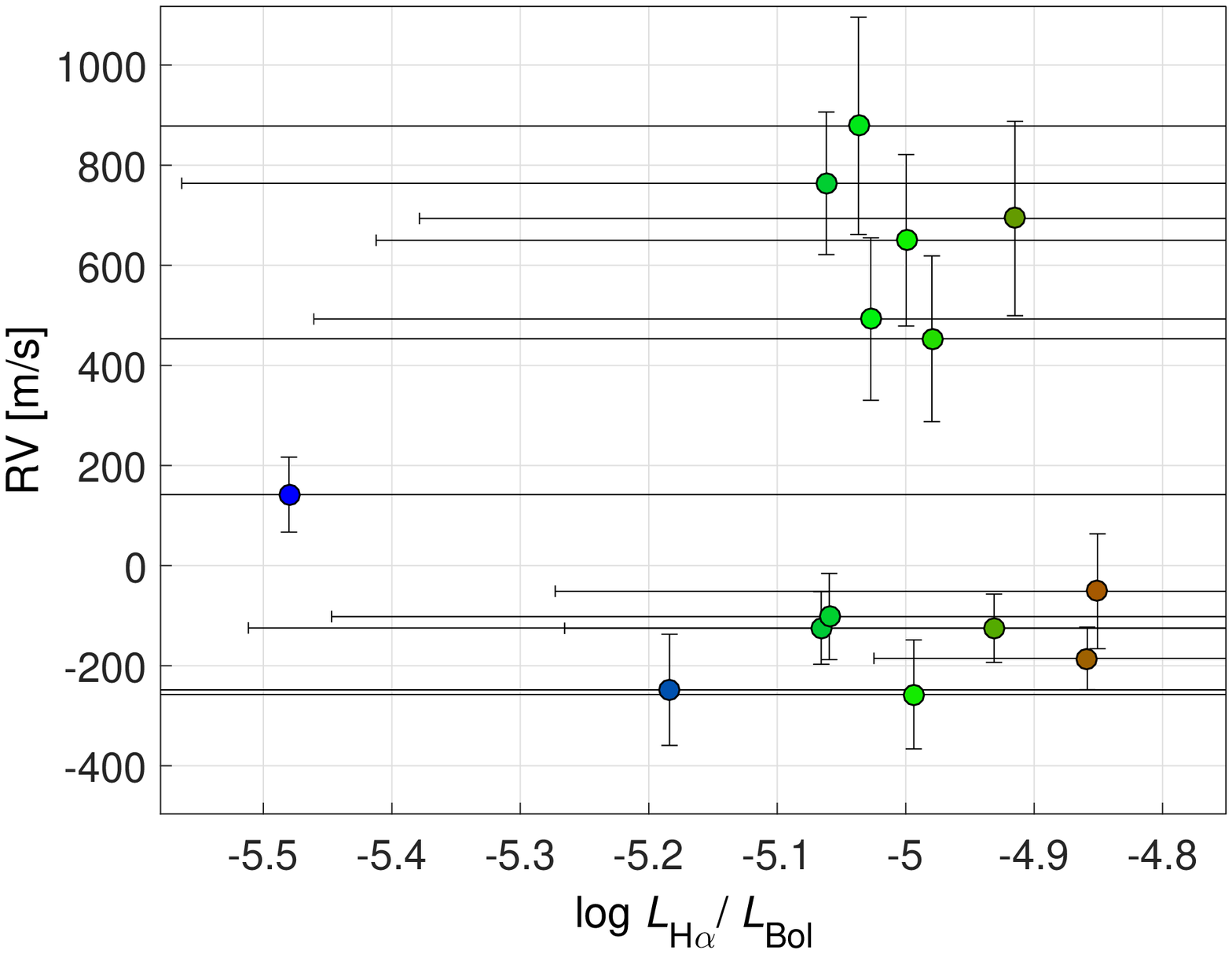}}
\endminipage


\minipage{0.33\textwidth}
{\includegraphics[width=\linewidth]{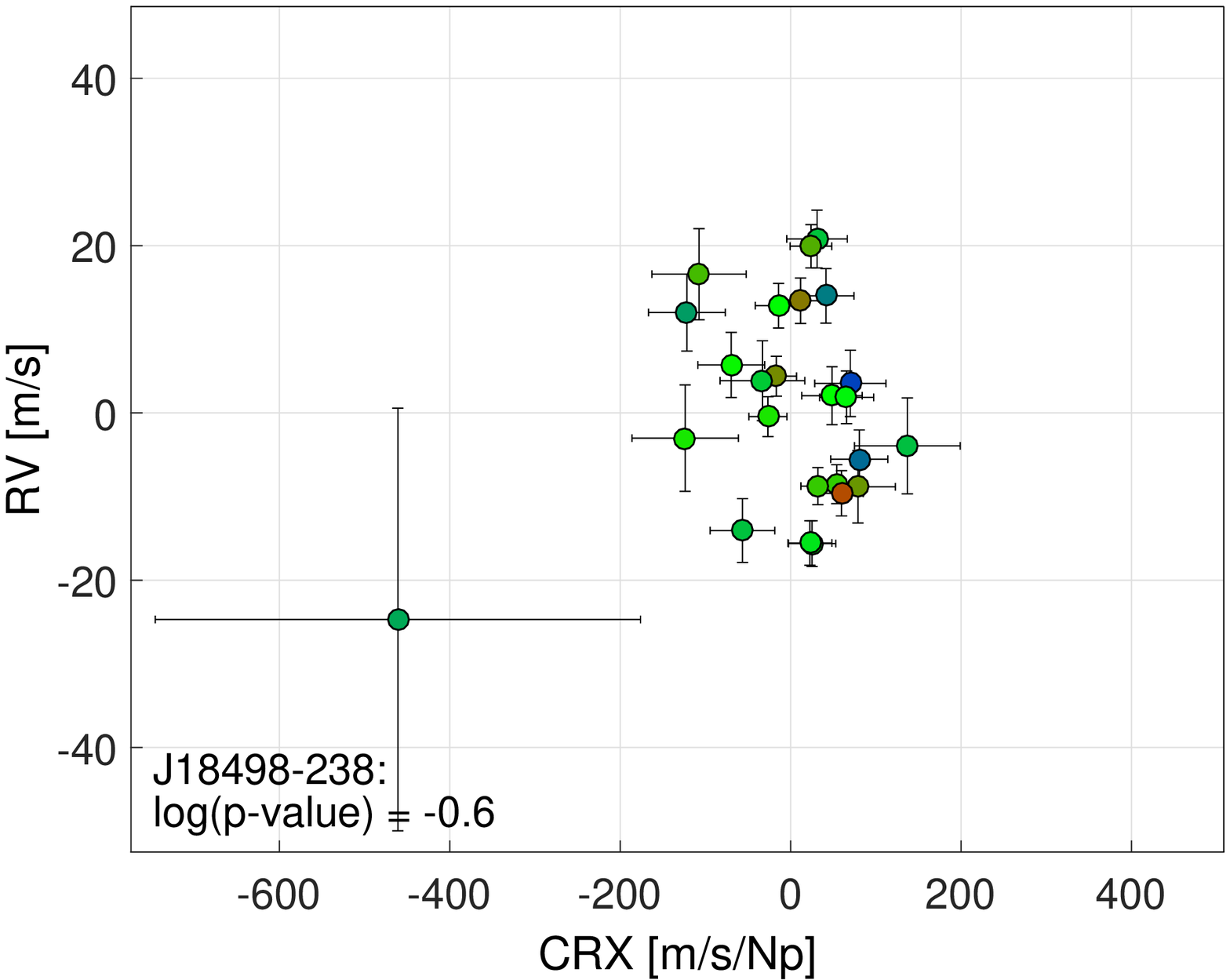}}
\endminipage\hfill
\minipage{0.33\textwidth}
{\includegraphics[width=\linewidth]{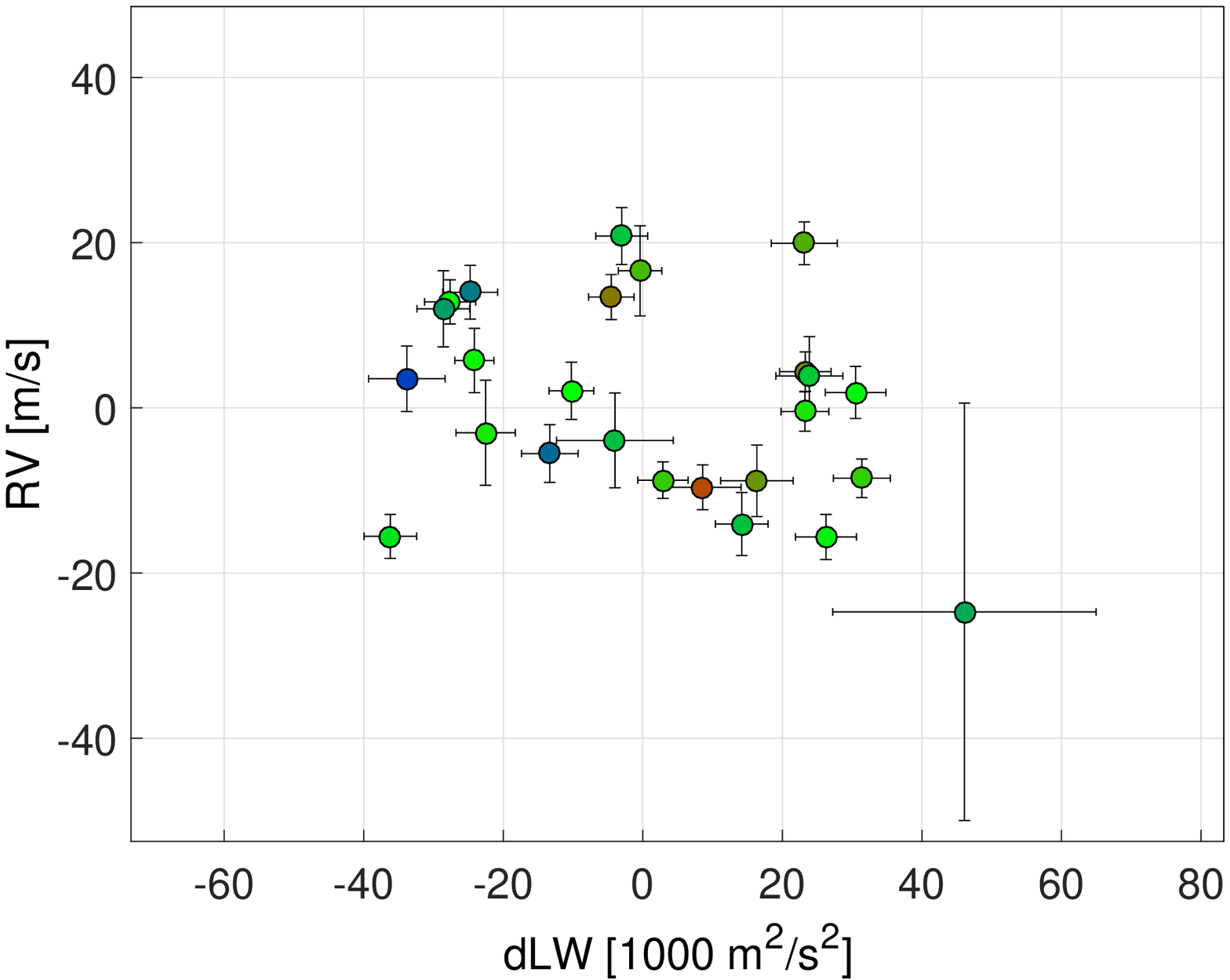}}
\endminipage\hfill
\minipage{0.33\textwidth}
{\includegraphics[width=\linewidth]{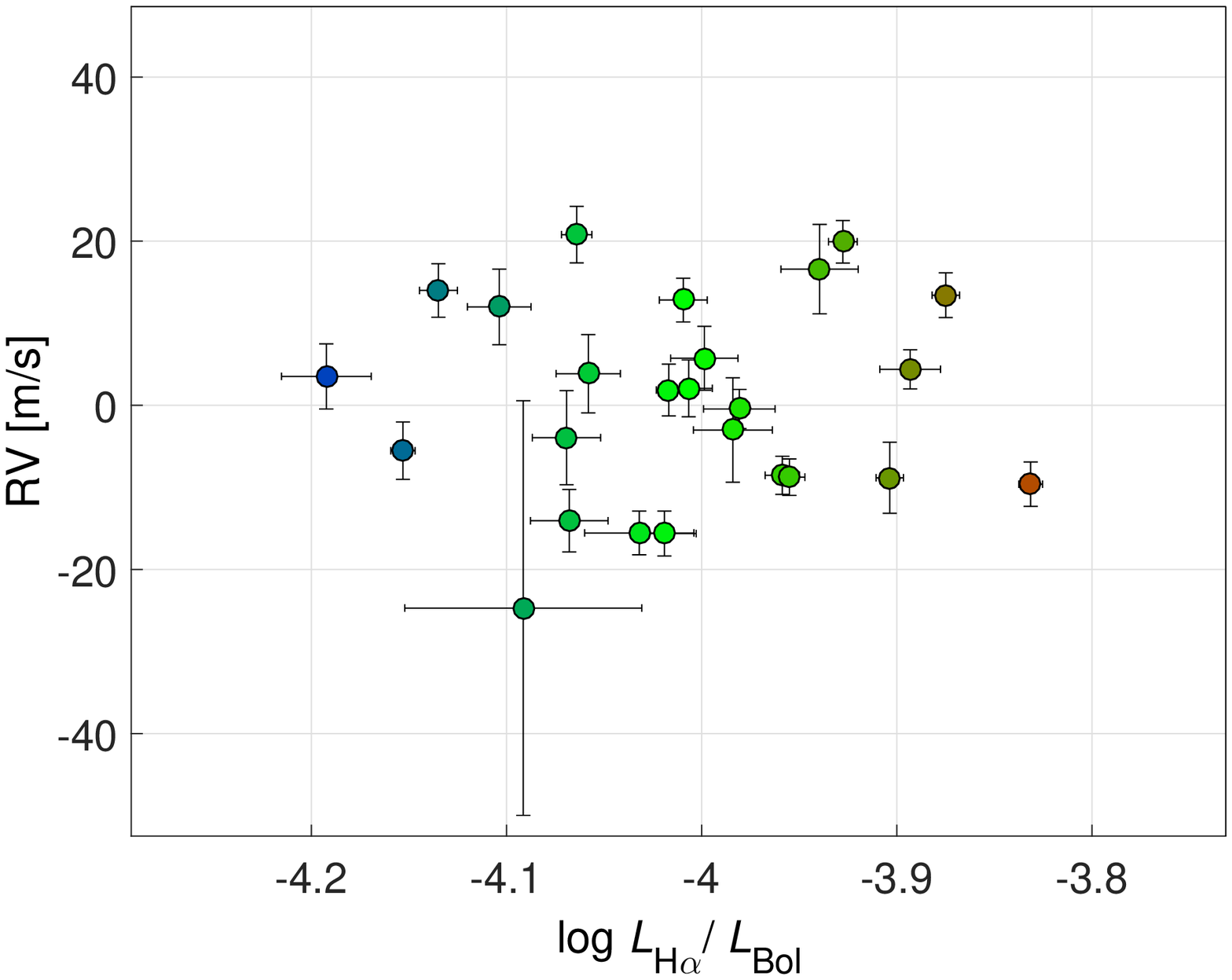}}
\endminipage


\minipage{0.33\textwidth}
{\includegraphics[width=\linewidth]{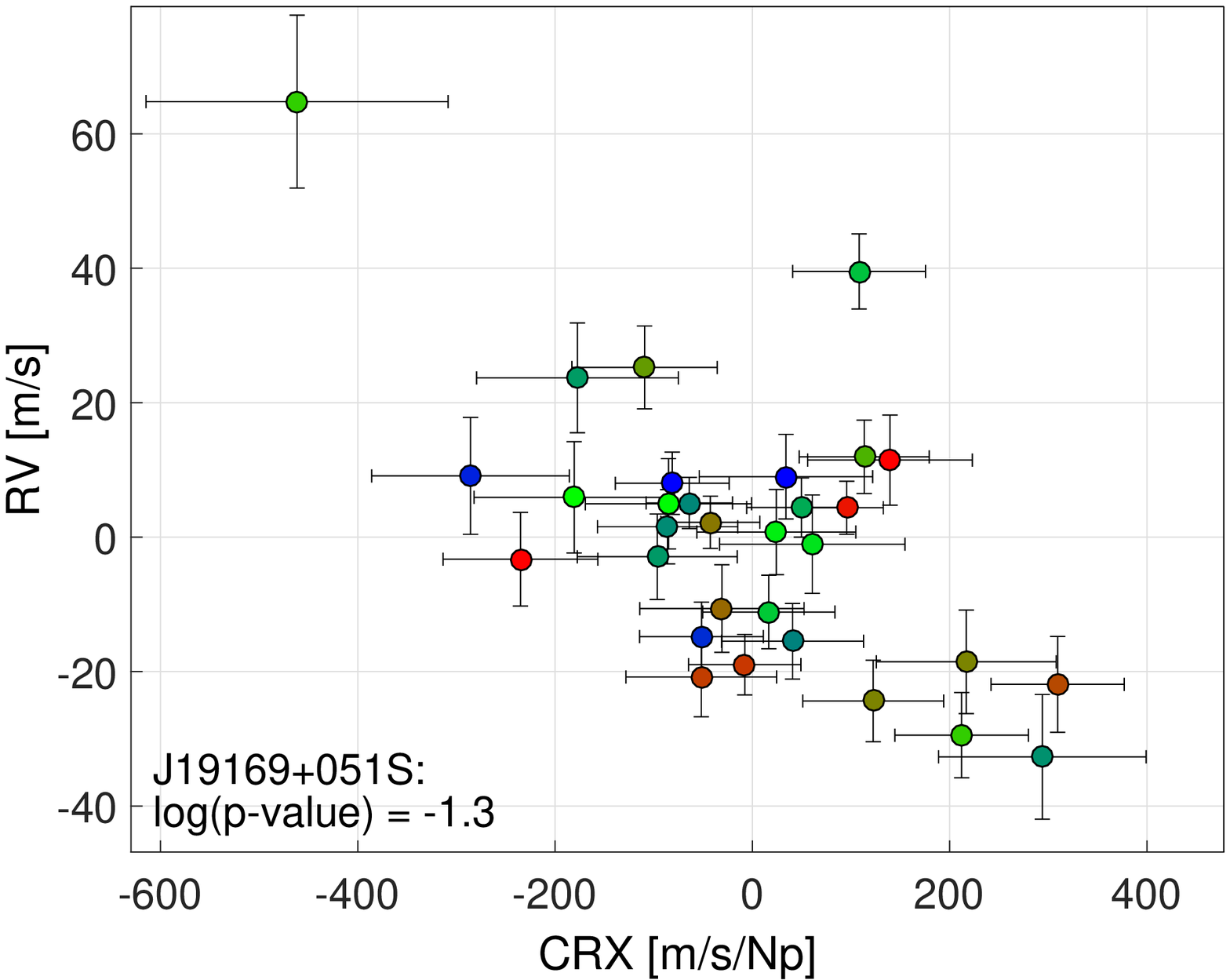}}
\endminipage\hfill
\minipage{0.33\textwidth}
{\includegraphics[width=\linewidth]{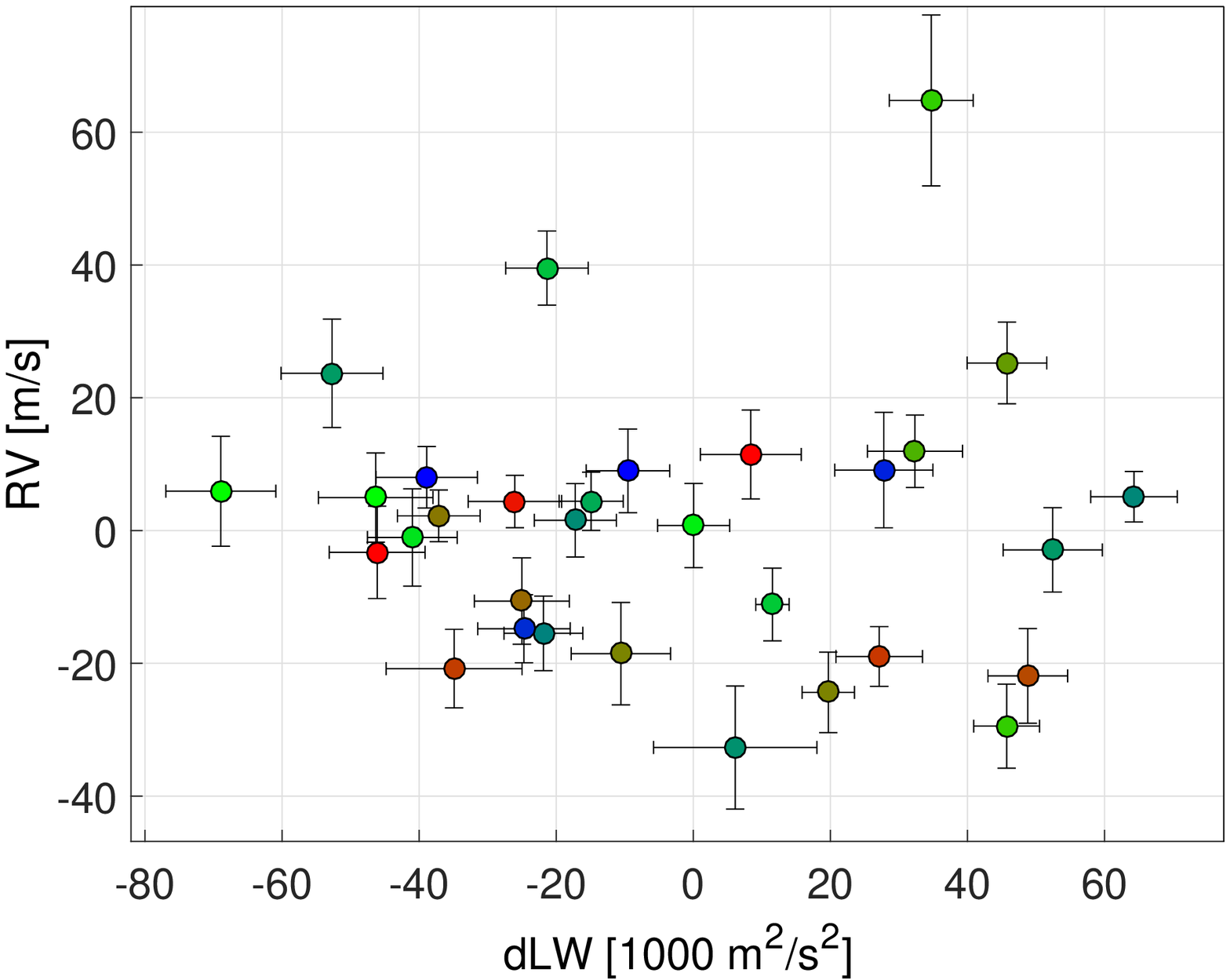}}
\endminipage\hfill
\minipage{0.33\textwidth}
{\includegraphics[width=\linewidth]{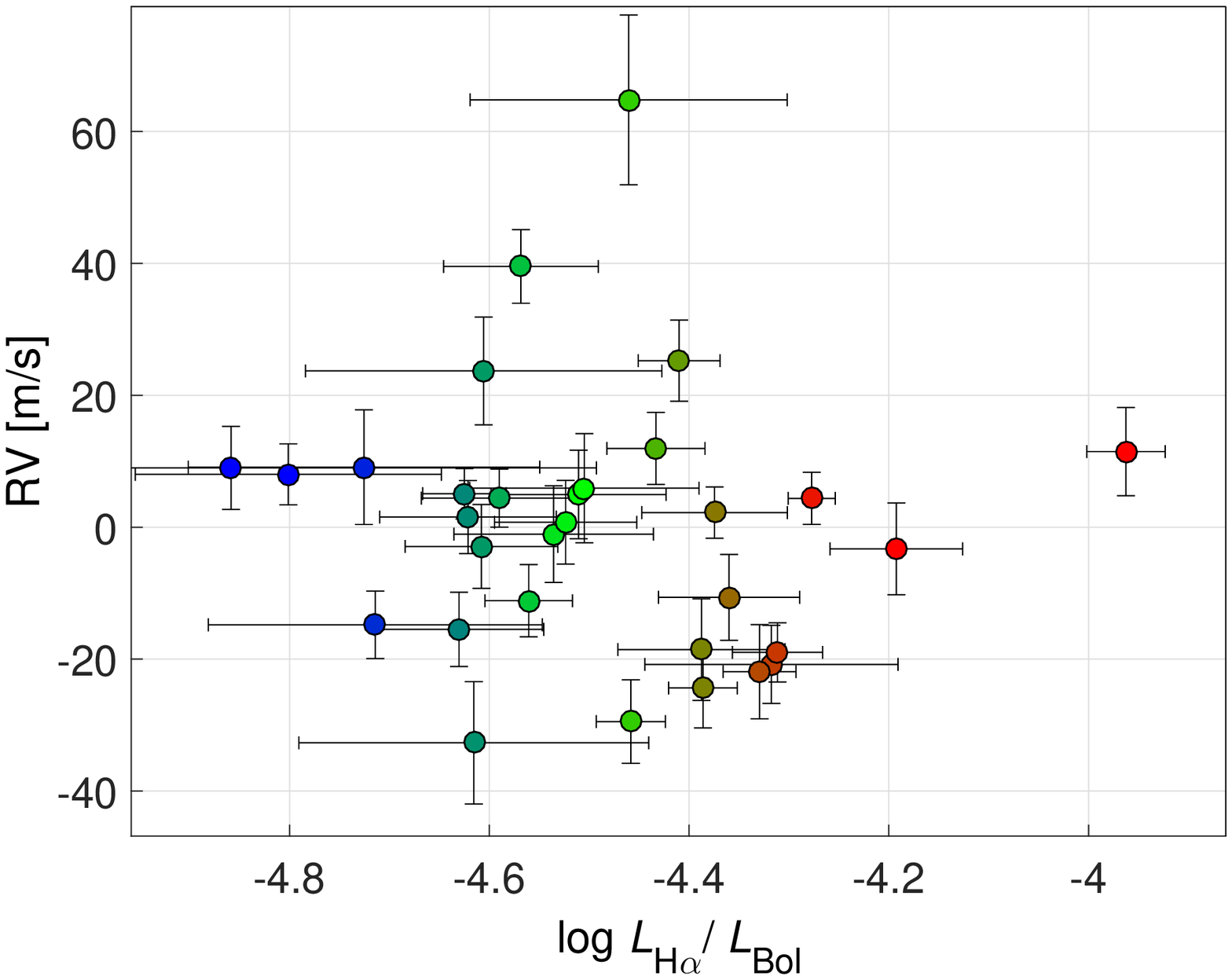}}
\endminipage


\minipage{0.33\textwidth}
{\includegraphics[width=\linewidth]{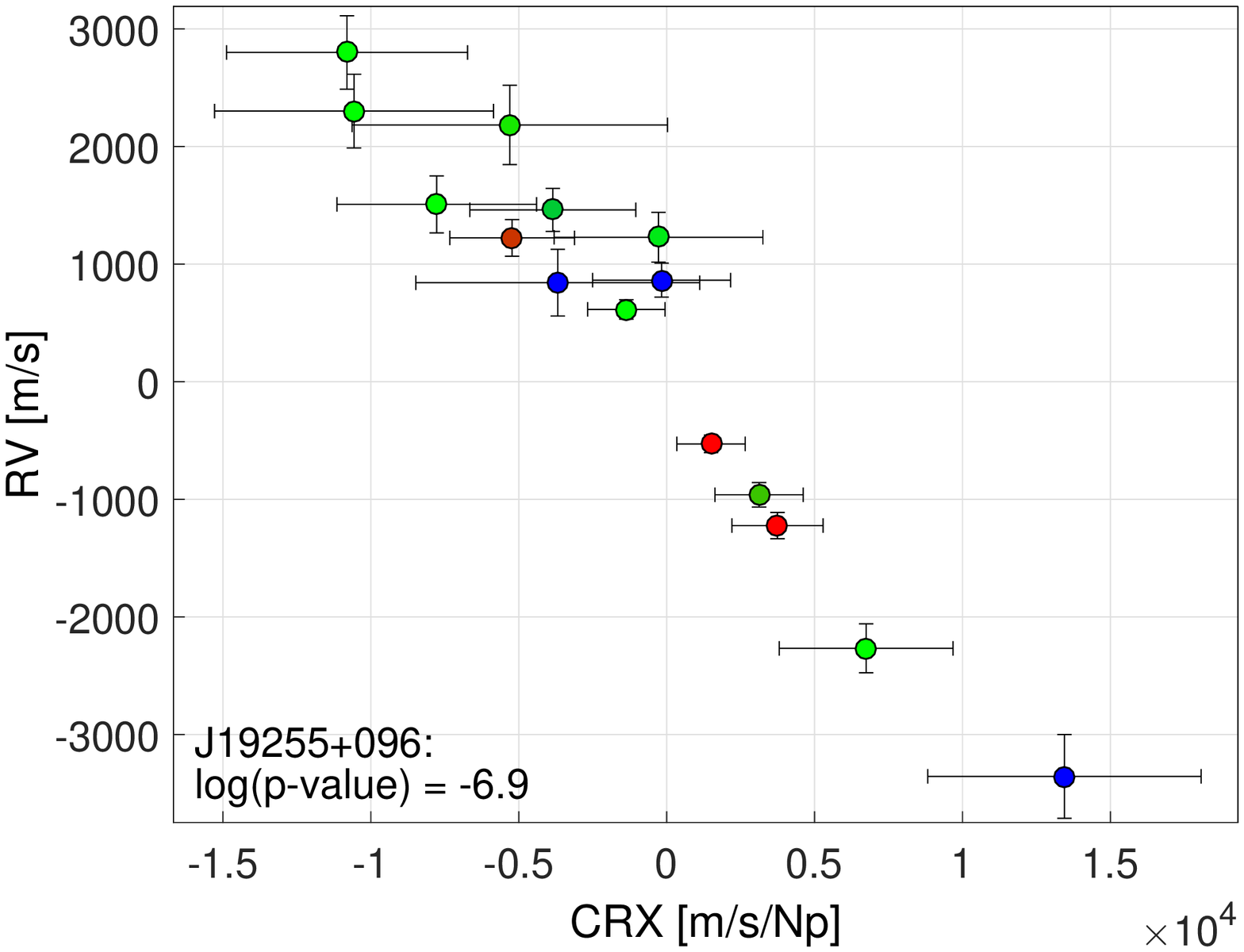}}
\endminipage\hfill
\minipage{0.33\textwidth}
{\includegraphics[width=\linewidth]{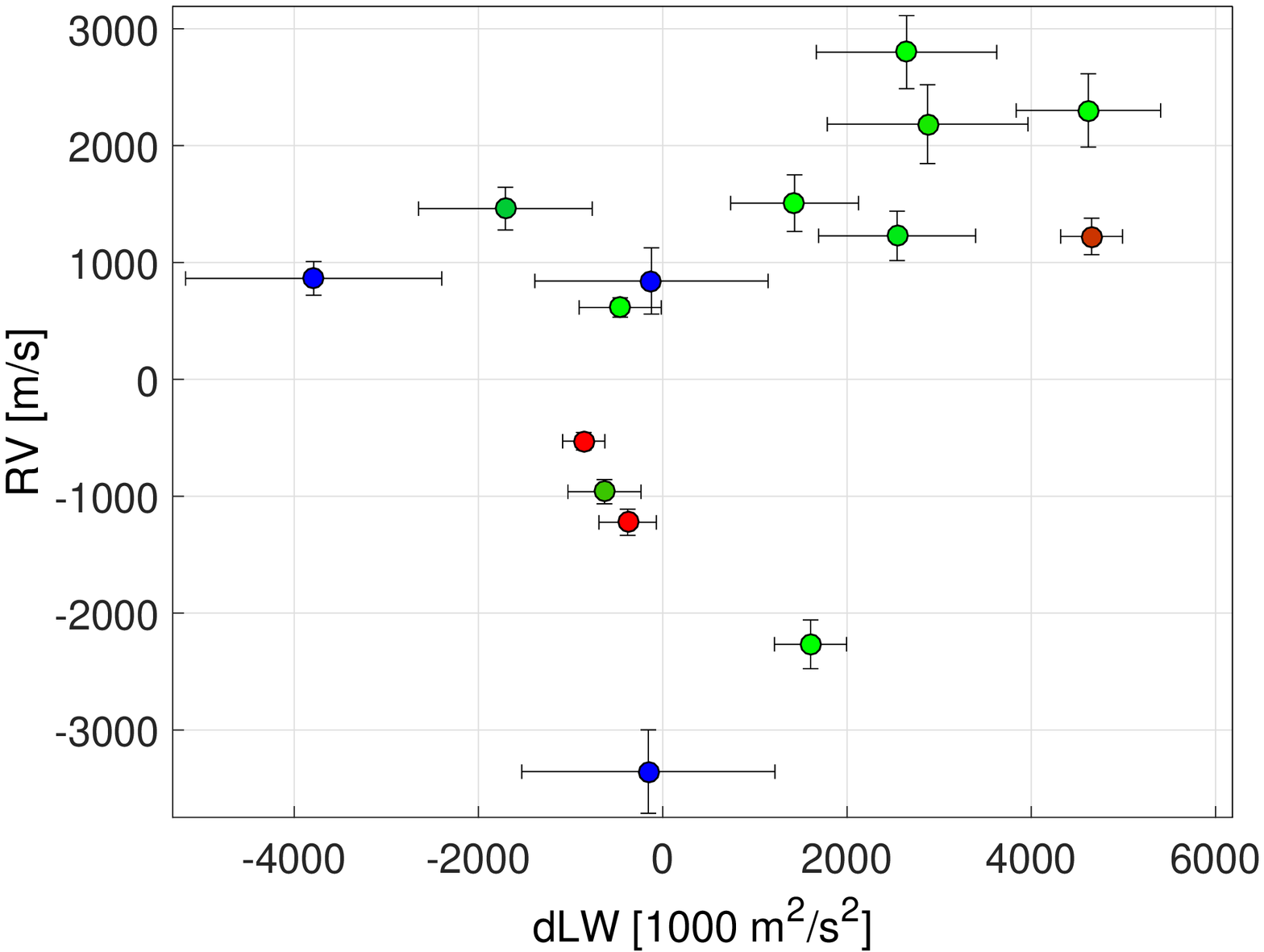}}
\endminipage\hfill
\minipage{0.33\textwidth}
{\includegraphics[width=\linewidth]{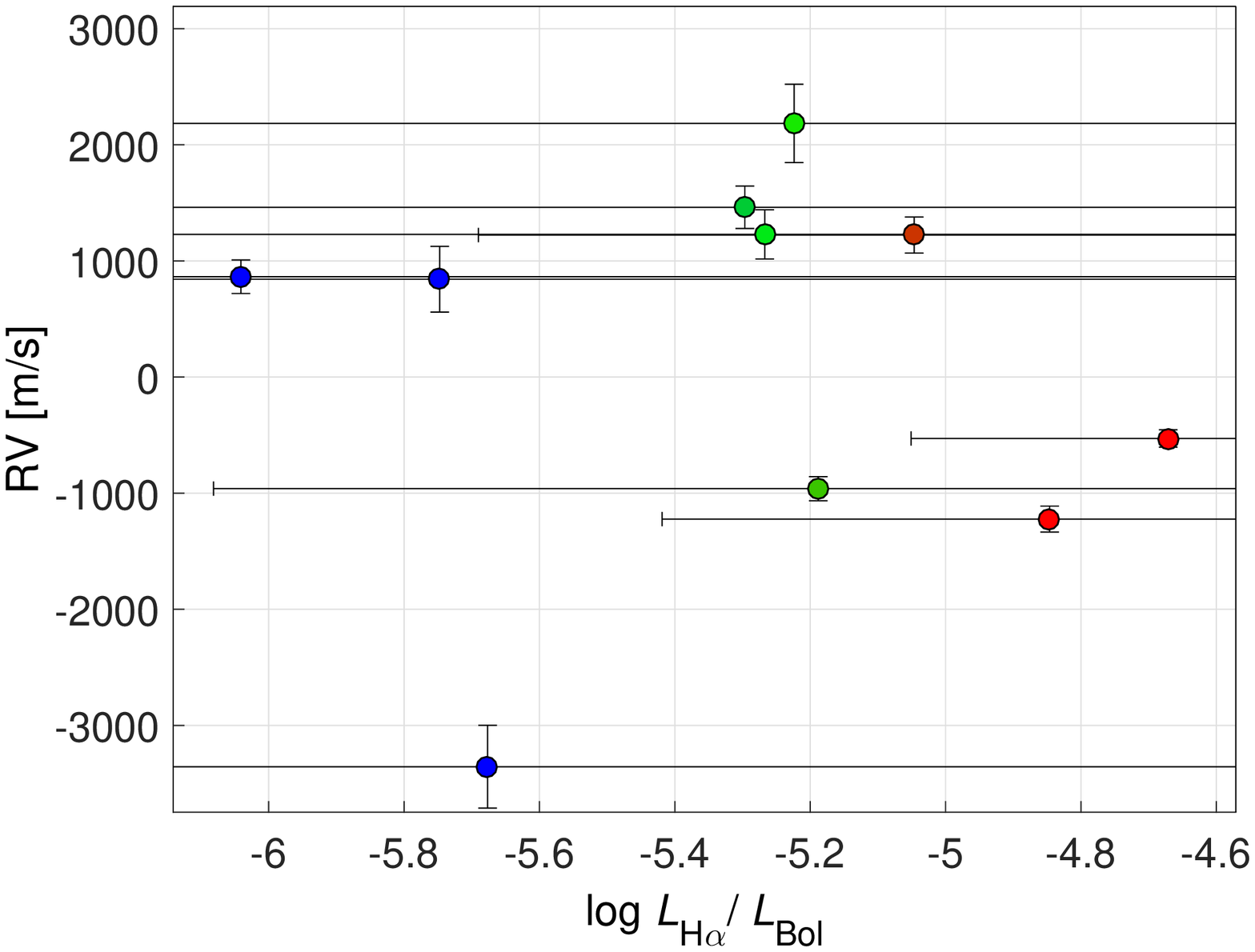}}
\endminipage
\caption{Continued.}
\label{figA1}
\end{figure*}

\addtocounter{figure}{-1}

\begin{figure*}[!htp]
\minipage{0.33\textwidth}
{\includegraphics[width=\linewidth]{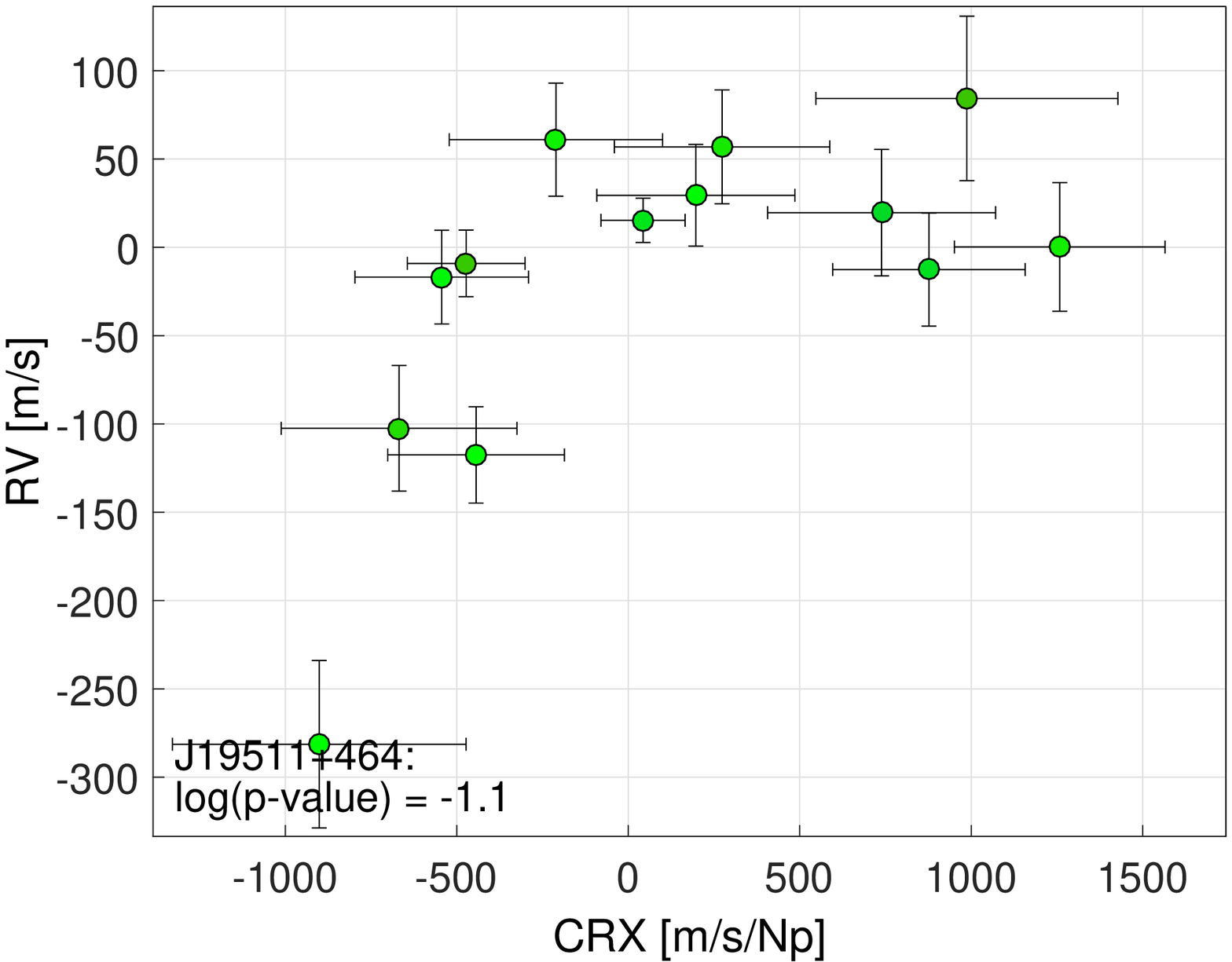}}
\endminipage\hfill
\minipage{0.33\textwidth}
{\includegraphics[width=\linewidth]{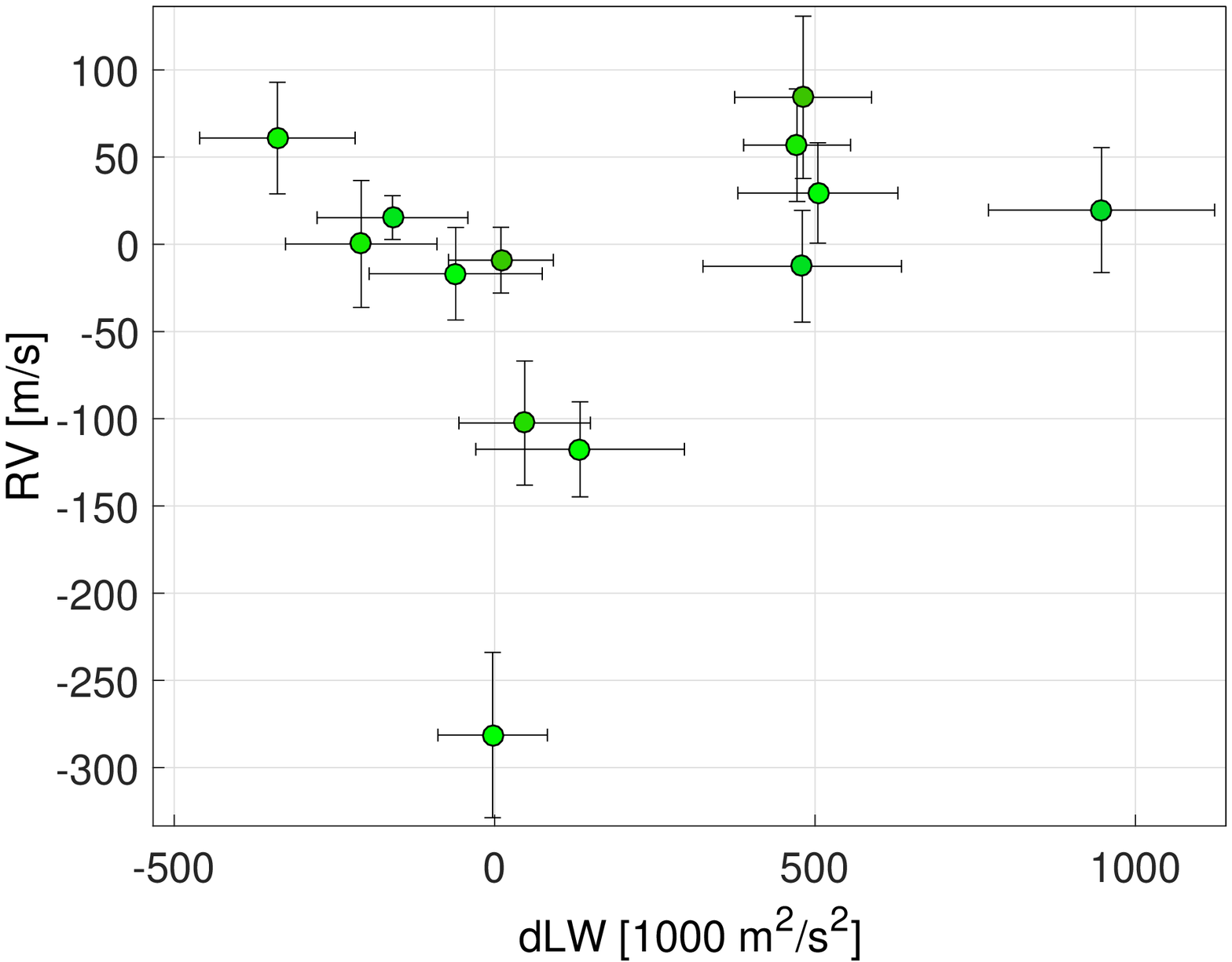}}
\endminipage\hfill
\minipage{0.33\textwidth}
{\includegraphics[width=\linewidth]{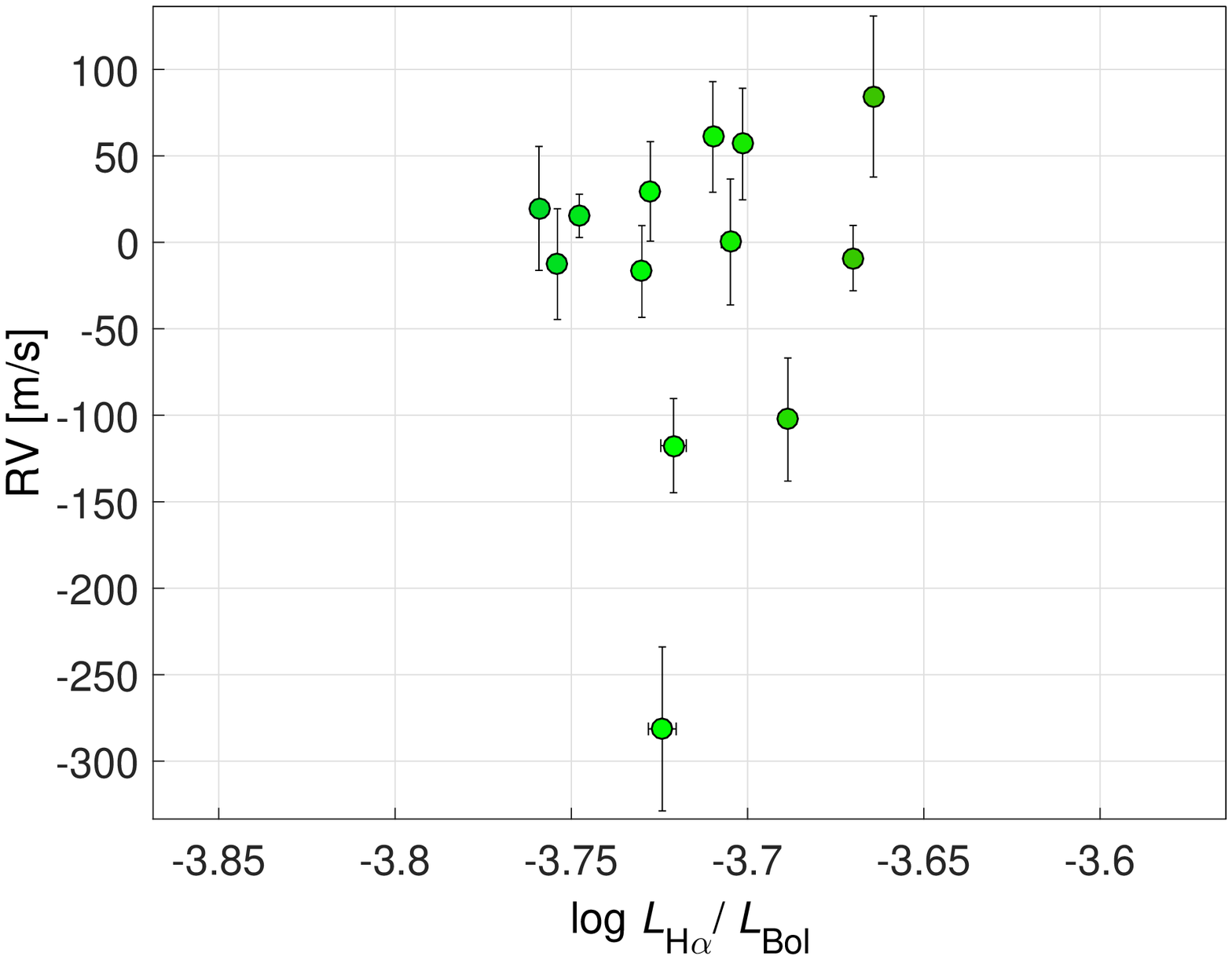}}
\endminipage


\minipage{0.33\textwidth}
{\includegraphics[width=\linewidth]{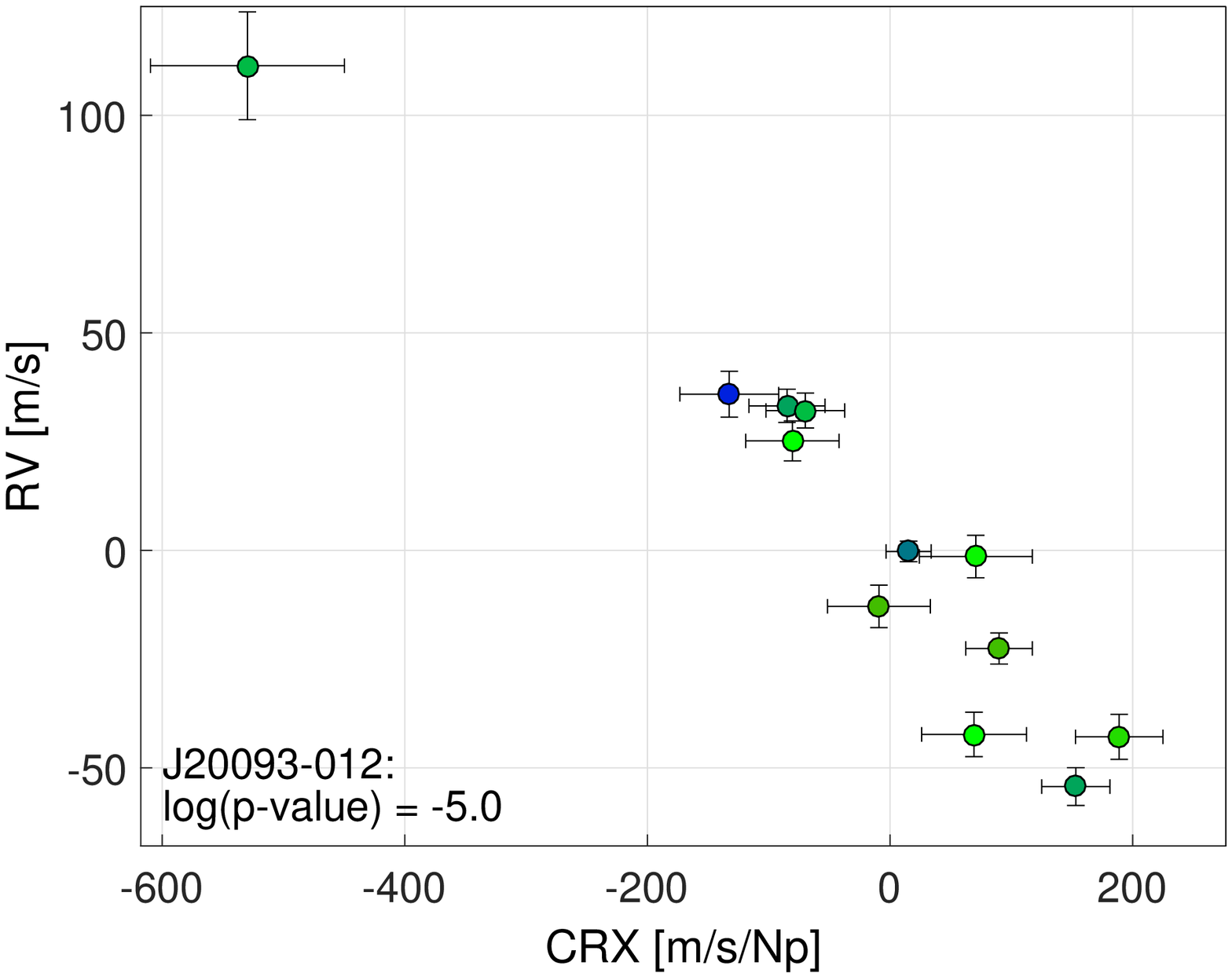}}
\endminipage\hfill
\minipage{0.33\textwidth}
{\includegraphics[width=\linewidth]{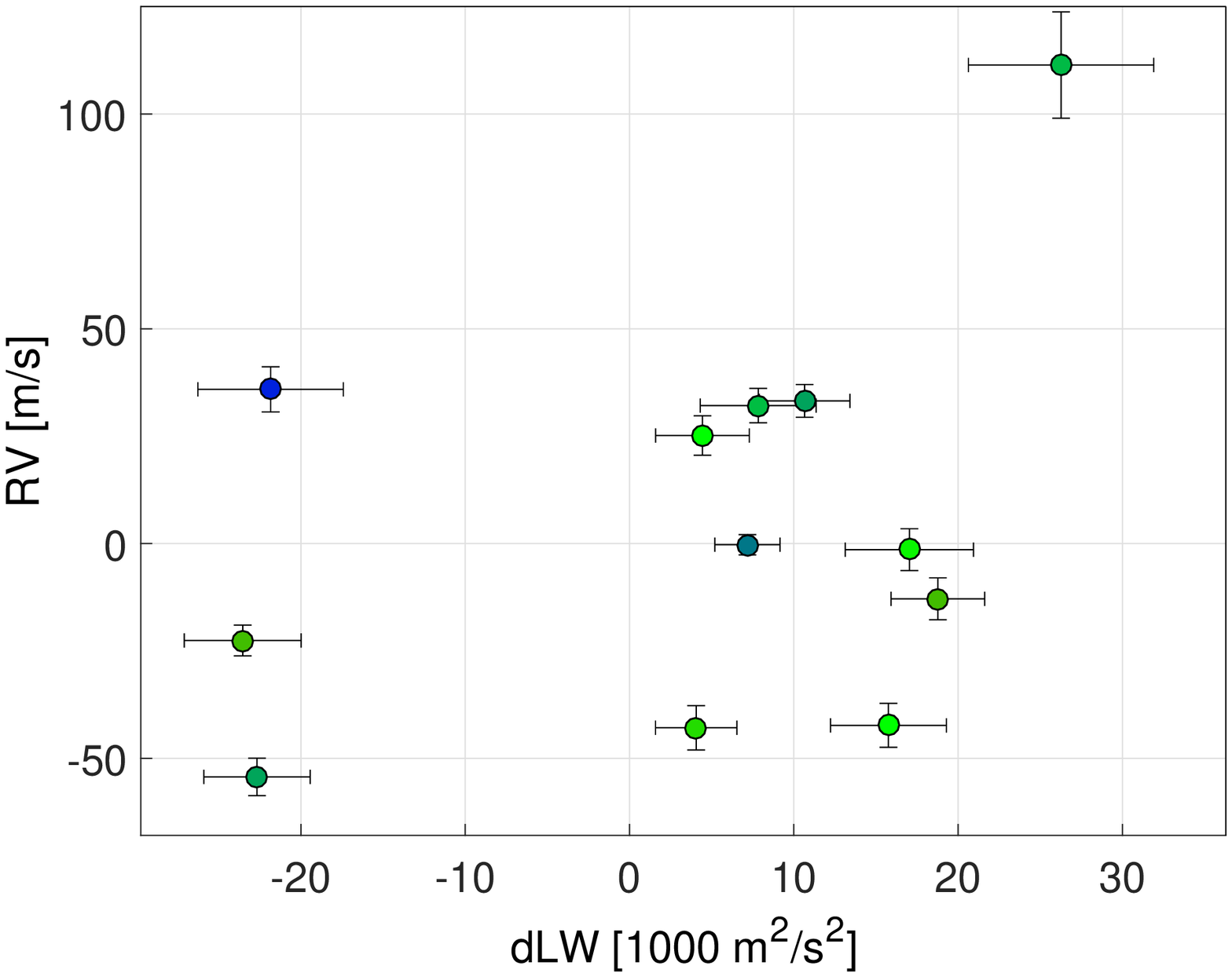}}
\endminipage\hfill
\minipage{0.33\textwidth}
{\includegraphics[width=\linewidth]{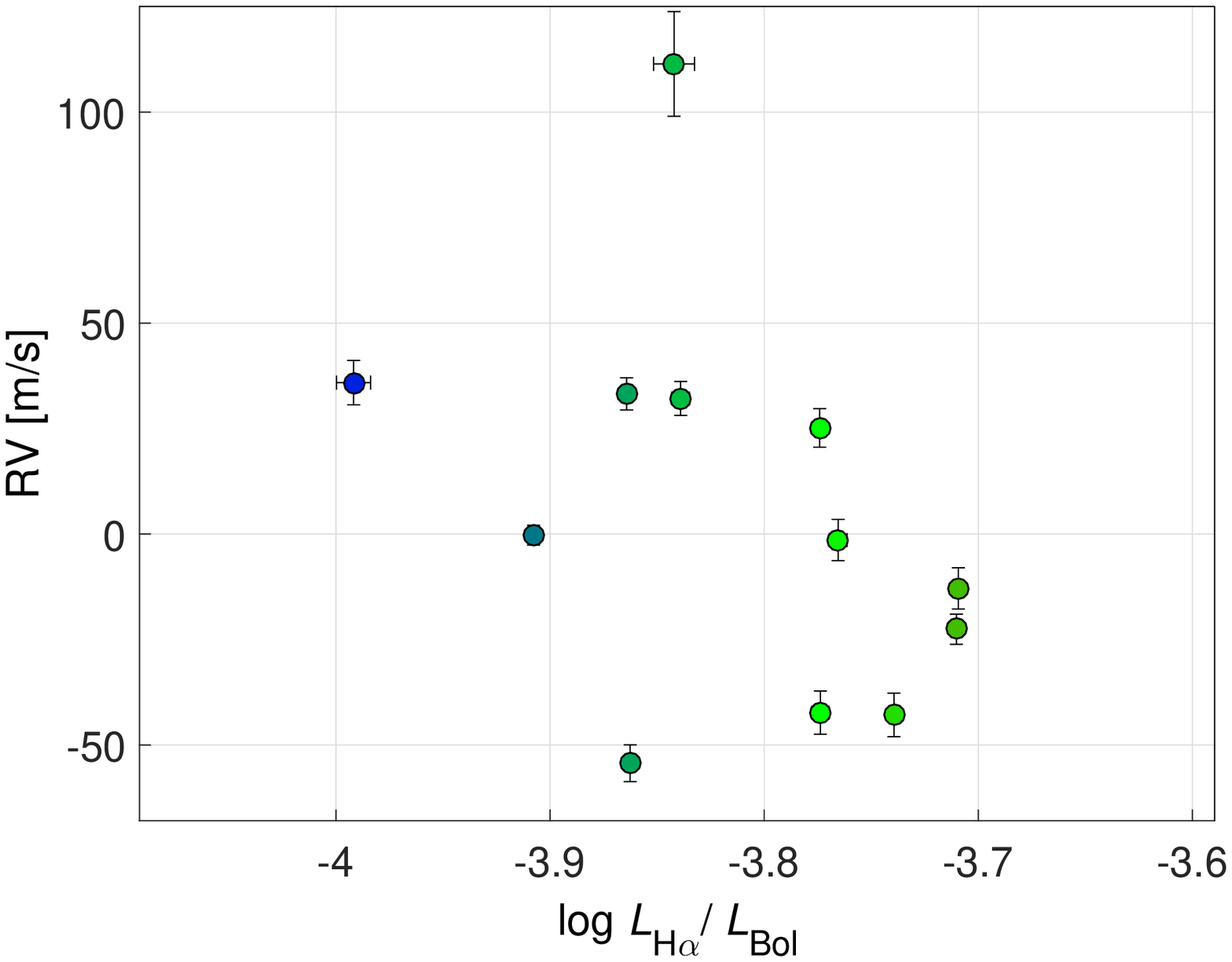}}
\endminipage


\minipage{0.33\textwidth}
{\includegraphics[width=\linewidth]{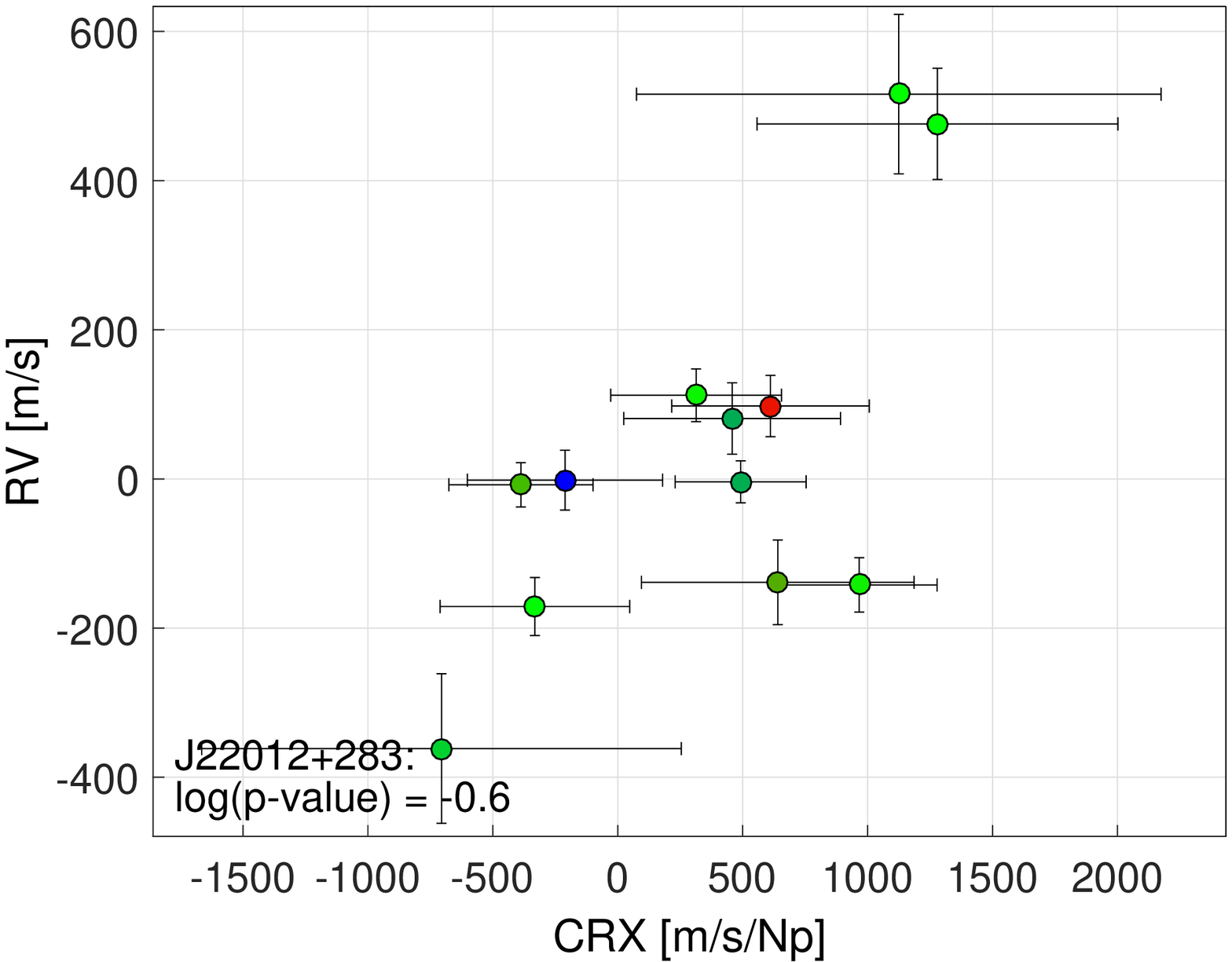}}
\endminipage\hfill
\minipage{0.33\textwidth}
{\includegraphics[width=\linewidth]{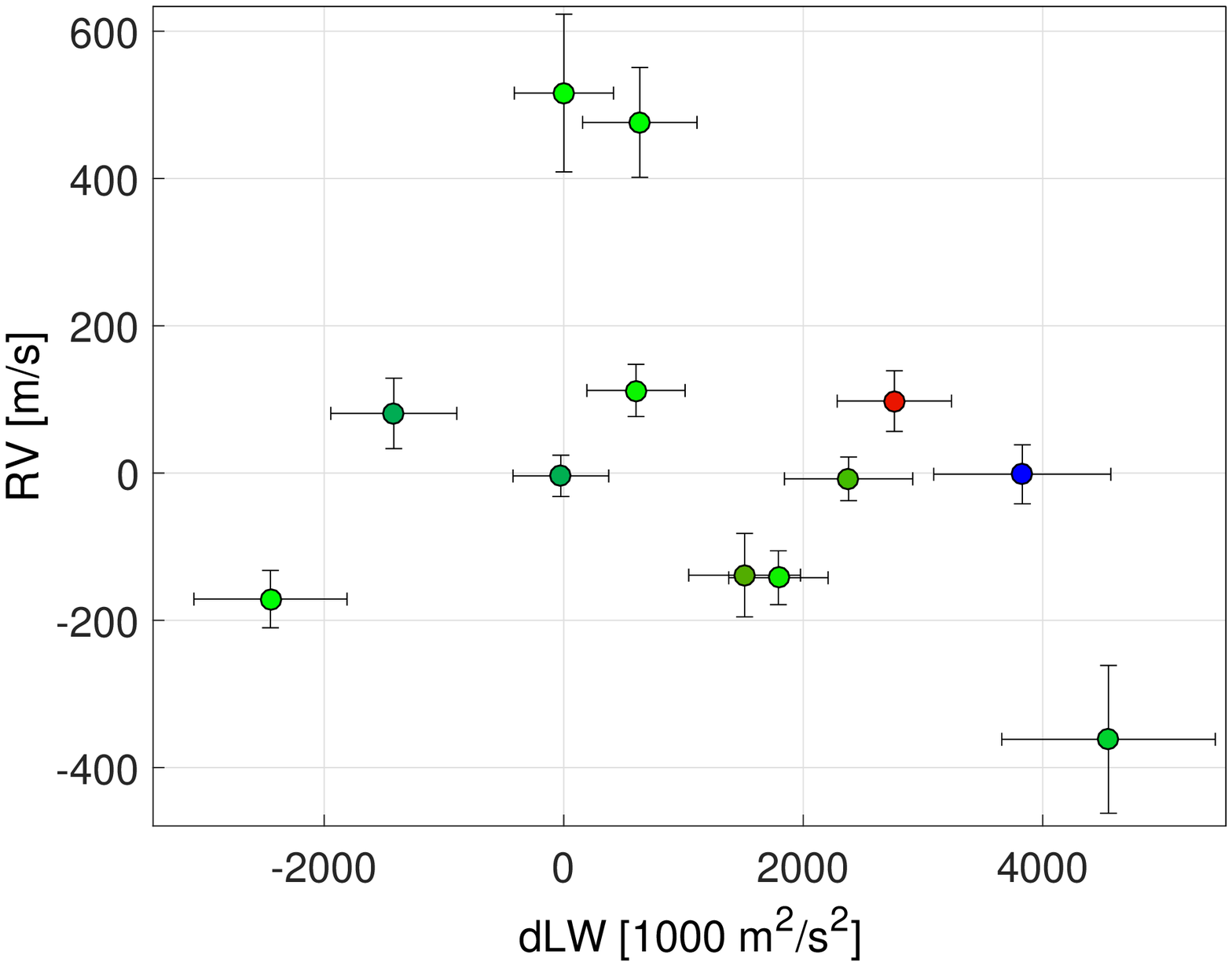}}
\endminipage\hfill
\minipage{0.33\textwidth}
{\includegraphics[width=\linewidth]{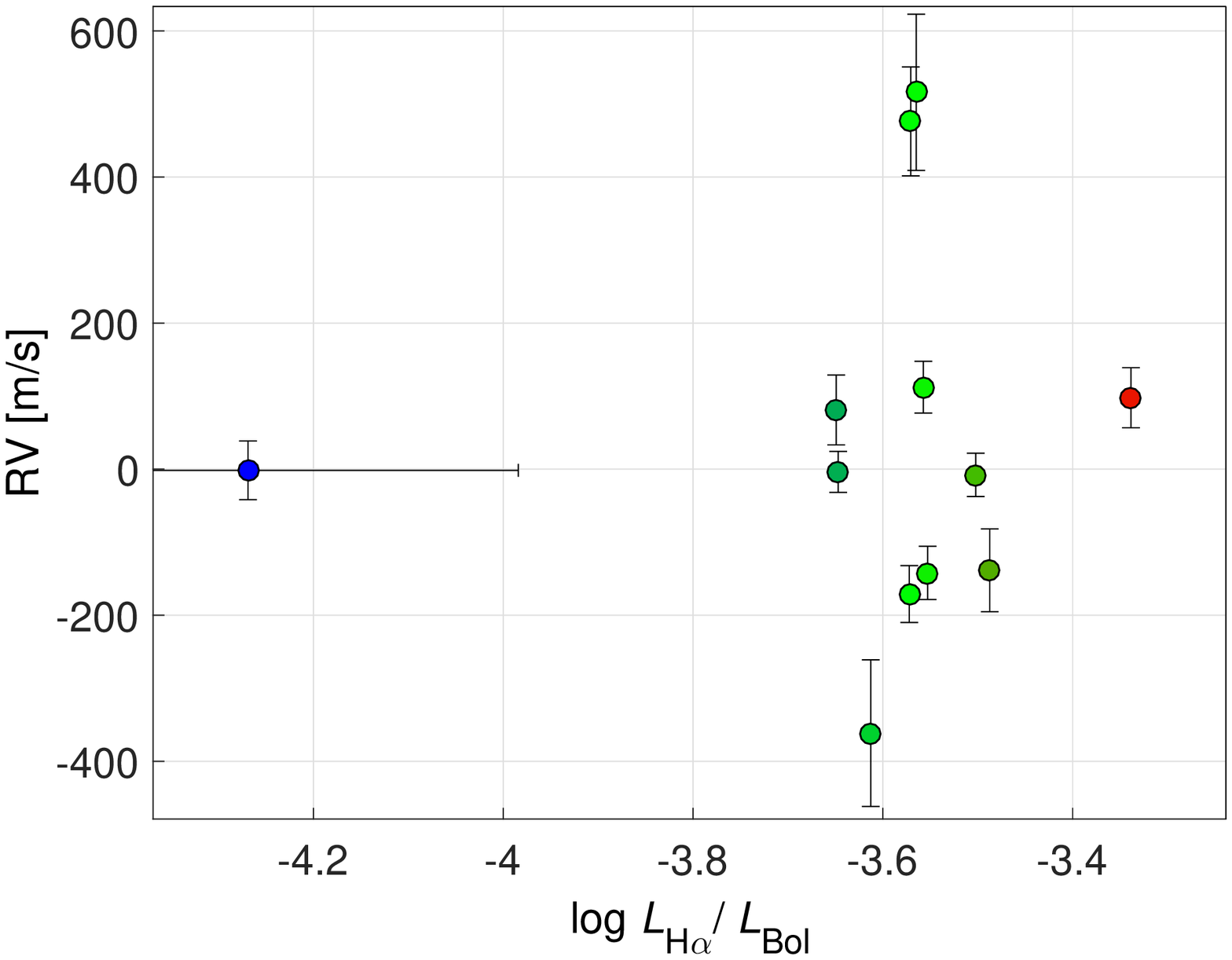}}
\endminipage


\minipage{0.33\textwidth}
{\includegraphics[width=\linewidth]{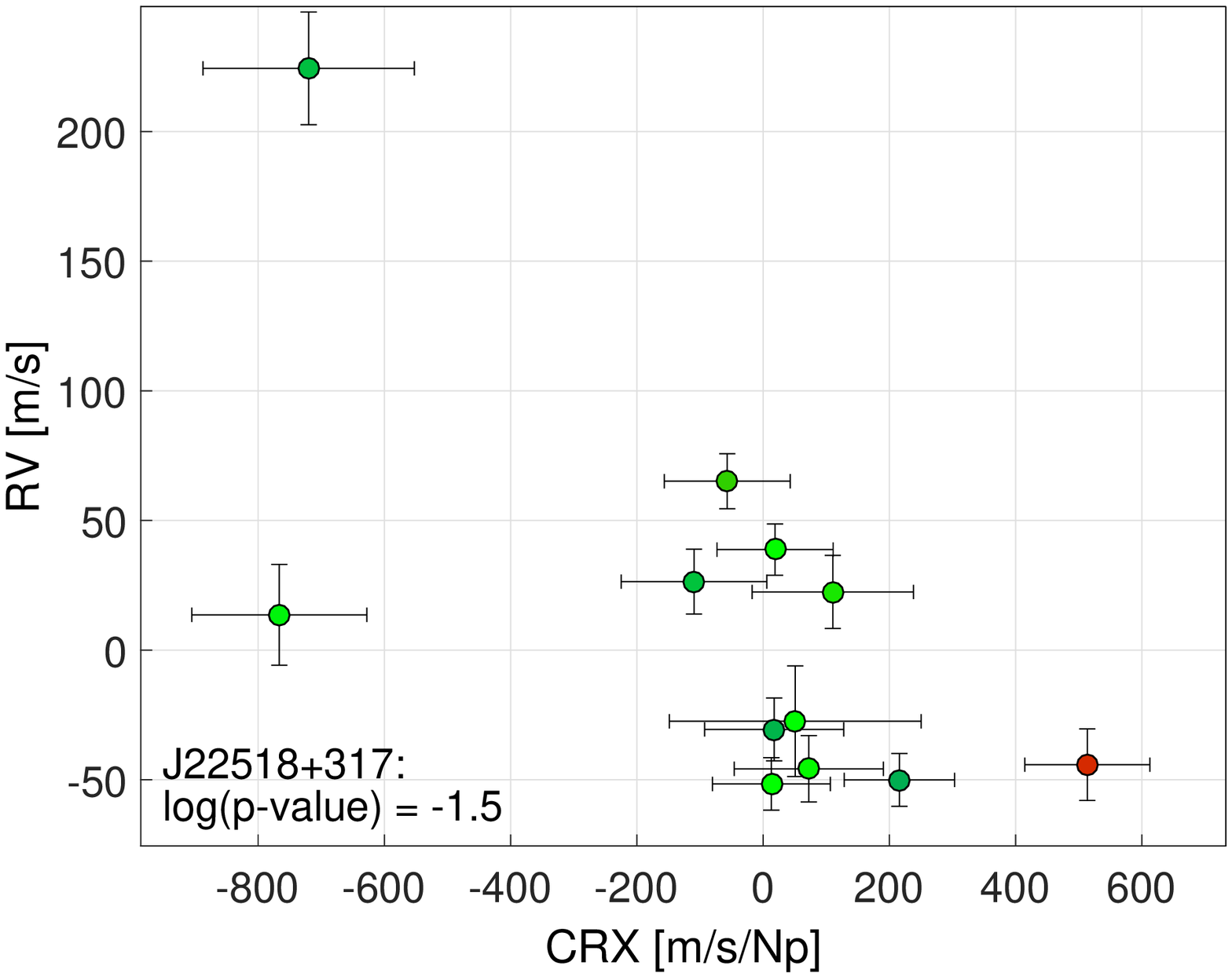}}
\endminipage\hfill
\minipage{0.33\textwidth}
{\includegraphics[width=\linewidth]{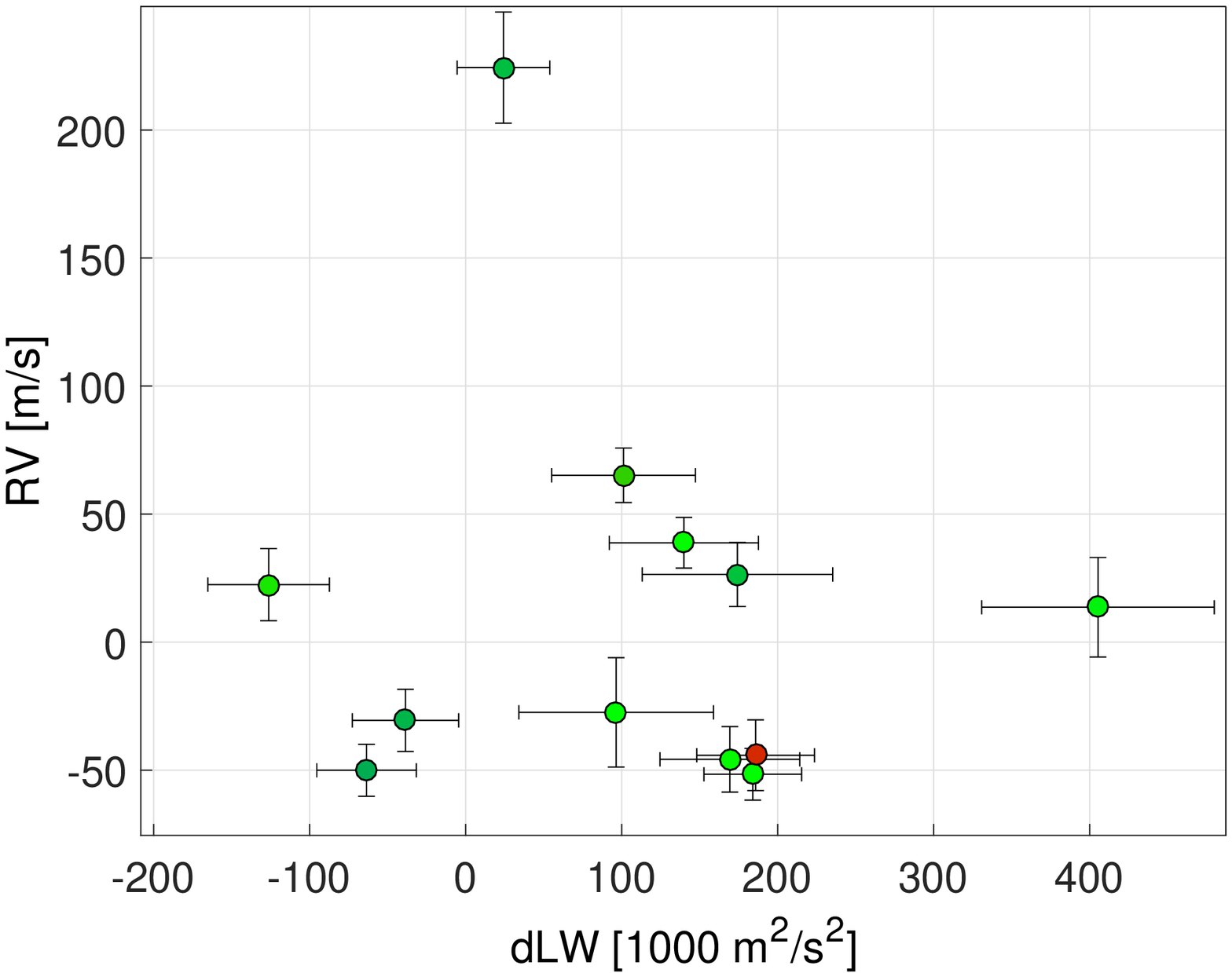}}
\endminipage\hfill
\minipage{0.33\textwidth}
{\includegraphics[width=\linewidth]{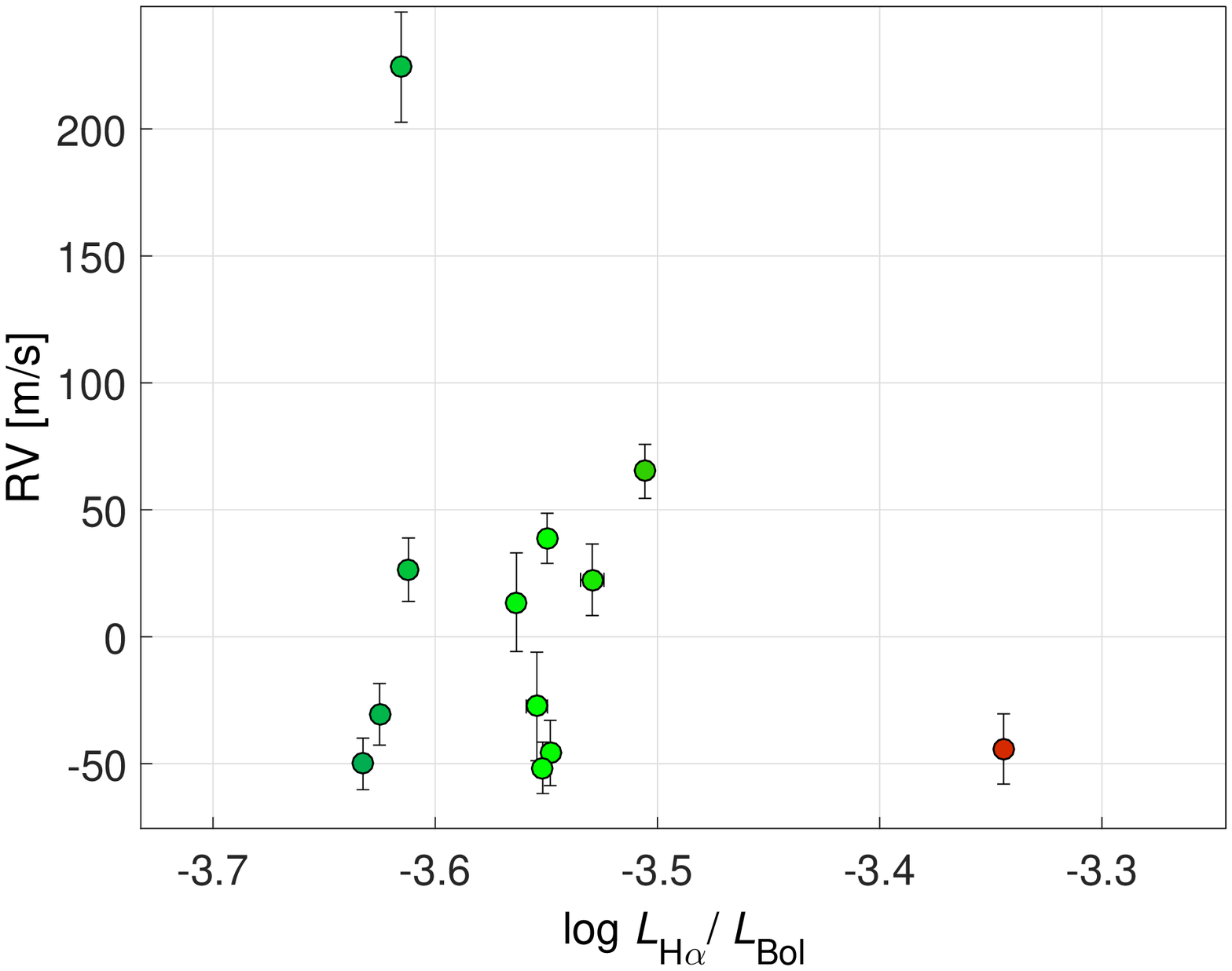}}
\endminipage


\minipage{0.33\textwidth}
{\includegraphics[width=\linewidth]{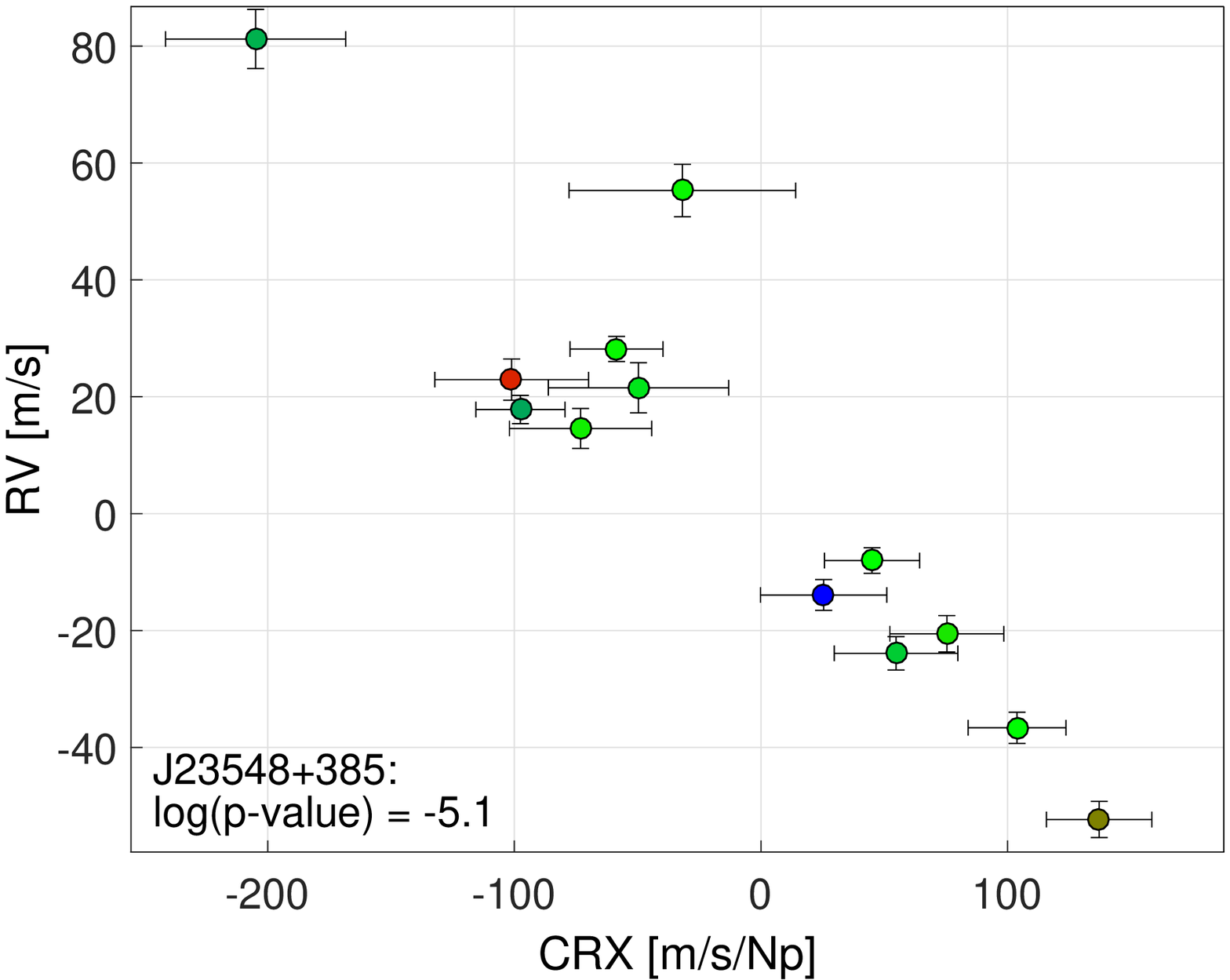}}
\endminipage\hfill
\minipage{0.33\textwidth}
{\includegraphics[width=\linewidth]{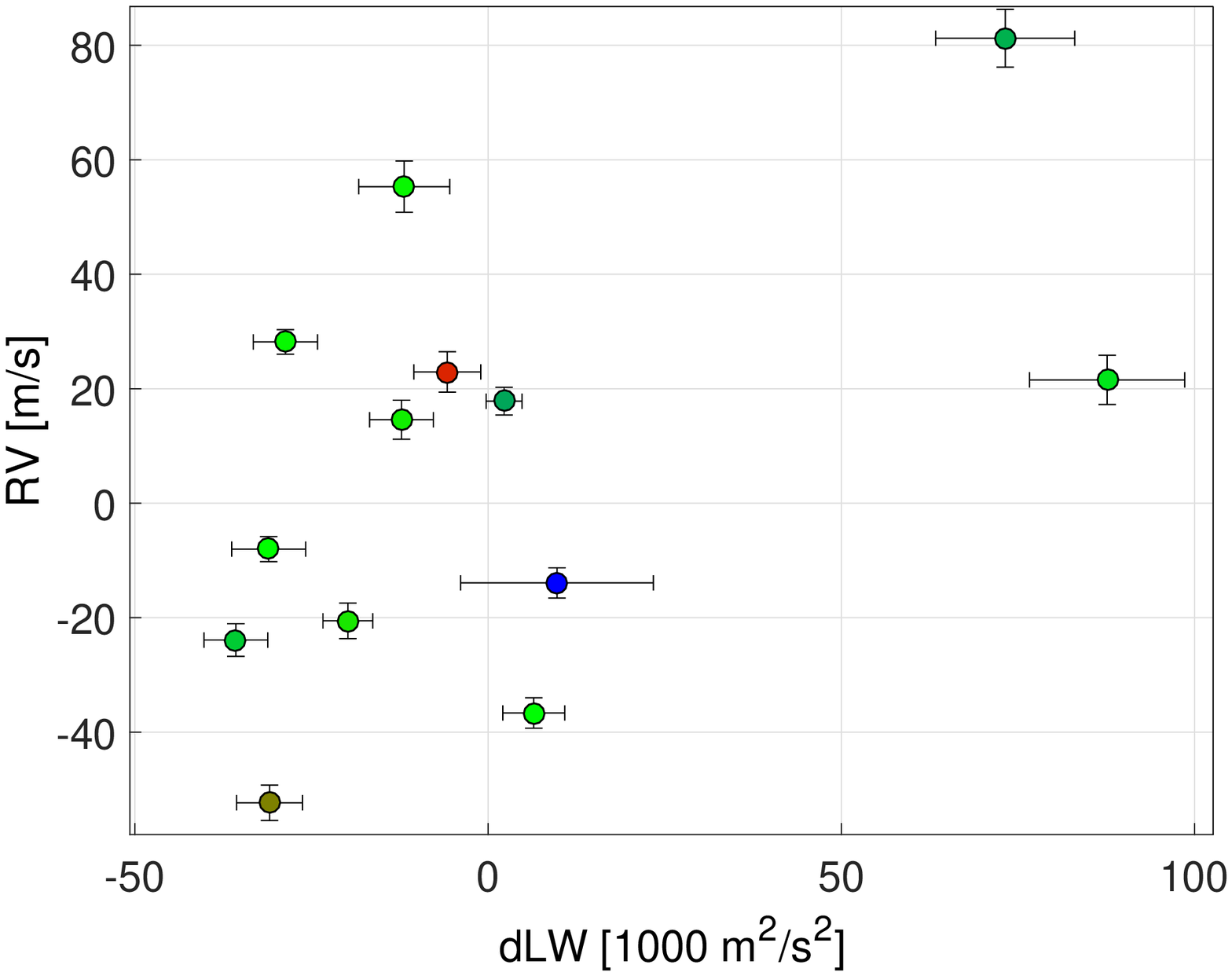}}
\endminipage\hfill
\minipage{0.33\textwidth}
{\includegraphics[width=\linewidth]{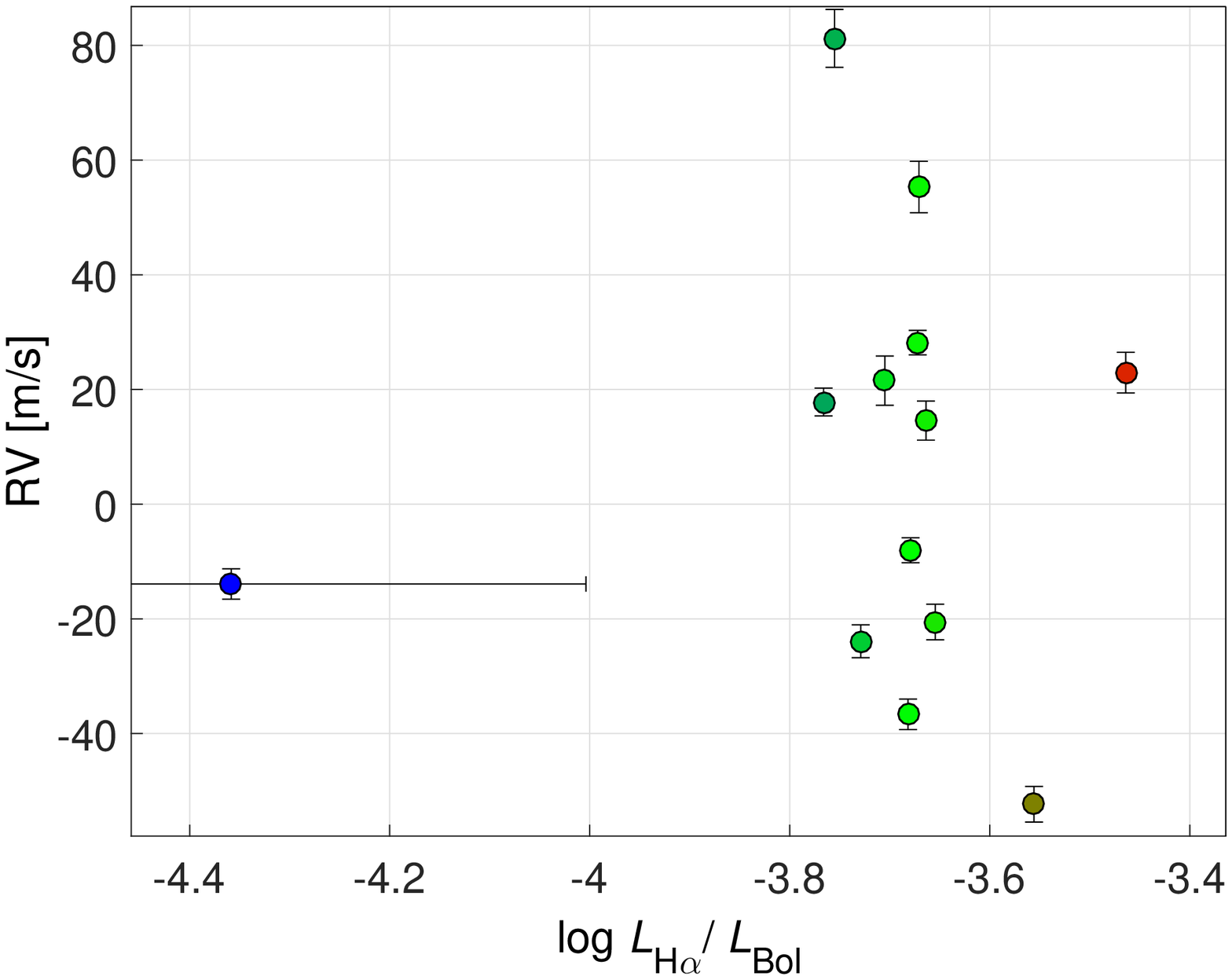}}
\endminipage
\caption{Continued.}
\label{figA1}
\end{figure*}

\end{appendix}

\end{document}